\documentclass[12pt]{iopart}
\usepackage{float}
\usepackage{morefloats}
\usepackage{datetime}
\usepackage{iopams}
\usepackage{epsf}
\usepackage{epsfig}
\usepackage{color}
\usepackage{psfrag}
\usepackage{verbatim}
\usepackage{setstack}
\usepackage{amsopn}
\usepackage{appendix}
\usepackage{graphicx}
\usepackage{graphics}
\usepackage{subfig}
\usepackage[margin=3.3cm]{geometry}

\usepackage{float}
\usepackage{tikz}

\usepackage{array}
\usepackage{tabularx}
\usepackage{multirow}
\usepackage{booktabs}
\usepackage{ctable}

\newcolumntype{?}{!{\vrule width 2pt}}

\newcommand{\be}{\begin{eqnarray}}	
\newcommand{\ee}{\end{eqnarray}}

\newcommand{\comments}[1]{}   

\DeclareFontFamily{OMX}{yhex}{}
\DeclareFontShape{OMX}{yhex}{m}{n}{<->yhcmex10}{}
\DeclareSymbolFont{yhlargesymbols}{OMX}{yhex}{m}{n}
\DeclareMathAccent{\wideparen}{\mathord}{yhlargesymbols}{"F3}

\begin{document}

\eqnobysec

\title{Coarsening and percolation in the kinetic $2d$ Ising model with spin exchange updates and the voter model}
\author{Alessandro Tartaglia, Leticia F. Cugliandolo, \\ and Marco Picco %
}

\address{Laboratoire de Physique Th\'eorique et Hautes Energies, UMR 7589,\\
Sorbonne Universit\'e et CNRS,\\
4 Place Jussieu, 75252 Paris Cedex 05, France
}

\today

\begin{abstract}
We study the early time dynamics of bimodal 
spin systems on $2d$ lattices evolving with different microscopic stochastic updates.
We treat the ferromagnetic Ising model with locally conserved order parameter (Kawasaki dynamics), the same model
with globally conserved order parameter (nonlocal spin exchanges), 
and the voter model. 
As already observed for non-conserved order parameter dynamics (Glauber dynamics), 
in all the cases in which the stochastic dynamics satisfy detailed balance, the critical percolation state
persists over a long period of time before usual coarsening of domains takes over and eventually takes the system to equilibrium.
By studying the geometrical and statistical properties of time-evolving spin clusters we are able to identify
a characteristic length $\ell_p(t)$, different from the usual length $\ell_d(t) \sim t^{1/z_{d}}$ that
describes the late time coarsening, 
that is involved in all scaling relations in the approach to the critical percolation regime.
We find that this characteristic length depends on the particular microscopic dynamics and the lattice geometry.
In the case of the voter model, we observe that the system briefly passes through a critical percolation state, to later
approach a dynamical regime in which the scaling behaviour of the geometrical properties of the ordered domains can be ascribed
to a different criticality.

\end{abstract}

\newpage
\tableofcontents

\vspace{2cm}

\newpage

\section{Introduction}
\label{sec:intro}

Take a system characterised by an equilibrium phase transition between 
a symmetric disordered phase and a symmetry-broken ordered one with two or more competing equilibrium states.
In typical classical realisations the adimensional control parameter is temperature over some energy scale inherent to the 
system's Hamiltonian. Coarsening is the process whereby such a system, 
initially prepared in a state drawn from the equilibrium distribution in the symmetric phase is let evolve,
typically under a certain relaxation dynamics that mimics the coupling to an environment, towards a new equilibrium state in the symmetry broken phase.
The relaxation dynamics produce local ordering: because of the build-up of correlations of the order parameter
field between distant space points, well-defined spatial domains of the ordered phase first emerge and then grow in time.

The theory of coarsening or 
phase ordering kinetics~\cite{Bray94,Puri09-article,CorberiPoliti} is based on the dynamic scaling hypothesis.
This hypothesis states that at long times the system enters a scaling 
regime regulated by a single growing length, denoted by $\ell_d(t)$ (where the subscript $d$ stands for ``dynamical''), such that the 
domain structure is statistically invariant when distances are measured with respect to it.
The way in which the length $\ell_d$ grows is 
determined by mesoscopic mechanisms and defines dynamic universality classes.
In the absence of frustration and/or quenched disorder
$\ell_d$ typically grows algebraically $\ell_d(t) \simeq t^{1/z_d}$, and 
the best known cases are the nonconserved order parameter (NCOP) class or model A 
with $z_d=2$, and the locally conserved order parameter (LCOP) class or model B with $z_d=3$, in the classification introduced in 
Ref.~\cite{Hohenberg-Halperin}. In order to apply the scaling hypothesis,
measuring times are asked to be longer than a microscopic 
time-scale, $t_0$, and observation distances $r$ are required to be such that 
$r_0 \ll r\ll L$ with $r_0$ a microscopic length-scale and $L$ the linear size of the system.

Usually, the analysis of coarsening phenomena is based on the investigation of the order parameter space-time 
correlation function or, equivalently, the dynamic structure factor. 
However, as already seen in previous works, see {\it e.g.}~\cite{BlCuPiTa-17}, the time-evolving domain structure contains additional useful information
that has not been much exploited so far. In this work we specifically analyse the statistical and geometrical properties
of the time-evolving spin clusters with the aim of establishing a connection between the growth process and percolation in $d=2$.

For the models with microscopic dynamics satisfying detailed balance, we prove the existence of a transient between the initial fully disordered state and a long-lived 
state in which the domain structure has critical percolation scaling properties. During the time regime
in which we observe approach to this critical-percolation-like behaviour we are able to identify a new characteristic length scale
that we denote $\ell_p(t)$ (where the subscript $p$ stands for ``percolation''),
different from the usual length associated to coarsening, $\ell_d(t) \sim t^{1/z_{d}}$,
and has the following meaning.
At time $t$, one is able to observe critical percolation properties in the domain pattern
over lengths $r$ such that $\ell_d(t) \ll r < \ell_p(t)$, while over lengths $r > \ell_p(t)$ the pattern appears as non-critical.
On length scales shorter than $\ell_d(t)$, instead, the system is already equilibrated.
When $\ell_p(t)$, which is growing in time, becomes comparable to the size $L$ of the lattice,
the system has reached a state where percolation criticality has extended over all distances.
Moreover, at this time, a ``stable'' pattern of percolating domains appears, 
where stable means that its topology (number of percolating spin clusters and their direction of percolation) is fairly
invariant under the microscopic dynamics.
The only further effect of the latter, during the late coarsening regime,
is to dissolve small non-percolating clusters and make interfaces smoother, 
until a very late time $t_{\mathrm{eq}} \sim L^{z_d}$
when the system gets close to the final equilibrium state.
(Even longer time-scales can be 
needed to equilibrate a sample at very low temperatures due to metastability.)

\begin{figure}[h]
\begin{center}
   \includegraphics[scale=1.25]{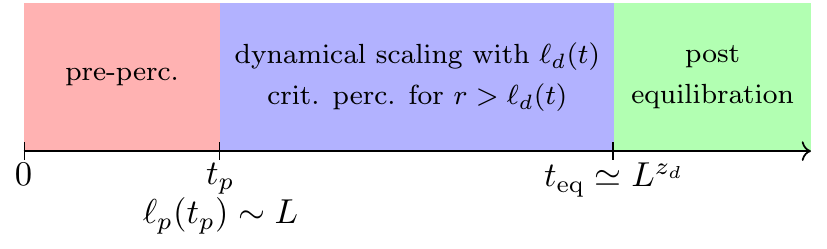}
\end{center}
\caption{\small
The separation of time scales for a $2d$IM evolving under stochastic spin dynamics.
For $t < t_p$, critical percolation properties are observed over distances $r$ up to $\ell_p(t)$.
The time $t_p$ is determined through the relation $\ell_p(t_p) \simeq L$.
For $t \ge t_p$, critical percolation properties extend over the entire system
(but can be observed for distances $r > \ell_d(t)$),
and scaling relations hold with the exponents of $2d$ critical percolation after rescaling all lengths by $\ell_d(t)$.
Around $t_{\rm eq} \sim L^{z_d}$ the system fully equilibrates. After this time, one observes equilibrium properties over all scales.
}
\label{fig:separation-of-scales}
\end{figure}

The onset of the critical-percolation-like regime occurs at the typical time $t_p$ that depends on the size of the system $L$
through the relation $\ell_p(t_p) = L$. We stress the fact that $t_p\ll t_{\mathrm{eq}} = L^{z_d}$ for all choices of the microscopic dynamics
and lattice geometry, so that this percolating regime can be observed and studied.
In particular, in the case of non-conserved order parameter 
dynamics on $2d$ lattices with even coordination number,  the characteristic length scale
$\ell_p$ behaves as~\cite{BlCuPiTa-17}
\begin{equation}
 \ell_p(t) \sim \ell_d(t) \, t^{1/\zeta}
 \; , 
\label{eq:ell_p}
\end{equation}
where the exponent $\zeta$ depends on the type of dynamics and the lattice geometry.
For example, on the square lattice $\zeta \simeq 0.5$. Instead,  $\ell_p(t)$ is non-algebraic, namely, $\ell_p(t) \sim \ell_d(t) \mathrm{e}^{\alpha t}$,
on the honeycomb lattice~\cite{BlCuPiTa-17}.
The separation of time scales induced by the presence of the additional growing length $\ell_p(t)$ 
is schematically represented in Fig.~\ref{fig:separation-of-scales}.

An approach to critical percolation was already observed and partially studied for Kawasaki dynamics~\cite{TaCuPi16,Takeuchi16}
and the voter model~\cite{TaCuPi15} (a purely dynamical model that we define below) on $2d$ spaces.
In this paper, we expand our analysis of the critical-percolation-like regime  and we measure 
the associated characteristic length $\ell_p$ in these models with greater precision. We also study the approach to 
critical percolation and later coarsening dynamics in the case of globally conserved order parameter (GCOP) dynamics.
Moreover, we make a distinction between the first time at which the system attains a critical percolation structure, denoted by $t_{p_1}$ in the text,
from the time after which the critical percolation structure becomes stable, in the sense explained above, the proper $t_p$ time. 
The results presented in this paper have to be compared to the ones in~\cite{BlCuPiTa-17} for the case of NCOP dynamics.

Concretely, we study the time evolution of a classical spin configuration, placed on a $2d$ lattice with finite size,
updated with different kinds of stochastic rules. The initial spin configuration corresponds to the $T\to\infty$
equilibrium state of the Ising model, that is, a spin $\pm 1$ is placed on each site with equal probability and independently of each other.
The stochastic update rule that produces the microscopic dynamics
(and can be encoded mathematically in terms of probability transition rates from one spin configuration to another)
is basically one of two types: single spin-flip  or spin-exchange. The first one consists in inverting the spin
at just one lattice site, the second one exchanges the spins at two sites. The transition probabilities of both moves
depend only on the value of the spins at the neighbouring sites, so that the update rules depend only locally on the spin configuration.
More specifically, the Kawasaki dynamics is realised by exchanging antiparallel spins at two nearest-neighbour lattice sites with
probability transition rates satisfying detailed balance.
To simulate the GCOP dynamics, we use the same spin-exchange rule but with the two chosen
sites being at any distance on the lattice. We will refer to this microscopic dynamics as
nonlocal Kawasaki dynamics.

The voter dynamics, instead, consist of single spin-flip moves and can be thought of as a linear Glauber model.
This is due to the fact that the spin-flip transition rates depend linearly on the effective magnetic field produced by the spin of the neighbouring sites on 
the target one. In our work we consider the symmetric voter model, with no control parameter playing the role of the temperature
as in the Metropolis Monte Carlo update rules.
The resulting dynamics lead the system towards one of two absorbing states:
either all sites have spin up or they all have spin down.
Although there is no notion of energy or temperature, the system reaches an absorbing state by means of the growth of ordered domains, 
and coarsening phenomena are observed and can be analysed with the same tools used in more common examples.

In the rest of the Introduction we explain these models in more detail and we outline the most important observables
that are measured and analysed in order to establish the existence and characterisation of the critical percolating regime.

\subsection{Models}
\label{subsec:models}

The model we consider is a classical spin system on a $2d$ lattice with finite size and periodic boundary conditions.
To each lattice site $i$ we associate a spin, $s_i = \pm 1$.

In the  ferromagnetic Ising model the energy is given by
\begin{equation}
H_J[\{s_i\}] = -J \sum_{\langle ij\rangle} s_i s_j
\label{eq:2dIM}
\end{equation}
with $J>0$ and the sum running over all pairs of nearest-neighbour lattice sites (each pair counted once).
In this paper, we will consider this model on a finite $2d$ lattice with periodic boundary conditions (PBC).

In the thermodynamic limit, 
this model undergoes a second order phase transition at a critical  temperature $T_c$ given by
$\beta^{\rm sq}_c J = J/(k_BT^{\rm sq}_c) = \ln{(1+\sqrt{2})} / 2 \simeq 0.4407$ on the square
lattice, $\beta^{\rm tr}_c J = J/(k_BT^{\rm tr}_c) = \ln{(3)} / 4 \simeq 0.2747$ on the triangular lattice, and
$\beta^{\rm honey}_c J = J/(k_BT^{\rm honey}_c) = \ln{(2+\sqrt{3})} / 2 \simeq 0.6585$ on the honeycomb lattice.

The initial condition is  taken to be a random spin configuration with no correlations, obtained by choosing $s_i =+ 1$
or $s_i = -1$ with probability $1/2$ on each lattice site. Under a mapping to occupation numbers,
$n_i = (s_i+1)/2$, such a configuration corresponds to a realisation of random site percolation with $p=1/2$. It is therefore right at
the critical percolation point for the triangular lattice and below the critical percolation points in the other 
two cases since $p_c^{\rm tr}=1/2$, $p_c^{\rm sq} \approx 0.5927$ and $p_c^{\rm honey} \approx 0.6970$.

In order to mimic an instantaneous quench in the temperature and the following relaxation towards an equilibrium state,
we adopt the usual Monte Carlo method.
The system can ``jump'' from a spin configuration $\{s_i\}$ to another one, $\{s^{\prime}_i\}$, 
with a rate $W( s, s^{\prime} )$ that must satisfy the well-known detailed balance condition,
\begin{equation}
  W( s, s^{\prime} ) \, P_{\rm eq} (s) \ = \ W( s^{\prime}, s ) \, P_{\rm eq} (s^{\prime})
 \label{eq:detailed-balance} 
\end{equation}
where $P_{\rm eq}(s)=\exp{ \left[ - H_{J}\left( \{ s_i \} \right) / ( k_{B} T )  \right] } / Z(T)$ is the canonical equilibrium
probability density for a given spin configuration $s = \{s_i\}$, with $T$ being the target temperature, $k_B$ the Boltzmann constant and
$Z(T)$ the canonical partition function at temperature $T$. The above condition ensures
that the dynamics enforced by the transition rates $W( s, s^{\prime} )$ bring the system towards
a stationary state with the desired probability distribution $P_{\rm eq}$ of canonical equilibrium at $T$.
A common choice for the transition rates $W$ is  
\begin{equation}
 W( s, s^{\prime} ) = \min{ \left( 1, \, \frac{P_{\rm eq} (s^{\prime})}{P_{\rm eq} (s)} \right)} \ ,
 \label{eq:metropolis-rule}
\end{equation}
known as the Metropolis rule. Note that, in this form, the function $W(s,s^{\prime})$ depends on the initial ``starting''
spin configuration, $s = \{ s_i \}$, and on the ``arrival'' one, $s^{\prime} = \{ s^{\prime}_i \}$, only through the energy cost
$\Delta E (s,s^{\prime}) = H_{J} \left( \{ s^{\prime}_i \} \right) - H_{J} \left( \{ s_i \} \right) $ for the system to go from 
$s$ to $s^{\prime}$. 

If one uses the transition rates defined by Eq.~(\ref{eq:metropolis-rule}) with the condition that the arrival spin configuration $s^{\prime}$
must either be the starting one itself, $s$, or differ from it only by the value of the spin at one single lattice site, the resulting process
is the one studied in~\cite{BlCuPiTa-17}. In this paper we focus on another type of relaxation dynamics,
namely, spin-exchange rules that conserve the total magnetisation of the system.
The upgrades consist simply in letting the starting spin configuration $s$ pass to a new one $s^{\prime}$ that differs from it only by 
having two lattice sites exchange their spins.

In the last part of the paper we will also consider a case of NCOP dynamics, the voter model.
In particular, we will deal with the symmetric rule: in a single move the system is allowed to flip the spin on a randomly chosen lattice site with
a transition rate $W( s, s^{\prime} )$ that is simply proportional to the fraction of nearest-neighbour sites with anti-parallel spins.
In this case there are no equilibrium states that the system can relax to, because there is no energy function like Eq.~(\ref{eq:2dIM}) associated
to the spin configuration. Therefore, one cannot interpret the transition rates as the ones given by the 
detailed balance rule in Eq.~(\ref{eq:detailed-balance}). Moreover, there is no control parameter 
playing the role of a temperature so there is no proper quench. 
Nevertheless, the system, which is initially prepared in a disordered state,
evolves towards one of two absorbing states, namely all lattice sites having the same spin, either $+1$ or $-1$,
and it achieves it by growing ordered domains.

Finally, we mention that a Monte Carlo step, consisting of $L^2$ attempts at making an infinitesimal transition, 
is the implicit time-unit in all our presentation.

\subsection{Observables}
\label{subsec:observables}

Here we give a list of the observables that allow us to characterise the coarsening process
that the system is subjected to, and the percolation geometrical and statistical properties that arise during the dynamics.

A crucial observable in our analysis is the characteristic dynamical length, denoted by $\ell_d$.
According to the dynamical scaling hypothesis, at sufficiently late times, a coarsening system develops a
domain mosaic the morphology of which is (statistically) scale-invariant with a spatial characteristic scale given by $\ell_d$. This means that
the typical domain structure is independent of time if all lengths are rescaled by $\ell_d$. In this sense,
$\ell_d(t)$ can be seen as the typical domain size at time $t$ since the relaxation dynamics has begun, or as the average
separation between ordered domains of the same phase. 

The scaling hypothesis is usually probed with the two-body dynamical correlation function, 
$C(\boldsymbol{r},t) = \langle  s(\boldsymbol{x}+ \boldsymbol{r},t) \, s(\boldsymbol{x},t) \rangle$, which should
take the scaling form
\begin{equation}
  C(\boldsymbol{r},t) \sim f\left( \frac{r}{\ell_d(t)}\right)
 \label{eq:corr_function_scaling}
\end{equation}
at late times $t$ and for distances $r$ much longer than the lattice spacing and much shorter than the linear system size $L$. 
In the context of numerical simulations, 
the characteristic length $\ell_d(t)$ is  usually estimated as the distance $r$ over which the correlation $C$ falls to a selected 
portion of its maximum value, or as the square root of 
$R^{2}(t) = ( \int^{+\infty}_{0} \mathrm{d} \boldsymbol{r} \, r^2 \, C(\boldsymbol{r},t) ) / ( \int^{+\infty}_{0} \mathrm{d} \boldsymbol{r} \, C(\boldsymbol{r},t) )$.

A more practical way of measuring the characteristic length $\ell_d$ is by means of the excess-energy associated to the domain walls,
\begin{equation}
 \epsilon(t) = \frac{ E_{\mathrm{eq}} - E(t)}{E_{\mathrm{eq}} }
 \; , 
 \label{eq:excess-energy}
\end{equation}
with $E(t)= \langle \mathcal{H} ( \{ \sigma_i (t) \} ) \rangle $ the average energy at time $t$ (where $\mathcal{H}$
is given by Eq.~(\ref{eq:2dIM}) and $\langle \dots \rangle$ represents the average over many independent realisations of 
the dynamics and the initial configuration)
and $E_{\mathrm{eq}}$ the energy of the equilibrium state or stationary state that the system is approaching (eventually in the limit
$t \rightarrow + \infty$). The excess-energy $\epsilon$ is proportional to the density of defects which, in our case, are 
the domain walls. In the case of the ferromagnetic Ising model, one can easily see that $\epsilon(t)$ coincides with the average
fraction of \textit{unsatisfied} lattice bonds in excess with respect to the reference equilibrium state, where an 
unsatisfied lattice bond, for a given spin configuration, is a bond between two sites having opposite spin values.
The typical domain radius, which is proportional to $\ell_d$, is roughly given by the inverse of this quantity~\cite{Bray94}.
In particular we introduce an \textit{excess-energy growing length}, $\ell_G(t)$, which is given by 
$\ell_G(t) = 1/\epsilon(t) = E_{ \mathrm{eq} }/(E_{\mathrm{eq}} - E(t))$.

In the paramagnetic initial state, $E(0) \simeq 0$ and $\ell_{G}(0) \simeq 1$.
As the system approaches thermal equilibrium at $T$ below the critical temperature $T_c$, 
the growing length increases, reflecting the fact that the density of defects (domain walls) decreases.
For a finite system of linear size $L$ subject to PBC, 
with equal number of $\pm$ spins in the initial state and conserved order parameter dynamics, 
the growing length can be at most $L/2$ since the optimal configuration is one
with two domains of ordered spins separated by two walls with length $L$.

Throughout our paper we will use the numerically estimated $\ell_G(t)$ as the measure of $\ell_d(t)$, except for the voter model,
that has no energy function.
Still, in the voter model,
an $\epsilon(t)$ can be defined as the average fraction of unsatisfied bonds, that is to say, links between nearest-neighbour sites with antiparallel spins.
Nevertheless, we found that, in this case, $\ell_G(t) = 1/\epsilon(t)$ is not the right estimate of $\ell_d(t)$ 
because of the particular stochastic spin-flip dynamics that make the voter model an example of domain growth in absence of surface tension.
Instead, we use the more traditional method of extracting the characteristic length $\ell_d(t)$ from the correlation function $C(r,t)$, or
we directly use the theoretical law $\ell_d(t) \sim t^{\frac{1}{2}}$~\cite{TaCuPi15,Krapivsky92,Krapivsky92b,FrachebourgKrapivsky96}.

Our analysis will focus on the geometrical and statistical properties of the ordered domains in the course of the dynamics, 
so a proper definition of \textit{domain} or \textit{spin cluster}  is in order: 
a \textit{spin cluster} is a subset of the lattice whose sites have all the same spin value and are connected by paths of 
satisfied bonds. The \textit{area} of a spin cluster is simply the number of sites that belong to it. 
A \textit{domain wall} or \textit{cluster hull} is a closed and non-self-intersecting path on the \textit{dual} lattice
(graphically, the dual lattice is constructed from the original one by connecting the centers of the original lattice plaquettes)
which is constructed by joining bonds (on the dual lattice) that intersect unsatisfied bonds on the original lattice.
The interface between a certain spin cluster and its neighbouring clusters of opposite orientation can be composed of many hulls.

\begin{figure}
\begin{center}
\subfloat[]{\includegraphics[scale=0.25]{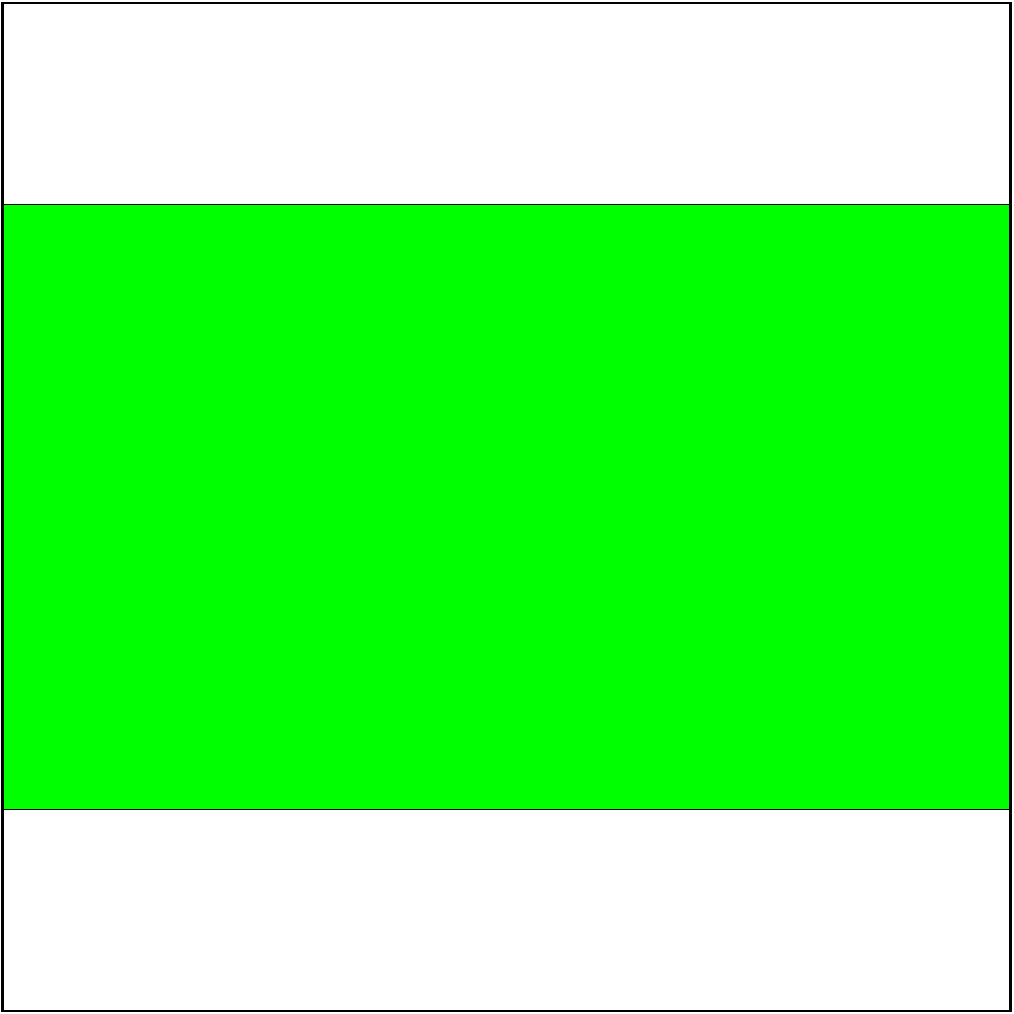}}\quad%
\subfloat[]{\includegraphics[scale=0.25]{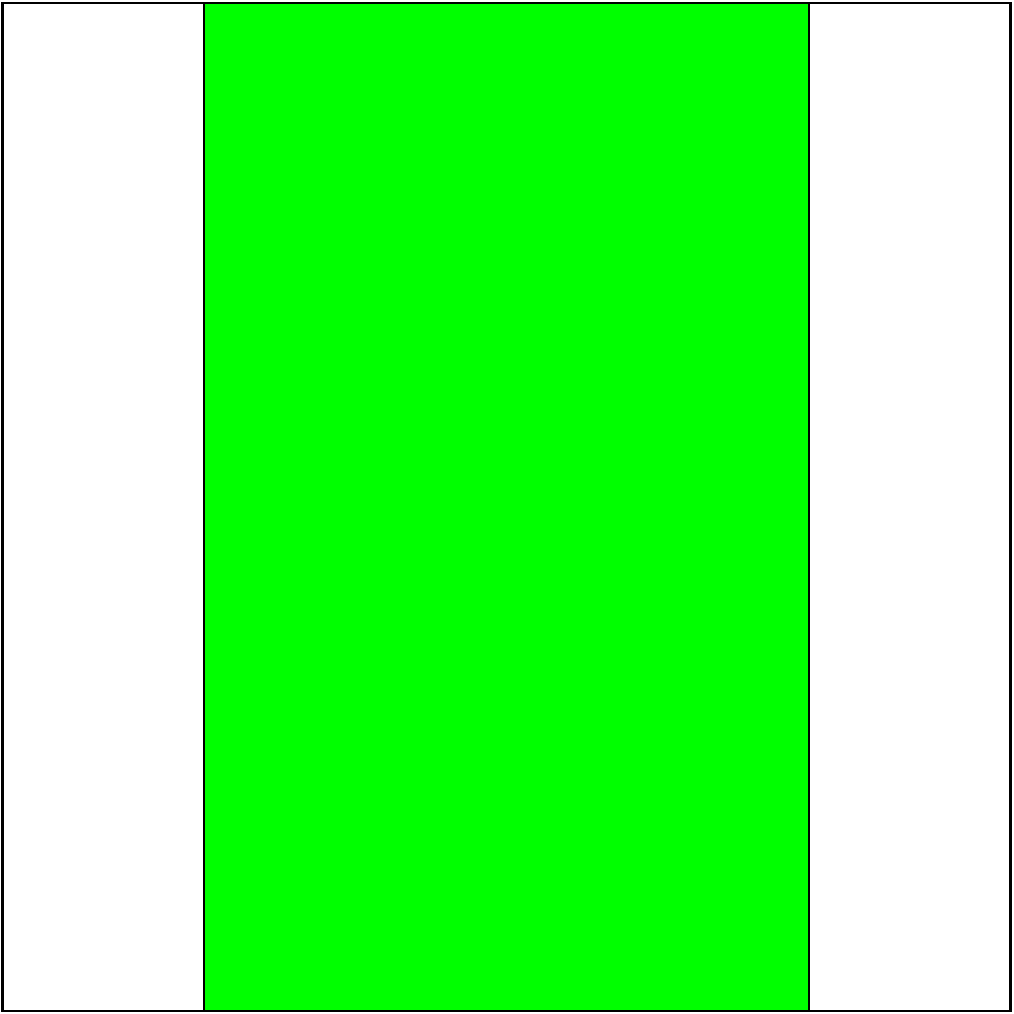}}\quad%
\subfloat[]{\includegraphics[scale=0.25]{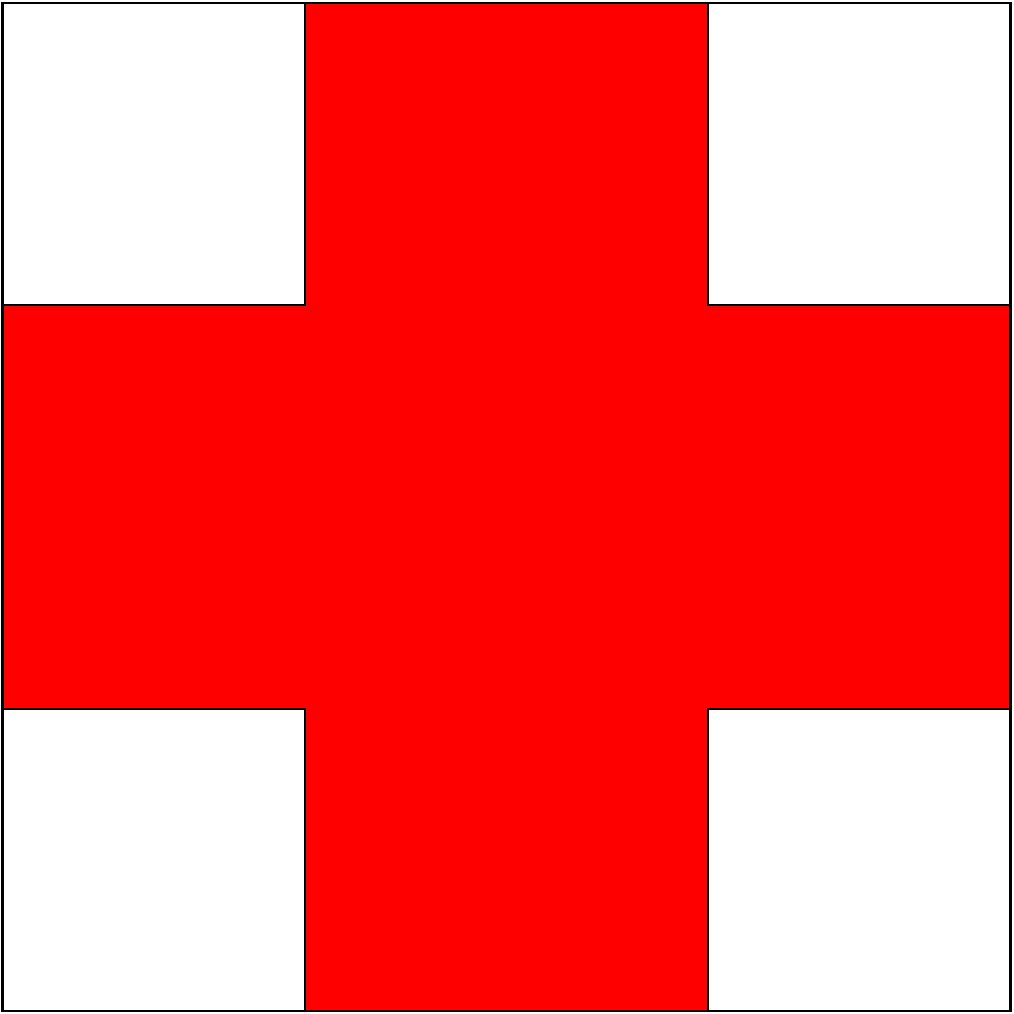}}\quad%
\subfloat[]{\includegraphics[scale=0.25]{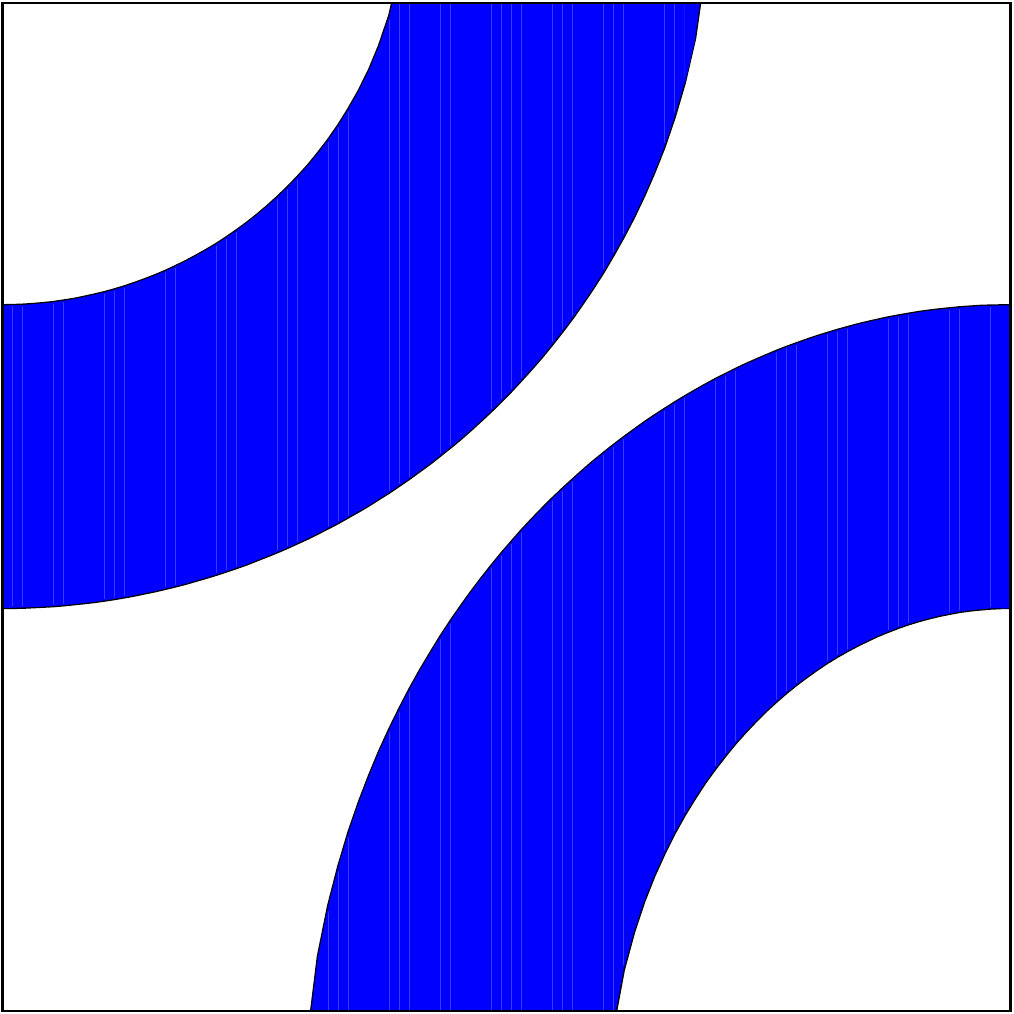}}\quad%
\end{center}
\caption{\small Sketches of wrapping clusters on a lattice with unit aspect ratio and PBC ({\it i.e.},~on a torus).
The panels show, in order, clusters spanning the system horizontally (a), vertically (b), both horizontally and vertically (c), and 
diagonally (d). In the first, second and fourth cases, the topology of the colored cluster implies 
the existence of a white percolating cluster next to it. On the contrary, in the third case the cluster percolating in both directions
forbids the existence of other percolating clusters.
}
\label{fig:sketch}
\end{figure}

We must now give a proper definition of {\it percolating spin configuration} on a finite-size $2d$ lattice and distinguish different possibilities. 
In all the cases that we consider, the model is defined on a torus, with toroidal and poloidal directions depicted as
horizontal and vertical directions when picturing the torus as a $2d$ sheet, see Fig.~\ref{fig:sketch}. 
A closed and non-self-intersecting path on a $2d$ lattice with PBC, when viewed as a curve on the torus,  can either be 
shrinked continuously to a single point without ``cutting'' the torus or it cannot, in which case it
winds around the torus ``body'' in at least one direction. We refer to the two types as a \textit{nonwrapping hull} and 
a \textit{wrapping hull}, respectively.
A spin cluster is said to \textit{wrap} around the lattice or to \textit{percolate} if one can construct a closed path inside the cluster that is
wrapping around the torus. Then, a spin configuration $\{ \sigma_i\}$ on the lattice is said to percolate if there is 
at least one spin cluster that percolates.
Schematic examples of the different topologies of the wrapping clusters are depicted in Fig.~\ref{fig:sketch}.
We identify four distinct situations:
\begin{itemize}
 \item a spin configuration with no wrapping cluster;
 \item a spin configuration that contains at least one cluster wrapping along one direction only (that is to say, horizontal or vertical stripes) (a) and (b);
 \item a spin configuration that contains a unique cluster wrapping in both directions, but without any wrapping walls
 (we refer to this case as a ``cross'' topology configuration) (c).
 \item a spin configuration that contains at least one cluster wrapping in both directions, with wrapping cluster walls (also called diagonal stripe configuration) (d);
\end{itemize}
These are the only possible cases in two dimensions and are mutually exclusive. The labels refer to the panels in Fig.~\ref{fig:sketch}.

To each one of the wrapping configurations described above we can associate a probability:
the probability of there being a cluster wrapping in both directions with a cross topology, $\pi_{\mathrm{hv}}$, the probabilities
of there being a cluster wrapping only horizontally or only vertically, $\pi_{\mathrm{h}}$ and  $\pi_{\mathrm{v}}$ respectively, and
the probability of there being a cluster wrapping in both directions in what we call a diagonal stripe configuration, $\pi_{\mathrm{diag}}$.
The probability that the spin configuration has no wrapping cluster is then given by 
$\pi_0 = 1 - \pi_{\mathrm{hv}} - \pi_{\mathrm{h}} - \pi_{\mathrm{v}} - \pi_{\mathrm{diag}}$.
We expect that, at a sufficiently long time after the quench, the system spends a very long period of time
in a regime in which geometrical and statistical properties of spin clusters have the same scaling behaviour of clusters
of occupied sites in critical site percolation on the same lattice.
As the system approaches this regime, the time-dependent wrapping probabilities should become equal to the corresponding ones
in $2d$ critical percolation. Actually, one must take care of the fact that in a percolation problem one considers just the
cluster of occupied sites, while in the spin models that we study here we have to consider clusters of both spin signs at equal footing.
With this consideration in mind, one can compute the values that the probabilities $\pi_{\mathrm{hv}}$, $\pi_{\mathrm{h}}$, $\pi_{\mathrm{v}}$ and
$\pi_{\mathrm{diag}}$ should have in the critical-percolation-like regime. In the case of a lattice of unit aspect ratio  
with PBC
$\pi^{(p)}_{\mathrm{hv}} \simeq 0.6190 $, $\pi^{(p)}_{\mathrm{h}} = \pi^{(p)}_{\mathrm{v}} \simeq 0.1694 $ and $\pi^{(p)}_{\mathrm{diag}} \simeq 0.0418$~\cite{Pi94},
with the superscript $(p)$ indicating that these are limit values that are approached in the critical-percolation-like regime.

Since we will present data relative to the dynamics on the honeycomb lattice, we mention here that,
because of the way in which we constructed this lattice in the numerical simulations,
the aspect ratio vertical/horizontal is equal to $\sqrt{3}$~\footnote{For practical purposes we constructed
the honeycomb lattice from the square lattice by removing half of the vertical lattice bonds.
More precisely, say that a site on a finite slice of the square lattice is indicated by the two indices $i$ and $j$, 
representing the row and column indices, respectively, with $1 \le i, j \le L$. Then, we constructed the corresponding
finite slice of the honeycomb lattice by removing the ``vertical'' bond between the sites $(i,j)$ and $(i+1,j)$ if
$i+j$ is an even number and we kept the bond otherwise, for each pair $(i,j)$.}. 
On this lattice, the critical percolation wrapping probabilities are given by~\cite{BaKrRe09,PruMol04}
$\pi^{(p)}_{\mathrm{hv}} \simeq 0.5120 $, $\pi^{(p)}_{\mathrm{h}} \simeq 0.4221$, 
$ \pi^{(p)}_{\mathrm{v}} \simeq 0.0408 $ and $\pi^{(p)}_{\mathrm{diag}} \simeq 0.0125$.

Other interesting observables are the area of the largest cluster and the length of its interface.
In $2d$ critical percolation the largest cluster is a fractal object, and both its area, $A_c$, and interface length, $l_c$,
are related to its linear size $l$ by
\begin{equation}
A_c \propto l^{D_A} \; , 
\qquad\qquad
l_c \propto l^{D_\ell}
\; , 
\end{equation}
with $D_A$ the bulk fractal dimension and $D_\ell$ the interface fractal dimension. 
These dimensions can be exactly computed for the critical point of the $q$-state Potts model in two dimensions
for $0 < q \le 4$ (where $q=2$ for the Ising model and $q\rightarrow 1$ for percolation)
through a Coulomb gas formulation~\cite{SaDu87}. 
One can introduce a parameter $\kappa$, related to $q$ through $\sqrt{q} = - 2 \cos{\left( 4 \pi/\kappa\right)}$,  
which determines the universality class of the model near criticality. The fractal dimensions are given in terms of $\kappa$ by
\begin{equation}
 D_A = 2 - \frac{\beta}{\nu} = 1 + \frac{3\kappa}{32} + \frac{2}{\kappa}
 \; , 
 \qquad\qquad
 D_\ell = 1+ \frac{\kappa}{8}
 \; , 
 \label{eq:fractal_dimensions}
 \end{equation}
where $\beta$ is the critical exponent of the order parameter and $\nu$ the one of the equilibrium correlation length.
For critical percolation $q=1$ or $\kappa=6$~\cite{Smirnov-01} and thus 
\begin{equation}
D_A = \frac{91}{48} \simeq 1.8958 \; , 
\qquad \qquad 
D_\ell = \frac{7}{4} = 1.75
\; .
\label{eq:fractal_dim_perc}
\end{equation}
We will show the time evolution of the quantities
$A_c$ and $l_c$ and see whether there is a certain period of time in which they have the fractal dimensions of critical percolation.

A quantity which plays a central role in percolation problems is the \textit{number density of cluster areas}, 
$N(A)$, that is to say, the number of clusters of occupied sites of size $A$ per unit area of the system. At the critical
percolation point, $N(A)$ is given by a power law
\begin{equation}
N(A) \, \sim \, A^{-\tau_A}
\; , 
 \label{eq:NA_crit_perc}
\end{equation}
with $\tau_A $ a characteristic exponent (also called Fisher exponent) related to $D_A$, the fractal dimension of the incipient percolating
cluster at the percolation critical point, by \cite{Stauffer94} 
\begin{equation}
\tau_A = 1+\frac{d}{D_A} = \frac{187}{91} \approx 2.0549
\; ,
\label{eq:Fisher_exp}
\end{equation} 
where $d$ is the dimensionality of the lattice, $d=2$ in our case.

We introduce a dynamical cluster size number density, $\mathcal{N}(A,t,L)$, that represents the number of spin clusters
(of both spin signs) with area $A$, per unit area of the lattice, at the time $t$ after the relaxation dynamics has started,
for a lattice of linear size $L$.
In general, we expect ${\mathcal N}(A,t, L)$ to be given by the sum of two contributions
\begin{equation}
{\mathcal N}(A,t, L) \simeq N(A,t) + N_p(A,t,L)
\; .
\label{eq:NA_general}
\end{equation}
$N(A,t)$ represents the contribution of the clusters that do not percolate or,
more precisely, the clusters that do not span across a length which is comparable to the lattice linear size $L$,
while $N_p(A,t,L)$ is the contribution of the clusters that do span the lattice, in general represented by percolating clusters,
and it therefore depends on $L$.
In other words, in the infinite lattice size limit, $\mathcal{N}(A,t, L)$ would be given by just $N(A,t)$, while
$N_p(A,t,L)$ is the term that we must add to take into account finite-size effects.

Our conjecture is that, at a characteristic time $t_p$, at which the system reaches the critical-percolation-like
regime, the term $N_p$ depends on $A$ and $L$ only through $A/L^{D_A}$ reflecting the scaling behavior of the incipient percolating
cluster at critical percolation, while the contribution $N(A,t)$ given by non-percolating clusters
gets an algebraic decay similar to the one at critical percolation, namely 
$N(A,t) \ \sim \ C(t) \, A^{-\tau_A}$, with $C(t)$ a particular time-dependent prefactor to be determined.
If the system of spin clusters were in equilibrium at exactly the critical percolation point, 
one would then have~\cite{SiArBrCu07,SiSaArBrCu09}
\begin{equation}
N(A) \, \simeq \, 2 \ c_d \ A^{-\tau_A}
\; , 
\label{eq:NA_eq_crit}
\end{equation}
with the constant $c_d$ given by $c_d = (\tau_A-2) (\tau_A-1)/2\approx 0.0290$~\cite{SiArBrCu07}.

An approximate expression for the time-dependence of the number density of non-percolating domains, $N(A,t)$,
was derived in \cite{ArBrCuSi07,SiArBrCu07} for the $2d$ NCOP dynamics
by assuming that, after a sufficiently long time, the domain growth can be reduced to the motion of independent domain walls
governed by a curvature-driven mechanism.
Essentially, this argument leads to
\begin{equation}
N(A,t) \, \simeq \, \frac{2c_d \, [ \lambda_d(t-t_p + t_0) ]^{\tau_A-2}}{[ A + \lambda_d(t-t_p + t_0) ]^{\tau_A}}
\qquad\qquad 
t \geq  t_p 
 \; ,
  \label{eq:NA_NCOP_dynamics}
\end{equation}
where  $c_d$ is the same constant as in Eq.~(\ref{eq:NA_eq_crit}), 
$\lambda_d $ is a parameter related to the diffusion coefficient of the domain walls, and $t_0$ is a 
characteristic cutoff time such that $\lambda_d t_0 = 1$.
If one assumes that, at a certain time $t_p$, the system has reached a critical percolation state,
$\tau_A$ is the Fisher exponent of $2d$ critical percolation.
After time $t_p$, $N(A,t)$ satisfies scaling with respect to the usual coarsening length $\ell_d(t) \sim t^{1/2}$.
In particular, Eq.~(\ref{eq:NA_NCOP_dynamics}) suggests that the typical area of the domains scales dynamically as $\mathcal{A}(t) \sim t$
in the coarsening regime ($t \gg t_p$).

Similar arguments were used in \cite{SiSaArBrCu09} to derive an approximate expression for $N(A,t)$ 
in the case of LCOP coarsening, yielding
\begin{equation}
N(A,t) \, \simeq \, \ell_d(t)^{-4} \, \frac{2c_d \, \left( \frac{A}{\ell_d(t)^2} \right)^{\frac{1}{2}}}
{ \left[ 1 + \left( \frac{A}{\ell_d(t)^2} \right)^{\frac{3}{2}} \right]^{\frac{2 \tau_A +1}{3}}}
 \quad,
  \label{eq:NA_COP_dynamics}
\end{equation}
where $\ell_d(t)$ is the characteristic length associated to coarsening (or typical domain radius)
which, for LCOP dynamics, should
behave asymptotically as $\ell_d(t) \sim t^{\frac{1}{3}}$. Again, this result is supposed to hold for times $t$ after the 
system has already reached a critical percolation state.
In this case, Eq.~(\ref{eq:NA_COP_dynamics}) suggests that the typical area of the domains scales dynamically as $\mathcal{A}(t) \sim t^{2/3}$
in the regime $t \gg t_p$ (if one assumes that $\ell_d(t) \sim t^{1/3}$, as expected for COP dynamics).
Note that both Eq.~(\ref{eq:NA_COP_dynamics}) and Eq.~(\ref{eq:NA_NCOP_dynamics}) 
can be rewritten as
\begin{equation}
N(A,t) \, \simeq \, \ell^{-4}_d(t) \, f\left( \frac{A}{\ell_d(t)^2} \right)
 \; ,
  \label{eq:NA_dynamics_scaling}
\end{equation}
with the scaling function 
\begin{eqnarray}
f(x)\, = \left\{
\begin{array}{ll}
\displaystyle{2c_d \ ( 1 + x)^{-\tau_A}}
\;\;\;\;\;\;\;\;
& 
\mbox{for NCOP dynamics}
\vspace{0.25cm}
\\
\displaystyle{2c_d \ x^{\frac{1}{2}} \ ( 1 + x^{\frac{3}{2}} )^{- \frac{2 \tau_A + 1}{3}} } 
\;\; \;\;\;\;\;\;
&  
\mbox{for LCOP dynamics}
\end{array}
\right.
\label{eq:NA_scaling_functions}
\end{eqnarray}%
assuming also that $\ell_d(t) \simeq \left[ \lambda_d ( t - t_p + t_0) \right]^{1/z_d}$ and $t - t_p \gg t_0$.
Note that, in the limit $x \rightarrow \infty$, both versions behave as $f(x) \sim 2 c_d \ x^{-\tau_A}$, and
thus we can use the approximation $N(A,t) \simeq 2 c_d  \ \left[\ell_d(t)\right]^{2(\tau_A - 2)} \ A^{-\tau_A}$ for $A \gg \ell^2_d(t)$.

Self-similarity (\textit{fractality}) and the associated scaling laws are the characteristic features of the critical
behaviour of statistical mechanics models undergoing second order phase transitions. In particular, the fractal dimensions
or \textit{scaling dimensions} of the geometrical objects appearing in these models
can be used to determine the universality class to which the critical behaviour of the model belongs to.
The scaling behaviour in the continuum limit of many $2d$ stochastic processes, 
as critical percolation, the critical Ising model, self-avoiding random walks, {\it etc.}, can be described by the
Schramm-Loewner evolution (known also as stochastic Loewner evolution or SLE).
An SLE with parameter $\kappa$, or $\mathrm{SLE}_{\kappa}$, is essentially a family of conformally invariant 
random planar curves~\cite{Schramm-99,Schramm-11,Lawler-07}, with the parameter $\kappa$ controlling how much the
curve ``turns''. It has been shown~\cite{Smirnov-01,Camia-06} that
the parameter $\kappa$ is the same as the one in the aforementioned Coulomb gas formulation of the $q$-state Potts model, and it
is linked to the central charge $c$ of the associated conformal field theory: for example, $\mathrm{SLE}_{3}$
describes the critical Ising model, $\mathrm{SLE}_{6}$ describes critical percolation, etc.

The fractal dimension of the curves of $\mathrm{SLE}_{\kappa}$ is given by $D_{\ell}(\kappa)= 1 + \kappa/8$.
Thus, by measuring $D_{\ell}$ it is possible to obtain $\kappa$ and, consequently, the type of criticality of the model.
Our approach will be then to study the fractality of the percolating hulls (or domain walls)
by analysing their scaling with $L$.
This is done, for example, by studying the scaling behaviour of the interface of the largest spin cluster.

Another method that we will use is the analysis of the \textit{winding angle} of the long cluster hulls.
The winding angle $\theta(x)$ between two points $P$ and $P'$ on a planar curve, such that the arc with extremes $P$ and $P'$
has length $x$, is defined as the incremental angle that the tangent to the curve is rotated by when moving from one point
to the other one. For conformally invariant stochastic planar curves belonging to $\mathrm{SLE}_{\kappa}$,
$\theta(x)$ as a function of the arc length $x$ behaves as~\cite{WiWi03,SaDu87}
\begin{equation}
\langle \theta^2(x) \rangle = \mbox{cst} +{4 \kappa \over 8+\kappa } \ln{x} \; ,
\label{eq:winding_angle_critical_hulls}
\end{equation}
with $\langle \dots \rangle $ the average over all possible realisations of the stochastic process.
In the dynamic problem we will deal with the time-dependent winding angle 
$\theta(x,t)$ measured for the hulls of the spin clusters.
If the system were to attain a critical-percolation-like state or another type of
$2d$ criticality (in general, any Potts model criticality) during the relaxation dynamics, we would then be able to observe 
$\langle \theta^2(x,t) \rangle$ satisfying the law expressed by Eq.~(\ref{eq:winding_angle_critical_hulls})
after having rescaled the curvilinear distance $x$ by a proper dynamical length.
In this way it is possible to distinguish between different types of criticalities. For example, we will see
that in the case of the voter model (see Sec.~\ref{subsec:winding-angle-vm}), the system reaches a critical percolation state
that does not persist and a new criticality is approached at late times.
The measurement of $\langle \theta^2 \rangle$ allows us to distinguish this new criticality from the one of percolation
with much more precision than what can be done with the analysis of the fractal dimension of the spin clusters.
Moreover, $\langle \theta^2\rangle$ gives an alternative  estimate of the typical time $t_p$.

\begin{figure}[h]
\begin{center}
   \includegraphics[scale=0.6]{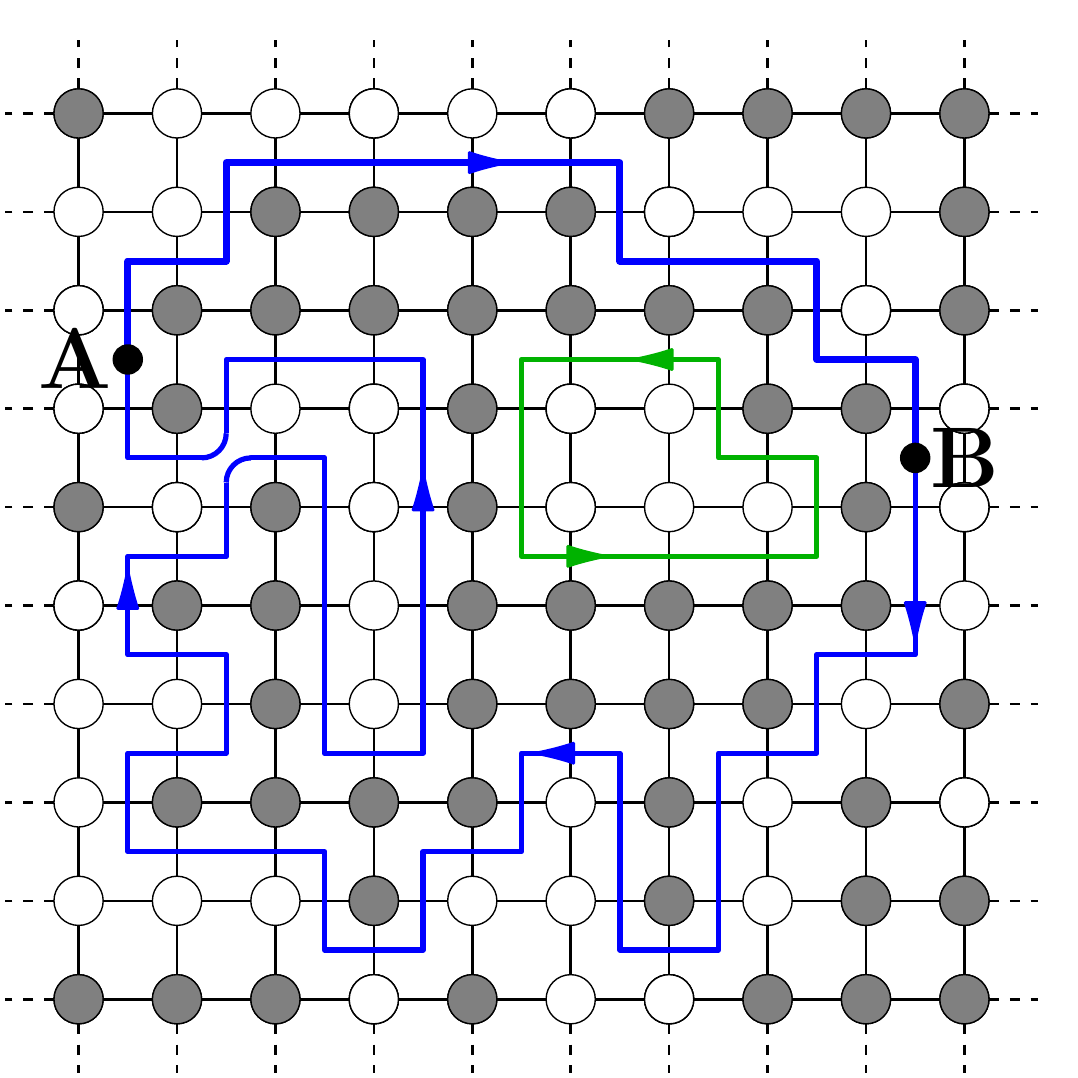}
\end{center}
\caption{\small
A schematic representation of the way we measure the winding angle for domain walls of spin clusters on a
square lattice. The figure presents a $10 \times 10$ square lattice mesh, where circles represent lattice sites.
Gray circles correspond to $+1$ spins, while white ones to $-1$ spins. We highlight two domain walls forming the interface of a ``gray'' spin cluster:
the blue one is the \textit{external} hull, \textit{i.e.} the hull that encloses the spin cluster; the green one is an \textit{internal} hull.
The arrows drawn on the hulls indicate the direction along which they are travelled in order to compute the winding angle, clockwise for external hulls,
anti-clockwise for internal ones.
As an example, the winding angle corresponding to the arc $\wideparen{AB}$ of length $13$ (number of unsatisfied lattice bonds traversed by the wall) 
on the blue domain wall is $ \Delta \theta_{\wideparen{AB}} = + \pi $.
The total winding angle for the external domain wall is $2 \pi$, while it is $-2 \pi$ for the internal one.
}
\label{fig:winding-angle-sketch}
\end{figure}

However, Eq.~(\ref{eq:winding_angle_critical_hulls}) is satisfied in the continuum space limit, while the models  are defined
on a lattice.
In Fig.~\ref{fig:winding-angle-sketch} we show a schematic representation of the definition of $\theta$ on a lattice.
To compute the winding angle for an arc of domain wall of length $n$ (with $n$ a positive integer), 
we start from one of the two extrema and we walk along the interface of the spin cluster (defined on the dual lattice) 
incrementing the tangent angle $\theta$ by a quantity $\Delta \theta$ each time the curve turns.
Notice that on the square lattice
$\Delta \theta$ can only take the values $0$, $\pi/2$ or $-\pi/2$.
The winding angle is then given by
\begin{equation}
 \theta(n) = \sum^{n}_{k = 1} \Delta \theta_{k}
 \label{eq:winding_angle_definition_lattice}
\end{equation}
with $\Delta \theta_{k}$ the increment corresponding to the $k$-th step of the walk. We always use the convention that
the domain wall is travelled in the direction such that the interior of the domain is on the right side of the walker, and
that right turns corresponds to $\Delta \theta > 0$, while left turns to $\Delta \theta < 0$. With these convention, one clearly sees
that the external hull of a domain has total winding angle (that is to say, the winding angle from one point to itself)
equal to $2 \pi$, while hulls that are in the interior have total winding angle equal to $- 2 \pi$. 
There might also be hulls that have zero total winding angle. These are hulls that wrap around the system.
For this reason, the total winding angle can be used to determine whether an interface wraps or not.
We also mention that the quantity in Eq.~(\ref{eq:winding_angle_definition_lattice}) must not depend on the starting point $P$ on the arc. 
Thus, we measure $\theta(n)$ for each point of the domain wall used as $P$ and we  then compute the average.

In the context of percolation theory, another useful tool is the 
{\it pair connectedness function}, $g(r)$, which is defined as the probability that
two lattice sites separated by a distance $r$ belong to the same cluster. 
In $2d$ critical percolation, the behaviour of $g(r)$ for
large $r$ ($r \gg r_0$, with $r_0$ the lattice spacing) is known~\cite{Stauffer94,ChristensenMoloney,Saberi15}
\begin{equation}
 g(r) \sim r^{-2\Delta_\sigma}, \quad	r\gg r_0
\end{equation}
where $\Delta_\sigma=2-D_A$ and $D_A$ is the fractal dimension of the critical percolation clusters.
In order to assess the presence of a critical-percolation-like regime in a coarsening process,
we introduce an analogous quantity. On a square lattice:
\begin{equation}
 g(r,t) = \frac{1}{4 L^{2}} \sum_{i} \sum_{i_{r}} \langle \gamma_{i,i_{r}}(t) \rangle 
\end{equation}
where the first summation runs over all lattice sites, the second one over the four sites $i_{r}$ that are located at distance $r$ from site $i$
along the horizontal and vertical directions, and $\gamma_{i,j}(t)=1$ if the sites $i$ and $j$
belong to the same spin cluster at time $t$, and equals $0$ otherwise.

\section{Local Kawasaki dynamics}
\label{sec:local-Kawasaki}

The local Kawasaki rules~\cite{Kawasaki66a,Kawasaki66b} are used to mimic 
phase separation in systems in which hydrodynamic effects can be neglected~\cite{Bray03}. 
In short, these dynamics describe  the process whereby a binary mixture
of components A and B, initially in a homogeneous phase,
demix leading to the coexistence of two phases: one rich in A
and the other in B. The system, initially in an unstable spatially
uniform state, performs a coarsening process to approach
its thermodynamically stable phase-separated state. Typical 
realisations are phase separation in binary alloys, high-viscosity fluids, and
polymer blends. The dynamics of phase separation has seen a revival of interest recently, 
in the context of experimental~\cite{De-etal14,Tojo-etal} and numerical~\cite{Hoffman-etal14,KudoKawaguchi,Takeuchi15,Takeuchi16,ShitaraBirBlakie17,WilliamsBlakie17,Takeuchi18} 
studies of separating binary mixtures of Bose gases.
In the cases of our interest, the control parameters are the temperature and the 
relative concentration of the two species.

The behaviour of the system at {\it late} times is well understood. In the long times limit the system approaches a dynamic scaling regime
described by an extension of the
Lifshitz-Slyozov-Wagner (LSW) theory~\cite{LifshitzSlyozov59,Wagner61}, in which the typical 
domain radius grows as~\cite{Hu86}
\begin{equation}
\ell_d(t) \simeq t^{1/z_d}
\qquad
\mbox{with}
\qquad
 z_d=3
 \end{equation}
(whereas for the NCOP dynamics the growing length is also given by a power law but the exponent is $z_d=2$). 
Numerical results in favour of this law were published in~\cite{Hu86,Amar88,Rogers88,Godreche-et-al-04,Krzakala-05} for spin exchange models 
although the time-dependence of the growth-law can be more complex in particle or polymer phase separating systems,
see {\it e.g.}~\cite{Reith12} and references therein.

It was noticed in~\cite{SiSaArBrCu09} that the low-temperature evolution 
of a 50:50 mixture after a quench from infinite temperature shares many points in common with the 
one of NCOP dynamics. On the one hand, an initial approach to 
critical percolation was noticed, although the time-scale needed to reach this 
state was not studied in detail. On the other hand, the number density of 
finite size cluster areas and domain wall lengths were studied numerically and 
they were both found to satisfy dynamic scaling with respect to the dynamic 
growing length $\ell_d(t) \simeq t^{1/z_d}$ with $z_d=3$, the one that characterises the 
scaling properties of the space-time correlation functions. 

We are interested in studying the early stages of the dynamical process 
(in contrast to the asymptotic LSW regime) and, in particular, the way in which the system approaches a state with a 
stable pattern of percolating domains. We confirm that this occurs for balanced mixtures whereas 
different behaviour is found for asymmetric ones~\cite{Takeuchi15}. A shorter account of the results in this 
Section appeared in~\cite{TaCuPi16}.

For NCOP dynamics the approach to the critical-percolation-like state 
is characterised by a characteristic length-scale, $\ell_p(t)$~\cite{BlCuPiTa-17}. The results that we will present 
in this Section, see also~\cite{TaCuPi16},
 suggest that the same mechanisms are at work for the local Kawasaki dynamics of binary mixtures with 
 equal concentration of the two species, with 
 \begin{equation}
 \ell_p(t) \simeq \ell_d(t) \, t^{1/\zeta} \simeq t^{1/z_p}
  \label{eq:ell_p_Ka}
 \end{equation}
 where the values of the exponents $\zeta$ and $z_p$ are expected to be different from the ones found for NCOP dynamics
 and may depend on the lattice geometry.
 In the latter case, reasoning in terms of the exponent $z_p$
 gives satisfactory results since the dynamical length-scale associated to coarsening, $\ell_d(t)$,
 has power law behaviour $\ell_d(t) \sim t^{1/z_d}$ with $z_d=2$, over a long time-interval
 which includes also the regime where the domain structure has critical percolation scaling properties.
 Therefore,  $\ell_p(t)$ has a power law dependence on time too, so that $z_p$ and $\zeta$ are related by
 $z_p = z_d \ \zeta/(\zeta + z_d)$.
 Instead, for Kawasaki dynamics the power law behaviour $\ell_d(t) \sim t^{1/z_d}$ with $z_d=3$
 establishes only very late when most of the features associated to critical percolation
 have been washed away.
 Accordingly, the characteristic length $\ell_p$ will still be related to $\ell_d$ in the way expressed by Eq.~(\ref{eq:ell_p}),
 but $\ell_d$ will be represented by the preasymptotic growing length $\ell_G$, defined in Sec.~\ref{subsec:observables}
 using the excess energy, that we can easily measure at any time.
 
\subsection{Formulation}
 
Throughout this Section we will use the spin language, in which 
up and down Ising variables correspond to the presence of the A and B 
species on a given site. All simulations are done on $2d$ lattices with PBC and
most at equal concentration of up and down spins.

As already explained in Sec.~\ref{subsec:models}, the Kawasaki rule consist in single spin-exchange configuration updates
with transition probabilities given by the Metropolis Monte Carlo method.
At each time step we randomly choose a pair of nearest-neighbour lattice sites. 
If the two sites are occupied by the same kind of spin, their state remains unchanged.
On the other hand, if the two sites have antiparallel spins, their spins are exchanged with a probability
given by the usual form $W( \Delta E) = \min{(1,\exp(-\beta \Delta E))} $, 
with $ \Delta E$ the change in energy provoked by the spin-exchange event.
Denoting by $s_{i}$ the spin at lattice site $i$ and by $\mathcal{N}(i)$ the set of sites that are nearest-neighbours 
of $i$, for a spin-exchange event involving the nearest-neighbour sites $i$ and $j$,  $\Delta E$ takes the form
\begin{equation}
 \Delta E = J \, (s_{i}-s_{j}) \left( \sum_{k \in \mathcal{N}(i) \backslash \{j\} } s_{k} - \sum_{h \in \mathcal{N}(j) \backslash \{i\} } s_{h} \right)
 \; . 
\end{equation}%
Since the local Kawasaki dynamics get blocked at zero temperature, most of the results presented in this Section
are obtained at finite but small sub-critical temperature.

\subsection{Snapshots}

\vspace{0.5cm}

\begin{figure}[h!]
\begin{center}
  \subfloat[$t=0$]{\includegraphics[scale=0.33]{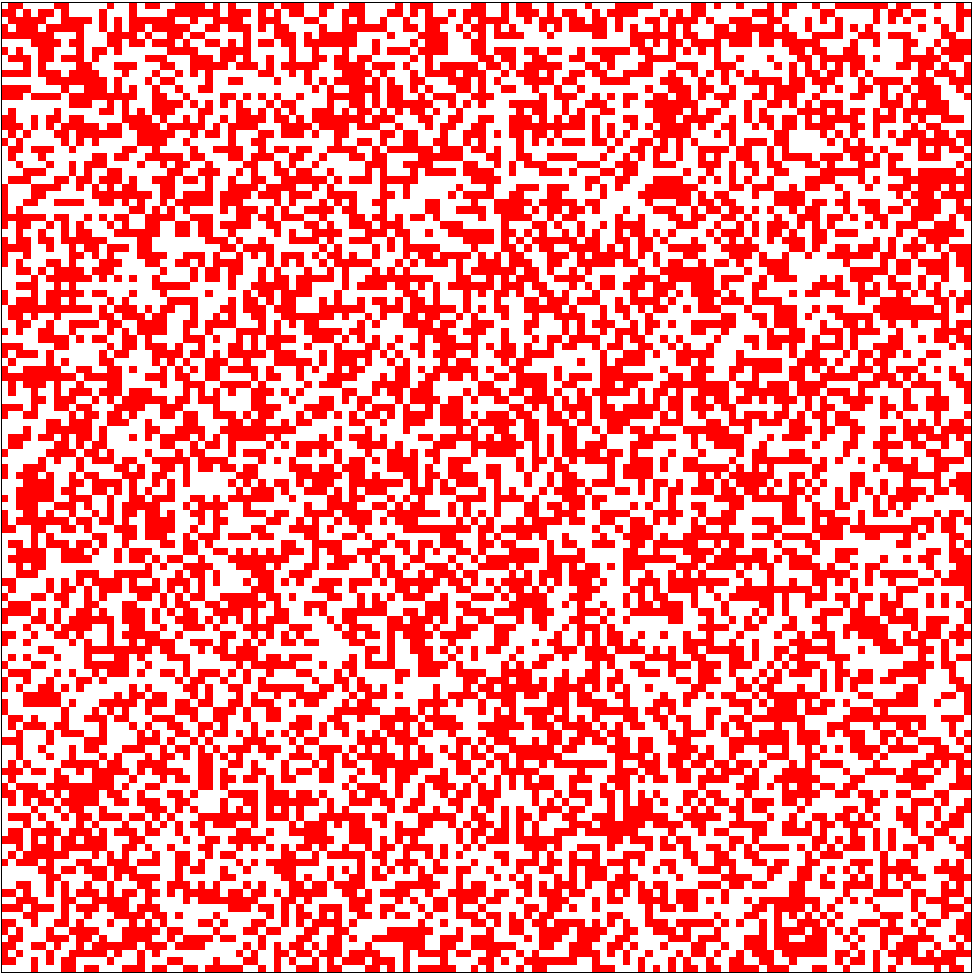}}\quad%
  \subfloat[$t=2$]{\includegraphics[scale=0.33]{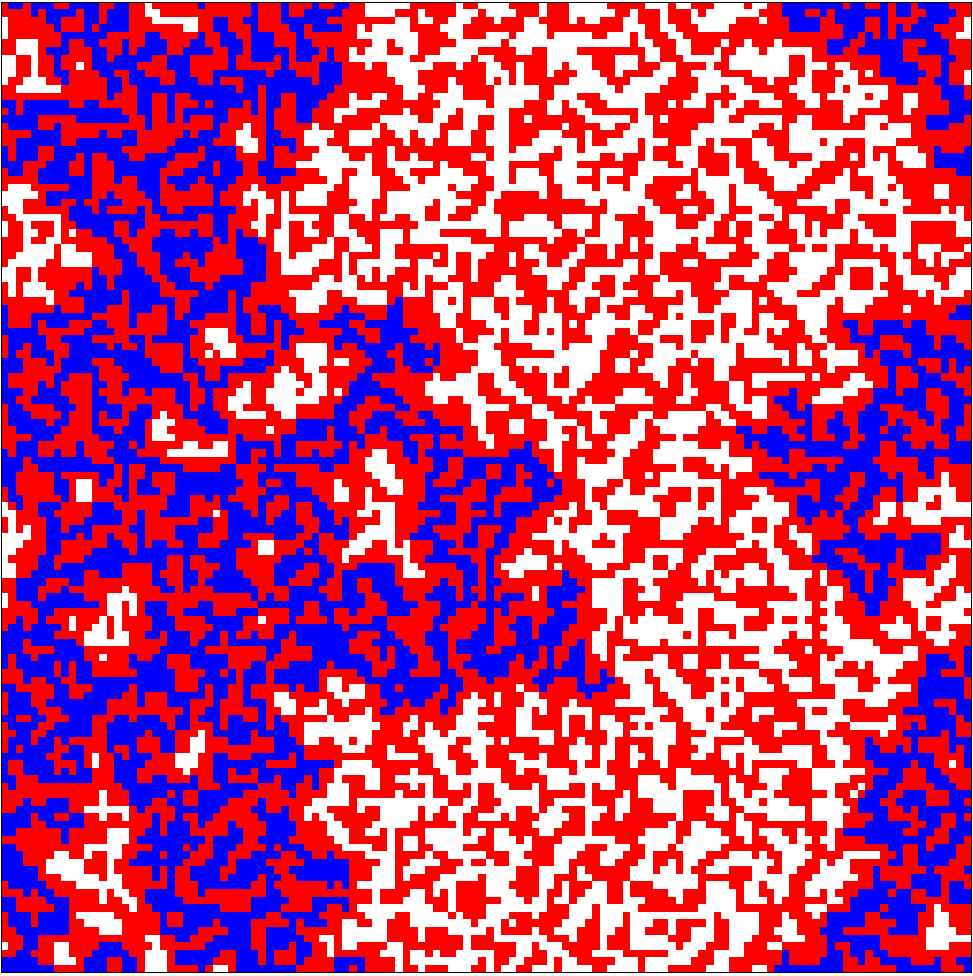}}\quad%
  \subfloat[$t=4$]{\includegraphics[scale=0.33]{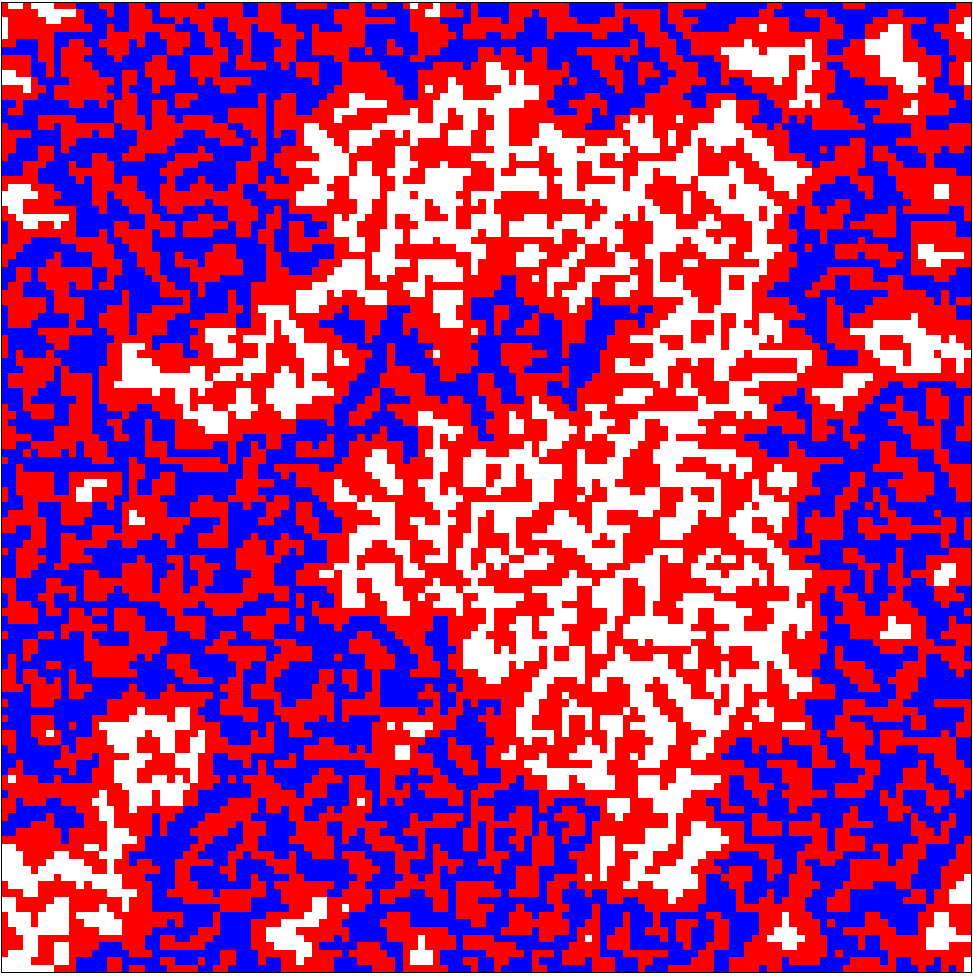}}\quad%
  \subfloat[$t=8$]{\includegraphics[scale=0.33]{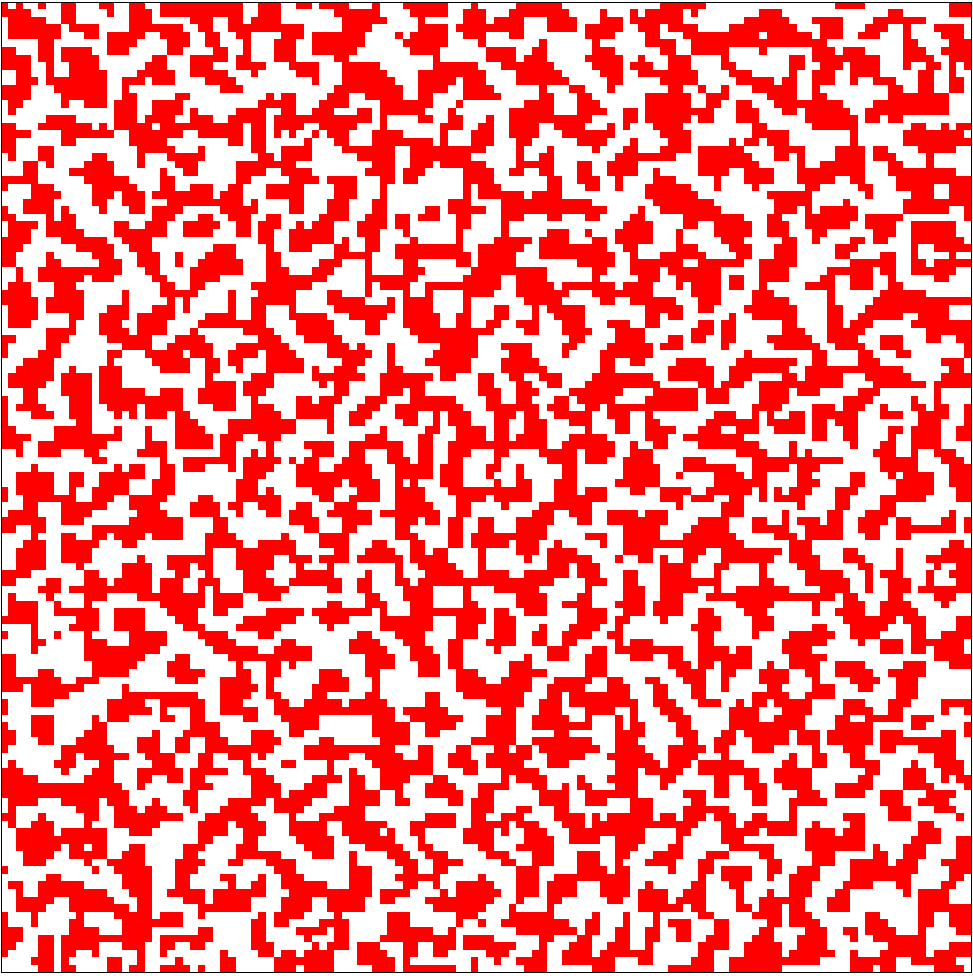}}

  \subfloat[$t=128$]{\includegraphics[scale=0.33]{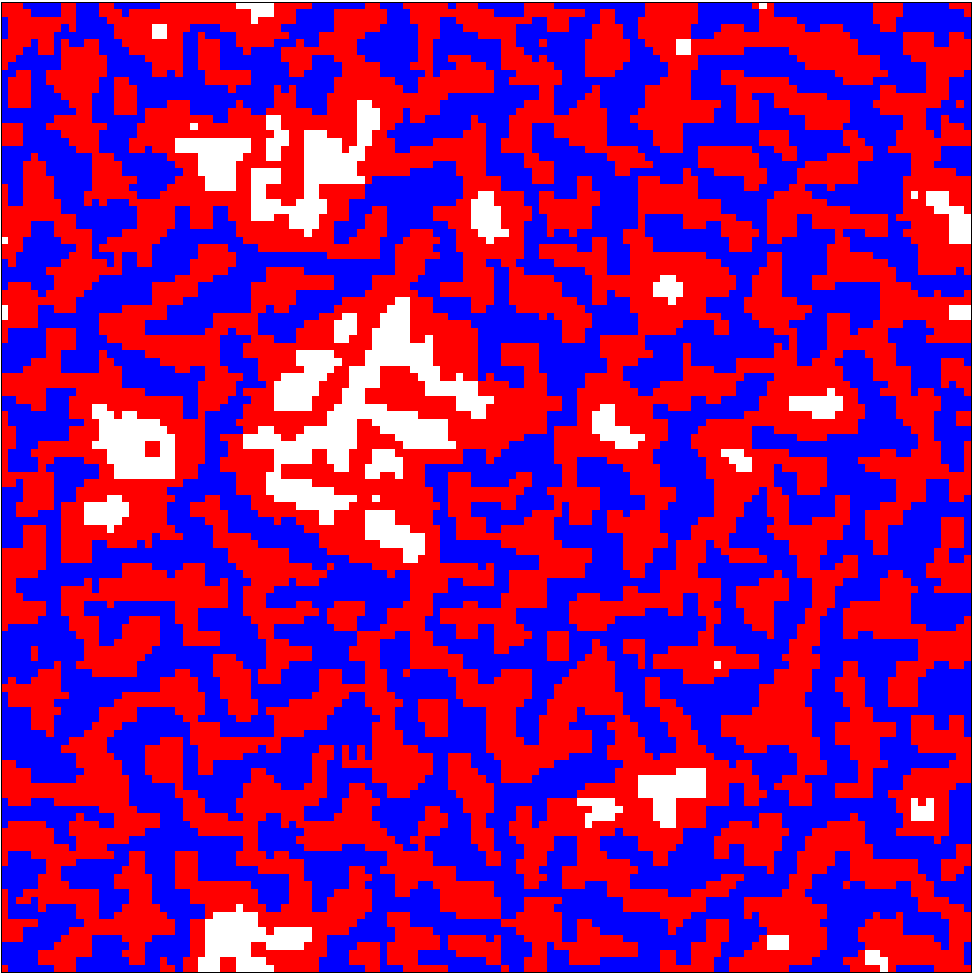}}\quad%
  \subfloat[$t=256$]{\includegraphics[scale=0.33]{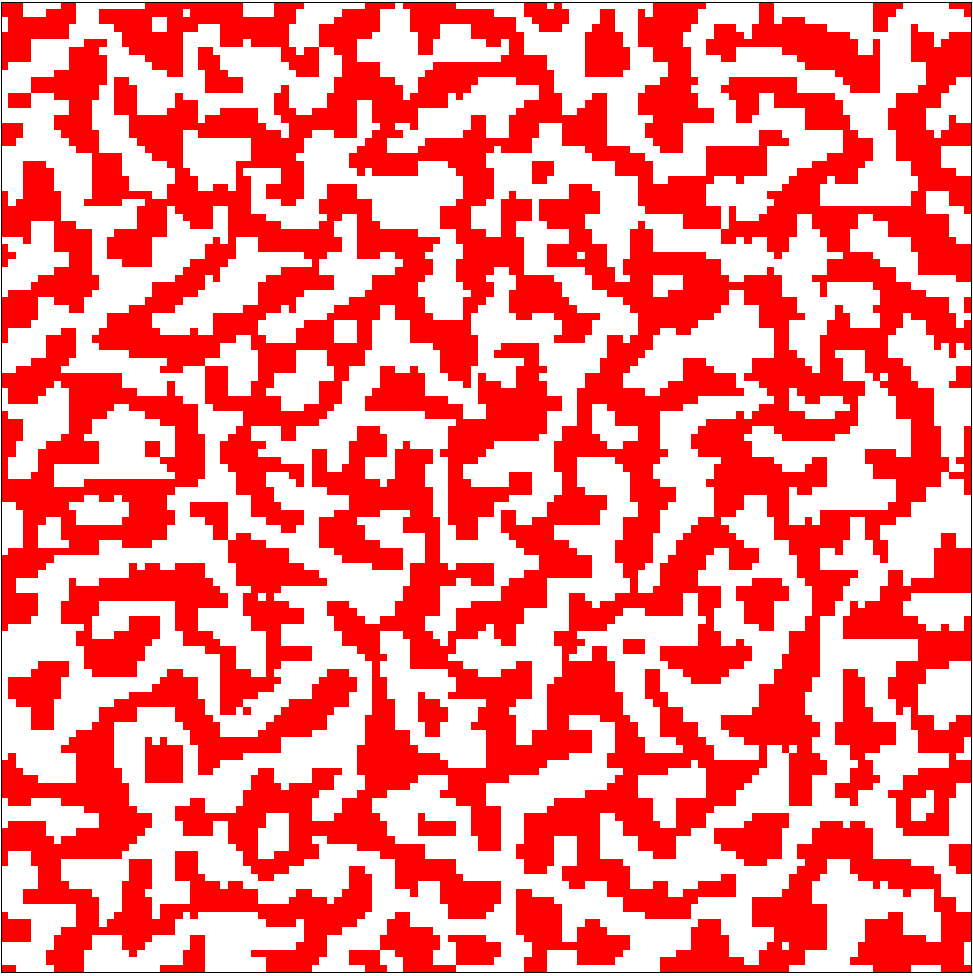}}\quad%
  \subfloat[$t=512$]{\includegraphics[scale=0.33]{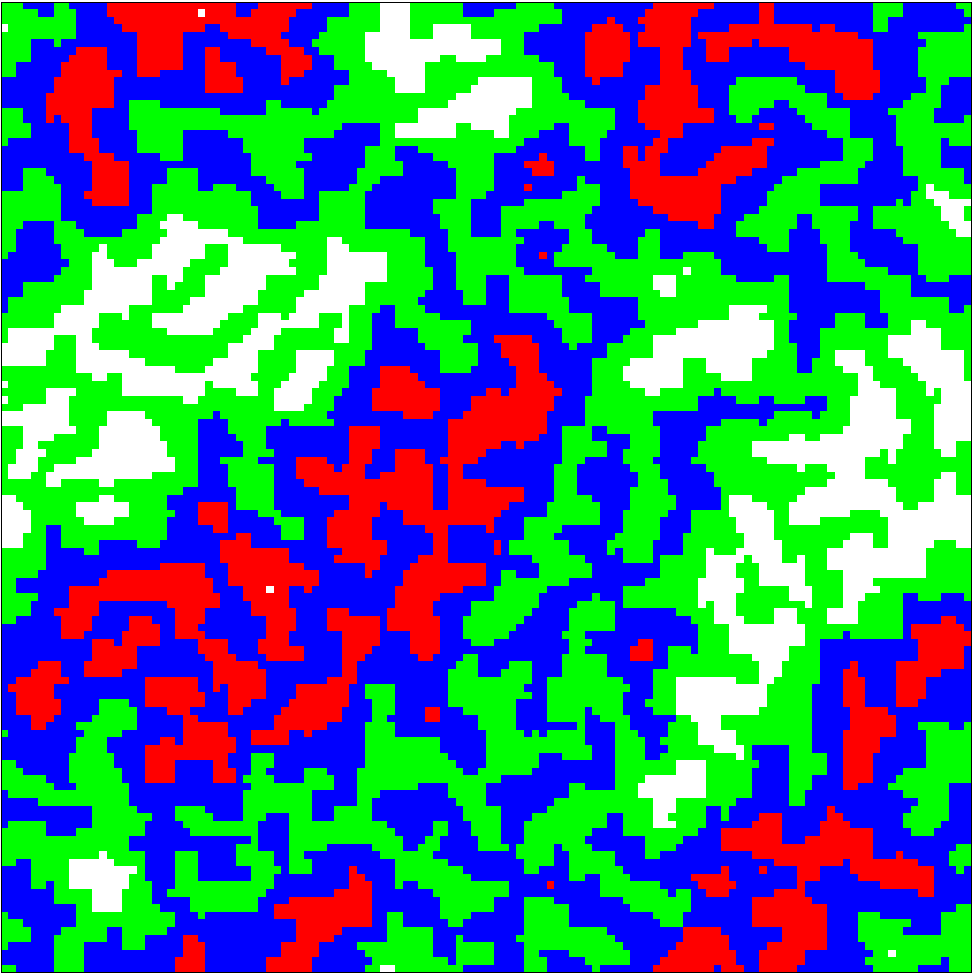}}\quad%
  \subfloat[$t=1024$]{\includegraphics[scale=0.33]{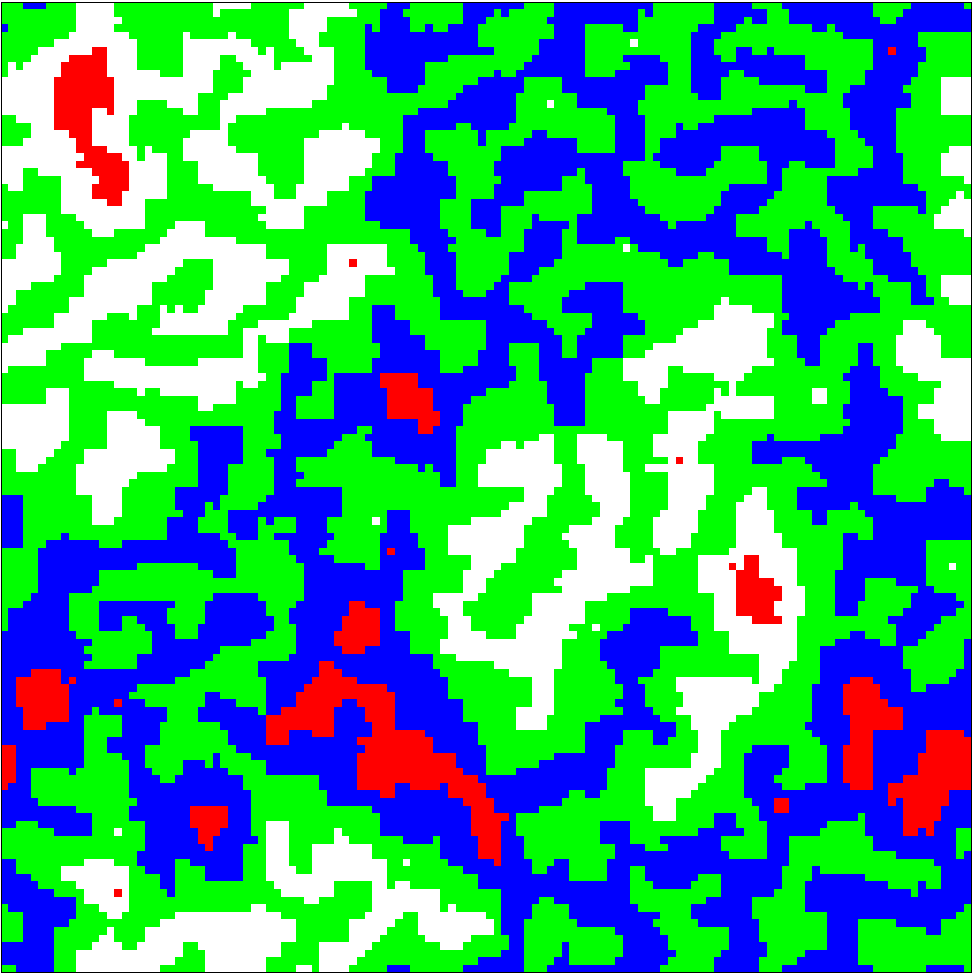}}
\end{center}
\caption{\small Some snapshots of a spin configuration of the $2d$IM 
evolving under local Kawasaki dynamics, following a sudden quench from
infinite temperature to $T_c/2$, on a square lattice with linear size $L=128$ and PBC.
The concentration of the two species of spin is the same.
Red sites and white sites represent $+1$ and $-1$ spins, respectively. Clusters that 
wrap around the system are highlighted in different colours, 
green for spin $+1$ wrapping clusters, blue for $-1$ wrapping clusters. 
The times selected are, from left to right, $t=0, \, 2, \, 4, \, 8$  in the upper row, 
and $t=128, \, 256, \, 512, \, 1024$  in the lower row.}
\label{KaSnap}
\end{figure}

In Fig.~\ref{KaSnap} we show some snapshots of the spin configuration evolving under 
the local Kawasaki update rule, after  a sudden quench of the Ising model from
infinite temperature to $T_c/2$. The spins are placed on a square lattice with linear size $L=128$ 
and PBC. The concentration of up and down spins is a half. These are shown with 
red and white dots, respectively. The panels show the evolution of the 
characteristic domain pattern. The clusters that percolate in at least one direction are highlighted in green and blue,
respectively. Interestingly enough, already at $t=2$, a percolating cluster appears but it then breaks.
In later time snapshots, say at $t \geq 512$,  {\it two} large clusters of opposite spin value are interlaced and
both percolate along the vertical direction.

\vspace{0.25cm}

\begin{figure}[h!]
\begin{center}
  \subfloat[$t=256$]{\includegraphics[scale=1.33]{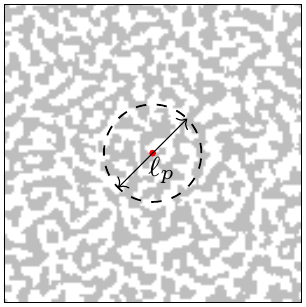}}\quad%
  \subfloat[$t=512$]{\includegraphics[scale=1.33]{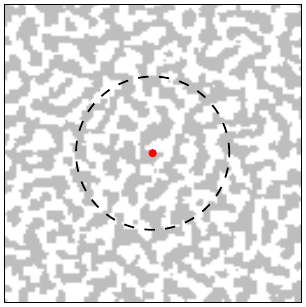}}\quad%
  \subfloat[$t=1024$]{\includegraphics[scale=1.33]{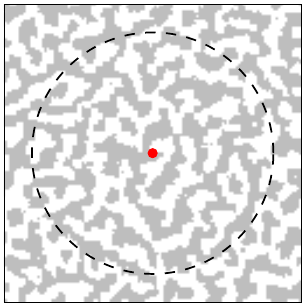}}\quad%
\end{center}
\caption{\small
A schematic representation of the time evolution of the characteristic length $\ell_p$ associated to critical-percolation-like  behaviour,
presented for a spin configuration evolving under local Kawasaki dynamics, following a sudden quench from
infinite temperature to $T_c/2$, on a square lattice with linear size $L=128$ and PBC.
The dashed circle has diameter equal to $\ell_p$ at the time $t$ ($t$ indicated below each snaphot). %
The red disk has instead diameter $\ell_d(t)$ and represents the typical size of areas that are equilibrated at the target temperature of the quench.
The coarsening length $\ell_d(t)$ has been estimated by means of numerical simulations (by averaging over many realisations) and
we assumed $\ell_p(t) \simeq \ell_d(t) t^{1/\zeta}$ with $\zeta=2$. See main text for more details.
}
\label{KaSnap-lp}
\end{figure}

In order to let the reader have a clearer understanding of the separation of length scales 
produced by the existence of the additional characteristic length $\ell_p(t)$ associated to critical-percolation-like behaviour,
we present a scheme in Fig.~\ref{KaSnap-lp} that compares $\ell_p(t)$ with the usual coarsening length $\ell_d(t)$.
In this figure, we show a sequence of snapshots of the spin configuration of a $2d$IM on a square lattice,
evolving under Kawasaki dynamics at temperature $T_c/2$ (with equal concentration of the two species of spin).
As one can see, the length $\ell_d$ (represented by the red spot) grows very slowly with time, approximately
as $\ell_d(t) \sim t^{0.2}$ in this range of times (see next Section and Fig.~\ref{LKaGL}).
The growth of equilibrated domains is thus barely noticeable. Compared to it, $\ell_p$ has a very rapid growth.
In fact, we assumed that $\ell_p(t) \sim \ell_d(t) t^{1/\zeta}$, with $\zeta \simeq 2$.
This relation is so far a conjecture based on the results obtained for the NCOP dynamics in our previous work~\cite{BlCuPiTa-17}.
In the following Sections, we will provide the methods and the observables used to probe this relation and estimate the exponent $\zeta$.

For the moment, it suffices to know that $\ell_p(t)$ represents the typical distance up to which one is able to observe critical percolation properties
in the domain pattern, at time $t$. 
At the time $t_p$ such that $\ell_p(t_p) \sim L$, the critical percolation features are extended to the whole system.
However, the reader should not get confused by Fig.~\ref{KaSnap-lp}. 
The length $\ell_p$ shown is not computed for the single realisation of the dynamics.
Instead, it is extracted from the finite-size scaling analysis of the observables
related to the geometrical and statistical properties of the spin clusters, averaged over many realisations,
as we will explain in the following Sections.

\subsection{The excess-energy growing length}

\begin{figure}[h]
\begin{center}
        \includegraphics[scale=0.5]{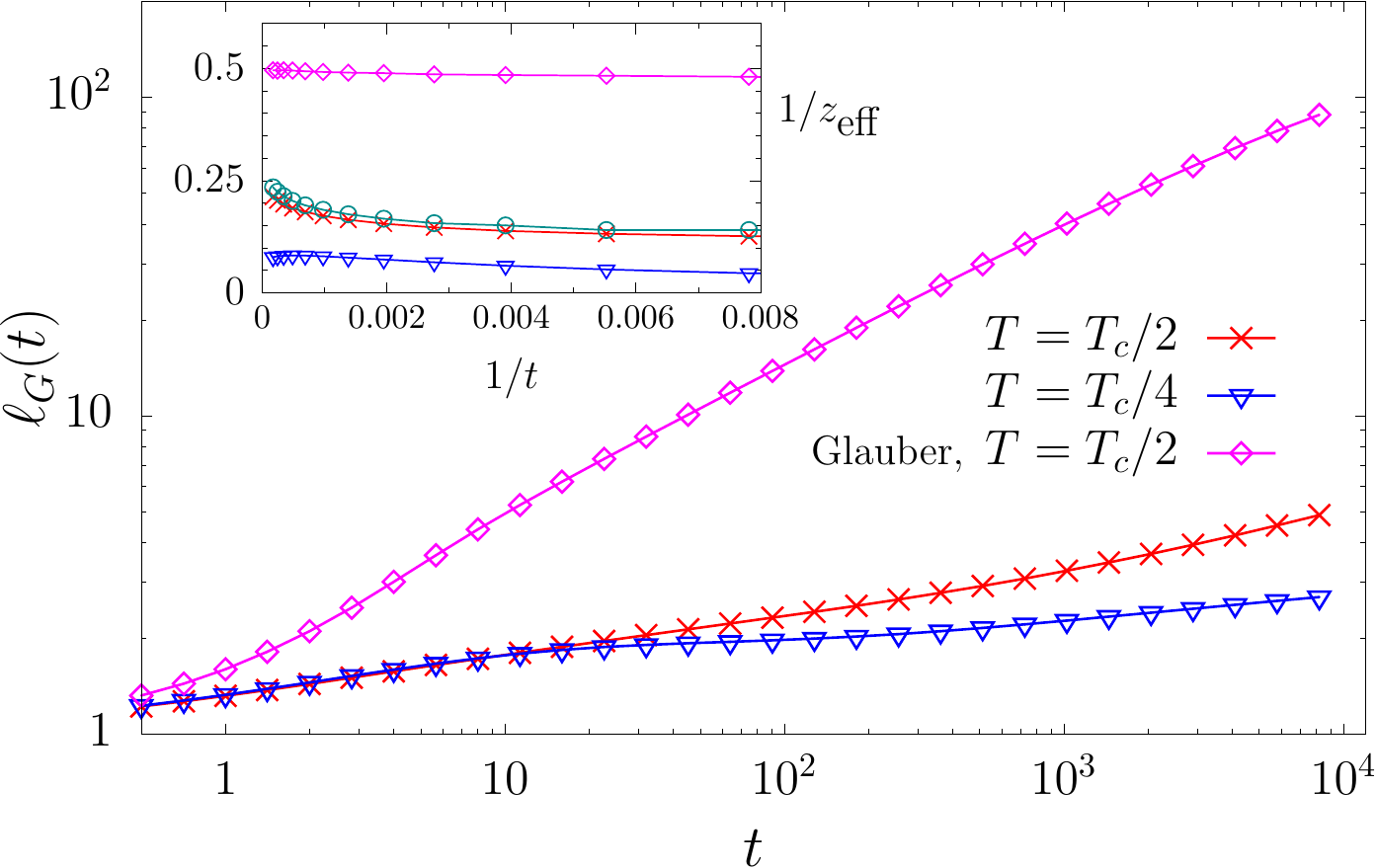}
        \includegraphics[scale=0.5]{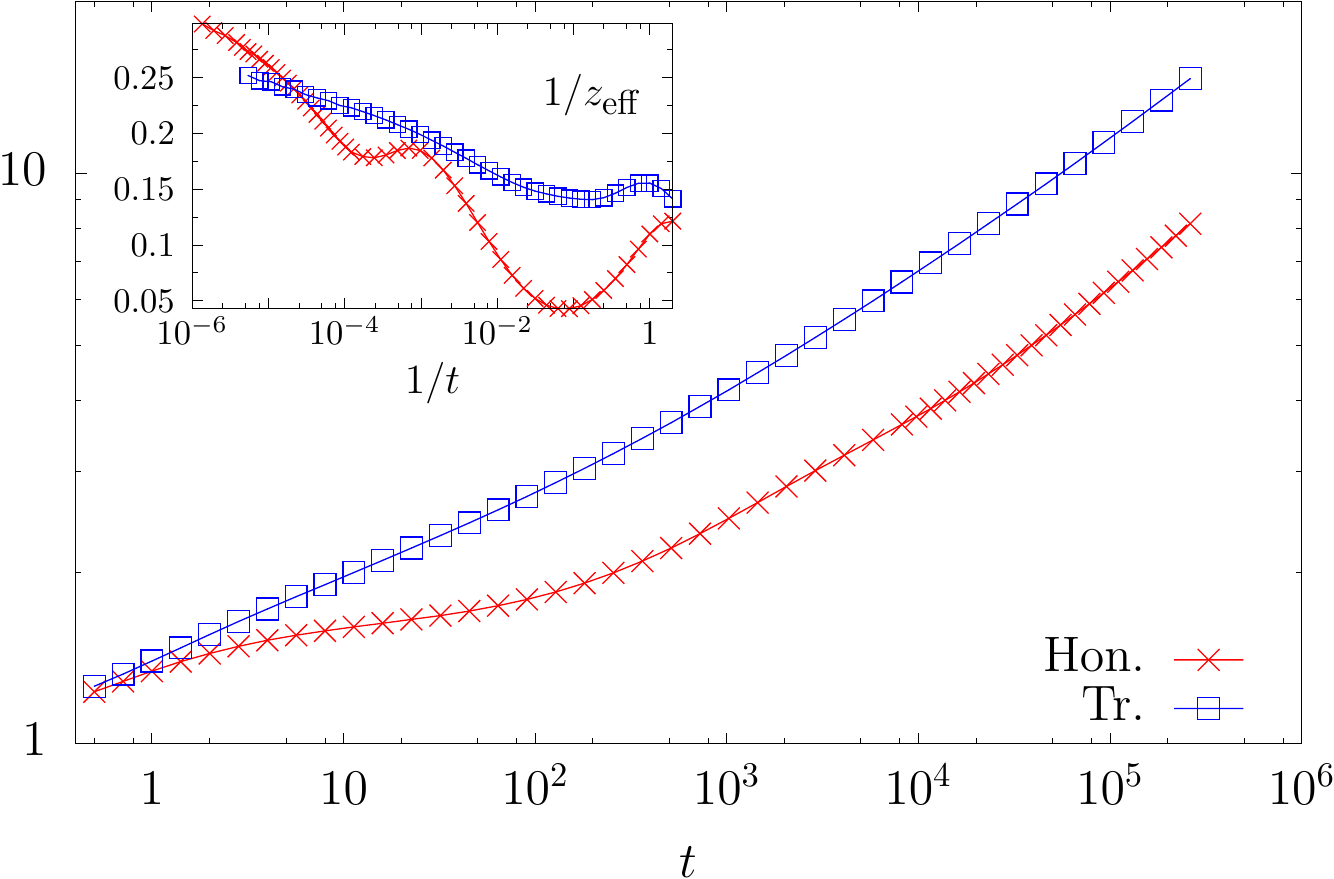}
\end{center}
\caption{\small Local Kawasaki dynamics of the $2d$IM with balanced densities of the two species.
The excess-energy growing length $\ell_G(t) = 1 / \epsilon(t)$ 
against time following a sudden quench. Left panel: model defined on a square lattice with linear size $L=640$ quenched
to $T=T_c/2$ (red curve) and $T=T_c/4$ (blue curve) and, for comparison, 
data for NCOP dynamics at $T_c/2$ (purple curve). 
Right panel: growing length for Kawasaki dynamics on honeycomb and triangular lattices, 
with linear size $L=320$, quenched to $T_c/2$.
In the insets, we show the effective growth exponent, $1/z_{\mathrm{eff}}(t)$, computed as the 
logarithmic derivative of the function $\ell_G(t)$, plotted as a function of $1/t$.
In the inset of the left panel, we also include the effective growth exponent estimated from 
the scaling of $\langle \theta^2 \rangle$ (circles).
}
\label{LKaGL}
\end{figure}

We study here the characteristic growing length $\ell_G(t)$ defined  as the inverse of the the excess-energy, Eq.~(\ref{eq:excess-energy}). 
In Fig.~\ref{LKaGL} we show  $\ell_G(t)$ for the local Kawasaki dynamics on square, honeycomb and triangular lattices,
at temperatures $T_c/2$ and $T_c/4$ for the square lattice and $T_c/2$ in the other two cases.

\vspace{0.5cm}

First, we note that the values reached by $\ell_G$
are notably smaller than the ones obtained with NCOP dynamics also shown in the left panel (labeled Glauber). Second, the 
data are temperature independent before $t \simeq 10$. At around this time,  the 
data at $T_c/4$ slow down and deviate from the ones at higher temperature.  Nowhere in the time span shown
in the figure a stable algebraic increase of $\ell_G$ establishes. 
The evolution of the effective exponent $1/z_{\rm eff}(t)$, computed as the logarithmic derivative of 
$\ell_G(t)$, is followed in the inset. In the time-window used, at $T_c/2$, $1/z_{\rm eff}(t)$  varies 
between $0.1$ and $0.25$, approximately, implying that $z_{\rm eff}(t)$ goes from $10$ to $4$. 
The measurement seems to slowly approach the
expected value for the dynamic exponent $z_d =3$~\cite{Hu86} but it is still far from it. 
The deviation is even worse for $T_c/4$ and the time-dependence of the effective exponent  $z_{\rm eff}$ is non-monotonic.
Indeed, we reckon that much longer simulations are needed to observe the asymptotic coarsening regime with the correct dynamical exponent.
However, we stress here that the relatively short times
at which we measured $\ell_G(t)$ are going to be the relevant ones for the study of the approach to the critical-percolation-like state. 
Contrary to what happens in the case of NCOP dynamics, the asymptotic $z_d=3$ has not established yet and time 
scales of at least two orders of magnitude longer are needed to see $z_{\rm eff}$ converge to $z_d=3$~\cite{Amar88,Rogers88}.
Another interesting issue is that the local Kawasaki dynamics on the honeycomb lattice do not block as do the 
NCOP rules, see the snapshots in~\cite{BlCuPiTa-17}, and the values of $z_{\rm eff}$ are similar to the ones for the square lattice.

The fact that the effective exponent varies so much in time and depends on temperature 
suggests to use the characteristic growing length $\ell_G(t)$ itself as a representative of $\ell_d(t)$ in all scaling relations,
to analyse the pre-asymptotic regime with the eventual approach to critical percolation. 
The difficulties involved in using the exponent $z_d$ and its effective evaluation
are confirmed by the analysis of other observables.

\subsection{Wrapping probabilities}
\label{subsec:WP-Kawasaki}

The wrapping probabilities $\pi_{\rm h}$, $\pi_{\rm v}$, $\pi_{\rm hv}$, $\pi_{\rm diag}$
defined in Sec.~\ref{subsec:observables} can be useful to determine the typical time required  to develop
a percolative structure and its scaling with the system linear size $L$.

\begin{figure}[h!]
\begin{center}
        \includegraphics[scale=0.56]{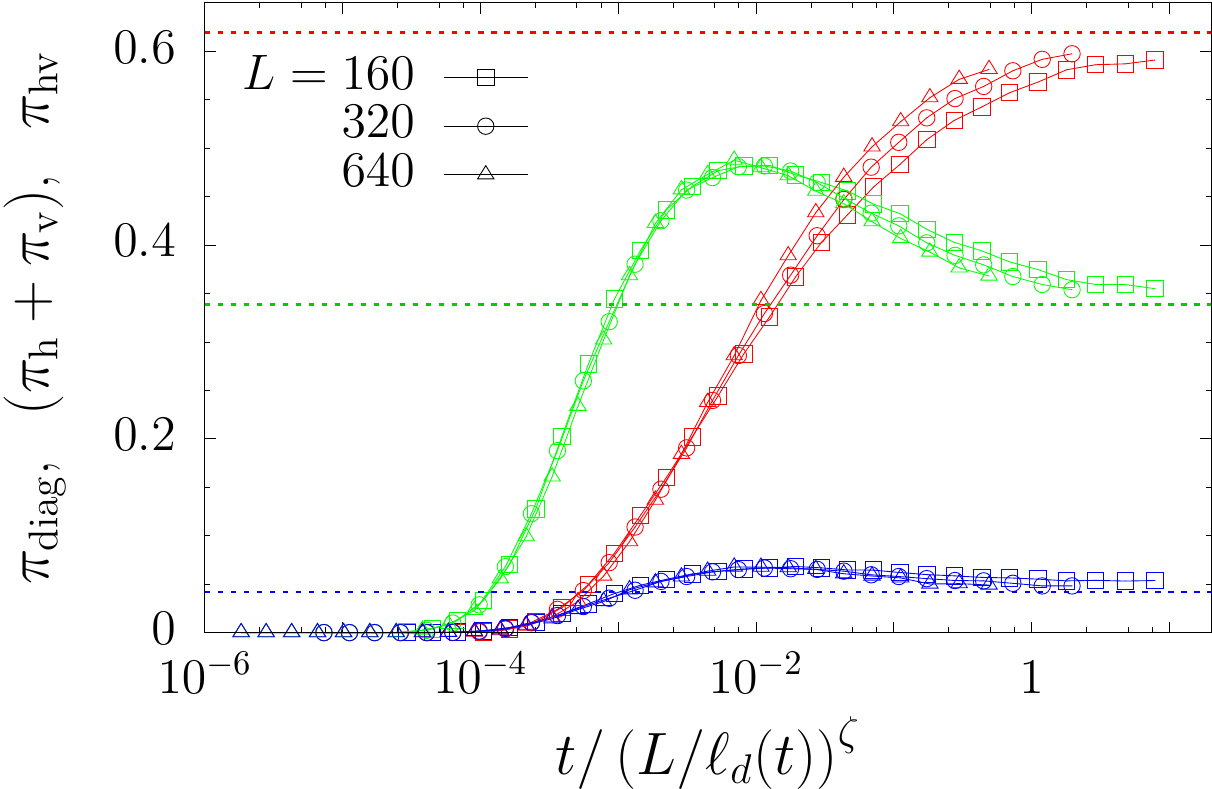}
        \quad%
        \includegraphics[scale=0.5]{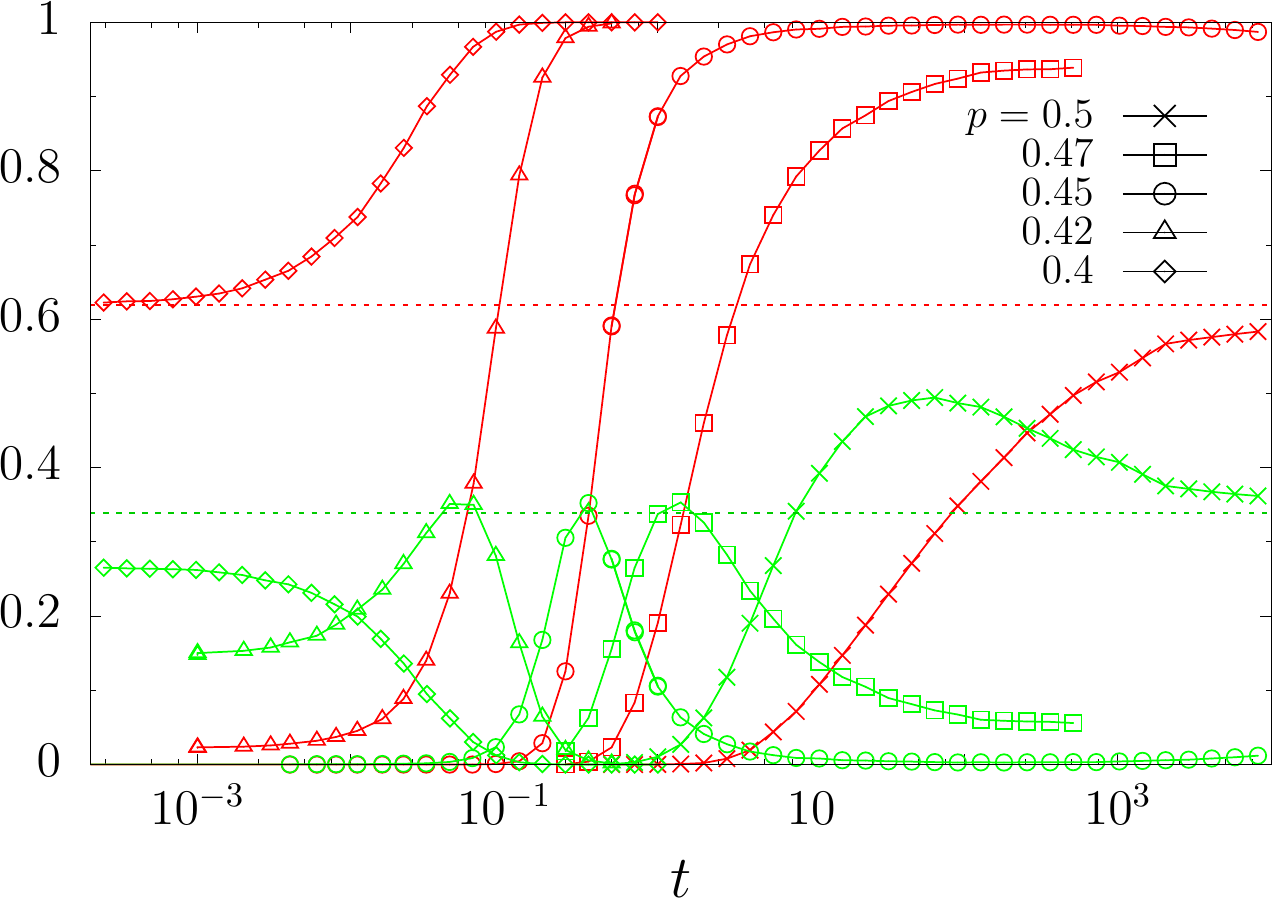}
\end{center}
\caption{\small Probability that a cluster wraps in a certain direction for the
local Kawasaki dynamics of the Ising model on a square lattice with PBC, quenched to $T_c/2$.
Red curves: clusters wrapping in both principal directions of the lattice ($\pi_{{\rm hv}}$).
Green curves: clusters wrapping in only one principal direction, either horizontally or vertically ($\pi_{{\rm h}} + \pi_{{\rm v}} $).
Blue curves: clusters wrapping in a diagonal direction ($\pi_{{\rm diag}}$).
The horizontal dotted lines are the exact values of the wrapping probabilities for critical percolation in $2d$.
In the left panel, data for equal concentration of up and down spins, and different values
of $L$, are plotted against the scaling variable 
$ t / \left( L/ \ell_d(t) \right)^{\zeta}$, with $\ell_d(t)=\ell_G(t)$ the characteristic length obtained 
from the inverse of the excess-energy.
The value of the exponent $\zeta\simeq 2.00$ is chosen to make the datasets corresponding to different $L$ collapse.
In the right panel, a system with linear size $L=160$  is
prepared initially at $t=0$ with a fixed concentration $p$ of $+1$ spin, for
$p=0.5$, $0.47$, $0.45$, $0.42$ and $0.4$, and it is then let evolve. 
}
\label{LKaWrapping_b2}
\end{figure}

What we should observe is that, at the time $t_p$, the time-dependent wrapping probabilities for the dynamical model
converge to the corresponding values of $2d$ critical percolation, and expect that they remain fixed at these values
for a very long period of time, more precisely, until the time at which the system fully equilibrate. 
In the thermodynamic limit, we expect this plateau to last indefinitely.
Our measurements suggest that this scenario is actually realised, see the data reported in Fig.~\ref{LKaWrapping_b2}.
Moreover, we should be able to perform a finite-size scaling by using the characteristic length $\ell_p(t)$, which was defined
as the length scale up to which the critical percolation properties can be observed, at time $t$.
We assume that, for $\ell_p(t) < L$, the $\pi$s depend on time $t$ and $L$, the system linear size, through the 
ratio $\ell_p(t)/L$,
\begin{equation}
 \pi_{\alpha}(t,L) \sim \tilde{\pi}_{\alpha} \left( \frac{\ell_p(t)}{L} \right) \; ,
 \label{eq:wrapping-prob-fss}
\end{equation}
with subscript $\alpha$ indicating the ``type'' of wrapping (horizontal, vertical, etc.) and $\tilde{\pi}_{\alpha}$ such that
$\tilde{\pi}_{\alpha}(x) \simeq \pi^{(p)}_{\alpha}$ for $x \gtrsim 1$, with $\pi^{(p)}_{\alpha}$ the corresponding wrapping probability in $2d$ critical percolation.
The time $t_p$ is such that $\ell_p(t) = L$ and it thus marks the onset of the critical-percolation-like scaling regime.
As stated in Sec.~\ref{sec:intro}, we assume that $\ell_p(t)$ has the form $\ell_p(t) = \ell_d(t) \ t^{1/\zeta}$, with $\zeta$ an exponent to be determined.

In the left panel of Fig.~\ref{LKaWrapping_b2}, 
we show the result of this scaling for
the wrapping probabilities in the case of the Kawasaki dynamics on a square lattice.
The data are plotted against the scaling variable $t / \left( L / \ell_d(t) \right)^{\zeta}$, 
which is completely equivalent to using the scaling variable $\ell_p(t) / L = ( \ell_d(t) \ t^{1/\zeta} ) / L$.
The value of the exponent $\zeta$ is chosen to make the datasets corresponding to different $L$ collapse on the same master curve.
The value that gives the best collapse is $\zeta \simeq 2.00$.
Indeed, we see that, for sufficiently large system sizes, the data
approach the exact values of critical percolation (represented by the horizontal dotted lines).
The scaling is rather good when the scaling variable is smaller than $10^{-2}$,
while finite size effects are more pronounced later. 
These results can be compared to the ones for NCOP dynamics shown in Ref.~\cite{BlCuPiTa-17}.

The time needed to approach the critical percolation wrapping probability values can be explicitly evaluated.
Concretely, from Fig.~\ref{LKaWrapping_b2} 
we can use $t_p / \left( L/\ell_d(t_p)\right)^{\zeta} = 1$
as a criterium to measure $t_p$. In this way we find
$t_p \simeq 233, \, 684, \, 1924, \, 5212, \, 13763$ for $L = 40, \, 80, \, 160, \, 320, \, 640$, respectively.

\begin{figure}[h!]
\vspace{0.5cm}
\begin{center}
        \includegraphics[scale=0.65]{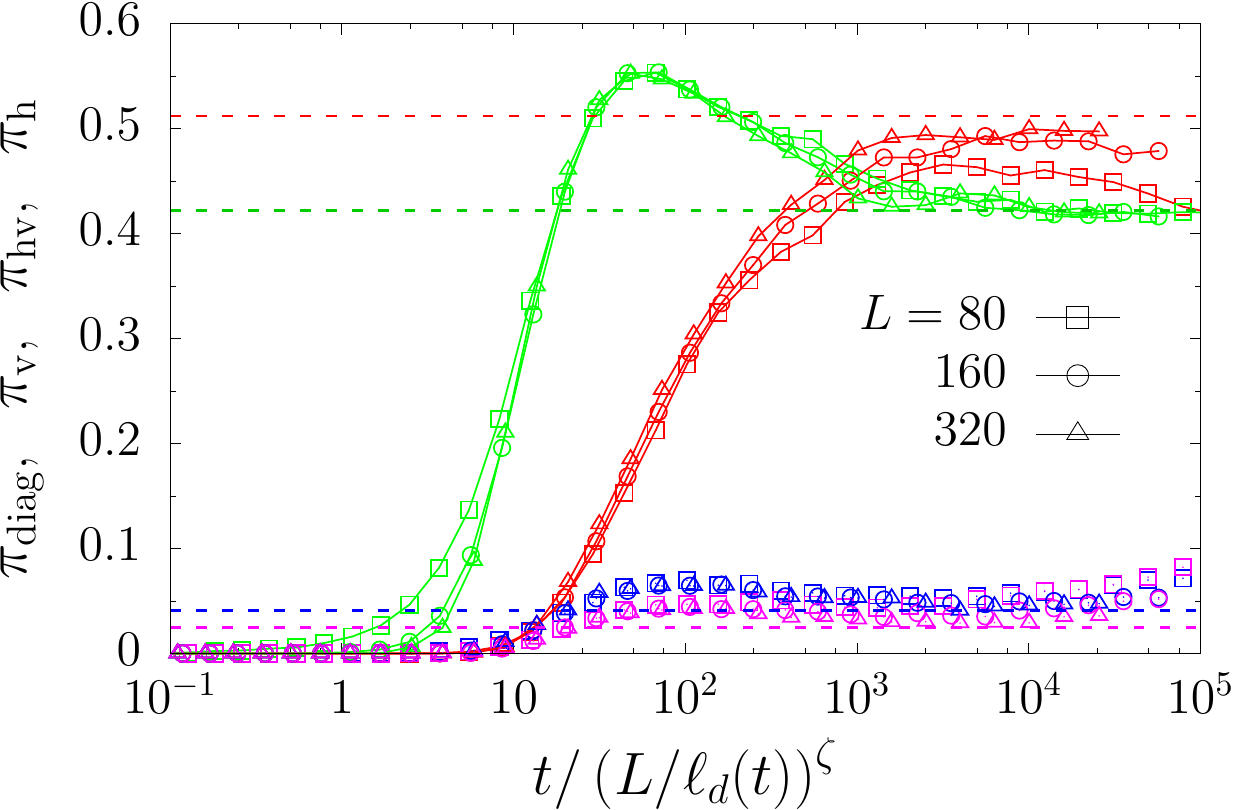}
\end{center}
\caption{\small
Local Kawasaki dynamics of the Ising model on a honeycomb lattice with PBC
and balanced densities of the two species, following a sudden quench to $T_c/2$, 
for different lattice linear sizes given in the key. We show the
probability that a cluster wraps in a certain direction at time $t$,
against the rescaled time
$t /\left( L /  \ell_d(t) \right)^{\zeta}$, where $\ell_d(t)=\ell_G(t)$ is the characteristic length obtained from the inverse excess-energy and the
value of the exponent $\zeta$ is chosen to make the data for different $L$ collapse one onto the other.
Red data points correspond to $\pi_{\mathrm{hv}}$, green ones to $\pi_{\mathrm{h}}$, blue ones to $\pi_{\mathrm{v}}$ and purple ones to $\pi_{\mathrm{diag}}$.
The value $\zeta \simeq 1.15$ yields approximately the best collapse.
The horizontal dashed lines correspond to the expected values at critical percolation
for a rectangular sheet of aspect ratio $\sqrt{3}$.}
\label{LKaWrapping_b2-Hon_new}
\end{figure}

In Fig.~\ref{LKaWrapping_b2-Hon_new} we show the same type of scaling on the honeycomb lattice.
In this case we plot the probabilities $\pi_{\rm h}$ and $\pi_{\rm v}$ separately, since 
the lattice that we use~\cite{BlCuPiTa-17} has aspect ratio different from $1$. 
In this case, the value of $\zeta$ that gives the best collapse is  $\zeta \simeq 1.15$,
but the finite-size scaling is not as good as on the square lattice.
We reckon that, in this case, we have simulated the dynamics on smaller systems, thus finite-size effects are more pronounced.

It is also interesting to check the influence of an unbalance between the densities of the two species and, in particular, 
to investigate whether clusters retain the critical percolation properties during a certain time regime 
when the initial concentration of one of the two species is close to the percolation threshold $p_c$.

In the right panel of 
Fig.~\ref{LKaWrapping_b2} we display the time evolution of the various cluster wrapping probabilities for different concentrations of up spins,
 $p=0.4, \ 0.42, \ 0.45, \ 0.47, \ 0.5$, in the case of the Kawasaki dynamics on a square lattice. We remind here that the site percolation threshold
for the square lattice is $p_c \simeq 0.5927 $.
As one can see, when the concentration of one of the two species is larger than $p_c$ (for the minority phase, this corresponds to the 
condition $p < 1 - p_c$),
the system has a cluster of the majority phase that percolates in both Cartesian directions already in the initial configuration,
see the data for $p=0.4$.
Instead, for $1-p_c < p < 1/2$, the curve $\pi_{\mathrm{hv}}(t)$ starts off from zero and slowly increases in a monotonic way,
approaching $1$ asymptotically.

For $p = 0.4 < 1-p_c$, the probability of having a cluster wrapping along only one principal direction of the lattice, $\pi_{\mathrm{h}} + \pi_{\mathrm{v}}$ (green curves), 
starts off close to the corresponding critical percolation value and then decreases rapidly to zero. A similar behaviour is observed for $\pi_{\mathrm{diag}}$.
In the cases in which $ 1 - p_c < p < 1/2$, the curves $\pi_{\mathrm{h}} + \pi_{\mathrm{v}}$ and $\pi_{\mathrm{diag}}$, 
start from values below the corresponding critical percolation probabilities,
increase and approach them at a certain instant that depends on $p$, but then rapidly detach from it and decrease to zero.

The curves for $p =  1/2$ (crosses) are the only ones that approach asymptotically the $2d$ critical percolation values
(shown with horizontal dotted lines). 
This result is consistent with the observation made in~\cite{Takeuchi15}, where the segregating dynamics of a mixture of Bose-Einstein 
condensates was studied, and similar percolation phenomena were observed.

\subsection{Average squared winding angle}
\label{subsec:LK-winding-angle}

As stated in Sec.~\ref{subsec:observables}, the measurements of the average squared winding angle, $\langle \theta^2 \rangle$, are very useful
to determine the type of criticality that the system is eventually approaching during its evolution,
{\it via} a fit of the data to Eq.~(\ref{eq:winding_angle_critical_hulls}) with $\kappa$ the relevant fitting parameter.
In our previous work~\cite{BlCuPiTa-17} we showed that $\langle \theta^2 \rangle$, measured
for the hulls of spin clusters produced by NCOP dynamics, approaches the form given by Eq.~(\ref{eq:winding_angle_critical_hulls})
with $\kappa \simeq 6$, the value that corresponds to $2d$ critical percolation.
In the following, we perform the same analysis for the Kawasaki dynamics.

\vspace{0.5cm}

\begin{figure}[h]
\begin{center}
        \includegraphics[scale=0.57]{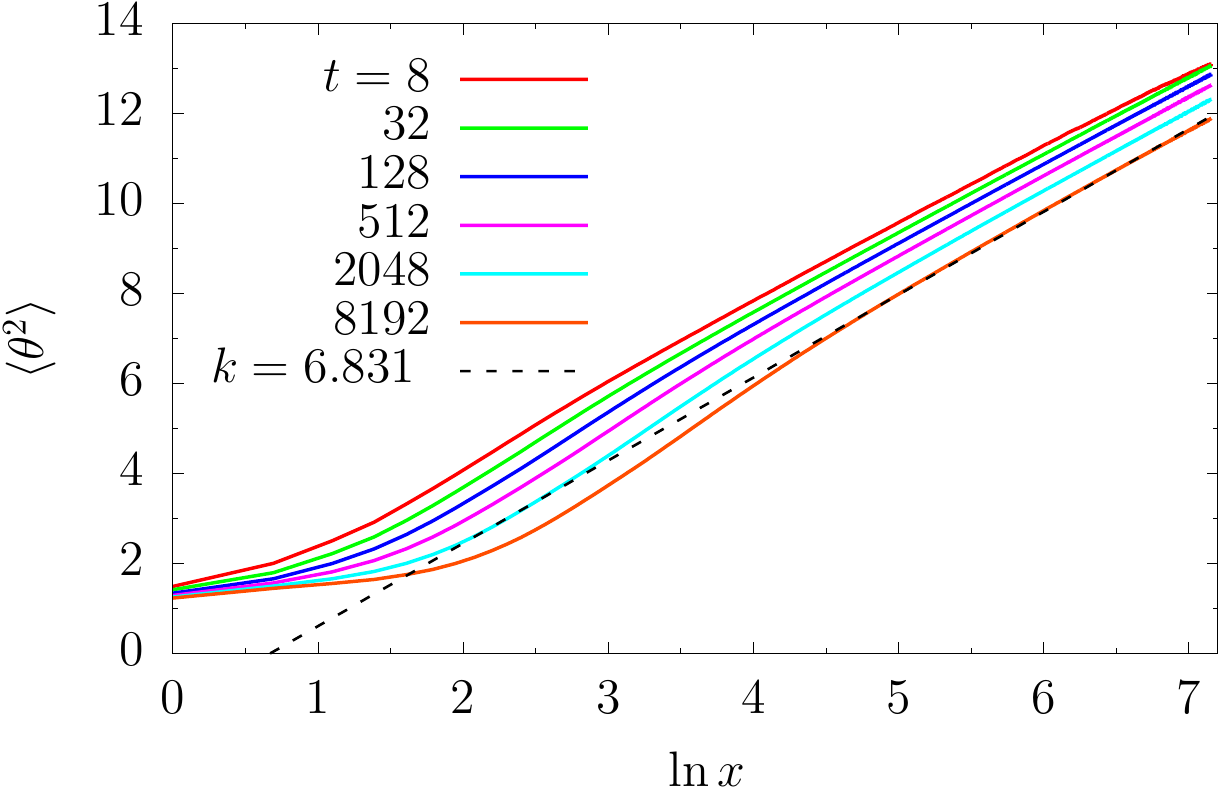}\quad%
        \includegraphics[scale=0.56]{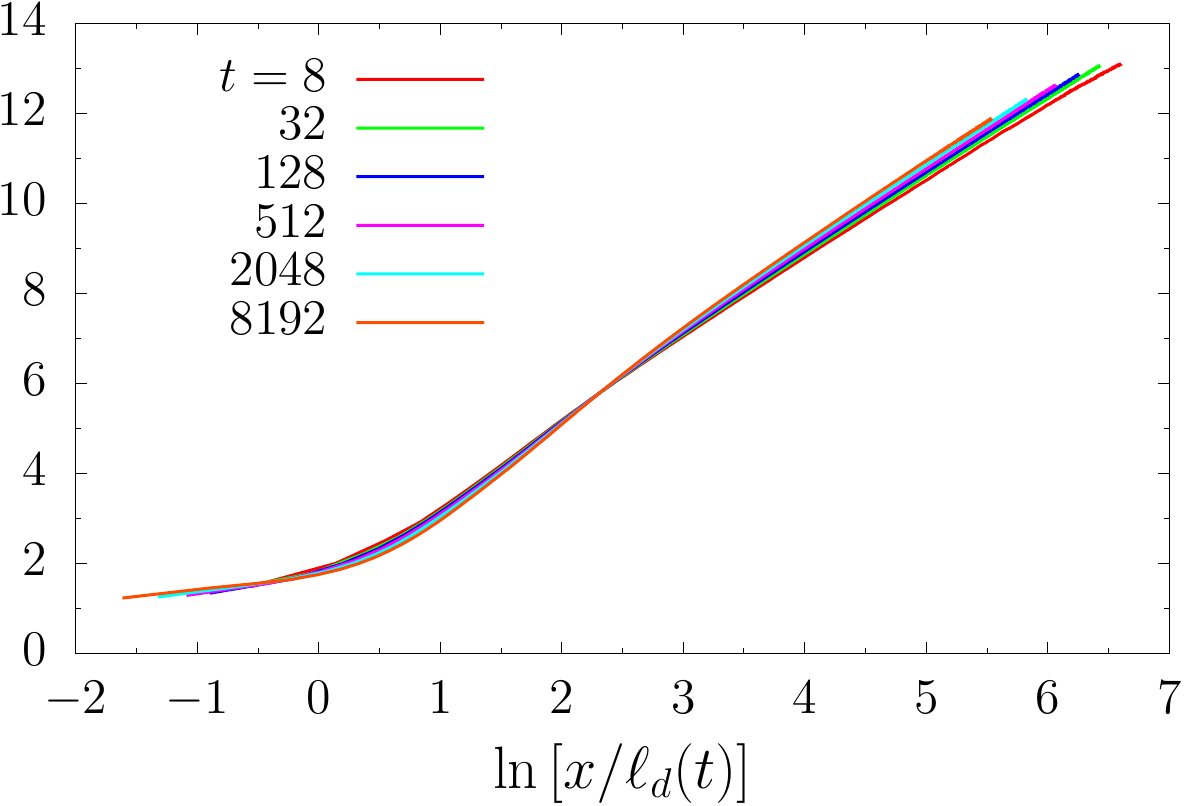}
\end{center}
\caption{\small 
Local Kawasaki dynamics of the Ising model on a square lattice with linear size $L=640$, 
following a quench to $T_c/2$. Equal concentration of up and down spins.
Left panel: the average squared winding angle $ \langle \theta^2 (x,t) \rangle $ for the
wrapping hulls that form the largest cluster interface against
the logarithm of the curvilinear coordinate $x$, at different times given in the key.
A linear fit to the data obtained for the latest time simulated ($t=8192$) is also shown in the left panel, yielding an 
SLE parameter $\kappa \simeq 6.8310$ (see Sec.~\ref{subsec:observables}).
Right panel: $ \langle \theta^2 (x,t)\rangle $ against the logarithm of $ x / \ell_d(t)$ with $\ell_d(t)=\ell_G(t)$ the growing 
length measured from the inverse of the excess energy.}
\label{LKaWA_b2}
\end{figure}

In Fig.~\ref{LKaWA_b2} we show the average squared winding angle for the hulls 
that form the interface of the largest spin cluster against the logarithm of the curvilinear coordinate $x$ (left panel) 
in the case of the Kawasaki dynamics on a square lattice with linear size $L=640$, at target temperature $T_c/2$. 
The interface of a cluster can be made of many components, as explained in Sec.~\ref{subsec:observables}, each one
being a closed path on the dual lattice constructed by joining sites (of the dual lattice) with bonds that intersect
``broken'' bonds between nearest-neighbour antiparallel spins on the original lattice.
To compute the quantity $ \langle \theta^2 \rangle $ shown in Fig.~\ref{LKaWA_b2} we considered only the hulls of the largest cluster
that are wrapping across the system, or said in another way, only those hulls that have zero total winding angle.
Of course, these hulls exist only if the largest cluster is wrapping and they come always in pairs.

We observe that the curves shown in the left panel of Fig.~\ref{LKaWA_b2} are very similar to the ones for 
NCOP dynamics (see~\cite{BlCuPiTa-17,BlCuPi12}). 
The dotted straight line is a fit to the logarithmic dependence of the data at $t=8192$ and yields 
an estimate of the SLE parameter $\kappa \simeq 6.8310$, relatively close to the one of critical 
percolation, $\kappa=6$.

In the right panel we show the same data plotted against
$\ln \left[ x/\ell_d(t) \right]$. The idea behind this scaling is that there is a separation of length scales characterised by
different fractal properties of the cluster hulls, with a crossover occurring at $\ell_d(t)$. If we measure the properties of the hulls
over a curvilinear distance $x$ smaller than $\ell_d(t)$ we are going to obtain the results of the equilibrium domain hulls, which are smooth.
Note that this statement is not exactly confirmed by our data since there is no sharp plateau for small values of $x$.
Nevertheless, $\langle \theta^2 \rangle$ seems to approach a plateau for $x \ll \ell_d(t)$ as $t$ increases.
Instead, for $x > \ell_d(t)$ the geometrical properties should be the ones of critical percolation
with $\langle \theta^2 \rangle$ given by Eq.~(\ref{eq:winding_angle_critical_hulls}).
This argument implies that, by rescaling the curvilinear distance $x$ by $\ell_d(t)$, data for different $t$ should collapse
on the same master curve.

It was not possible to collapse the data over the {\it whole range of times} available from the simulations by using the theoretical 
asymptotic power law behaviour of $\ell_d(t)$, $\ell_d(t) \propto t^{1/z_d}$, with a unique choice of 
the dynamical exponent $z_d=3$.
As already explained, in the range of times when the critical-percolation-like state is observed, the characteristic length scale
$\ell_d(t)$ has not yet acquired the asymptotic behaviour of the LSW theory for LCOP dynamics.
Thus, as done previously, $\ell_d(t)$ is taken to be the full time-dependent growing length obtained as the inverse of the 
excess energy, $\ell_G(t)$. As one can see from the right panel in Fig.~\ref{LKaWA_b2} the scaling thus achieved  
is very good.

On the other hand, we could estimate an effective growth exponent, or more precisely its inverse $z^{-1}_{\rm eff}(t)$,
by attempting pairwise collapse of curves $ \{ \langle \theta^2(x,t_i) \rangle \}_i $ corresponding to 
consecutive measuring times $t_{i}$ and $t_{i+1}$. 
We rescaled the distance $x$ by the factor $(t_{i+1}/t_i)^{\alpha}$ and we looked for the value of the exponent $\alpha$ that made the curve
$ \langle \theta^2(x,t_{i+1}) \rangle$ collapse onto the curve $ \langle \theta^2(x,t_{i}) \rangle$, when the former is plotted
against the rescaled distance. By performing this procedure for all $i$, we obtained an estimate $\alpha_i$ of the effective growth
exponent for each time interval $[t_i, t_{i+1}]$.
This estimate of  $z^{-1}_{\rm eff}(t)$ is included in the inset to Fig.~\ref{LKaGL} (as open green circles) 
and they show the same trend as the one extracted from $\ell_G(t)$.

\subsection{Largest cluster scaling}
\label{subsec:LC-Kawasaki}

\begin{figure}[h!]
\begin{center}
        \includegraphics[scale=0.55]{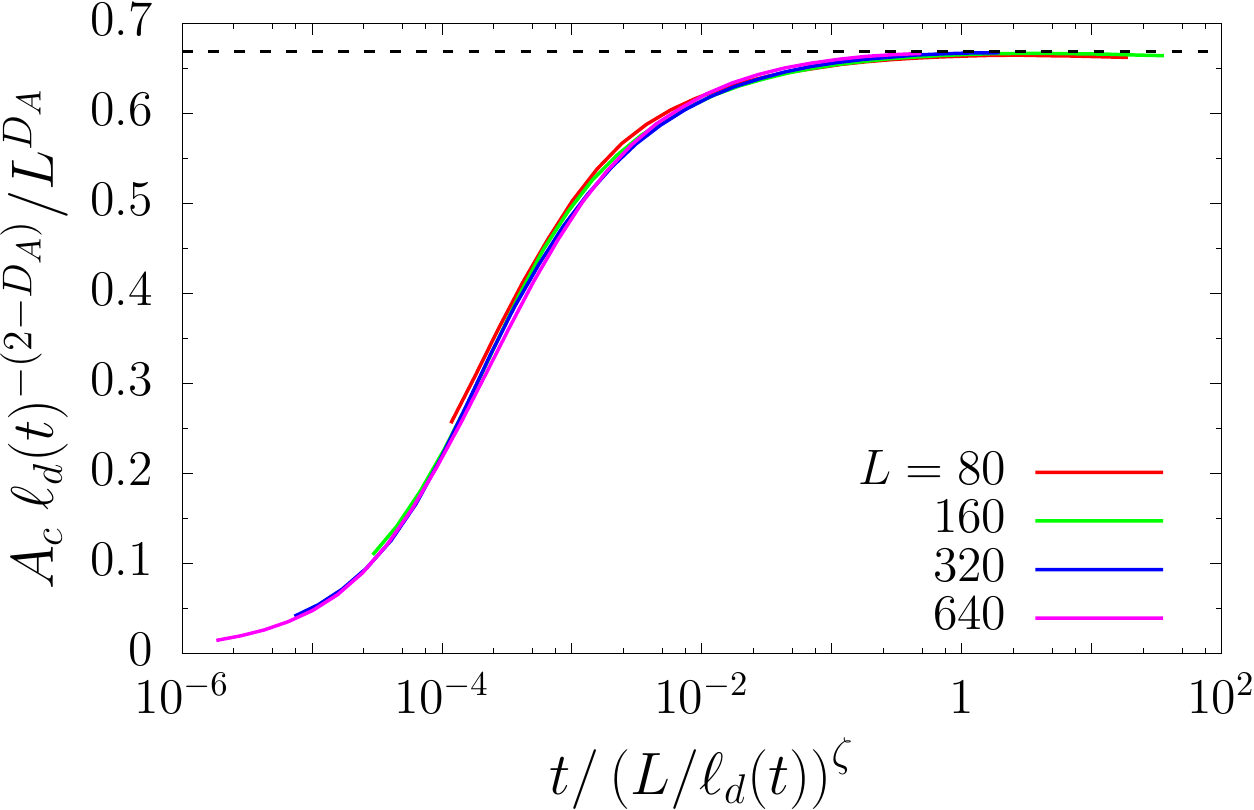}\quad%
        \includegraphics[scale=0.55]{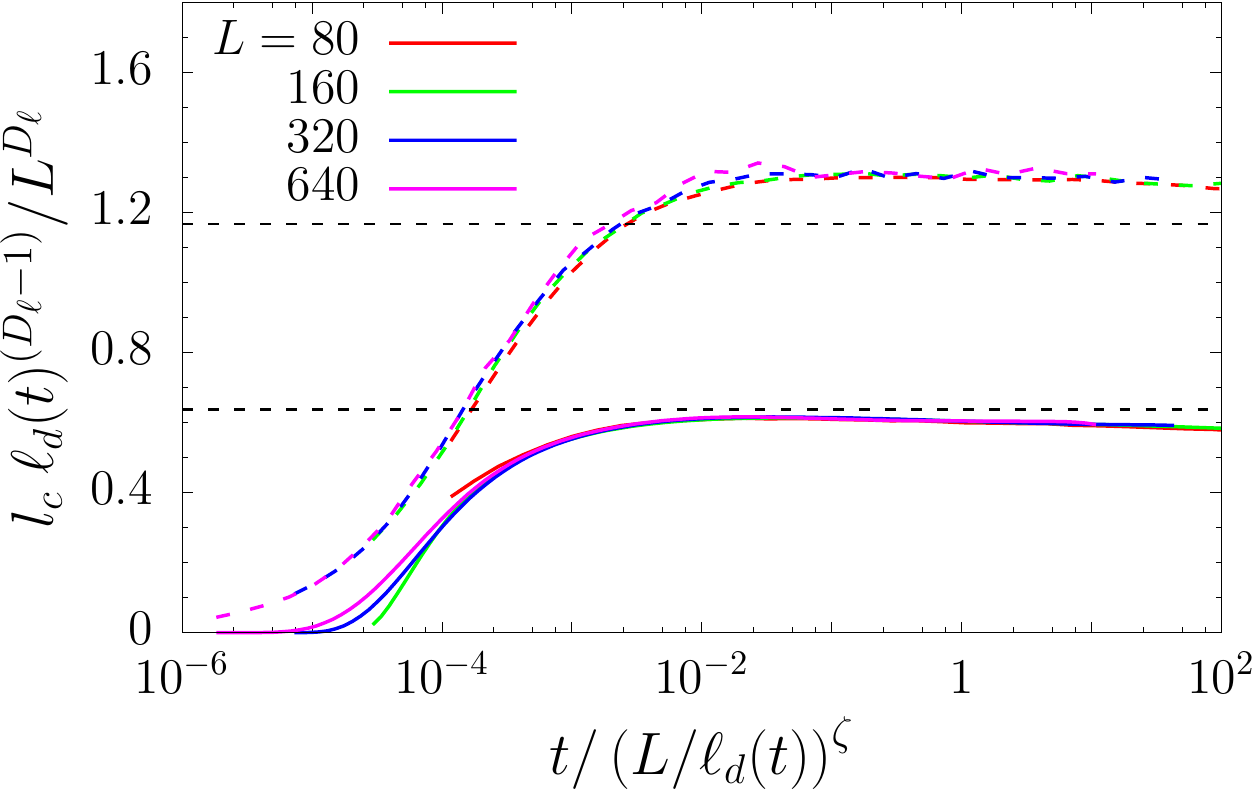}
\end{center}
\caption{\small
Local Kawasaki dynamics of the $2d$IM on the square lattice with PBC
and balanced densities of the two species, at temperature $T_c/2$, for different values of $L$, indicated in the key.
In the left panel we show $\left( A_c/L^{D_A} \right) \; \ell_d(t)^{-(2-D_A)}$, with $A_c$ the largest cluster size,
while in the right panel $\left( l_c/L^{D_{\ell}} \right) \; \ell_d(t)^{-(1-D_{\ell})}$, with
$l_c$ the length of the largest cluster interface. 
The interface length $l_c$ is computed separately for the non-wrapping hulls
(indicated by dashed lines) and the wrapping hulls (indicated by continuous lines),
if they exist. $\ell_d(t) = \ell_G(t)$ is the growing length extracted from the inverse of the excess-energy.
All quantities are plotted against the rescaled time $t / \left( L /  \ell_d(t) \right)^{\zeta}$, 
where the exponent $\zeta \simeq 2.00$ was chosen to make the datasets corresponding to different $L$ collapse on to each other.
The black dashed horizontal lines represent the value of the ratios $A_c/L^{D_A}$ (in the left panel) and 
$l_c/L^{D_{\ell}}$ (in the right panel) for critical site percolation on a square lattice (with PBC), computed by numerical simulations:
$A_c/L^{D_A} \simeq 0.6683$, while $l^{\mathrm{wrap}}_c/L^{D_{\ell}} \simeq 0.6383$ and $l^{\mathrm{non-wrap}}_c/L^{D_{\ell}} \simeq 1.1678$
for wrapping hulls and non-wrapping hulls, respectively.  
}
\label{LKaLC_b2_new}
\end{figure}

\begin{figure}[h]
\vspace{0.5cm}
\begin{center}
        \includegraphics[scale=0.55]{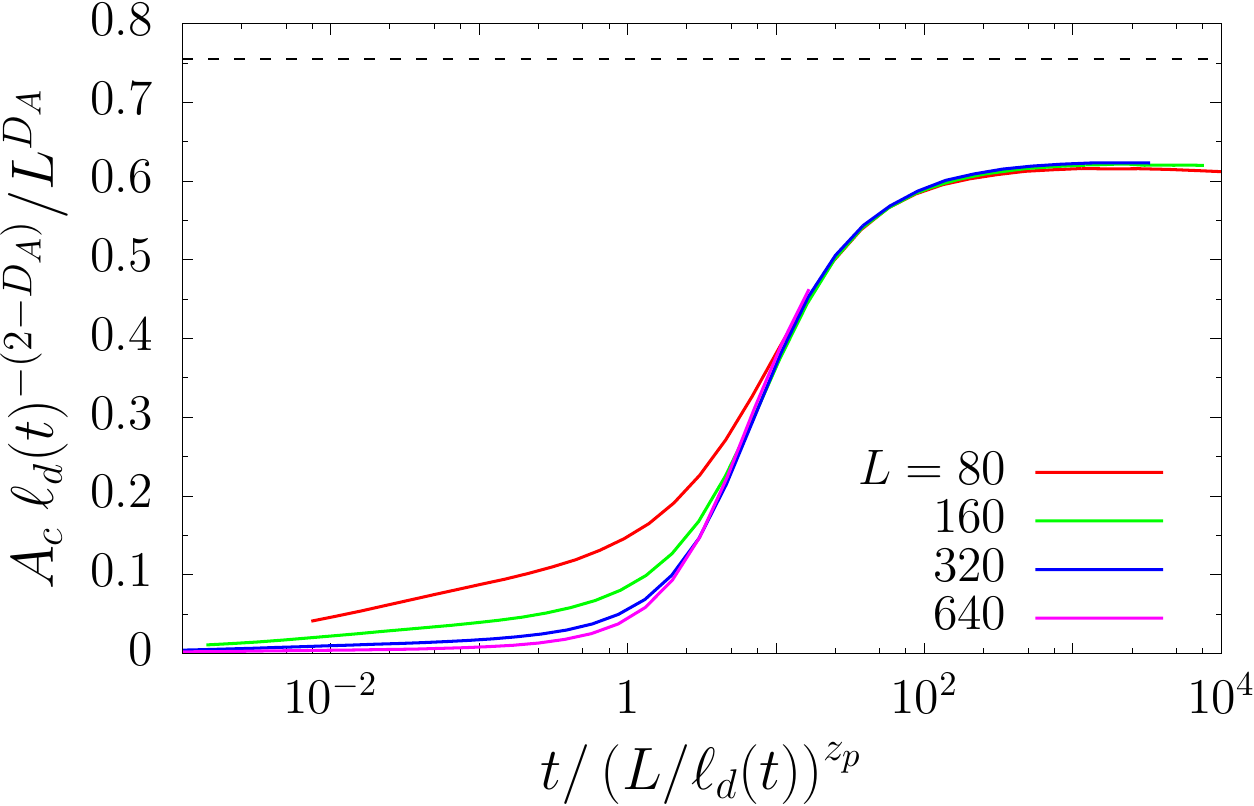}\quad%
        \includegraphics[scale=0.55]{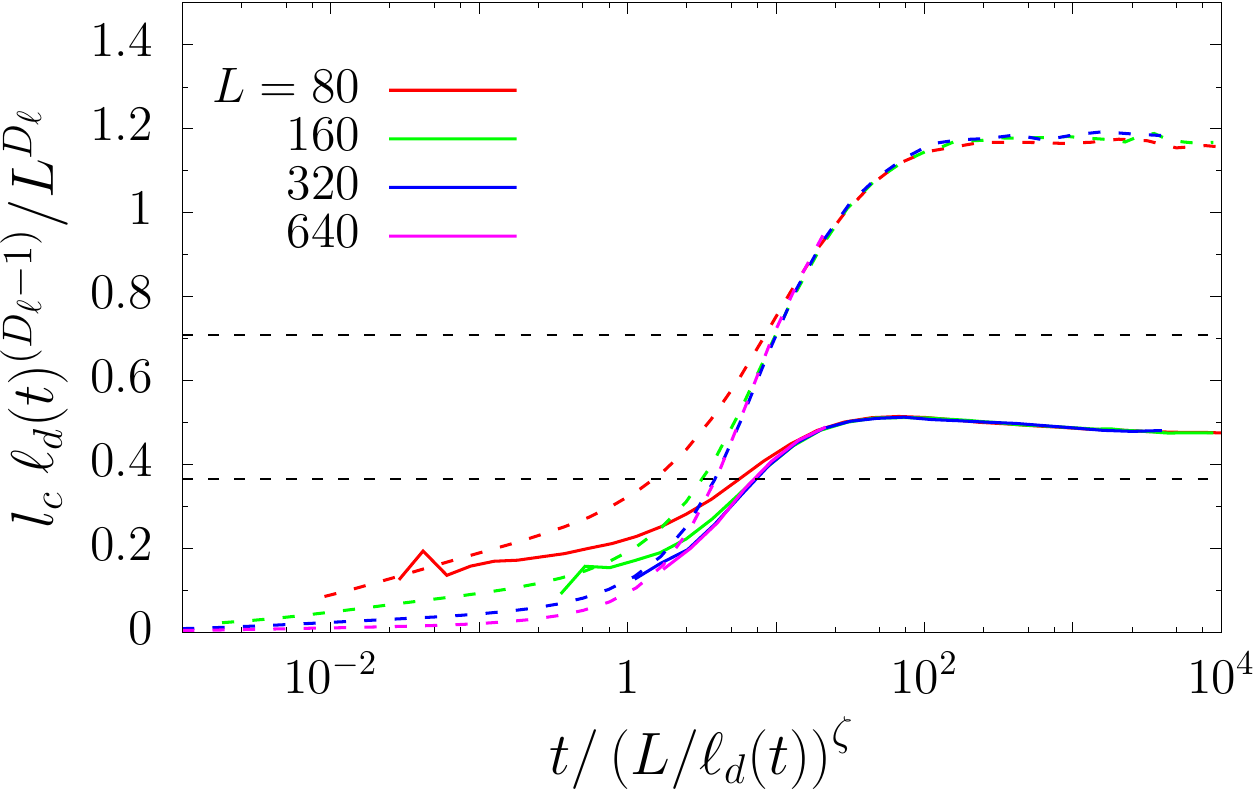}
\end{center}
\caption{\small
Local Kawasaki dynamics on the honeycomb lattice with PBC
and balanced densities of the two species, at temperature $T_c/2$, for different values of $L$, indicated in the key.
In the left panel we show $\left( A_c/L^{D_A} \right) \; \ell_d(t)^{-(2-D_A)}$,
while in the right panel $\left( l_c/L^{D_{\ell}} \right) \; \ell_d(t)^{-(1-D_{\ell})}$.
As in Fig.~\ref{LKaLC_b2_new}, the interface length $l_c$ is separated into two parts, 
one being the contribution coming from non-wrapping hulls (indicated by dashed lines), 
while the other one being the contribution due to wrapping hulls (indicated by continous lines), if they exist.
Here $\ell_d(t) = \ell_G(t)$ is the characteristic length obtained by the inverse of the excess-energy.
All the quantities are plotted against the rescaled time $t / \left( L /  \ell_d(t) \right)^{\zeta}$, 
where the exponent $\zeta$ was chosen to make the datasets corresponding to different $L$ collapse one onto the other.
The best collapse is obtained with $\zeta \simeq 1.15$, shown in the plots.
The black dashed horizontal lines represent the value of the ratios $A_c/L^{D_A}$ (in the left panel) and 
$l_c/L^{D_{\ell}}$ (in the right panel) for critical site percolation on a honeycomb lattice (with PBC), computed by numerical simulations:
$A_c/L^{D_A} \simeq 0.7554$, while $l^{\mathrm{wrap}}_c/L^{D_{\ell}} \simeq 0.3658$ and $l^{\mathrm{non-wrap}}_c/L^{D_{\ell}} \simeq 0.7090$
for wrapping hulls and non-wrapping hulls, respectively.
}
\label{LKaLC_b2-Hon_new}
\end{figure}

\begin{figure}[h!]
\vspace{0.5cm}
\begin{center}
        \includegraphics[scale=0.55]{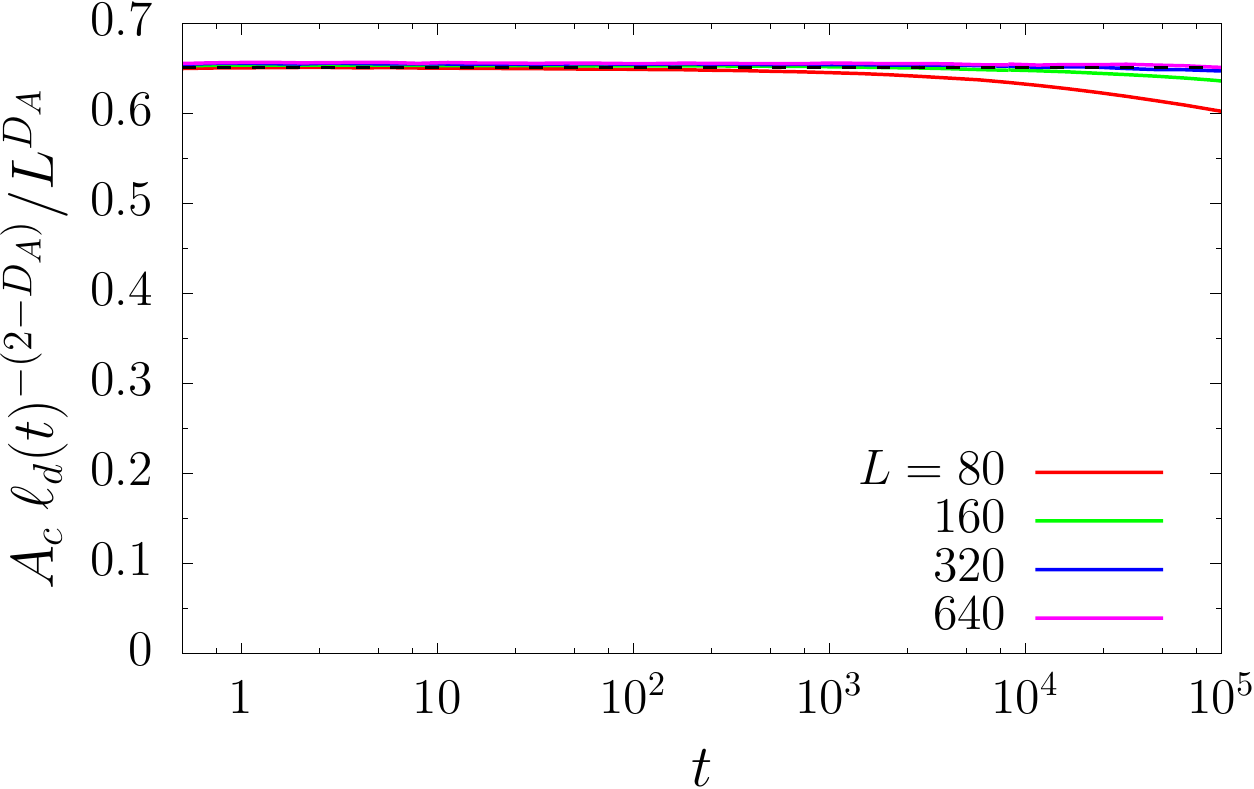}\quad%
        \includegraphics[scale=0.55]{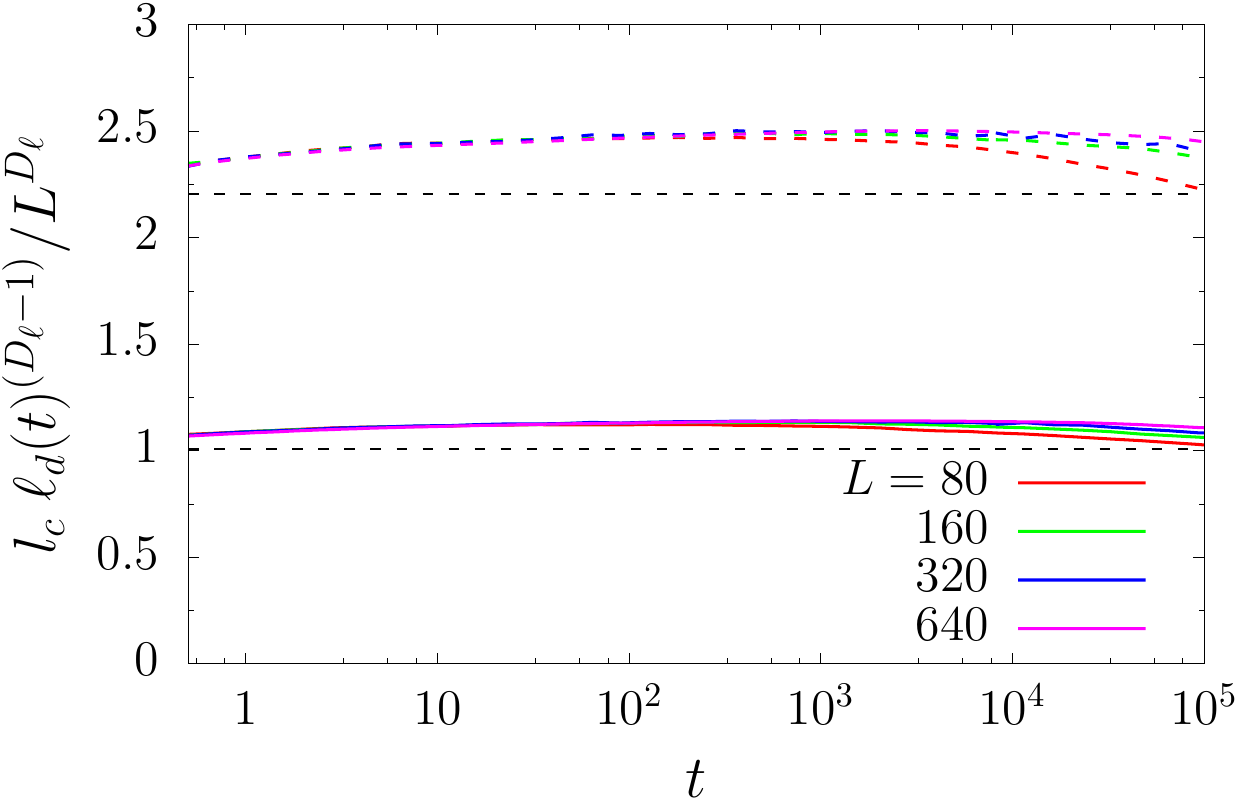}
\end{center}
\caption{\small
Local Kawasaki dynamics of the $2d$IM on the triangular lattice with PBC
and balanced densities of the two species, at $T_c/2$.
In the left panel we show $\left( A_c/L^{D_A} \right) \; \ell_d(t)^{-(2-D_A)}$,
while in the right panel $\left( l_c/L^{D_{\ell}} \right) \; \ell_d(t)^{-(1-D_{\ell})}$.
The interface length $l_c$ is separated into two parts, 
one being the contribution coming from non-wrapping hulls (indicated by dashed lines), 
while the other one being the contribution due to wrapping hulls (indicated by continous lines), if they exist.
Here $\ell_d(t) = \ell_G(t)$ is the characteristic length obtained by the inverse of the excess-energy.
All the quantities are plotted against time $t$.
The black dashed horizontal lines represent the value of the ratios $A_c/L^{D_A}$ (in the left panel) and 
$l_c/L^{D_{\ell}}$ (in the right panel) for critical site percolation on a triangular lattice (with PBC), computed with numerical simulations:
$A_c/L^{D_A} \simeq 0.6550$, while $l^{\mathrm{wrap}}_c/L^{D_{\ell}} \simeq 1.0075$ and $l^{\mathrm{non-wrap}}_c/L^{D_{\ell}} \simeq 2.2044$
for wrapping hulls and non-wrapping hulls, respectively. 
}
\label{LKaLC_b2-Tr_new}
\end{figure}

In this Section we report some results on the geometric and scaling properties of the largest cluster in the early time regime.
The data shown correspond to local Kawasaki dynamics at temperature $T_c/2$ on the square lattice (Fig.~\ref{LKaLC_b2_new}),
on the honeycomb lattice (Fig.~\ref{LKaLC_b2-Hon_new}) and on the triangular one (Fig.~\ref{LKaLC_b2-Tr_new}).
The quantity $A_c$ is the (averaged) size of the largest cluster, while $l_c$ is the (averaged) length of its interface.

We need to distinguish two classes of hulls.  A cluster that is wrapping across the system along only one direction (horizontal, vertical
or in a diagonal direction) always possesses two wrapping parts, while a cluster that is not wrapping or that wraps simultaneously in both
principal directions of the lattice does not possess any of these.
As already explained in Sec.~\ref{subsec:observables}, wrapping hulls are characterised by having zero total winding angle.

For all the cases that we show in this Section, we present the two contributions to the total length of the largest cluster interface, $l_c$,
the one coming from wrapping hulls and the one coming from non-wrapping hulls, as two separate quantities.
In general, they may have different scaling properties.

Following the same scaling arguments used in our previous work on the NCOP dynamics
of the ferromagnetic $2d$IM~\cite{BlCuPiTa-17},
the correct way of scaling the size of the largest cluster, $A_c(t,L)$, 
in order to take into account the effects of coarsening and percolation, is
\begin{equation}
  \frac{A_c(t,L)}{L^{D_A}}  \sim \, \ell_d(t)^{2-D_A} \, x_{A}\left( \frac{\ell_p(t)}{L} \right)
  \; , 
 \label{eq:Ac_scaling_2}
\end{equation}
with $D_A=91/48$ the fractal dimension of the largest cluster in $2d$ critical percolation,
$\ell_d(t)$ the characteristic length scale associated to the growth of domains, $\ell_p(t)$ the one
associated to the approach to the critical percolation state and $x_{A}$ an unknown function.
Analogously, the scaling for the largest cluster hulls is given by
\begin{equation}
   \frac{l_c(t,L)}{L^{D_{\ell}}}  \sim \, \ell_d(t)^{1-D_{\ell}} \, x_{\ell}\left(  \frac{\ell_p(t)}{L} \right)
 \label{eq:lc_scaling_2}
\end{equation}
with $D_{\ell}=7/4$ the fractal dimension of percolating domain hulls in $2d$ critical percolation
and $x_{\ell}$ an unknown function. As stated in Eq.~(\ref{eq:ell_p_Ka}), 
we suppose that $\ell_p(t)$ is given by $\ell_p(t) \simeq \ell_d(t) \, t^{1/\zeta}$, with $ \zeta $ an exponent to be determined.

The above relations are supposed to hold for sufficiently large system size (ideally in the limit $L \gg r_0$), for
$t\gg t_0$, with $t_0$ a microscopic time scale, and $t\ll t_{\mathrm{eq}}$,
with $t_{\mathrm{eq}}$ the time needed by the system to reach the equilibrium state imposed by the bath temperature.
Moreover, we expect that as $\ell_p(t) \rightarrow L$ the system reaches a critical-percolation-like state on the scale of $L$, and
for larger times both $x_{A}$ and $x_{\ell}$ converge to constants so that 
$ A_c(t,L)/L^{D_A}  \simeq \, C_A \, \ell_d(t)^{2-D_A} $ and $l_c(t,L)/L^{D_{\ell}}  \simeq \, C_{\ell} \, \ell_d(t)^{1-D_{\ell}}$,
with $C_A$ and $C_{\ell}$ some constants.

\vspace{0.5cm}

We find acceptable scaling of the quantities $ (A_c(t,L)/L^{D_A}) \, \ell_d(t)^{-(2-D_A)}$ and \\
$ (l_c(t,L)/L^{D_\ell}) \, \ell_d(t)^{-(1-D_{\ell})}$ 
as functions of the rescaled time $ t \left( L /  \ell_d(t) \right)^{\zeta} $ 
(equivalently, as functions of $\ell_p(t)/L$) where the value of the exponent $\zeta$
is again determined by looking for the best data collapse.
On the square lattice at temperature $T_c/2$,
the best collapse is found for $\zeta \simeq 2.00$ for both $A_c$ and $l_c$, see Fig.~\ref{LKaLC_b2_new}, 
as found for the scaling of the wrapping probabilities in Sec.~\ref{subsec:WP-Kawasaki}.
Note that we have also included
the numerical measurements of the ratios $A_c/L^{D_A}$ and $l_c/L^{D_{\ell}}$ for critical site percolation on the square lattice (with PBC).
As one can see, $ (A_c(t,L)/L^{D_A}) \, \ell_d(t)^{-(2-D_A)}$ is very close to the critical percolation value, $A_c/L^{D_A}\simeq 0.6683$.
For the rescaled $l_c(t,L)$ the agreement is not as good. 
The contribution from wrapping hulls reaches a plateau located slightly below 
the critical percolation value, $l^{\mathrm{wrap}}_c/L^{D_{\ell}} \simeq 0.6383$, while the contribution from 
non-wrapping hulls goes a little above the corresponding
critical percolation value, $l^{\mathrm{non-wrap}}_c/L^{D_{\ell}} \simeq 1.1678$.

The reason why the rescaled $l_c$ does not coincide with the measurement of the ratio
$l_c/L^{D_{\ell}}$ (for both wrapping and nonwrapping hulls) for critical percolation, at times when the LCOP dynamics
should have reached the so-called critical-percolation-like scaling regime, can be attributed to the presence
of a proportionality factor between the growing length $\ell_G$ computed as the inverse of the excess-energy
(which is used as estimate of $\ell_d$ here) and the true $\ell_d$, that is $\ell_d(t) \simeq \alpha \ell_G(t)$.
We expect this factor to be close to $1$ for the Kawasaki dynamics on the square lattice. However in the scaling
of $A_c$ and $l_c$ discussed above it appears as $\alpha^{2 - D_A}$ and $\alpha^{1 - D_{\ell}}$, respectively. 
In the case of the scaling of $A_c$ this factor does not have much affect since the exponent $D_A \simeq 1.8958$ is pretty close
to $2$, while it has in the case of the scaling of $l_c$ because the difference $D_{\ell} - 1$ is much larger in comparison.

In Fig.~\ref{LKaLC_b2-Hon_new} we show the same type of scaling plots for the dynamics on the honeycomb lattice and again, as for
the scaling of the wrapping probabilities, we find $\zeta \simeq 1.15$.
Notice, however, that the quality of the collapse is not as good as on the square lattice. More precisely, finite
size effects are more visible.
We also show the numerical values that $A_c/L^{D_A}$ and $l_c/L^{D_{\ell}}$ 
take at critical site percolation on this lattice with non-unit aspect ratio.
In this case there is no clear agreement between the values reached by the rescaled $A_c(t,L)$ and $l_c(t,L)$
for $ t \left( L /  \ell_d(t) \right)^{\zeta} \gg 1$ and the corresponding ones of critical percolation.
Also in this case the argument presented above explaining the discrepancy between the asymptotic values of the rescaled 
$A_c$ and $l_c$ and the measurements of the ratios $A_c/L^{D_A}$ and $l_c/L^{D_{\ell}}$ 
in critical percolation on the same lattice, applies. The factor of proportionality $\alpha$ between $\ell_d$ and $\ell_G$
may depend drastically on the particular lattice geometry considered and this can explain why, in the case of the dynamics
on the honeycomb lattice, the discrepancy is more pronounced than what we observe for the case of the square lattice.

A different discussion must be dedicated to the case of the triangular lattice.
In Fig.~\ref{LKaLC_b2-Tr_new} we present the rescaled largest cluster size (left panel) 
and length of its interface (right panel) against time $t$, for the Kawasaki dynamics on the triangular lattice
at temperature $T_c/2$, for different values of $L$.
As it was also observed for the NCOP dynamics on the triangular lattice~\cite{BlCuPiTa-17}, it is not possible to rescale
time as in \, $t \left( L /  \ell_d(t) \right)^{\zeta}$ with $\zeta>0$ in order to make the rescaled
data, $ (A_c(t,L)/L^{D_A}) \, \ell_d(t)^{-(2-D_A)}$ and $ (l_c(t,L)/L^{D_\ell}) \, \ell_d(t)^{-(1-D_{\ell})}$,
collapse. The reason for this is that the system is already at a critical percolation
state at $t=0$ (since the initial spin configuration is chosen to be fully uncorrelated and with concentration $p=0.5$ of the two species)
and there is no additional time scale to take into account other than the usual coarsening one, $\ell_d(t)$.
The rescaled largest cluster size, $ (A_c(t,L)/L^{D_A}) \, \ell_d(t)^{-(2-D_A)}$, coincides with the ratio $A_c/L^{D_A}$
of critical site percolation on the triangular lattice, namely $A_c/L^{D_A}\simeq 0.6550$, at $t=0$ (outside of the plot because of the logarithmic scale) 
and remains close to this value for a very long period of time 
with a very slow decrease, which is more visible for the smaller lattice sizes. 
In the case of the hull length $l_c$, both two contributions scale exactly as $L^{D_{\ell}}$ at $t = 0$
(where $\ell_d = 1$) and $ l_c(0,L)/L^{D_\ell} \simeq 1.0075$ for the wrapping hulls, while $ l_c(0,L)/L^{D_\ell} \simeq 2.2044$ for the non-wrapping ones
(again, not visible in the plot because of the logarithmic scale in the time axis).
Notice that, however, very early in the dynamics, the rescaled hull length, $ l_c(t,L)/L^{D_\ell}$, increases from the initial critical percolation value
and, for large values of $L$, it seems to reach a plateau.

\subsection{Pair connectedness function}
\label{subsec:pair-connectedness_LK}

We show here some results regarding the time evolution of the pair connectedness function for the
Kawasaki dynamics following a quench to $T_c/2$.
The definition of the pair connectedness $g(r)$ for a spin system is recalled in Sec.~\ref{subsec:observables}.

\begin{figure}[h]
\vspace{0.5cm}
\begin{center}
        \includegraphics[scale=0.55]{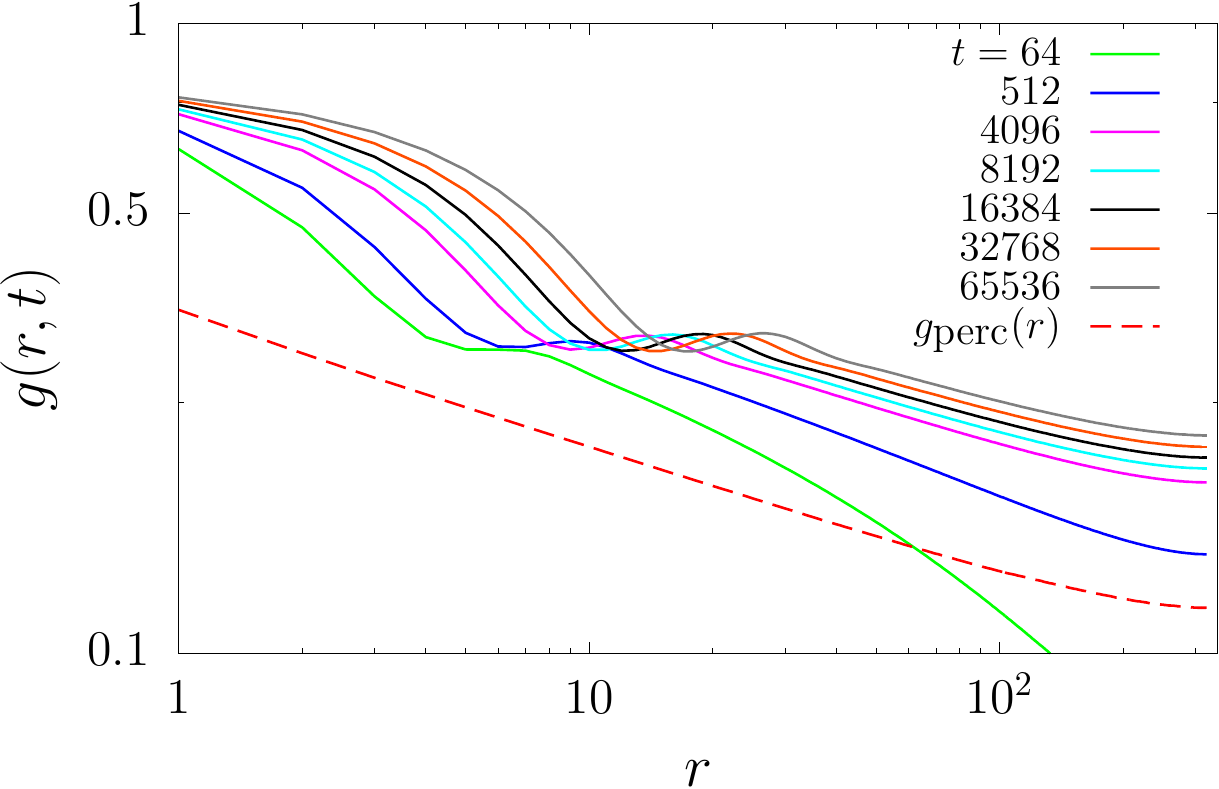}\quad%
        \includegraphics[scale=0.55]{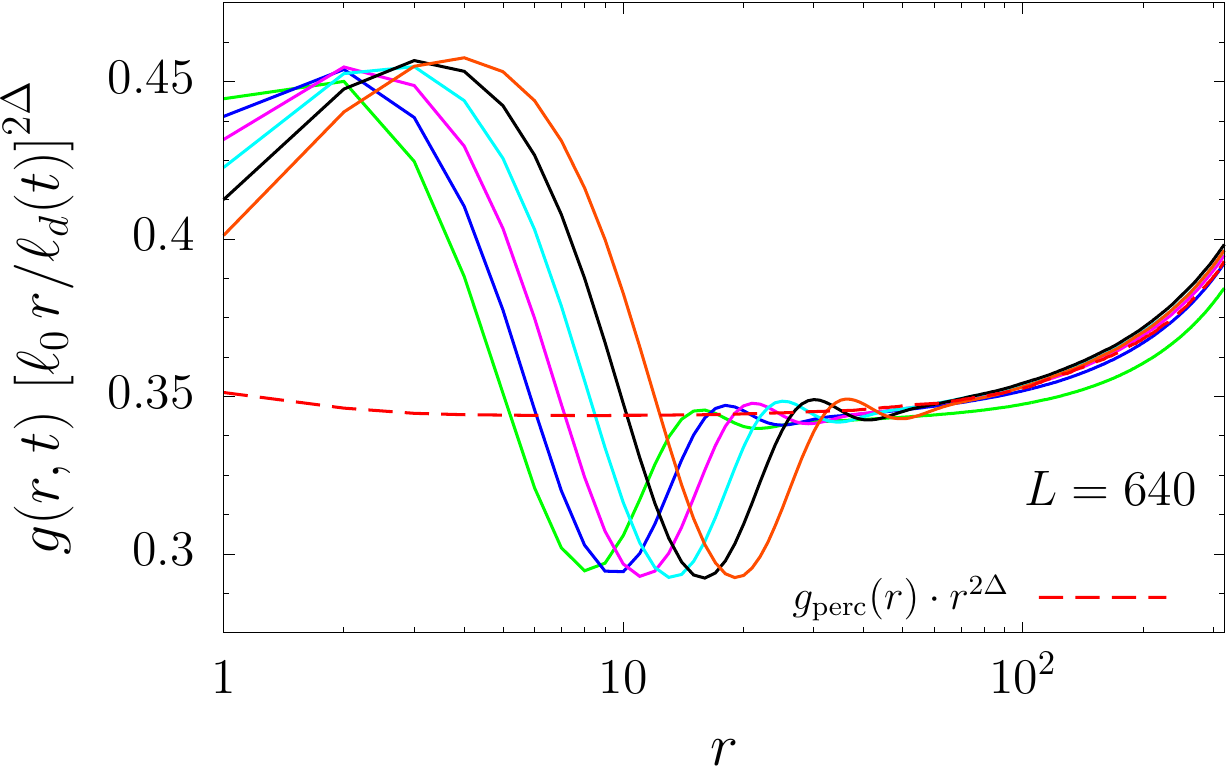}%
\end{center}
\caption{ \small 
Local Kawasaki dynamics on a square lattice with linear size $L=640$, 
at temperature $T_c/2$, with equal concentration of up and down spins. 
Comparison between the time evolving pair connectedness function and the one for critical site percolation 
shown with a dashed line. 
In the left panel we show $g(r,t)$, while in the right panel we plot 
$ g(r,t) \,  \left[ \ell_0 \, r/ \ell_d(t) \right]^{2 \Delta } $,
where $\Delta =2-D_A$, $\ell_d(t)=\ell_G(t)$ is the characteristic length obtained by the inverse of the excess-energy and
$\ell_0$ is an adjustable parameter, both against distance $r$, for different times indicated in the key.
The color code for the different times is the same in both panels. 
The value of the constant $\ell_0$ was chosen so that 
the data relative to the dynamical problem collapsed onto the data for critical percolation
in the region of the tail.
}
\label{LKaPC_b2}
\end{figure}

In the left panel in Fig.~\ref{LKaPC_b2}, 
we show $g(r,t)$ against distance $r$ at different times $t$ 
for the dynamics on a square lattice with $L=640$ and PBC. 
We compare it to the one of random site percolation, $g_{\mathrm{perc}}(r)$, on a square lattice with the 
same linear size and boundary conditions,
at the site occupation probability $p=0.5927$ that is approximately the threshold value $p_c$ (red dashed curve).
Notice that, because of the PBC, the maximum allowed value of $r$ is $L/2$.
One can clearly see that, for $t > 4096$, the long-distance behaviour (approximately, for $r \gtrsim 20$) of $g(r,t)$ is the same as the one
of the critical percolation pair-connectedness, $g_{\rm perc}$, apart from some time-dependent scaling factor.
This observation suggests that, in this range of times, the system is in the critical-percolation-like scaling regime. Accordingly, we expect
$g(r,t)$ to satisfy
\begin{equation}
 g(r,t) \simeq \mathrm{const.} \times \left( \frac{r}{\ell_d(t)} \right)^{- 2\Delta}
 \label{eq:pair-connectedness-scaling}
\end{equation}
for $r \gg \ell_d(t)$, with $\Delta = 2 - D_A$, $D_A = 91/48$ being the fractal dimension of the incipient percolating cluster in $2d$ critical percolation.

To test this claim, we plotted the quantity $ g(r,t) \,  \left[ \ell_0 \, r/ \ell_d(t) \right]^{2 \Delta} $ against $r$,
where $\ell_d(t) = \ell_G(t)$, in the right panel of Fig.~\ref{LKaPC_b2}.
The parameter $\ell_0$ was chosen so that the data relative to the dynamical problem collapsed onto the one for critical percolation.
This costant scaling factor is necessary because of the different normalization between $g$ and $g_{\rm perc}$.
The best collapse was achieved by using $\ell_0 \simeq 0.16$.

\begin{figure}[h]
\vspace{0.5cm}
\begin{center}
        \includegraphics[scale=0.6]{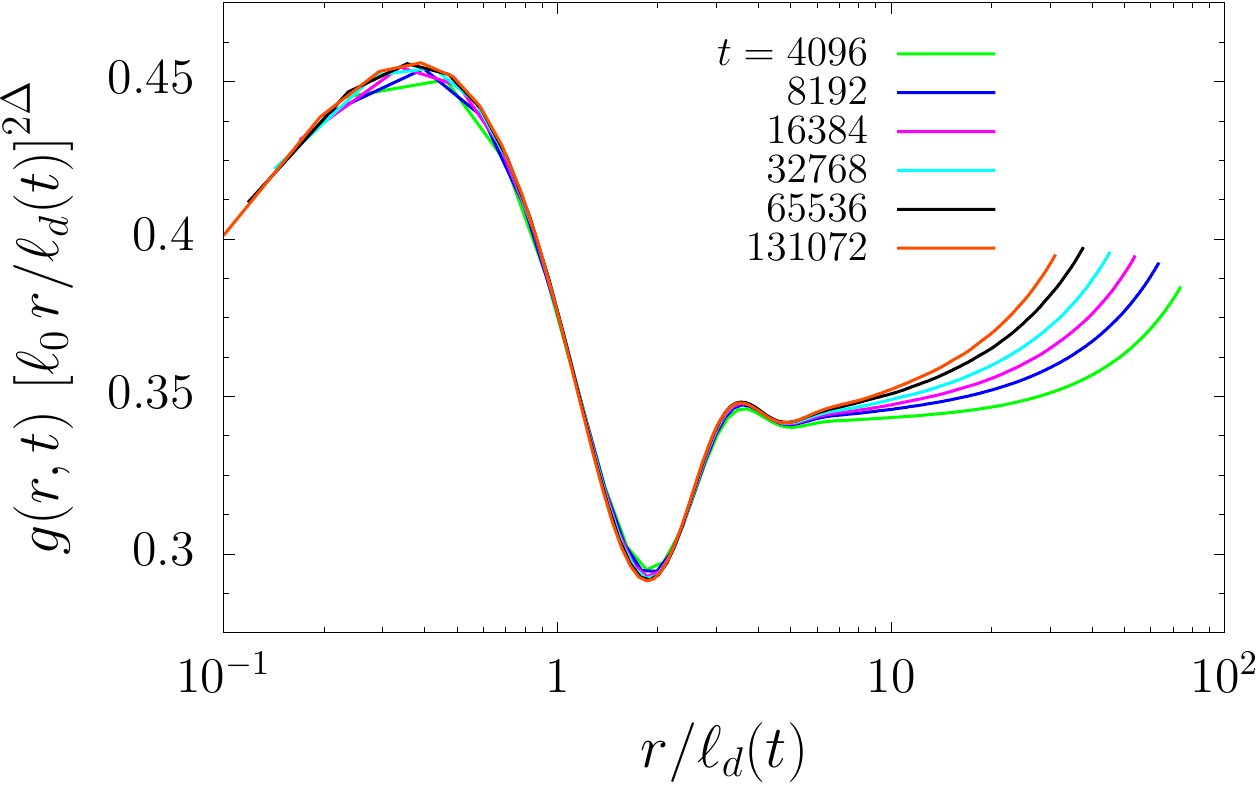}%
\end{center}
\caption{ \small 
Local Kawasaki dynamics on a square lattice with linear size $L=640$, at temperature $T_c/2$,
with equal concentration of up and down spins.  
In the vertical axis $ g(r,t) \,  \left[ \ell_0 \, r/ \ell_d(t) \right]^{2 \Delta} $,
where $\Delta = 2-D_A$, $\ell_d(t)=\ell_G(t)$ is the characteristic length derived from the inverse of the excess-energy 
and $\ell_0$ takes the same value as in Fig.~\ref{LKaPC_b2}. In the horizontal axis, 
the rescaled distance $r/\ell_d(t)$.}
\label{LKaPC_b2-2}
\end{figure}

The data relative to the dynamics collapse for sufficiently long distances.
Due to the PBC and finite system size, the rescaled dynamical and critical percolation pair-connectedness
do not approach a constant value at large values of  $r$.
The behaviour of $g(r,t)$ for small values of $r$ is a characteristic of the coarsening process induced by the microscopic dynamics
and signals the fact that at small length scales the structure of clusters is very different from the one of critical percolation. 
The same separation of length scales that was observed for the scaling behaviour of the average squared winding angle,
$\langle \theta^2 \rangle $, is also playing a role in the case of the pair connectedness function. 
To make it more evident, in Fig.~\ref{LKaPC_b2-2}, we plot 
the rescaled pair connectedness, $ g(r,t) \,  \left[ \ell_0 \, r/ \ell_d(t) \right]^{2 \Delta_\sigma} $, against
the rescaled distance $r/\ell_d(t)$. As one can see, the short-distance part of the rescaled pair 
connectedness ($r/\ell_d(t) \lesssim 1$) is
collapsing on the same master curve for different values of $t$.

\subsection{Number density of domain areas}
\label{subsec:num_dens_dom_Ka}

Following the same arguments exposed in our previous work on the NCOP dynamics~\cite{BlCuPiTa-17}, we assume that 
the number density $N$ of non-percolating spin clusters depends on $A$ and $t$ as
\begin{equation}
 N(A,t) 
  \simeq  
2 c_d^{\rm eff}(t) \;
\frac{ A^{1/2} }{ \left[ A^{3/2} +\ell^{3}_d(t) \right]^{ \frac{(2\tau_A + 1)}{3} }  }  
\; \Phi \left( \frac{A/\ell_d^{2-D_A}(t)}{\ell^{D_A}_p(t)} \right)
\label{eq:NA_COP_1}
\end{equation}
where we defined an effective normalisation constant
\begin{equation}
c_d^{\rm eff}(t)\equiv 2c_d \; [ \, \ell_d(t) \, ]^{2(\tau_A-2)}
\; . 
\end{equation}
This is the same form given by Eq.~(\ref{eq:NA_COP_dynamics}) with the addition of a pre-percolation scaling factor
$\Phi$ that depends on $A$ and $t$ through
\begin{equation}
\frac{A/\ell_d^2(t)}
{\left(\ell_p(t)/\ell_d(t)\right)^{D_A}} 
\end{equation}
with $\ell_p(t)$ the characteristic length which governs the regime of approach to the critical-percolation-like state.

The idea behind this scaling is that, at time $t$, on linear scales larger than $\ell_p(t)$, the system has not fully reached the critical percolation state yet, that
is, domains with size $A$ much larger than $\ell^{D_A}_p(t)$ have still some features of the initial fully disordered state. In this sense, 
$\ell_p(t)$ serves as a crossover scale between the percolation criticality and the high temperature disorder. Note that, in terms of areas, the corresponding
crossover scale is $\ell^{D_A}_p(t)$. Moreover, 
both $A$ and $\ell_p(t)$ must be rescaled by $\ell_d(t)$ with the corresponding scaling dimensions to take into account
the effects of usual coarsening (see Ref.~\cite{BlCuPiTa-17} for more details).

Finally, we expect the pre-percolation scaling function $\Phi$ to satisfy $\Phi(x) \rightarrow 1$ as $x\rightarrow 1$ so that we 
recover the expression given by
Eq.~(\ref{eq:NA_COP_dynamics}) when $A/\ell_d^2(t) \ll \left(\ell_p(t)/\ell_d(t)\right)^{D_A}  $, {\it i.e.},  for scales $A$ such that 
the criticality of percolation has already set in
at time $t$.

\subsubsection{Triangular lattice.}

\begin{figure}[h!]
\begin{center}
	  \includegraphics[scale=0.7]{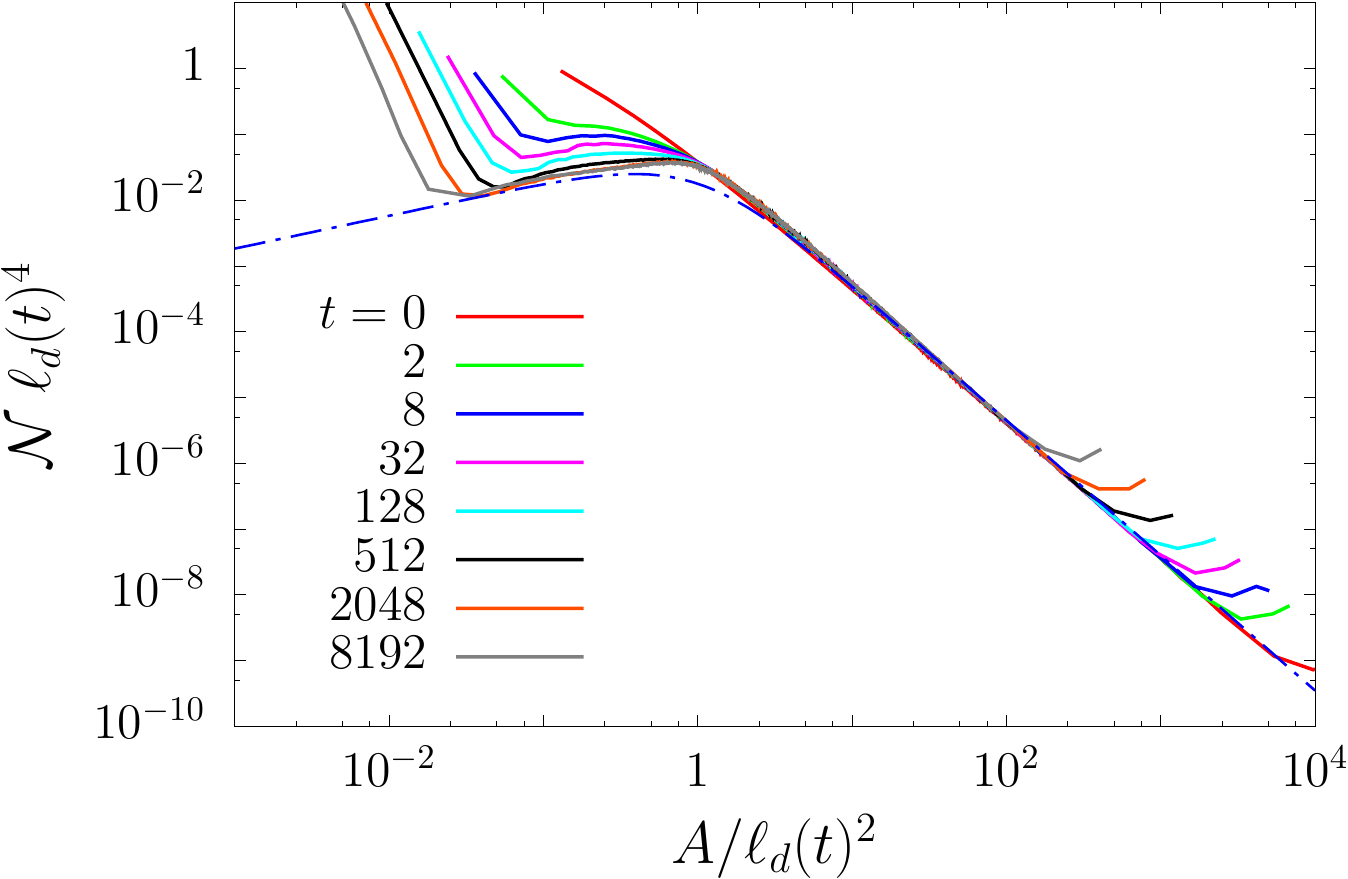}
\end{center}
\caption{\small Local Kawasaki dynamics of the Ising model
on a triangular lattice with linear size $L=640$, quenched to $T_c/2$, with equal concentration
of the two species. We show $\mathcal{N}(A,t) \, \ell^4_d(t)$ against the rescaled area $A/\ell^2_d(t)$, with
$\ell_d(t) = \alpha \ell_G(t)$ the characteristic length associated to the coarsening process, and $\alpha = 2.78$.
The value of $\alpha$ was chosen so that the datasets collapse onto the master curve
$f(x) = 2 \, c_d \, x^{1/2} \, \left( 1 +x^{3/2} \right)^{-(2 \, \tau_A+1)/3}$, represented by the blue dashed line.
}
\label{LKaNA-NWR_b2-Tr}
\end{figure}

On the triangular lattice the pre-percolating regime is absent since the initial spin configuration is already a
realisation of critical percolation, and Eq.~(\ref{eq:NA_COP_1}) should hold without the pre-percolation factor $\Phi$.
To check the scaling with $\ell_d(t)$, in Fig.~\ref{LKaNA-NWR_b2-Tr} we plotted the rescaled cluster size distribution
$\mathcal{N}(A,t) \, \ell^4_d(t)$ against the rescaled area $A/\ell^2_d(t)$ for a system of linear size 
$L=640$ evolving at $T=T_c/2$.
We take $\ell_d(t)$  to be proportional to the inverse of the excess-energy, 
$\ell_d(t) = \alpha \ell_G(t)$, with the proportionality constant $\alpha = 2.78$. This value is approximately
the one that gives us the best data collapse.

The master curve $f(x) = 2 \, c_d \, x^{1/2} \, \left( 1 + x^{3/2} \right)^{-(2\, \tau_A + 1)/3}$, 
the scaling function for the LCOP domain growth (see Eq.~(\ref{eq:NA_scaling_functions})), is shown with blue discontinuous line.
Deviations from the master curve are expected at very large values of the scaling variable $A/\ell^2_d(t)$,
that is to say, for domains with linear size comparable to $L$.
Deviations are also expected at small values of $A/\ell^2_d(t)$ because of the discreteness of the lattice.
The master curve from the data collapse slightly differs from the analytic form around the ``shoulder'', the point at which there is the 
crossover between the $\sqrt{A}$ behaviour for small domains and the power law
decay $A^{-(2\, \tau_A + 1)/3}$ for large domains.

\vspace{0.5cm}

\subsubsection{Pre-percolation scaling.}
\label{subsubsec:Ka_prepercolating}

As we did for the scaling of the largest cluster size and interface length and for
the wrapping probabilities, we assume that in the regime of approach to critical percolation
the relevant length scale is $\ell_p(t) \propto \ell_d(t) \, t^{1 / \zeta}$.

Neglecting for the moment the contribution of the large percolating clusters, 
the number density of domain areas has the following scaling behaviour,
\begin{eqnarray}
\frac{A^{\tau_A} \,  {\mathcal N}(A,t,L)}
{2 c^{\rm eff}_d(t)} \, 
\simeq 
\left\{
\begin{array}{ll}
\!\!
\displaystyle{\left(\frac{A}{\ell^2_d(t)}\right)^{\frac{2\tau_A+1}{2} } }
\;\;\;\;\;\;\;\;
& 
A \ll \ell^{2}_d(t) 
\vspace{0.25cm}
\\
\displaystyle{1} 
\;\; \;\;\;\;\;\;
&  
\ell^{2}_d(t) \ll A \ll \ell^{D_A}_p(t)
\vspace{0.25cm}
\\ 
\displaystyle{ \Phi \left( \frac{A/\ell^{2-D_A}_d(t)}{\ell^{D_A}_p(t)} \right) } 
&
 A \gtrsim \ell^{D_A}_p(t)
\end{array}
\right.
\label{eq:NA_COP_2}
\end{eqnarray}%

\begin{figure}[h!]
\begin{center}
        \includegraphics[scale=0.57]{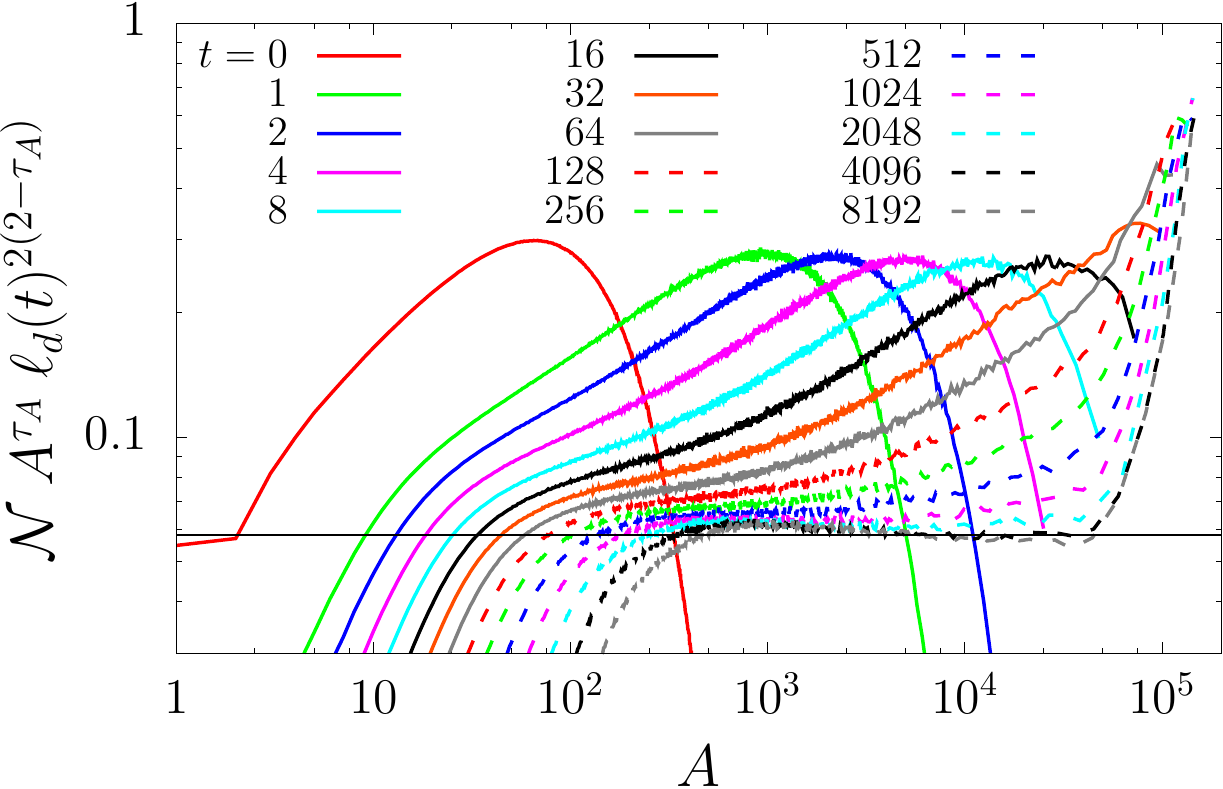}\quad%
        \includegraphics[scale=0.56]{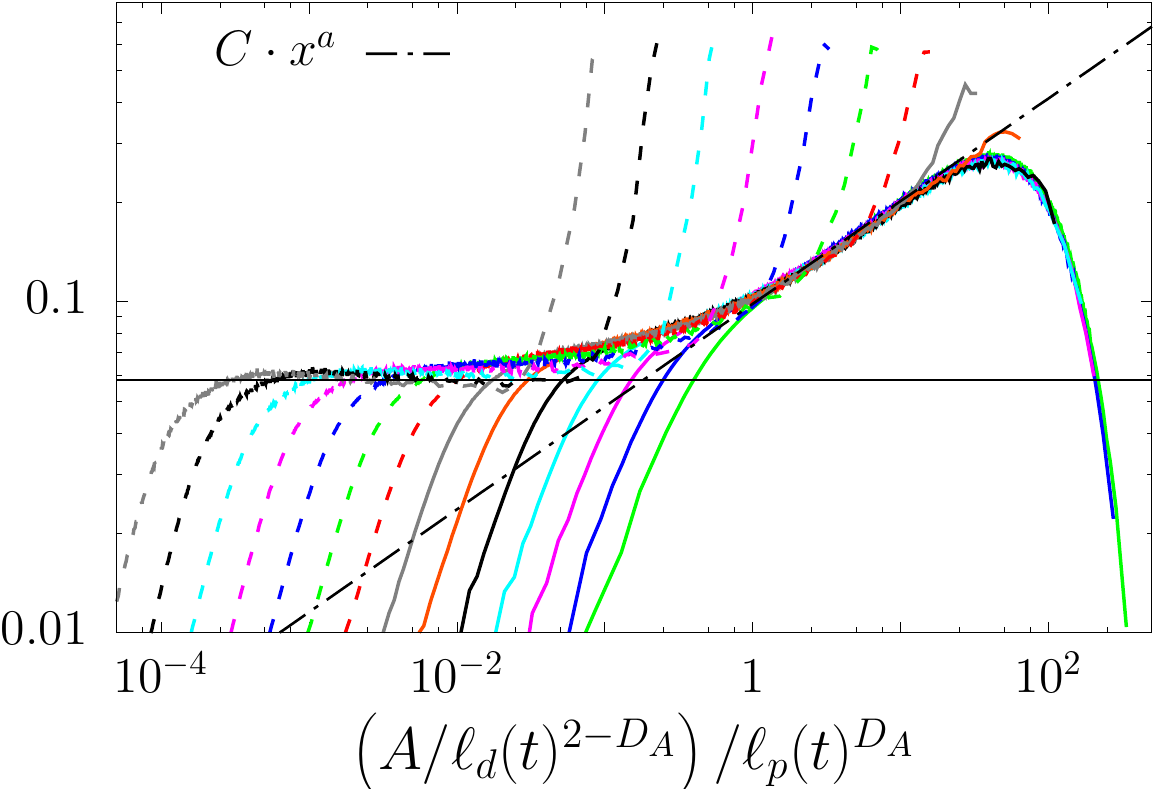}
\end{center}
\caption{\small Local Kawasaki dynamics of the Ising model on a square lattice with linear size $L=640$, quenched to $T_c/2$.
50:50 up-down spin mixture. Instantaneous number density of cluster areas, ${\mathcal N}(A,t, L)$,  
at different times given in the key in the left panel, plotted in the form
$A^{\tau_A} \;  \ell_d(t)^{2(2- \tau_A)} \, {\mathcal N}(A,t, L)$ against the area $A$,
with $\tau_A=187/91$ the Fisher exponent for critical percolation and $\ell_d(t) = \alpha \ell_G(t)$.
The proportionality constant $\alpha$ was tuned so that the plateau appearing in the data for  
the last time matched approximately the constant $2 c_d \approx 0.0580$, indicated by the black horizontal line.
In the right panel, the same quantity is plotted against the rescaled area
$ \left( A / \ell_d(t)^{2-D_A} \right) / \ell_p(t)^{D_A}$, with $\ell_p(t) = \ell_d(t) \, t^{1/\zeta}$, where
the exponent $\zeta$ was chosen to make the datasets corresponding to different times collapse in the scaling region
represented by the ``shoulder''. 
The best result is achieved by using $\zeta \simeq 2.00$.
The function $\Phi(x) = C \, x^{a}$ has been fitted to the data at $t=1$ in the interval $[1,50]$ 
yielding $a \simeq 0.310$.
}
\label{LKaNA-NWR_b2}
\end{figure}

The data for a square lattice with linear size $L=640$ are presented in Fig.~\ref{LKaNA-NWR_b2},
where $A^{\tau_A} \, \ell_d(t)^{2(2- \tau_A)} \;  {\mathcal N}(A,t,L)$ is plotted against $A$ 
in the left panel and against the rescaled area $(A/\ell_d^{2-D_A}(t))/\ell^{D_A}_p(t)$ in the right panel.
In both panels the critical percolation  Fisher exponent,
$\tau_A=187/91$,  and the fractal dimension of the percolating clusters at critical percolation, $D_A=91/48$, were used.

\vspace{0.5cm}

The coarsening characteristic length $\ell_d(t)$ has been taken to be 
proportional to $\ell_G(t)$, $\ell_d(t) = \alpha \ell_G(t)$.
The value of $\alpha$ was adjusted so that the plateau appearing in the data corresponding to the last time shown 
($t=8192$) matched approximately the constant $2 c_d$; we found $\alpha \simeq 3.64$.
The value of the exponent $\zeta$ was chosen to make the datasets corresponding to different times collapse in the scaling region
represented by the ``shoulder''. Consistently, the best result is achieved by using $\zeta \simeq 2.00$, 
the same value obtained for the scaling of the largest
cluster size and longest interface, and the wrapping probabilities.

The pre-percolation regime scaling function, $\Phi$,
takes approximately the form $\Phi(x) = C \,  x^{a}$, with $a>0$.
By fitting this function to the data at  $t=1$,
in the region $ [1,50] $
of the scaling variable $ \left( A / \ell_d(t)^{2-D_A} \right) / \ell_p(t)^{D_A}$, we found a value of 
$a$ compatible with the one obtained for the NCOP dynamics~\cite{BlCuPiTa-17}, namely $a\simeq 0.310$ 
(the fit is indicated by a black dashed line in the left panel of Fig.~\ref{LKaNA-NWR_b2}).

\begin{figure}[h!]
\begin{center}
        \includegraphics[scale=0.57]{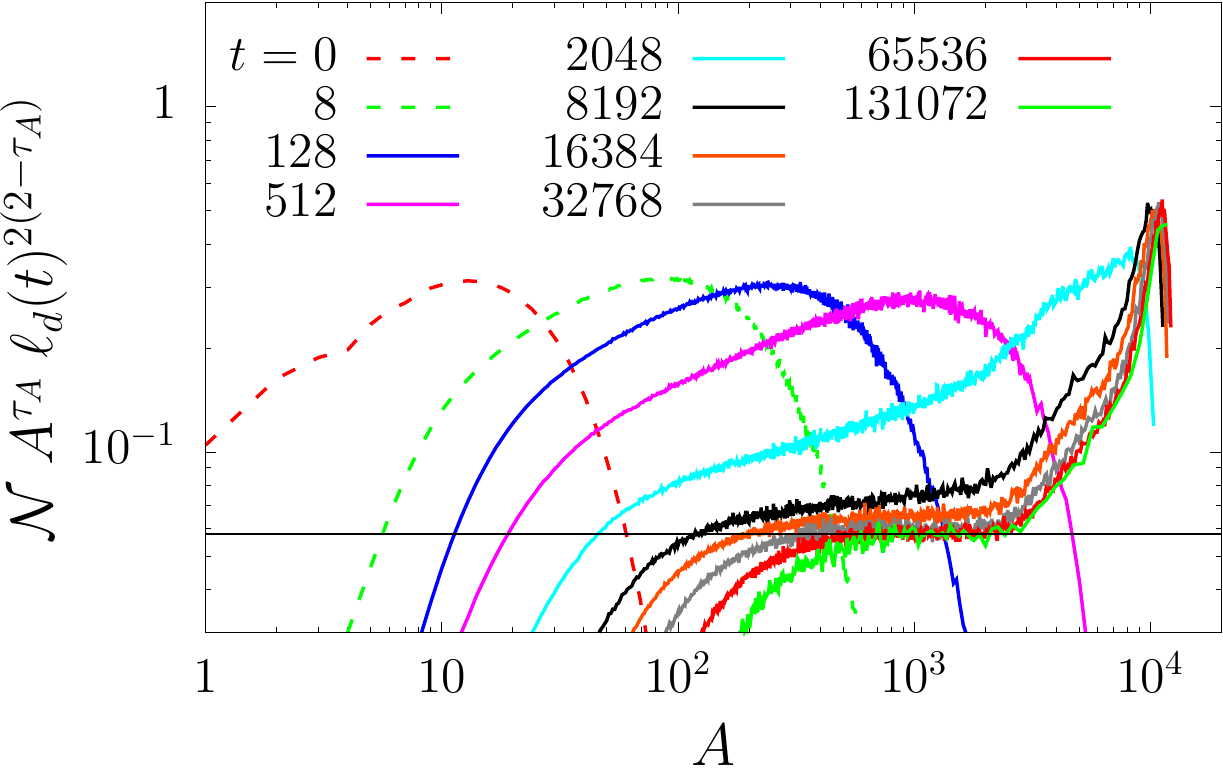}\quad%
        \includegraphics[scale=0.56]{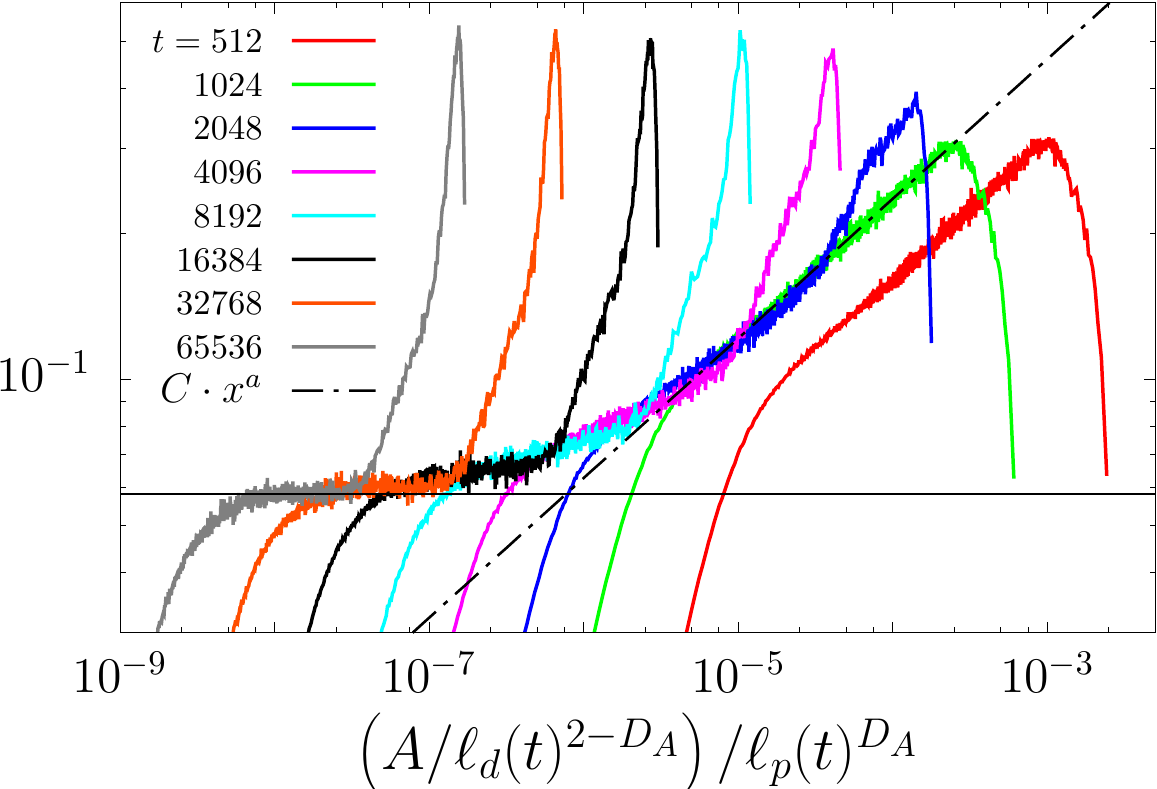}
\end{center}
\caption{\small 
Local Kawasaki dynamics of the Ising model
on a honeycomb lattice with linear size $L=160$, quenched to $T_c/2$.
50:50 up-down spin mixture. Same scaling plots as in Fig.~\ref{LKaNA-NWR_b2}.
Here again $\ell_d(t) = \alpha \ell_G(t)$, with the value of $\alpha$ tuned so that the plateau appearing in the data for  
the latest time shown, $t\simeq 1.31 \cdot 10^5 $, matched approximately the constant $2 c_d \approx 0.0580$, indicated by the black horizontal line.
The characteristic length $\ell_p(t)$ is given by $\ell_p(t) = \ell_d(t) \, t^{1/\zeta}$,
where the exponent $\zeta$ was chosen to be equal to the one used for the scaling of the wrapping probabilities, that is $\zeta = 1.15$.
In the right panel, the black dashed line represents a fit of the function $f(x) = C \, x^{a}$ to the data realtive to $t=1024$ in the interval
$[5 \times 10^{-5}, 10^{-4}]$ of the scaling variable, yielding $a \simeq 0.290$.
}
\label{LKaNA-NWR_b2-Hon}
\end{figure}

An analogous scaling  for the dynamics on the honeycomb lattice is presented in Fig.~\ref{LKaNA-NWR_b2-Hon}.
In the left panel,
at large values of $t$, the plateau corresponds to a range of areas obeying the
critical percolation statistics. In the right panel, 
$A^{\tau_A} \; \left[ \ell_d(t) \right]^{2(2- \tau_A)} \, {\mathcal N}(A,t, L)$ against the rescaled area $(A/\ell_d^{2-D_A}(t))/\ell^{D_A}_p(t)$ highlights
the pre-percolating regime.
Here again we set $\ell_p(t) = \ell_G(t) \, t^{1/\zeta}$, with the exponent $\zeta$ taking the value used for the scaling of 
the wrapping probabilities, namely $\zeta \simeq 1.15$. 
As one can see, the collapse is not as good as in the case of the square lattice, but 
the qualitative behaviour of the rescaled distribution is the same.
Moreover, a fit of the function $\Phi(x) = C \, x^{a}$ to the data relative to $t=1024$, in the interval
$[5 \times 10^{-5}, 10^{-4}]$ of the scaling variable $x = ( A/\ell_d(t)^{2-D_A} ) /\ell_p(t)^{D_A}$, yields $a \simeq 0.290$.

\vspace{0.5cm}

\subsubsection{Size distribution of percolating clusters.}

\begin{figure}[h!]
\begin{center}
        \includegraphics[scale=0.52]{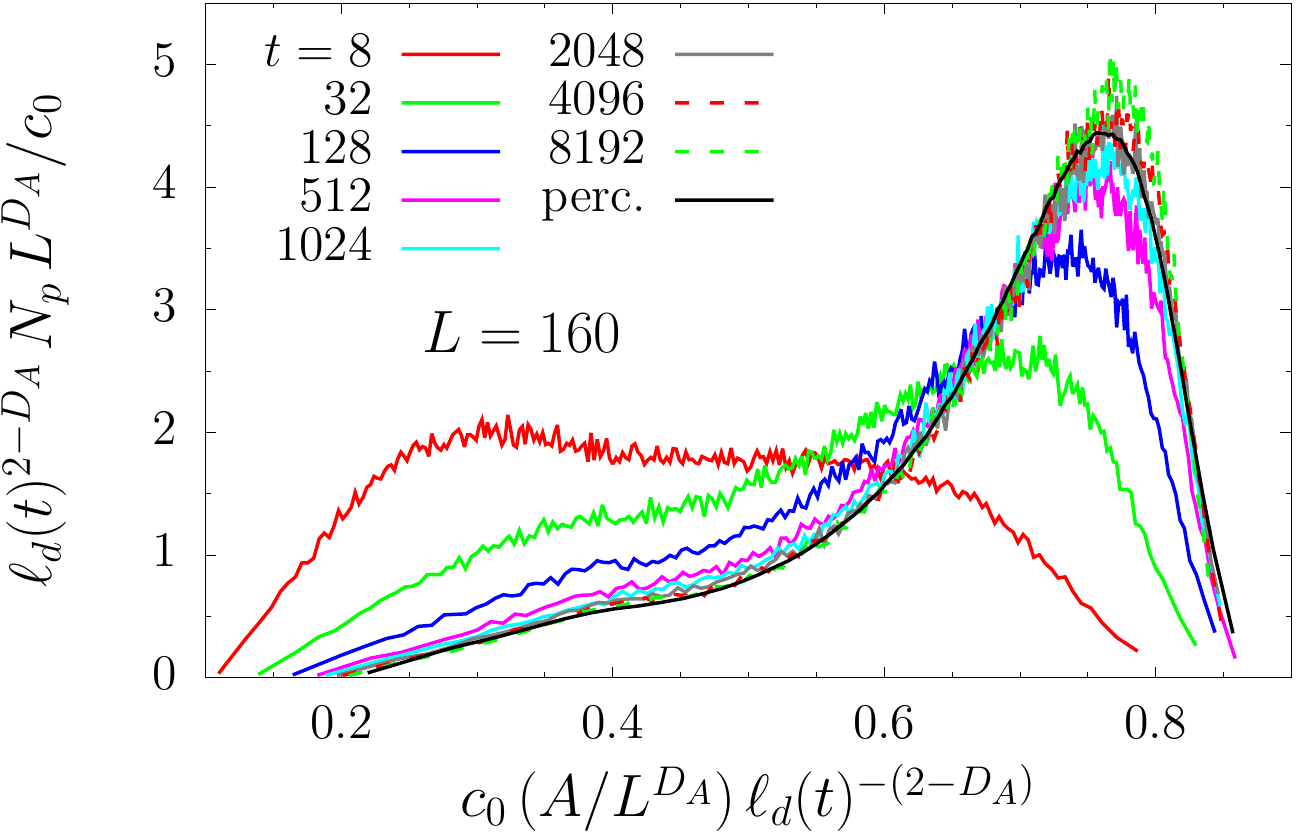}\quad%
        \includegraphics[scale=0.52]{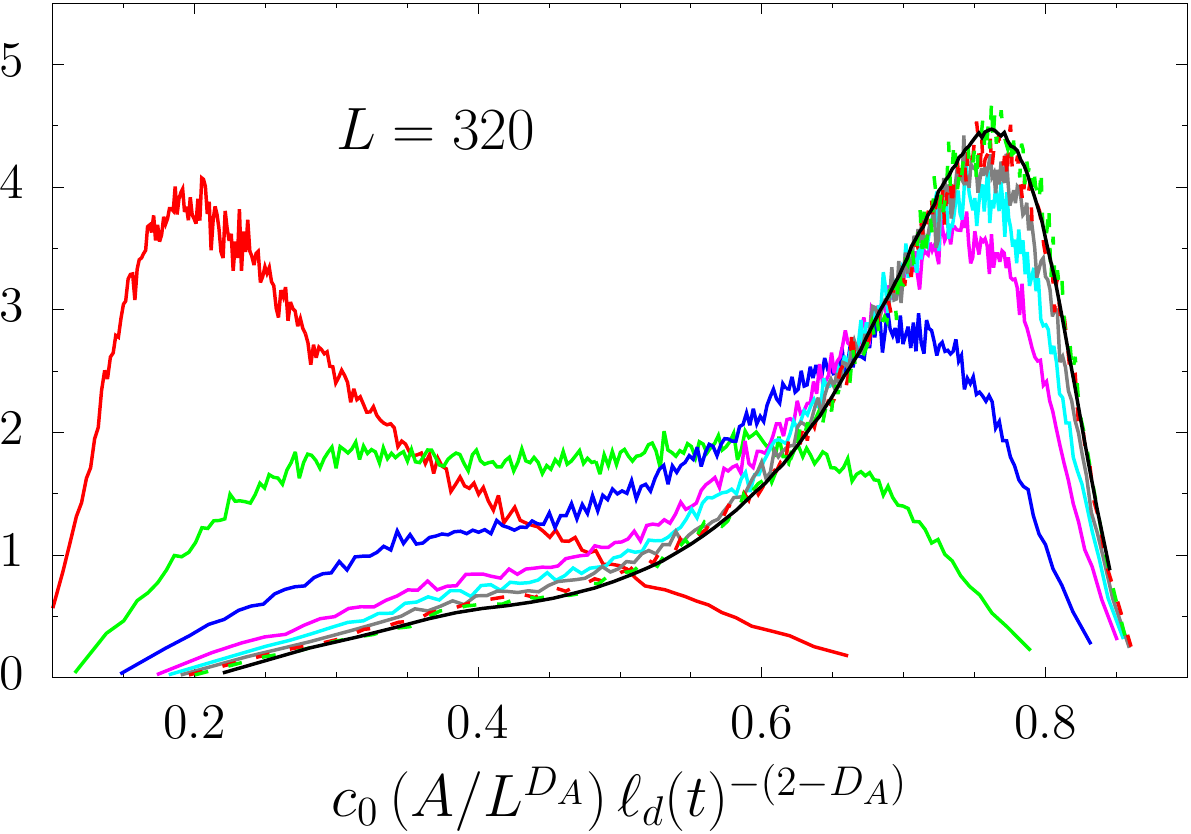}
\end{center}
\caption{\small The size distribution of the two largest clusters, $N_p(A,t,L)$, for
the Kawasaki dynamics at  $T_c/2$ of the IM on a square lattice with $L = 160$ (left) and $L=320$ (right). 
The color code for the measuring times (see the key in the left panel) is the same in the two plots.
The distribution is multiplied by $L^{D_A} \, \ell_d(t)^{2-D_A} / c_0$ and plotted against the rescaled area
$c_0 \, ( A/L^{D_A} ) \, \ell_d(t)^{-(2-D_A)}$, where $D_A$ is the fractal dimension of the percolating cluster in
$2d$ critical percolation, $\ell_d(t)=\ell_G(t)$ is the characteristic length obtained as the inverse of the excess energy, and
$c_0=1.18$. The size distribution of the largest cluster for random site percolation, at the threshold occupation
probability on a square lattice of corresponding size, is also shown with a black solid
line, multiplied by $L^{D_A}$ and plotted against $A/L^{D_A}$. The value of the constant $c_0$ was
chosen so that the rescaled distributions for the dynamical problem collapsed onto the
static one of critical percolation.
}
\label{LKaNA-LC-SLC_b2}
\end{figure}%

We present here an analysis of the quantity $N_p$ introduced in Sec.~\ref{subsec:observables},
that is the contribution given by the percolating clusters 
(or clusters whose size is comparable with the system size) to the full cluster size distribution $\mathcal{N}$.

The very few large clusters that survive the coarsening process after
a sufficiently long time are the ones that are used to define the characteristic time $t_p$. 
Around this time, these clusters span most of the lattice and their geometrical and statistical properties coincide with the ones
of the clusters occurring at critical site percolation on the same lattice, as seen in Sec.~\ref{subsec:LC-Kawasaki}.
Usually, at this time, only two large clusters, with opposite spin orientation, are percolating 
and become ``stable'' with respect to the microscopic dynamics.
This is the reason why $N_p$ can be effectively considered, for all practical purposes, as the size distribution of the two largest clusters in the system.

In critical percolation, the size distribution of the (incipient) percolating cluster, $N_p$, should satisfy the scaling
behaviour $N_p(A,L) \sim L^{-D_A} n_p \left( A/L^{D_A} \right)$, with $n_p$ a proper scaling function.
In the dynamical problem one has to take into account the effects of coarsening, and
we have seen that the largest cluster size needs to be rescaled by the factor $\ell_d(t)^{2-D_A}$ in order to reproduce the properties of critical percolation.
Given these considerations and the fact that $N_p$ is a probability density, we propose
\begin{equation}
  N_p(A,t,L) \sim L^{-D_A} \ell_d(t)^{-(2-D_A)} \, n_p \left( A/L^{D_A} \, \ell_d(t)^{-(2-D_A)} \right)
 \label{eq:scaling_NP}
\end{equation}
in the limit $L \rightarrow \infty$ and for $t$ approaching the characteristic time $t_p$. 
This scaling behaviour was confirmed by numerical results in~\cite{BlCuPiTa-17} for NCOP dynamics.
It implies that if one plots the quantity $L^{D_A} \, \ell_d(t)^{2-D_A} N_p(A,t,L)$ (for the dynamical problem) against the
rescaled size $ (A / L^{D_A} ) \, \ell_d(t)^{-(2-D_A)}$, the curves corresponding to different times $t$ and lattice linear size $L$ should approach
the same master curve as $t \rightarrow t_p(L)$. 
Furthemore, this master curve should coincide with the scaling function $n_p$ of critical percolation on the same lattice.

In Fig.~\ref{LKaNA-LC-SLC_b2} we show this scaling behaviour for the size distribution of the two largest spin clusters for the Kawasaki dynamics
on a square lattice at temperature $T_c/2$. 
Together with these data we also show the size distribution of the largest cluster, multiplied by $L^{D_A}$,
for random site percolation on a square lattice of the same size, at the threshold occupation probability ($p_c \simeq 0.5927 $), against $A/L^{D_A}$.
As one can see, for $L=160$ (left panel), the data relative to times $t=2048$ and $t=4096$ collapse approximately on the master curve represented by the critical percolation size distribution
(indicated by a black solid line). We can then conclude that the critical-percolation-like domain structure is attained at 
time $t_p \in [2048, 4096]$. Similarly for $L=320$ (right panel) we obtain that the data for $t=8192$ is the closest to match the critical percolation distribution.
However, in order to get the collapse of the data relative to the quench dynamics onto the critical  percolation ones we needed to include an additional
scaling factor $c_0 \simeq 1.18$, which seems to be independent of $L$.

\subsection{Summary}
 
In this Section we analysed the way in which the phase separation of a balanced mixture of two
species approaches a state with a stable pattern of critical percolating domains under local spin (or particle) 
exchange updates.
 
 First of all, we proved that extremely long time scales are needed to reach the LSW $t^{1/3}$  
 algebraic growth of the typical domain length of the coarsening phenomenon. Accordingly, we argued that 
 in the scaling analysis the full time-dependent $\ell_G(t)$ should be used as a representation of this 
 growing length.
 
For the three lattice geometries, by studying the  scaling properties of the wrapping probabilities,
the variance of the winding  angle,  the averaged largest area, the averaged longest hull, and the domain area distribution, 
we concluded that the characteristic length $\ell_p(t)$ associated to the critical-percolation-like structures 
behaves as 
\begin{equation}
\ell_p(t) \simeq \ell_G(t) \, t^{1/\zeta}
\nonumber
\end{equation}
with $\ell_G(t)$ the growing length extracted from the excess energy, and 
$\zeta \simeq 2.00$ on the square lattice and $\zeta \simeq 1.15$
 on the honeycomb lattice. The model on the triangular lattice behaves differently 
 since it is at the critical percolation point initially and no further rescaling is needed.  
 
As regards the statistical distribution of the very large areas,
we found a number of common features with the results of NCOP dynamics~\cite{BlCuPiTa-17}. 
For example, the scaling function describing the crossover between the algebraic tail in the number density of finite-size domains
and the weight of the domains that scale with the size of the system is, within numerical accuracy, the same
$\Phi(x) \propto x^a$ with $a \simeq 0.3$ in both cases.

After the time $t_p$ such that $\ell_p(t_p) = L$, the percolating cluster(s) become fatter and fatter
and a second ordering regime characterised by the expected growing length $\ell_d(t) \simeq  t^{1/z_d}$ with $z_d = 3$ should eventually 
establish. However, very long times are needed to reach this algebraic behaviour and these go beyond 
the accessible simulation time-window at sufficiently low temperatures.

We completed in this way the analysis of the local spin-exchange dynamics and we are now ready to 
analyse how is this picture modified (or not) when the spin exchanges do not respect the locality
condition.

\section{Nonlocal Kawasaki dynamics}
\label{sec:nonlocal-Kawasaki}

We now consider the nonlocal version of the spin exchange dynamics which was studied in the previous
section. We only use initial configurations with equal concentration of the two species and we let them evolve
under nonlocal exchanges of anti-parallel spins situated anywhere in the lattice (see Ch.~5 in~\cite{Barkema}). 
The moves are accepted with probability dictated by the Boltzmann weight at temperature $T$ and satisfying the usual
detailed balance condition. In other words, the transition rates for the spin exchange events are exactly
the same as the ones explained in Sec.~\ref{sec:local-Kawasaki},
but without the restriction that the lattice sites involved in the spin exchange must be nearest-neighbours.
The total magnetisation of the system remains a conserved quantity, but the local magnetisation
(that is, the average magnetisation over an arbitrarily small number of neighbouring lattice sites) does not.

We checked, by studying the magnetisation density at equilibrium, that is to say the order parameter for the 
paramagnetic-ferromagnetic transition, that this {\it a priori} artificial dynamics
capture the thermodynamic instability at the equilibrium critical temperature $T_c$ of the $2d$ Ising model 
on the corresponding lattice geometry (not shown).

The renormalisation group arguments proposed in~\cite{Bray91,Bray94b} suggest that, asymptotically, the 
global conservation law should become irrelevant and the dynamics be controlled by the 
$\ell_d(t)\simeq t^{1/2}$ growth law.
Studies of the phase ordering kinetics of coarse-grained field theories with the 
total magnetisation imposed (on average) with a time-dependent magnetic field also suggest 
$\ell_d(t) \simeq t^{1/2}$ at finite temperature~\cite{Sire95}. This result was derived in the large 
$N$ limit and with various approximation schemes of the Ginzburg-Landau-like description.
A simple argument exposed by Rutenberg in~\cite{Rutenberg96} signals a difference between the dynamics 
at $T=0$, where he claimed that the LCOP results should be found, $\ell_d(t) \simeq t^{1/3}$, and 
$\ell_d(t) \simeq t^{1/2}$ at $T>0$. Numerical simulations~\cite{Tamayo89,Tamayo91,Bray91,Annett92,Moseley92,Sire95,Rutenberg96}
have only explored relatively short time scales, $t < 10^4$, they are not decisive in their 
description of the growth law exponent, and this deficiency created 
some debate around this issue. An experiment using a $2d$ chiral liquid crystal with dynamics in this universality class 
was consistent with $\ell_d(t)\simeq t^{1/2}$~\cite{Sicilia-etal08}.

In this Section we present a rather complete analysis, reaching times many orders of magnitude longer than the ones accessed
before, $t\simeq 10^{20}$. We present results on sub-critical quenches of the $2d$IM defined on the three lattices studied, all with PBC. 
We start from the evaluation of the usual growing length from 
the decay of the excess energy and we then move on to the study of the dynamic geometric structure. 

\vspace{0.5cm}

\begin{figure}[h]
\begin{center}
  \subfloat[$t=0$]{\includegraphics[scale=0.33]{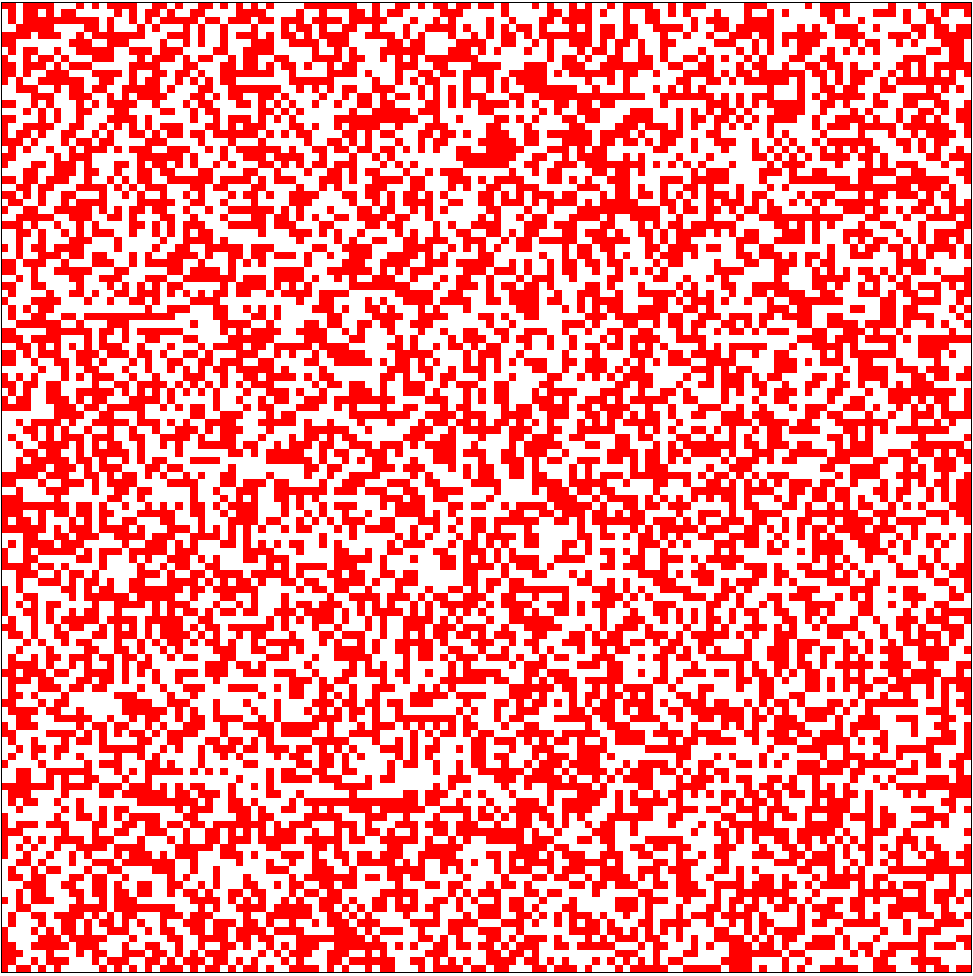}}\quad%
  \subfloat[$t=4107$]{\includegraphics[scale=0.33]{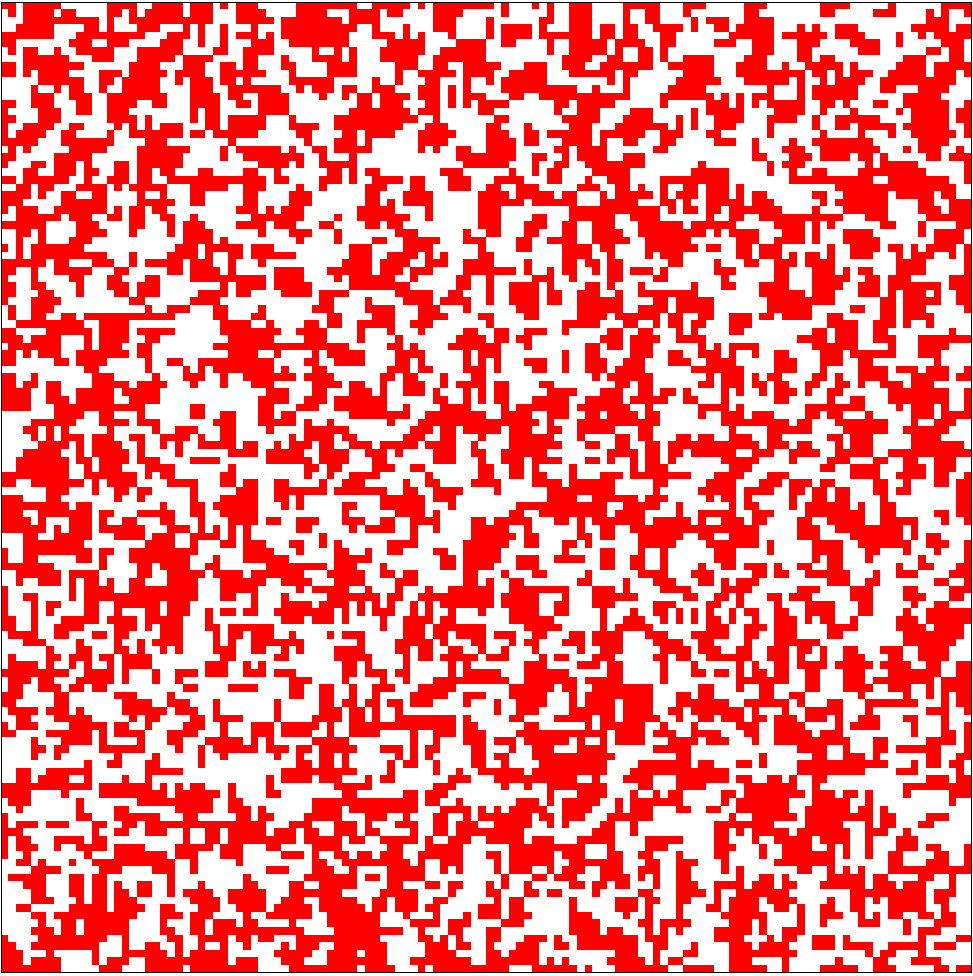}}\quad%
  \subfloat[$t=8202$]{\includegraphics[scale=0.33]{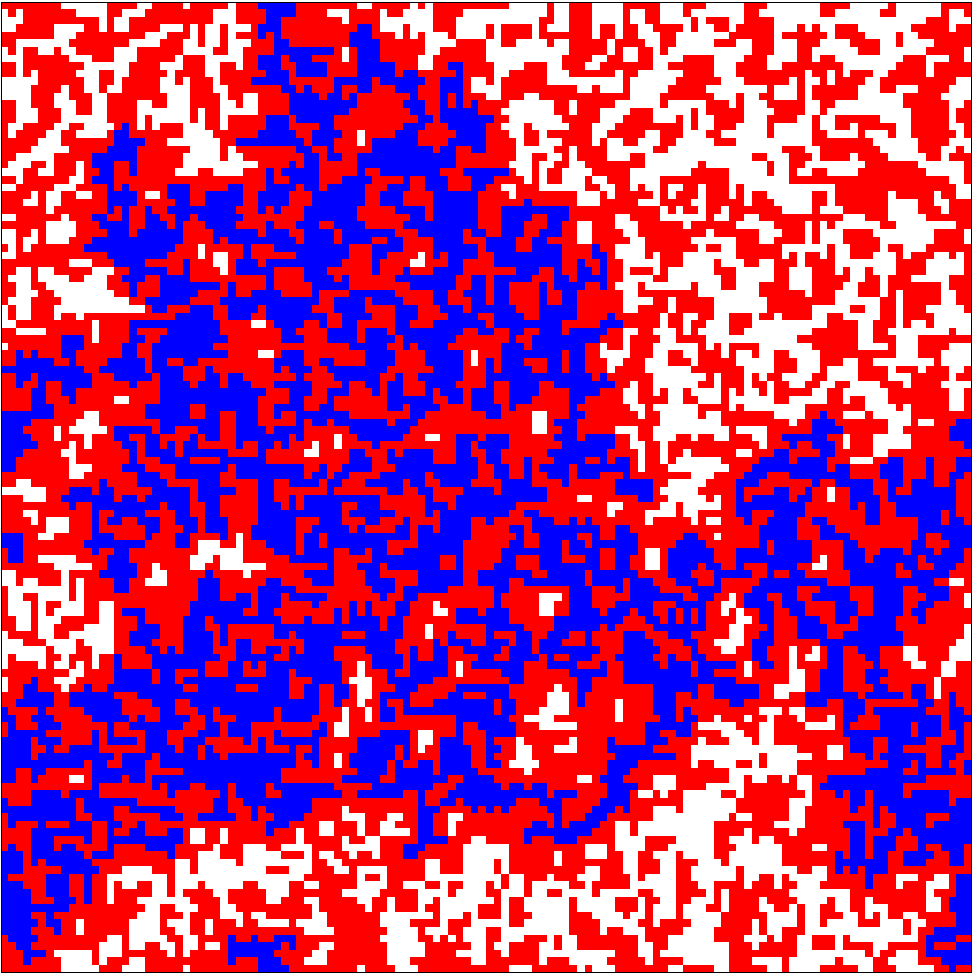}}\quad%
  \subfloat[$t=131450$]{\includegraphics[scale=0.33]{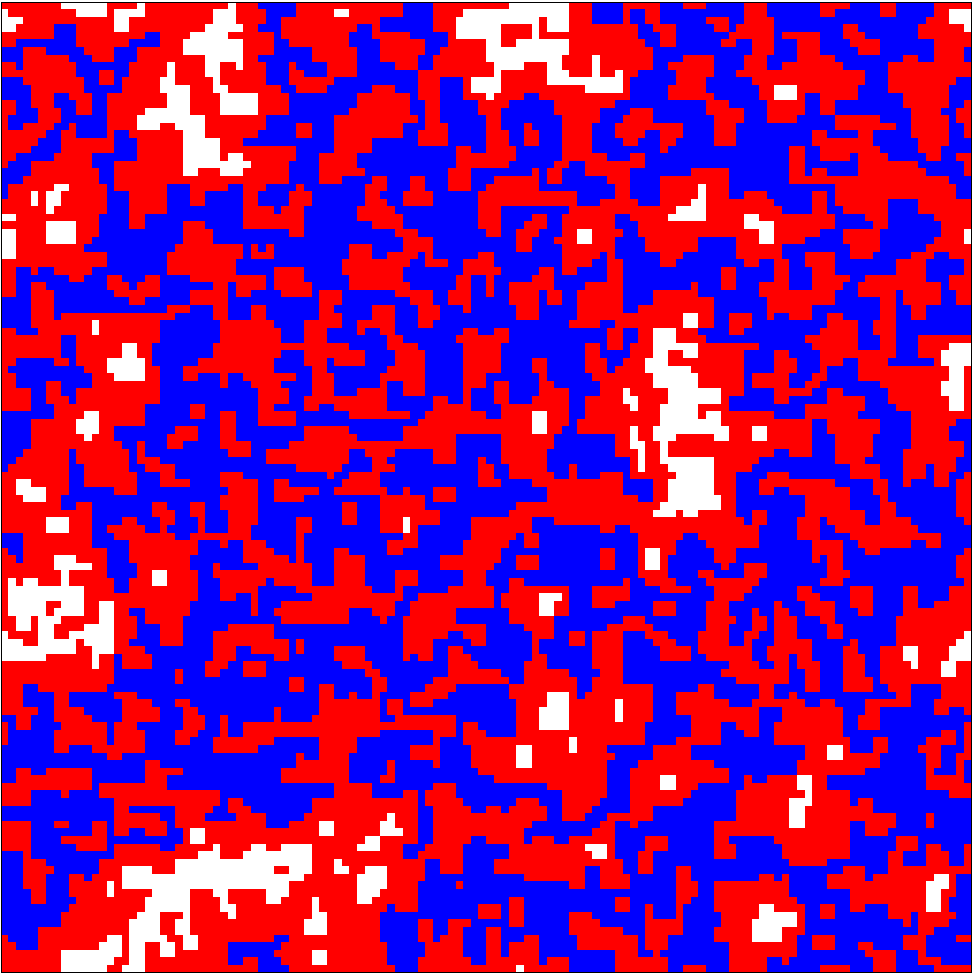}}

  \subfloat[$t=4.2 \times 10^6$]{\includegraphics[scale=0.33]{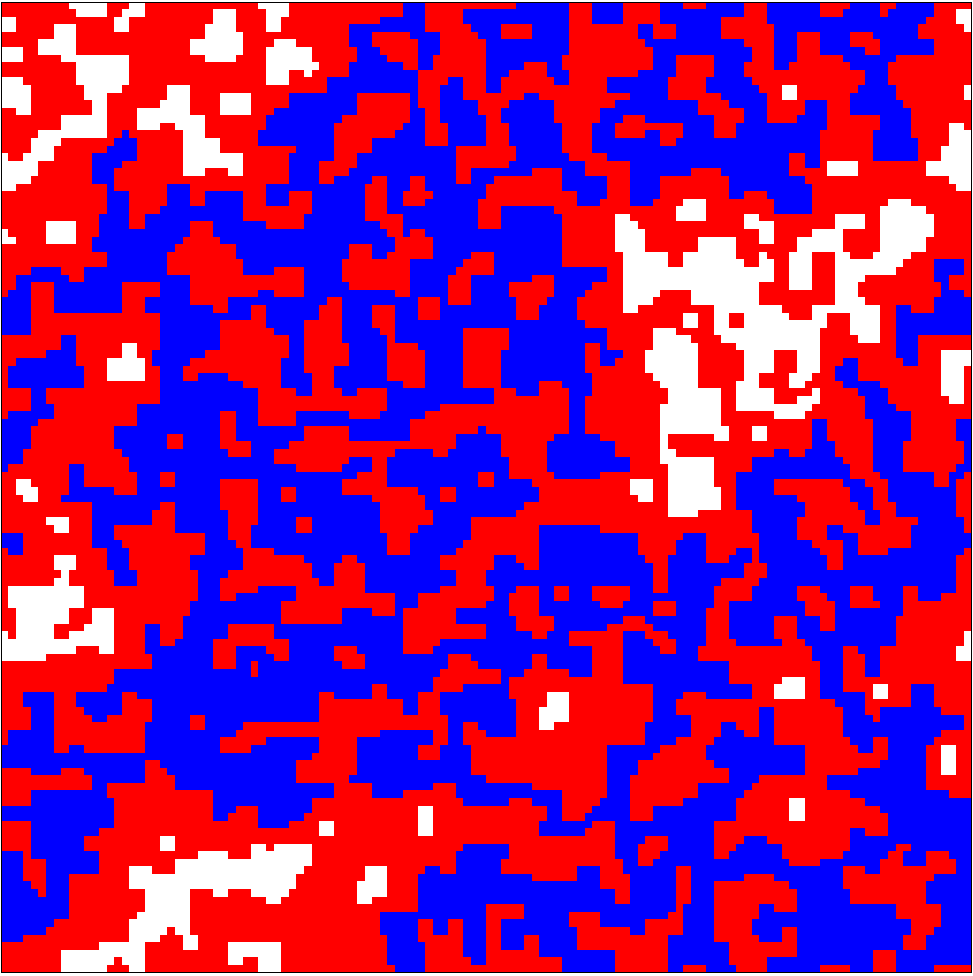}}\quad%
  \subfloat[$t=1.3 \times 10^8$]{\includegraphics[scale=0.33]{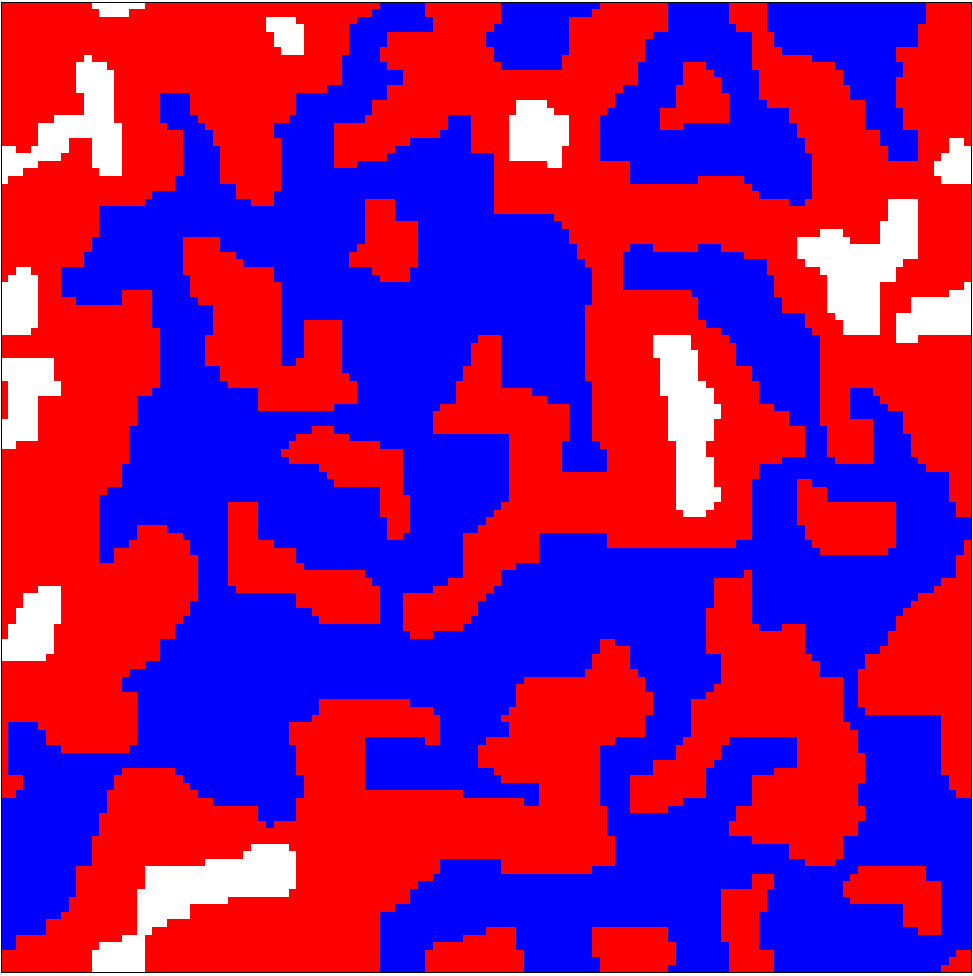}}\quad%
  \subfloat[$t=4.3 \times 10^9$]{\includegraphics[scale=0.33]{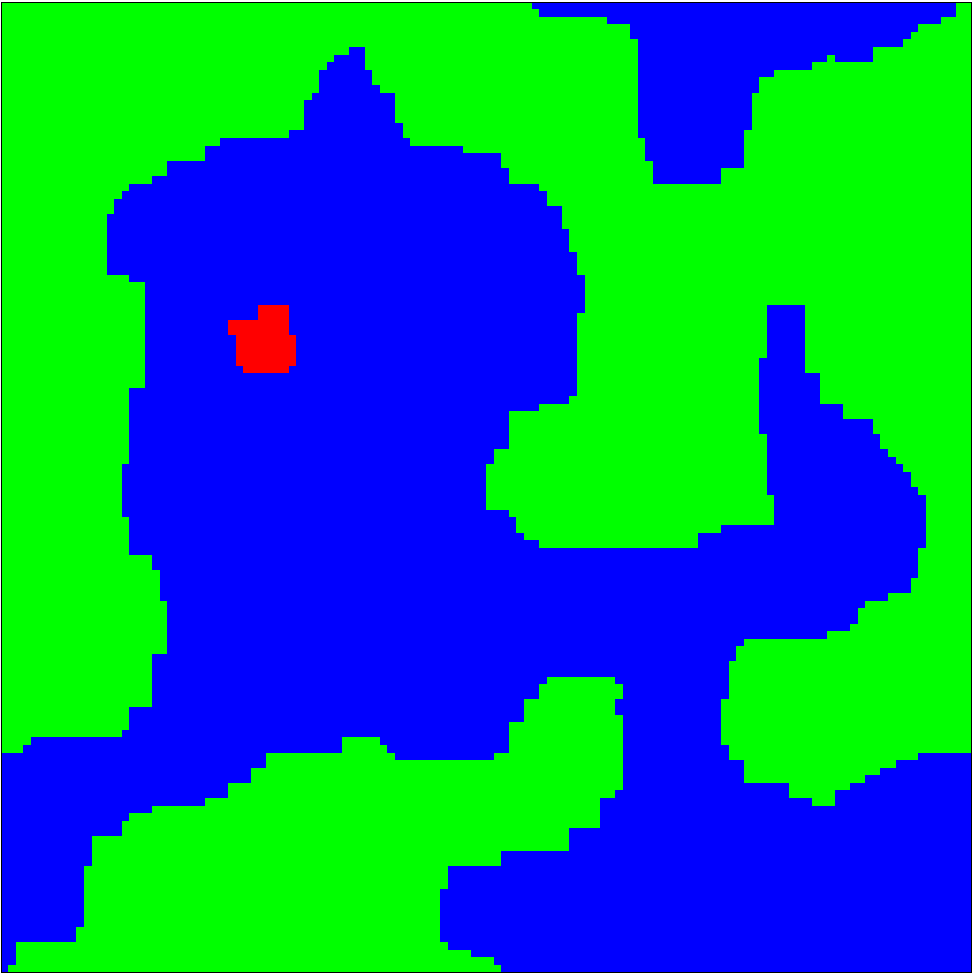}}\quad%
  \subfloat[$t=1.4 \times 10^{11}$]{\includegraphics[scale=0.33]{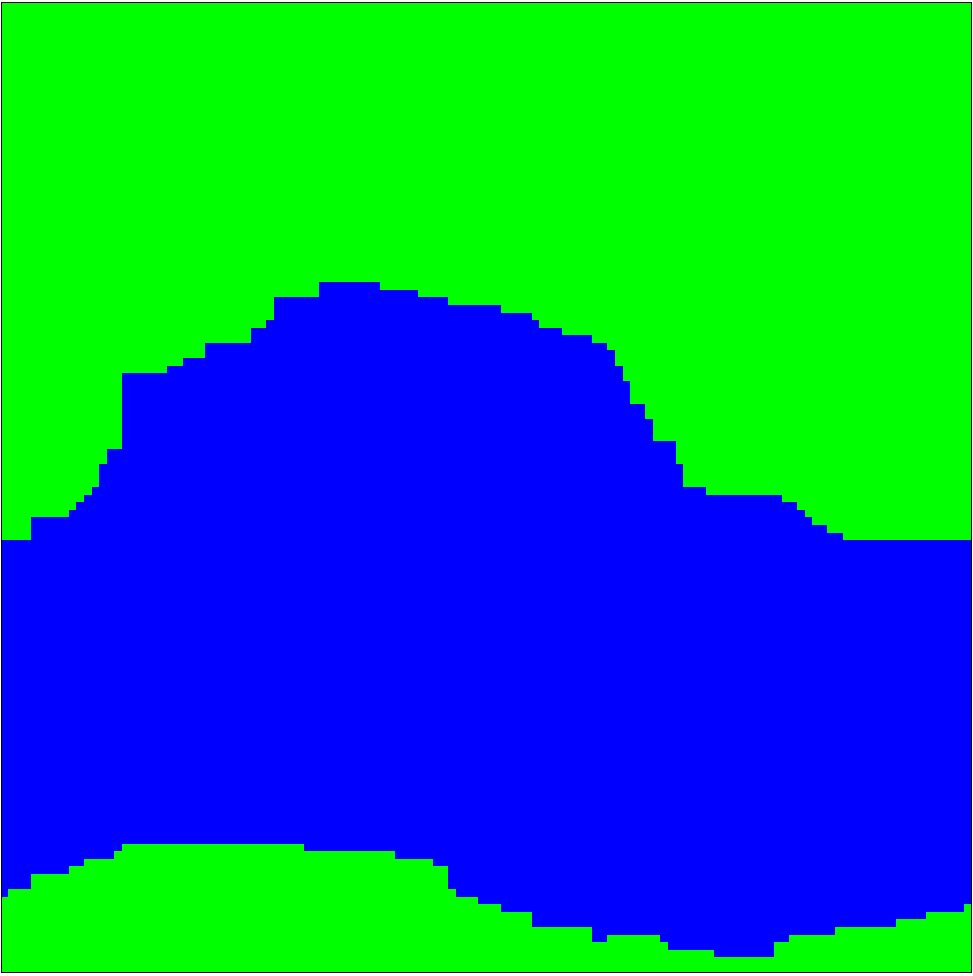}}
\end{center}
\caption{\small 
Snapshots of a spin configuration on a square lattice with size $128 \times 128$ 
and PBC, evolving at $T=T_c/4$ with nonlocal Kawasaki dynamics from  an infinite temperature initial condition.
The concentration of the two species  is the same.
Spins $s_i=-1$ are shown as red points while spins $s_i=+1$ are shown as white points. 
Percolating clusters of spins $s_i=-1$ 
are shown in green and percolating clusters of spins $s_i = +1$ in blue. The times at which the snapshots are 
taken are indicated below them.
}
\label{fig:snapshots-nlk}
\end{figure}

\subsection{Snapshots}

In Fig.~\ref{fig:snapshots-nlk} we show snapshots of the spin configurations on a square lattice with PBC evolving
under nonlocal Kawasaki dynamics at temperature $T_c/4$. 
We observe the presence of growing domains and percolating structures. 
Apart from the absolute time values, that are definitely much longer than the ones
explored using local Kawasaki spin-exchange updates, we do not see much of a difference 
in the overall morphology of the spin clusters with respect to the snaphots in Fig.~\ref{KaSnap}, 
at least at naked eye.

\subsection{The growing length}
\label{subsec:growing-length-NLK}

\vspace{0.5cm}

\begin{figure}[h]
\begin{center}
\includegraphics[scale=0.52]{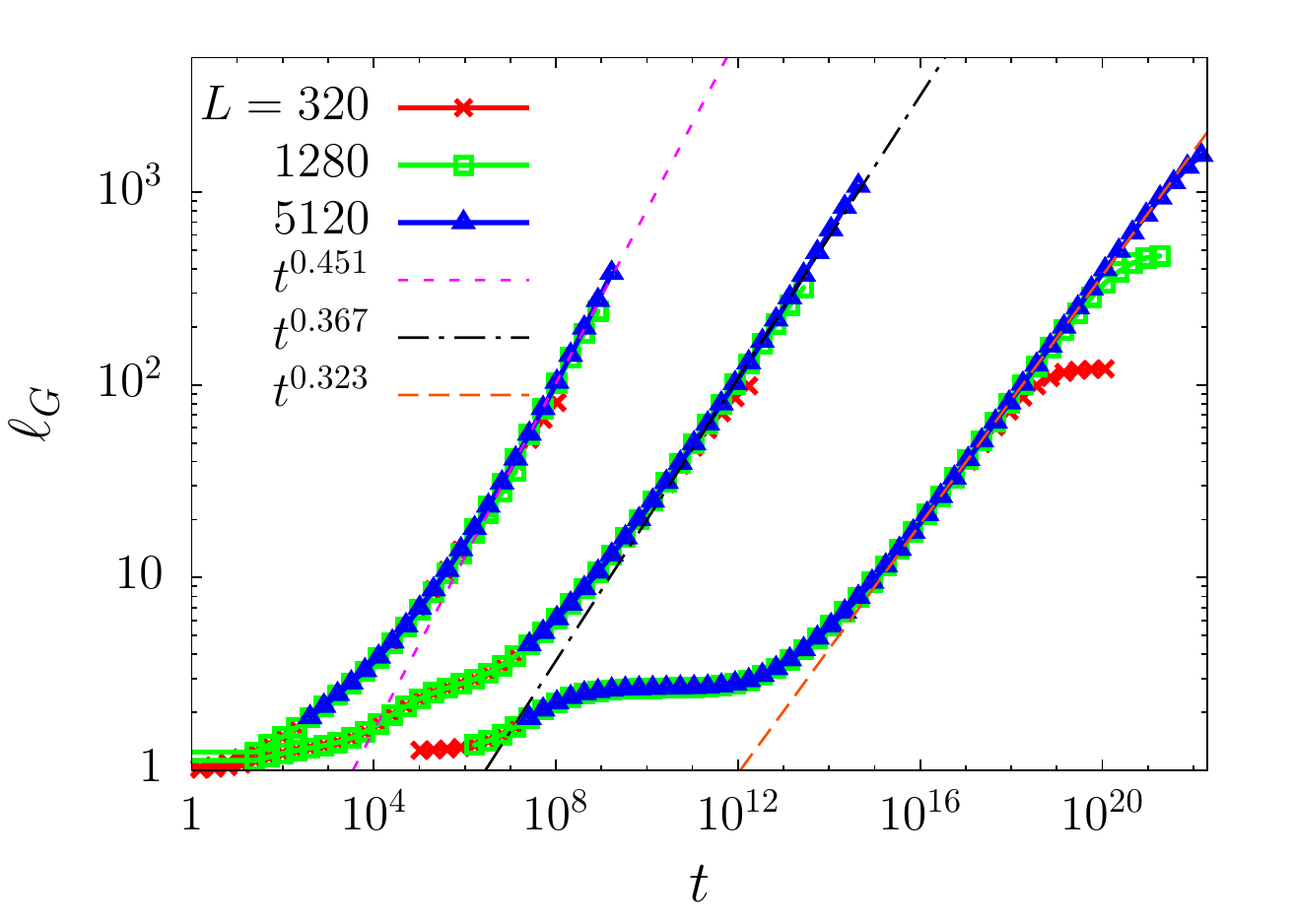}\quad%
\includegraphics[scale=0.52]{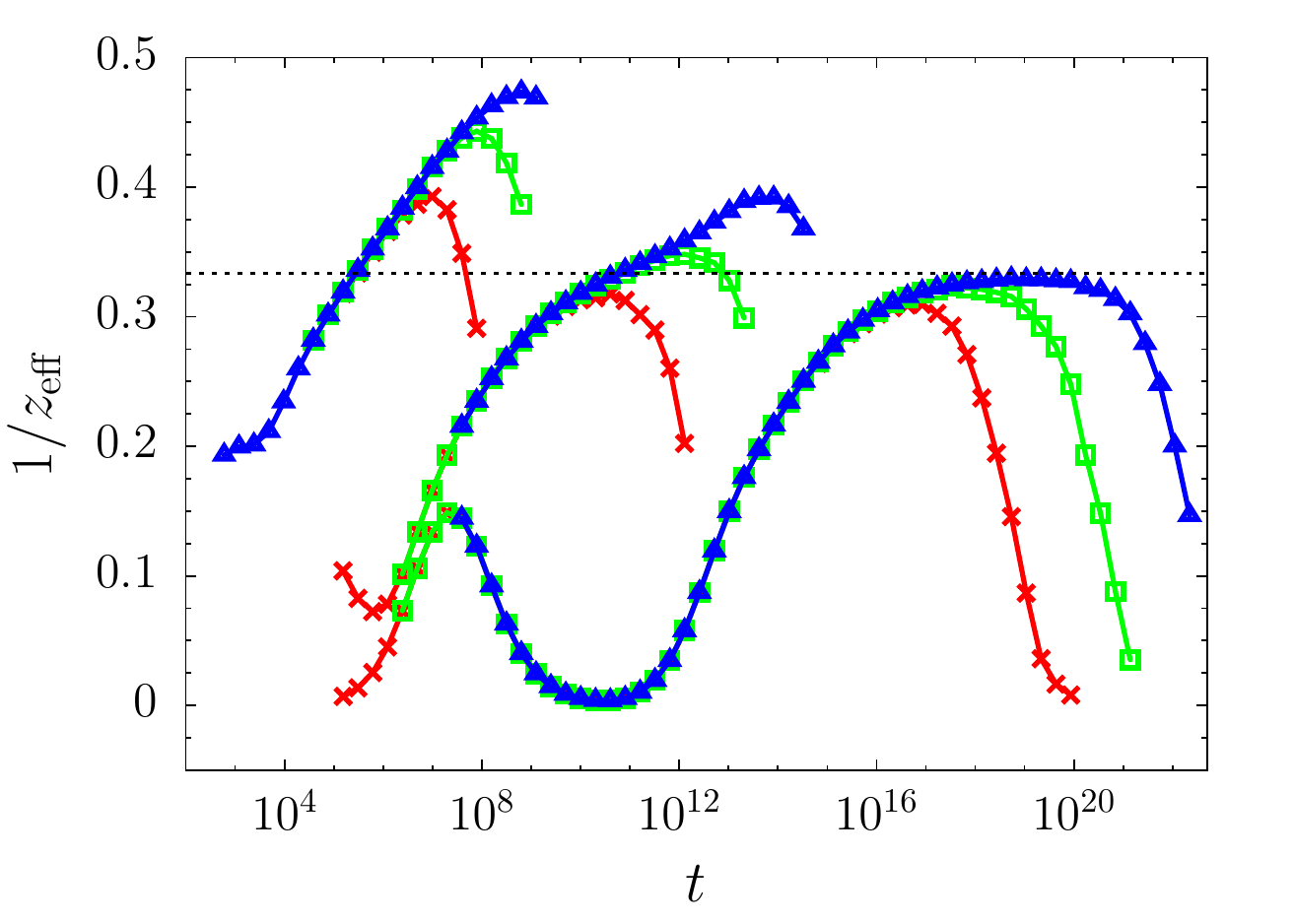}
\end{center}
\caption{\small Left: the excess-energy growing length $\ell_{G}$
as a function of time $t$ for the nonlocal Kawasaki dynamics on a square lattice with PBC at temperatures
$T = T_c/2$ (data on the left), $T = T_c/4$ (data in the middle) and $T=T_c/8$ (data on the right).
For each temperature, we show the data for three different values of $L$, the lattice linear size.
A power law $a \, t^{b}$ has been fitted to the data corresponding to $L=5120$
in the intermediate region (between the initial transient region and the final saturation one) 
and is represented by the straight dashed lines (see the key for the fitted exponents).
Right panel: corresponding effective growth exponents. 
They increase weakly with the linear size $L$ of the lattice.
The dotted horizontal line corresponds to $1/3$.
}
\label{FigGL}
\end{figure}

In the left part of Fig.~\ref{FigGL} we show $\ell_{G}(t)$
for the nonlocal Kawasaki dynamics on a square lattice with PBC, with linear sizes $L=320, \, 1280, \, 5120$. 
The leftmost curves correspond to the dynamics at temperature
$T=T_c/2$, the ones in the middle to $T=T_c/4$, and the ones on the right to $T=T_c/8$.
Note that $\ell_{G}(t)$ can be at most $L/2$ because of the global conservation of the number of spins of each species 
which makes sure that the spin configuration corresponding to minimal energy is the one with just two domains 
separated by a straight interface.
In each case, we observe the presence of an initial transient regime and a final saturation (which is clearly visible only
for the smaller $L$) and in both regimes the behaviour of $\ell_G(t)$ cannot be described by a simple power law.
Moreover, we notice the presence of a \textit{plateau} in the transient regime which becomes longer
as $T$ decreases with the effect of delaying when the actual power law behaviour sets in.

In the left panel we display a fit of the function $f(t) = C \, t^{1/z}$ to the data relative to $L=5120$
in the intermediate region, for each  temperature.
For the dynamics at $T=T_c/2$ we obtain $1/z \simeq 0.442$, a value which is close to the expected asymptotic result
for GCOP dynamics, $1/z_{d}=1/2$; for the evolution at
$T=T_c/4$  we obtain $1/z \simeq 0.367$; and, for the one at $T=T_c/8$ we find $1/z \simeq 0.322$, 
close to the value $1/z_{d}=1/3$ expected for the dynamics at $T=0$, according to Rutenberg~\cite{Rutenberg96}. 
 
In the right panel in  Fig.~\ref{FigGL} we show the effective growth exponent ${z_{\rm eff}^{-1}}$, 
defined as the logarithmic derivative of $\ell_G(t)$, for each one of the cases present in the left panel.
At $T_c/2$, the effective exponent $z_{\rm eff}^{-1}$ seems to converge towards $1/2$ as $L$ increases, as expected. 
At $T_c/8$ it spends approximately three time decades around $1/3$. At $T_c/4$, we observe an intermediate result.
At each temperature, 
the effective exponent reaches a maximum and then starts to fall off due to the final saturation of $\ell_G(t)$.
Note that, for fixed $T$, the larger is $L$ the larger is the maximum value reached by ${z_{\rm eff}^{-1}}$ and longer is the 
period of time
that ${z_{\rm eff}^{-1}}$ stays at this value.
These observations confirm that the expected value $z_{\rm eff}=2$ at finite temperature 
is reached for not too low temperature ({\it i.e.} $T=T_c/2$), as $L \rightarrow \infty$.
Instead, for sufficiently low temperature ({\it i.e.} $T=T_c/8$), the zero-temperature predicted result, $z_{\rm eff}=3$, 
would instead be mostly observed for large $L$.

\begin{figure}[h]
\begin{center}
\includegraphics[scale=0.7]{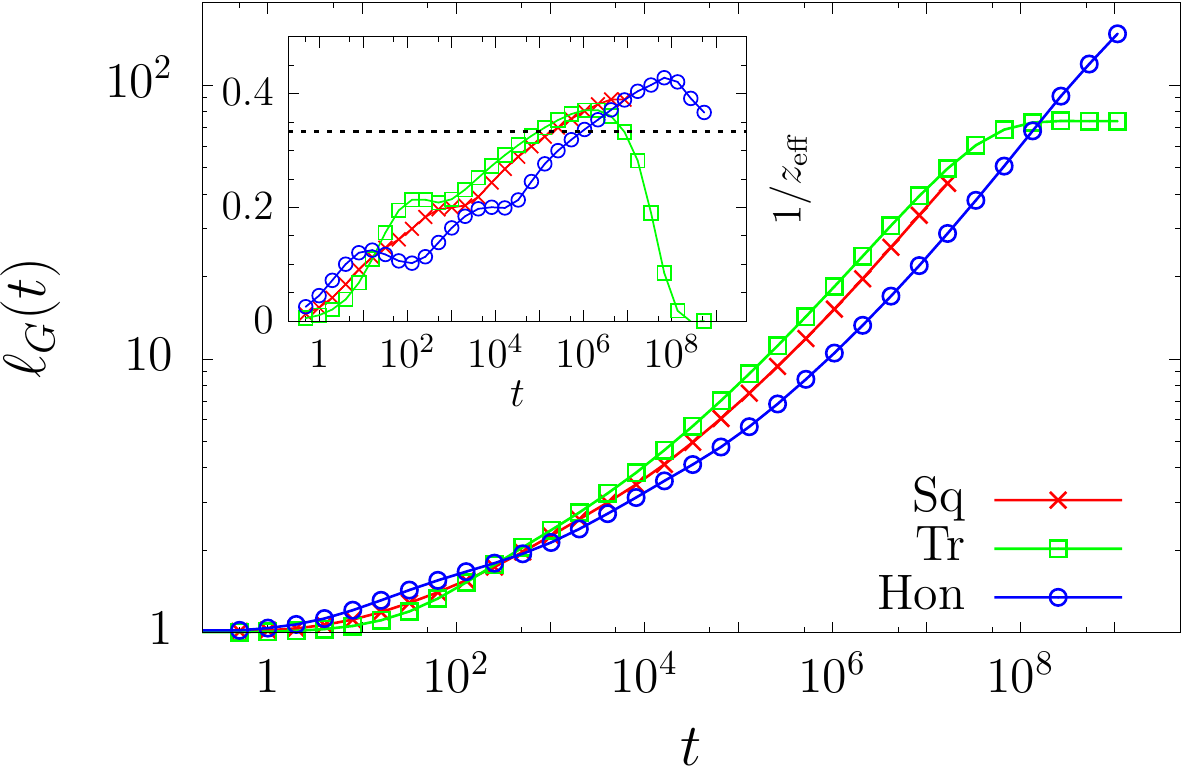}
\end{center}
\caption{\small The excess-energy growing length, $\ell_G(t)$, for the nonlocal Kawasaki dynamics on the square, 
triangular and honeycomb lattices
with linear size $L=320$ and PBC, at  $T=T_c/2$, starting from a fully disordered initial condition
with equal concentration of the two species. In the inset, the effective growth exponent, $1/z_{\rm eff}(t)$, as a function of time,
calculated from the logarithmic derivative of  $\ell_G(t)$. The dashed horizontal line is at $1/3$.
}
\label{NLK-GL_other}
\end{figure}

In Fig.~\ref{NLK-GL_other} we report the excess-energy growing length for the nonlocal Kawasaki dynamics 
on the honeycomb and triangular lattices, along with the analogous data for the square lattice.
In the three cases the working temperature is $T_c/2$ and the system linear size is $L=320$.
In the inset we show the effective growth exponent, $1/z_{\rm eff}$, calculated from the 
logarithmic derivative of the curves in the main plot, the same procedure used in the right panel in 
Fig.~\ref{FigGL}.
The behaviour of $\ell_G(t)$ for the dynamics on these three lattices is similar.

\vspace{0.5cm}

\subsection{Wrapping probabilities}
\label{subsec:WP-NLK}

\begin{figure}[h]
\begin{center}
\includegraphics[scale=0.52]{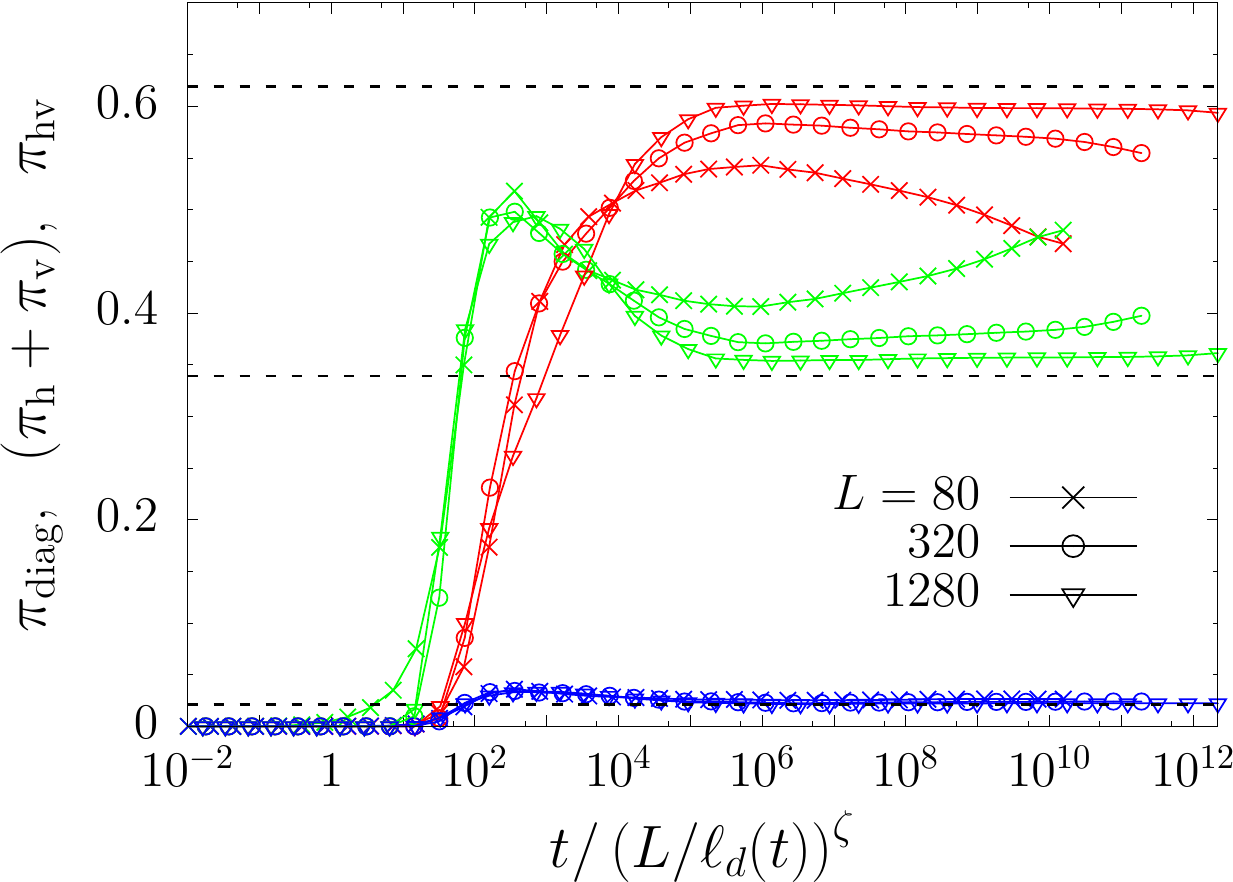}\quad%
\includegraphics[scale=0.52]{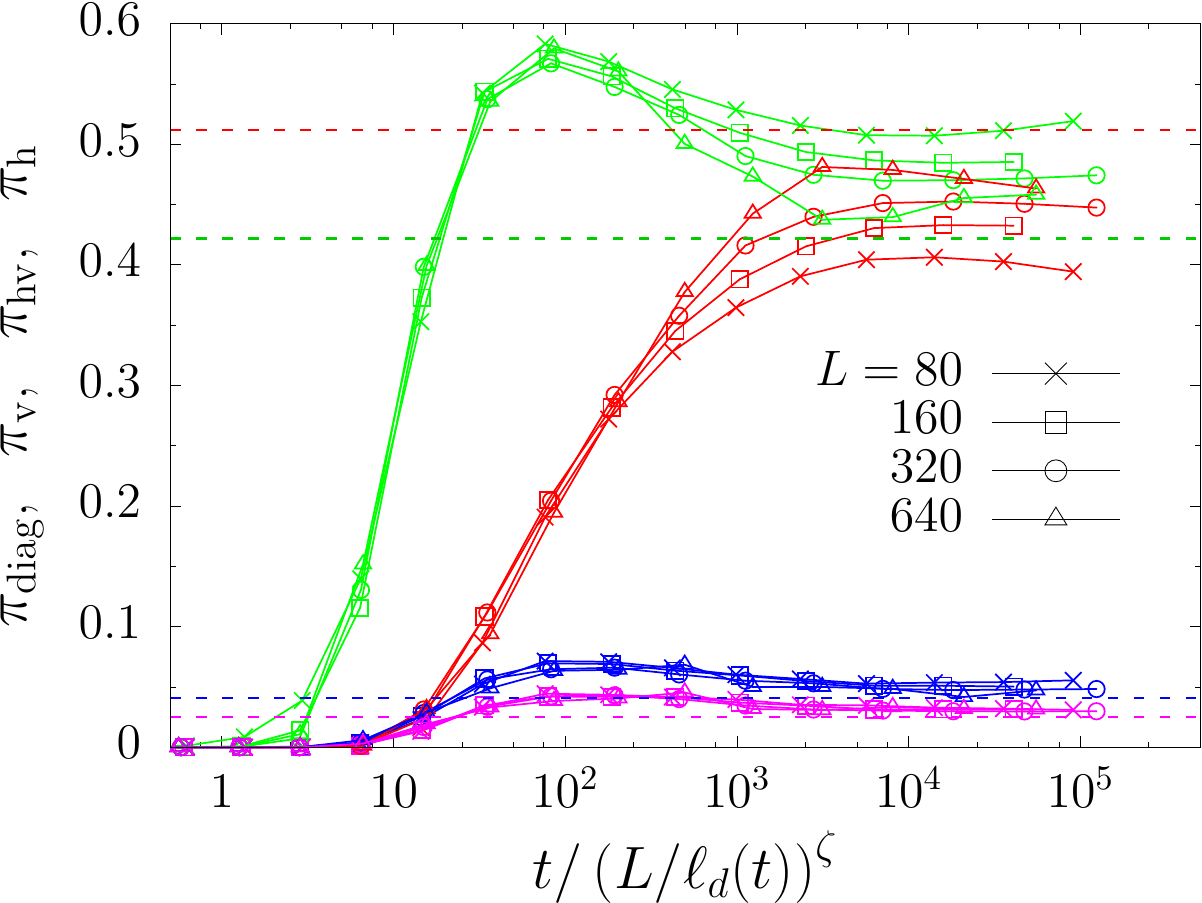}
\end{center}
\caption{\small
In the left panel we show the probability of having a cluster wrapping in both directions (red curves), $\pi_{\rm hv}$, either 
horizontally or vertically (green curves), $ ( \pi_{\rm h} + \pi_{\rm v} )$, and diagonally 
(blue curves), $\pi_{\rm diag}$, for
nonlocal Kawasaki dynamics on a square lattice with PBC, at  $T=T_c/4$, for different values of $L$ indicated in the key.
The data are plotted against the rescaled time $t / ( L / \ell_d(t) )^{\zeta}$, 
where $\ell_d(t)$ is taken to be  the inverse of the excess energy, $\ell_G(t)$,
and the value of the exponent $\zeta$ is chosen to obtain data collapse, $\zeta \simeq 1.15$.
The horizontal dashed lines are the wrapping probabilities in critical percolation on a torus
with unit aspect ratio.
In the right panel, the same wrapping probabilities on a honeycomb lattice,
with $\pi_{\rm hv}$ in red, $\pi_{\rm h}$ in green, $\pi_{\rm v}$ in blue and $\pi_{\rm diag}$ in purple. The best data collapse is also 
achieved with $\zeta \simeq 1.15$. The dashed horizontal lines are the exact values of
$\pi_{\rm hv}$, $\pi_{\rm h}$, $\pi_{\rm v}$ and $\pi_{\rm diag}$ at critical percolation on a lattice with aspect ratio $\sqrt{3}$.
}
\label{FigWP-NLK}
\end{figure}

In Fig.~\ref{FigWP-NLK} we display
the wrapping probabilities for the nonlocal Kawasaki dynamics on a square lattice (in the left panel) and on a honeycomb lattice
(in the right panel), for different values of $L$, and at temperature $T_c/4$ in both cases.
Notice that for the  honeycomb lattice the probabilities
$\pi_{\rm h}$ and $\pi_{\rm v}$ are shown separately, since our construction of this lattice does not have
unit aspect ratio, as explained before.

\vspace{0.5cm}

As for the local version of the dynamics, the $\pi$s are plotted against the rescaled time
$t / ( L / \ell_d(t) )^{\zeta}$, with $\ell_d(t)$ taken to be the excess-energy growing length $\ell_G(t)$,
and the exponent $\zeta$ determined numerically searching for data collapse.
This is achieved with $\zeta \simeq 1.15$ on the square and honeycomb lattices.

As the system size increases, the data approach asymptotic values that are
very close to the ones of $2d$ critical percolation on the corresponding lattice, shown with dashed horizontal lines.
Note that very large system sizes are needed to see the approach to these values (the curves for $L=1280$ are still a bit away from them).
We do not observe, in the time window explored by our simulations, the final disappearance of the diagonally percolating clusters.

Another feature is that, at sufficiently long times,  $\pi_{\rm hv}$ tends to decrease while
$\pi_{\rm h} + \pi_{\rm v}$ increases.
This behaviour is expected since, as the system evolves under the GCOP dynamics, 
configurations consisting of two large clusters of opposite phase both percolating along the same 
direction of the lattice become more and more likely (this should be true independently of the lattice geometry).

By using the criterium $t_p / ( L / \ell_d(t_p) )^{\zeta} \simeq 10^4$
for the nonlocal Kawasaki dynamics at $T= T_c/4$ on the square lattice, and 
$t_p / ( L / \ell_d(t_p) )^{\zeta} \simeq 10^3$ for the same dynamics on the honeycomb lattice
(that corresponds, approximately, to the value of the scaling variable $t / ( L / \ell_d(t) )^{\zeta}$
at which the wrapping probabilities get close to the critical percolation ones) we are able to find estimates of $t_p$,
namely $ t_p \simeq 4.91 \cdot 10^5, \ 2.11 \cdot 10^6, \ 8.46 \cdot 10^6 $ for $ L=80, \ 320, \ 1280 $, respectively, for the case of the square lattice, and
$ t_p \simeq 3.32 \cdot 10^4, \ 6.36 \cdot 10^4, \ 1.20 \cdot 10^5, \ 2.23 \cdot 10^5 $ for $L=80, \ 160, \ 320, \ 640 $, respectively, for the case of the honeycomb lattice.

\subsection{Averaged squared winding angle}
\label{subsec:NLK-winding-angle}

\begin{figure}[h]
\begin{center}
\includegraphics[scale=0.5]{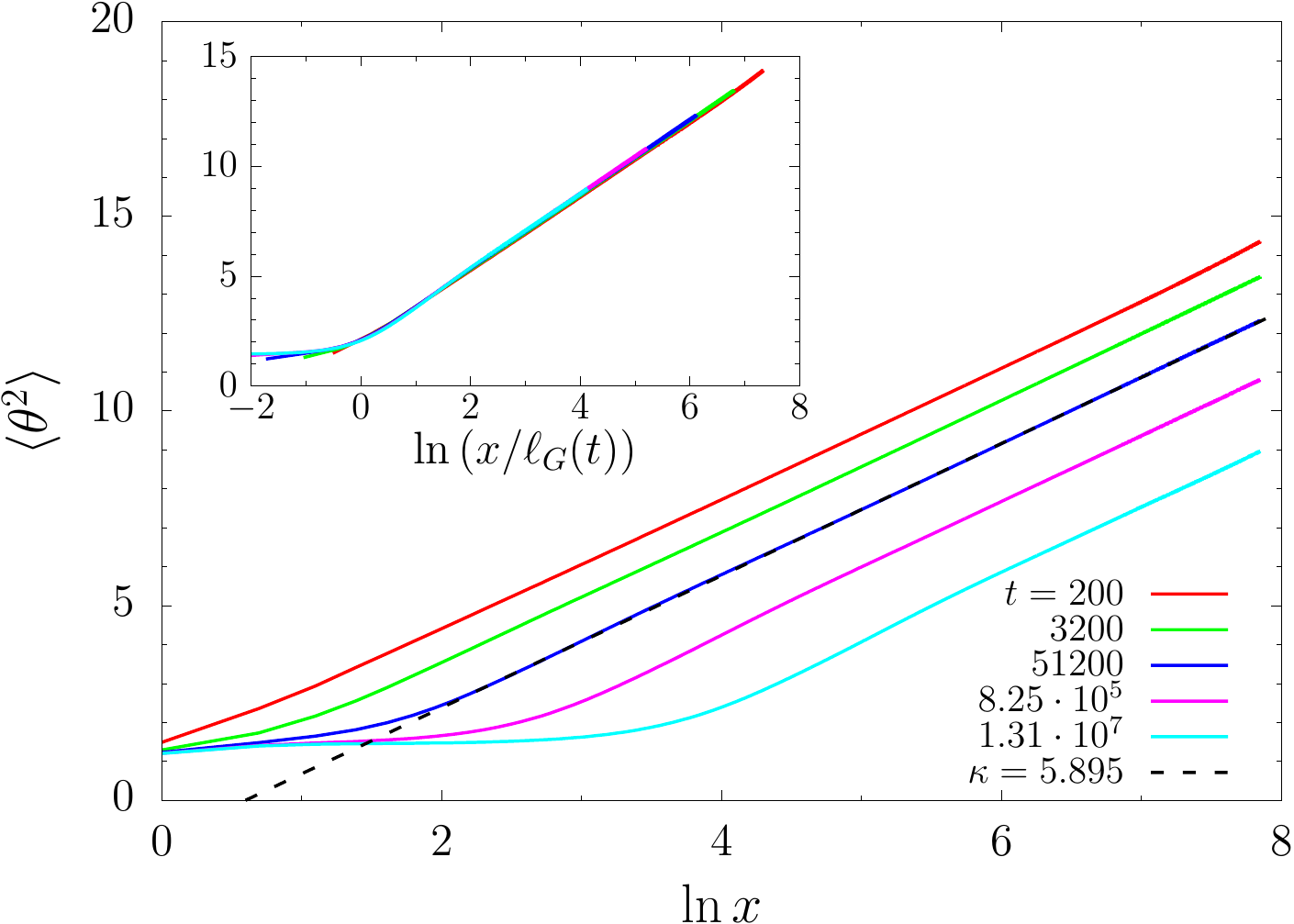}\quad%
\includegraphics[scale=0.5]{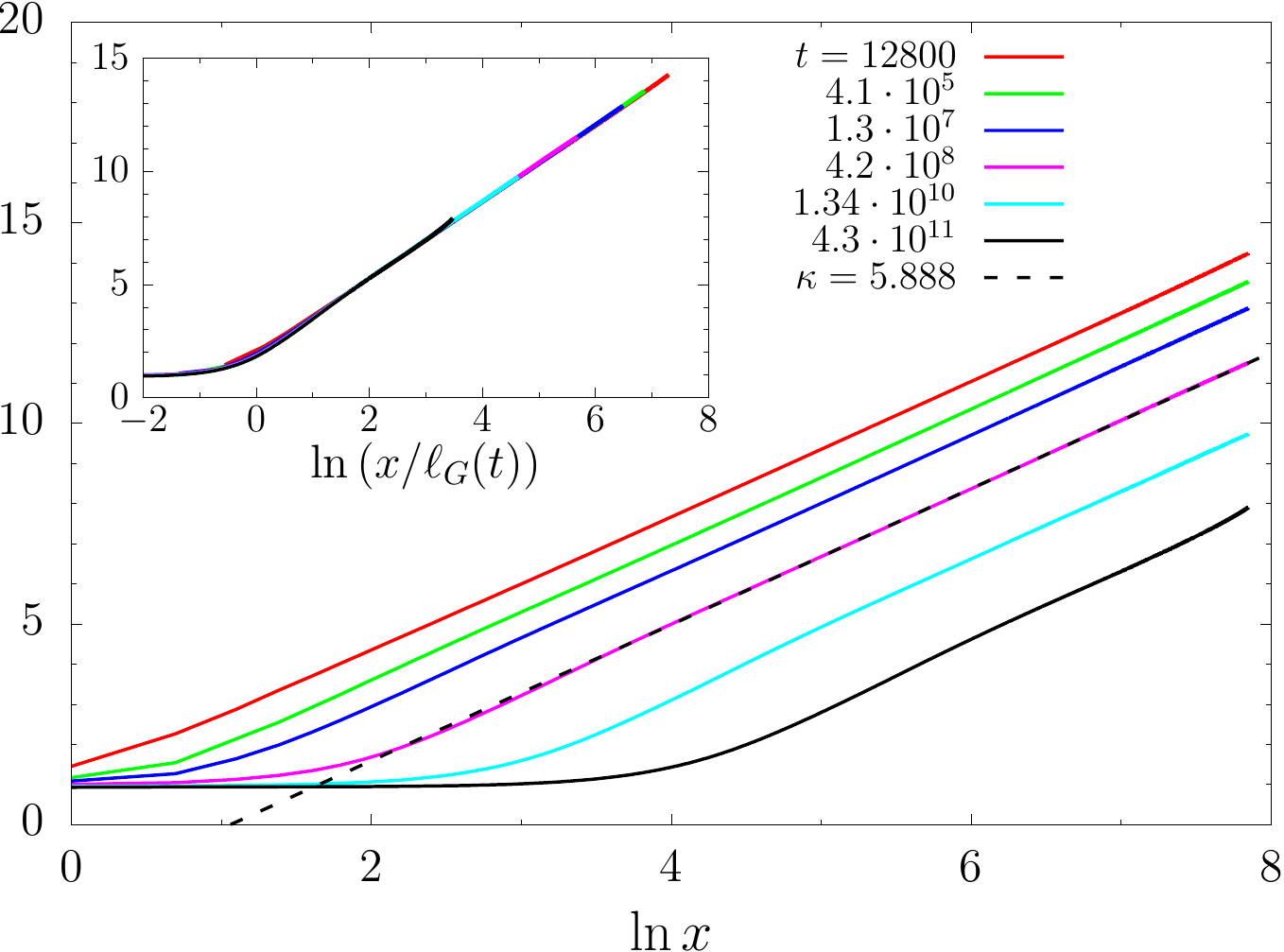}
\end{center}
\caption{\small Average squared winding angle, $\langle \theta^2 \rangle$, against $\ln{x}$ with $x$ the curvilinear distance
along a hull, for closed hulls (nonzero total winding angle). Model defined on a square lattice with $L=1280$
and nonlocal Kawasaki dynamics at $T = T_c/2$ (left panel) and $T=T_c/4$ (right panel). 
In the insets, the length $x$ is rescaled by $\ell_{G}(t)$, the characteristic length obtained as the inverse of the excess energy.
The black dashed lines are fits of the data (in the ``linear'' region) to
the function $f(x) = \mathrm{cst.} \ + \ 4\kappa / ( 8+\kappa ) \ln{x}$
yielding $\kappa\simeq5.895$ and $\kappa\simeq5.888$
for the quench to $T_c/2$ and the one to $T_c/4$, respectively. The times shown are the same in the main plots 
and the insets and are indicated in the keys.}
\label{FigW}
\end{figure}

\begin{figure}[h]
\begin{center}
\includegraphics[scale=0.56]{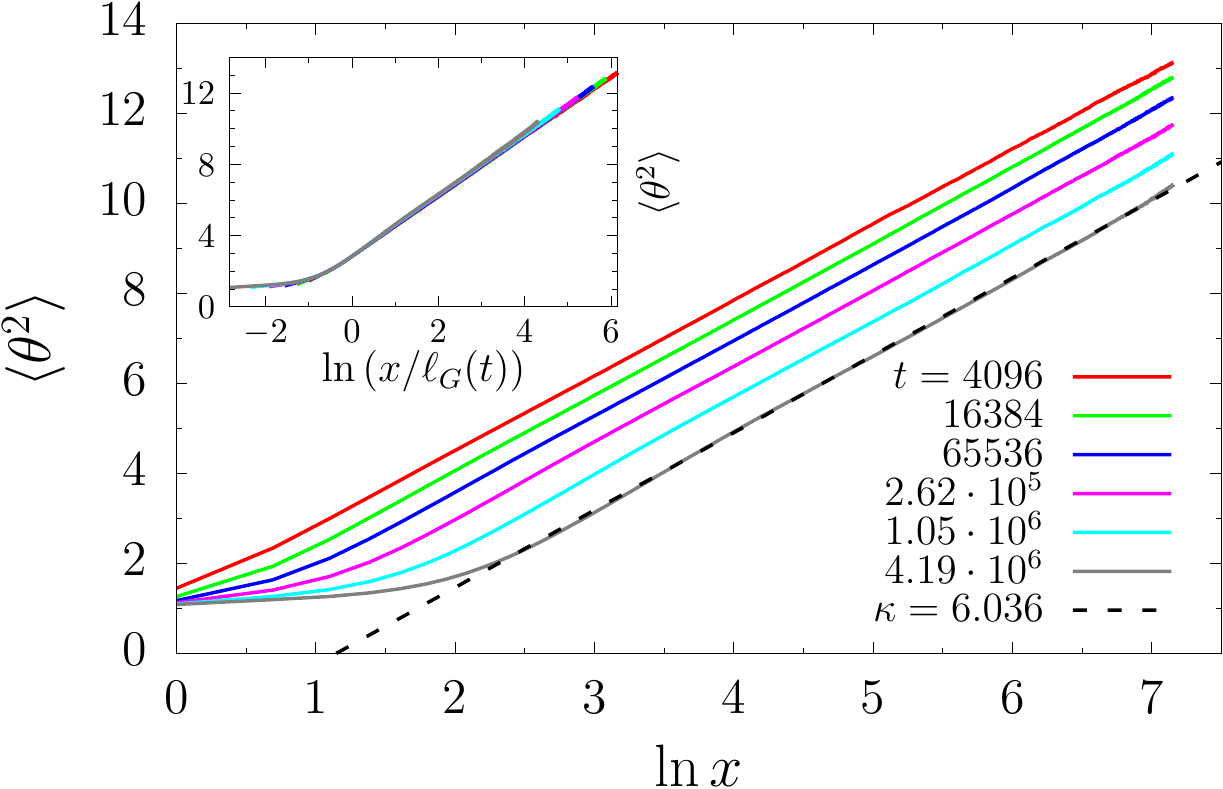}\quad%
\includegraphics[scale=0.56]{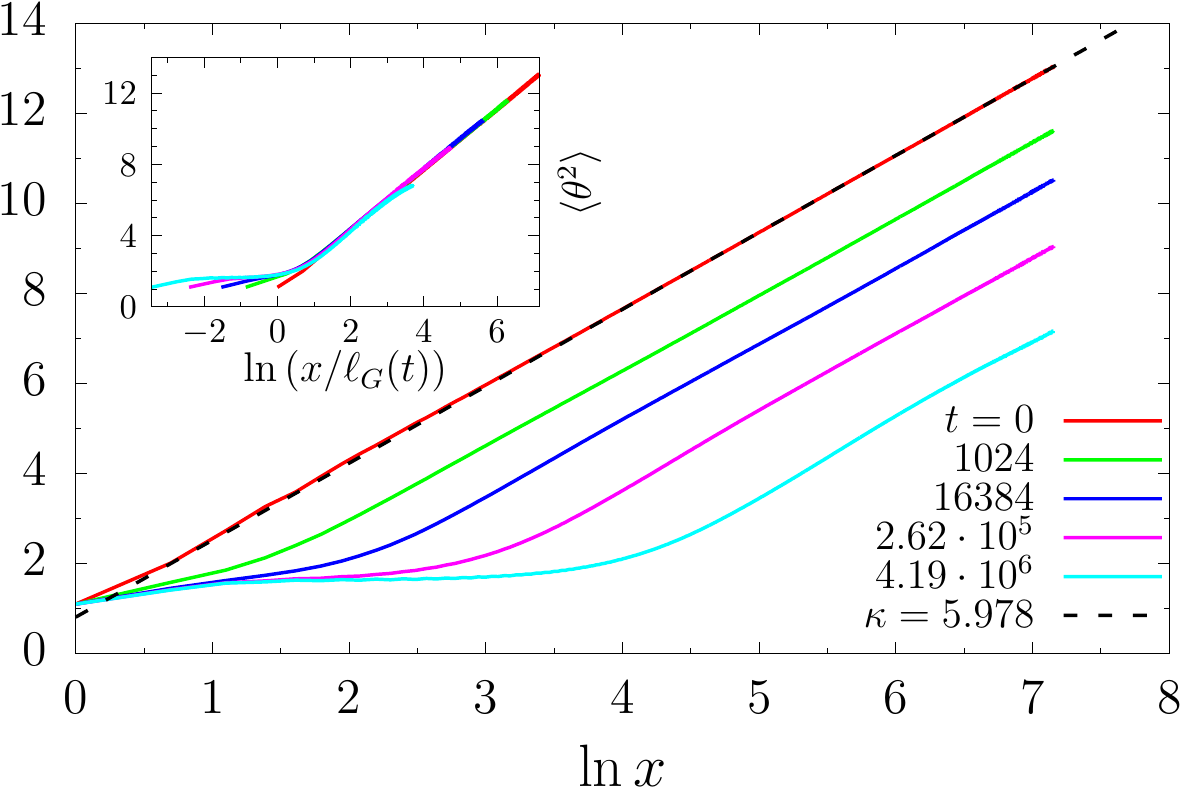}
\end{center}
\caption{\small Nonlocal Kawasaki dynamics at temperature $T_c/2$.
The average squared winding angle, $\langle \theta^2 \rangle$, against $\ln{x}$ 
with $x$ the curvilinear distance along a cluster hull, for wrapping hulls (zero total winding angle).
Dynamics on a honeycomb lattice (left panel) and on a triangular lattice (right panel), both with linear size $L=640$.
As in Fig.~\ref{FigW}, the black dashed lines represent a fit of the function 
$f(x) = \mathrm{cst.} \ + \ 4\kappa / ( 8+\kappa ) \ln{x}$ to the data in the ``linear'' region,
yielding $\kappa \simeq 6.036$ (honeycomb lattice) and $\kappa \simeq 5.978$ (triangular lattice).
In the insets  we report $\langle \theta^2 \rangle$ against
$\ln{(x/\ell_G(t))}$, with $\ell_G(t)$ the excess-energy growing length. The times shown are the same in the main plots 
and the insets and are indicated in the keys.
}
\label{NLK_WA_other}
\end{figure}

In  Fig.~\ref{FigW} we show $\langle \theta^2(x,t) \rangle$ as a function of the logarithm of the curvilinear distance $x$
along a cluster interface, for closed hulls (with nonzero total winding angle) on a square lattice. 
A very similar behavior is obtained for the case 
of wrapping hulls (with zero total winding angle) so we do not show these results.
The left panel displays data at $T_c/2$, while the right panel at $T_c/4$. 
This quantity is expected to behave as in Eq.~(\ref{eq:winding_angle_critical_hulls}).
A fit of the function $f(x) = \mathrm{const.} + 4\kappa / ( 8+\kappa ) \ln{x}$ to the data in the region where 
the dependence of $\langle \theta^2 \rangle$ on $\ln{x}$ is linear,
gives us an estimate of the $\mathrm{SLE}$ parameter: $\kappa \simeq 5.895$ for the dynamics at $T=T_c/2$ 
and $\kappa \simeq 5.888$ for the dynamics at $T=T_c/4$, results that are rather close to the expected value $\kappa=6$ for critical percolation hulls.
A similar analysis for the nonlocal Kawasaki dynamics on the honeycomb and triangular lattices at $T_c/2$ is shown in Fig.~\ref{NLK_WA_other}.
In the inset in each panel we present the same data as a function of $x/\ell_G(t)$,  with $\ell_G(t)$ being the 
excess-energy growing length.

\vspace{0.5cm}

\subsection{Largest cluster scaling}

\vspace{0.5cm}

\begin{figure}[h]
\begin{center}
 \includegraphics[scale=0.55]{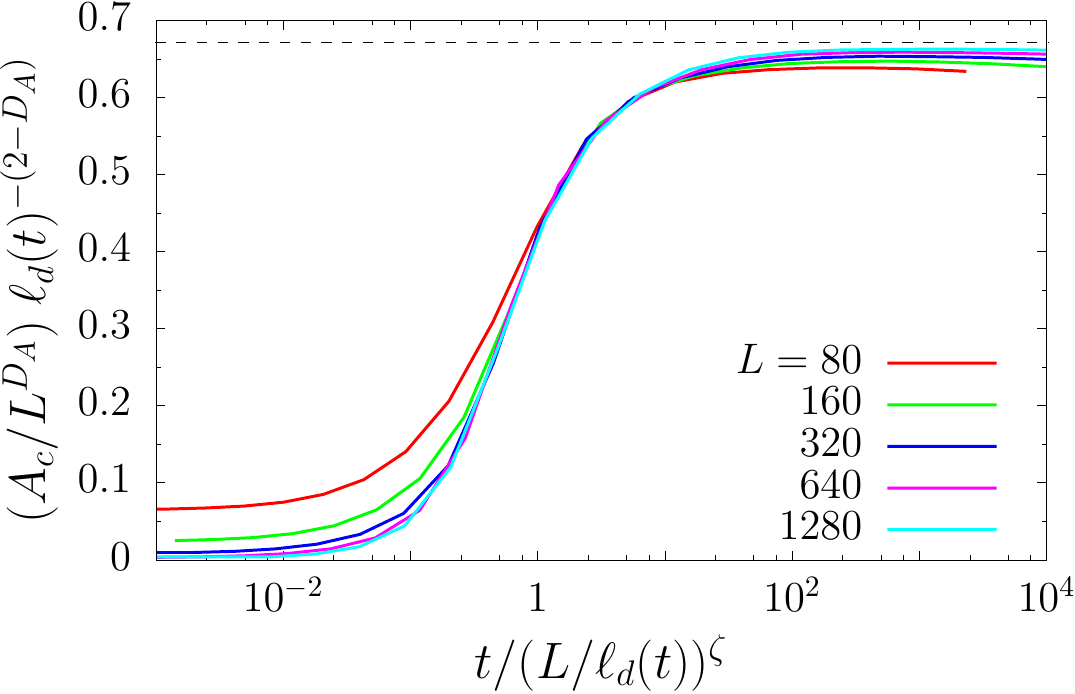}\quad%
 \includegraphics[scale=0.55]{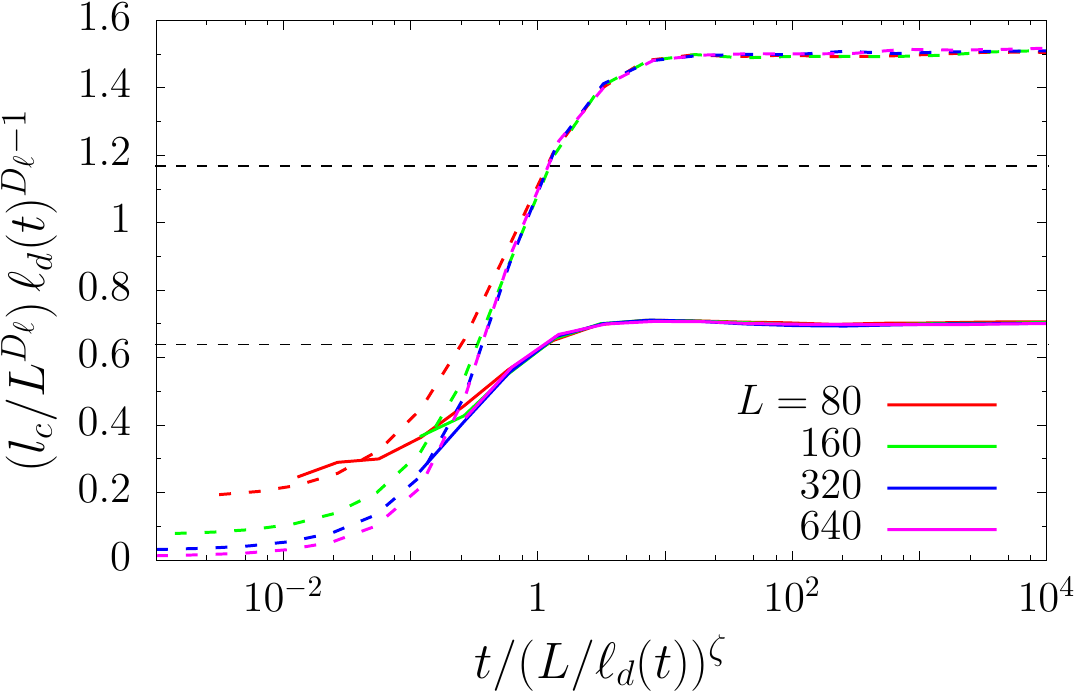}
\end{center}
\caption{\small
Nonlocal Kawasaki dynamics of the $2d$IM on the square lattice at  $T_c/2$.
In the left panel,  $ \left( A_c/L^{D_A} \right) \; \ell_d(t)^{-(2-D_A)}$, with $A_c$ the largest cluster size, while
in the right panel, $ \left( l_c/L^{D_{\ell}} \right) \; \ell_d(t)^{-(1-D_{\ell})}$,
with $l_c$ the length of the largest cluster interface.
Regarding $l_c$, we have separated the contribution of the wrapping hulls (zero total winding angle) shown with 
continuous lines, from the one 
of the non-wrapping hulls (nonzero total winding angle) shown with dashed lines. 
We took $\ell_d(t) = \ell_G(t)$, with $\ell_G(t)$
the  inverse excess-energy growing length.
All the quantities are plotted against the rescaled time $t / ( L / \ell_d(t) )^{\zeta}$
with $\zeta \simeq 1.16$ giving the best collapse for both $A_c$ and $l_c$.
The black dashed horizontal lines represent the value of the ratios $A_c/L^{D_A}$ (in the left panel) and 
$l_c/L^{D_{\ell}}$ (in the right panel) for critical site percolation on a square lattice (with PBC), computed with numerical simulations:
$A_c/L^{D_A} \simeq 0.6683$, while $l^{\mathrm{wrap}}_c/L^{D_{\ell}} \simeq 0.6383$ and $l^{\mathrm{non-wrap}}_c/L^{D_{\ell}} \simeq 1.1678$
for wrapping hulls and non-wrapping hulls, respectively.}
\label{NLK-LC}
\end{figure}

\begin{figure}[h]
\begin{center}
 \includegraphics[scale=0.55]{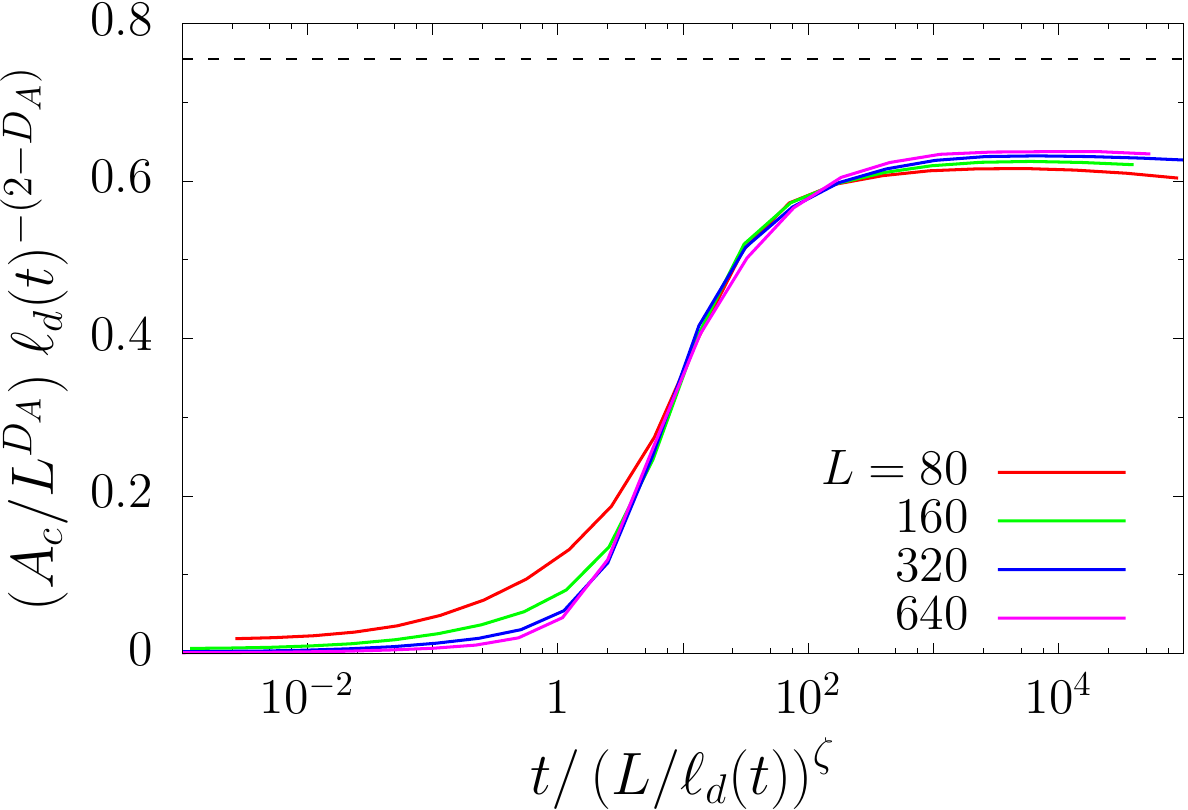}\quad%
 \includegraphics[scale=0.55]{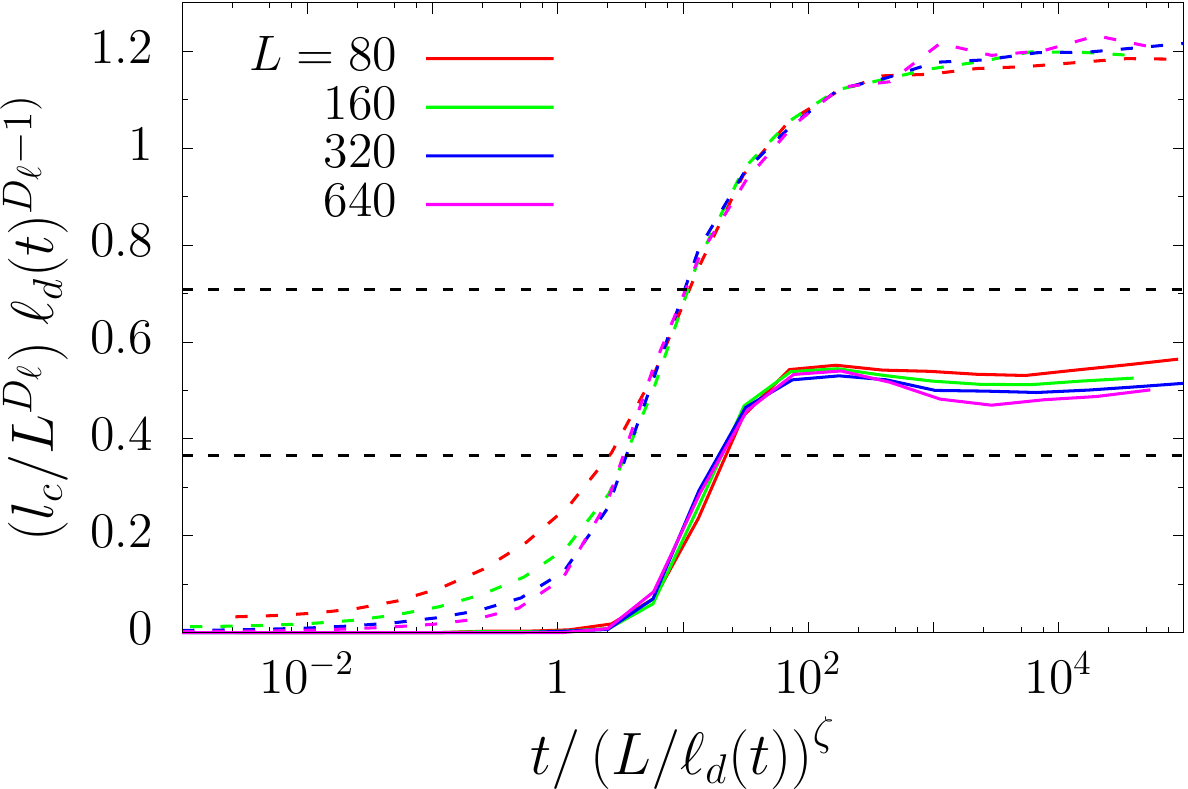}
\end{center}
\caption{\small
Nonlocal Kawasaki dynamics on the honeycomb lattice at  $T_c/2$.
In the left panel, $ \left( A_c/L^{D_A} \right) \; \ell_d(t)^{-(2-D_A)}$, with $A_c$ the largest cluster size, while
in the right panel, $ \left( l_c/L^{D_{\ell}} \right) \; \ell_d(t)^{-(1-D_{\ell})}$,
with $l_c$ the length of the largest cluster interface, separating the contribution due to wrapping hulls (zero total winding angle) from the one 
due to non-wrapping hulls (nonzero total winding angle). 
As in Fig.~\ref{NLK-LC}-right, the former is indicated with continuous lines, while the latter with dashed lines.
We took $\ell_d(t) = \ell_G(t)$, with $\ell_G(t)$
the characteristic length from the inverse of the excess-energy.
All the quantities are plotted against  $t / ( L / \ell_d(t) )^{\zeta}$, 
where $\zeta \simeq 1.17$ gives the best collapse in both cases.
The black dashed horizontal lines represent the value of the ratios $A_c/L^{D_A}$ (in the left panel) and 
$l_c/L^{D_{\ell}}$ (in the right panel) for critical site percolation on a honeycomb lattice (with PBC), computed with numerical simulations:
$A_c/L^{D_A} \simeq 0.7554$, while $l^{\mathrm{wrap}}_c/L^{D_{\ell}} \simeq 0.3658$ and $l^{\mathrm{non-wrap}}_c/L^{D_{\ell}} \simeq 0.7089$
for wrapping hulls and non-wrapping hulls, respectively. 
}
\label{NLK-LC_Hon}
\end{figure}

\begin{figure}[h]
\begin{center}
 \includegraphics[scale=0.55]{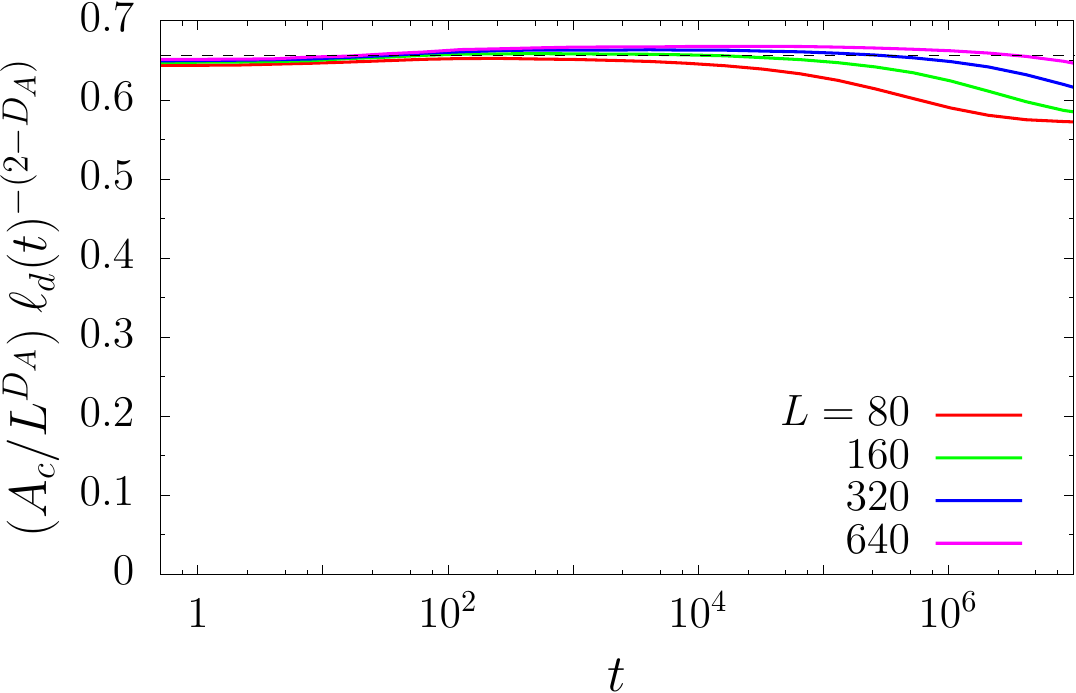}\quad%
 \includegraphics[scale=0.55]{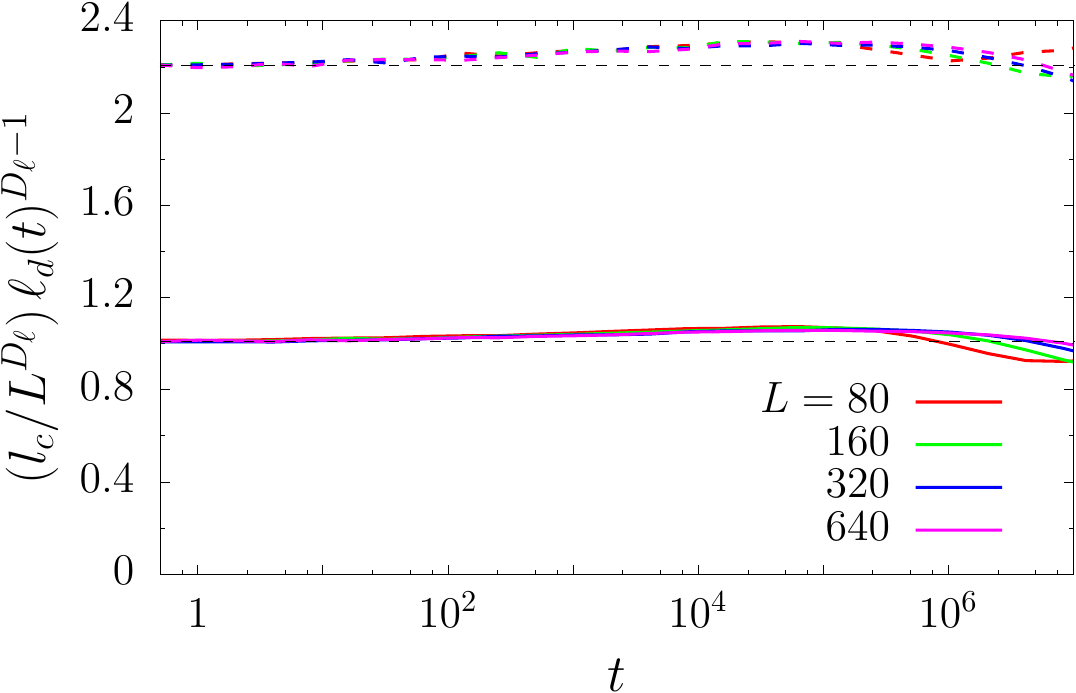}
\end{center}
\caption{\small
Nonlocal Kawasaki dynamics on the triangular lattice at $T_c/2$.
In the left panel,  $ \left( A_c/L^{D_A} \right) \; \ell_d(t)^{-(2-D_A)}$, with $A_c$ the largest cluster size, while
in the right panel, $ \left( l_c/L^{D_{\ell}} \right) \; \ell_d(t)^{-(1-D_{\ell})}$,
with $l_c$ the length of the largest cluster interface, separating the contribution due to wrapping hulls (zero total winding angle) from the one 
due to non-wrapping hulls (nonzero total winding angle). 
The former is indicated with continuous lines, while the latter is drawn with dashed lines, as in the previous figures.
We took $\ell_d(t) = \ell_G(t)$, with $\ell_G(t)$
the characteristic length obtained from the inverse of the excess-energy.
All the quantities are plotted against time $t$ as its rescaling is not necessary in this case.
The black dashed horizontal lines represent the value of the ratios $A_c/L^{D_A}$ (in the left panel) and 
$l_c/L^{D_{\ell}}$ (in the right panel) for critical site percolation on a triangular lattice (with PBC), computed with numerical simulations:
$A_c/L^{D_A} \simeq 0.6550$, while $l^{\mathrm{wrap}}_c/L^{D_{\ell}} \simeq 1.0075$ and $l^{\mathrm{non-wrap}}_c/L^{D_{\ell}} \simeq 2.2044$
for wrapping hulls and non-wrapping hulls, respectively.
}
\label{NLK-LC_Tr}
\end{figure}

We now discuss the scaling properties of the size of the largest spin cluster, $A_c$, and 
the length of its interface, $l_c$.
The data shown correspond to nonlocal Kawasaki dynamics at temperature $T_c/2$ on the square lattice (Fig.~\ref{NLK-LC}),
on the honeycomb lattice (Fig.~\ref{NLK-LC_Hon}) and on the triangular one (Fig.~\ref{NLK-LC_Tr}).
As in Sec.~\ref{subsec:LC-Kawasaki} for the local version of the constraint, we present the two contributions to the total length of the largest cluster interface, $l_c$,
the one coming from wrapping hulls (which exists only if the cluster is wrapping) and the one coming from non-wrapping hulls, as two separate quantities.

We would like to recover the same scaling relations expressed by Eqs.~(\ref{eq:Ac_scaling_2}) and~(\ref{eq:lc_scaling_2}).
Namely, the largest cluster size $A_c$ must be scaled as $ (A_c(t,L)/L^{D_A}) \, \ell_d(t)^{-(2-D_A)}$, 
while the length of its interface $l_c$ as $ (l_c(t,L)/L^{D_\ell}) \, \ell_d(t)^{-(1-D_{\ell})}$,
with $\ell_d(t)$ taken to be the excess-energy growing length $\ell_G(t)$ and
$D_A=91/48$ and $D_{\ell}=7/4$ the fractal dimensions of the largest cluster and 
the percolating hulls in $2d$ critical percolation, respectively.
Both quantities are plotted against the rescaled time $ t \left( L /  \ell_d(t) \right)^{\zeta} $.
Again, the value of the exponent $\zeta$
is determined by looking for the best collapse of the datasets corresponding to different $L$.

On the square lattice we find $\zeta \simeq 1.16$, while on the honeycomb lattice
$\zeta \simeq 1.17$, both values being compatible, within numerical accuracy, with the ones obtained studying the scaling
of the wrapping probabilities. On the triangular lattice, as already explained in Sec.~\ref{subsec:LC-Kawasaki},
the regime of approach to the critical percolation state is not present (since the spin configuration is already critical at $t=0$)
thus the scaling of time as $t \left( L /  \ell_d(t) \right)^{\zeta}$ fails with any $\zeta > 0$, signalling that there is no
additional time scale other than the usual one of coarsening. Thus in Fig.~\ref{NLK-LC_Tr}
the quantities $ (A_c/L^{D_A}) \, \ell_d(t)^{-(2-D_A)}$ and 
$ (l_c/L^{D_\ell}) \, \ell_d(t)^{-(1-D_{\ell})}$ are plotted against $t$.

As already done for the local version of the dynamics, we have also included
the value of the ratios $A_c/L^{D_A}$ and $l_c/L^{D_{\ell}}$ for critical site percolation on the corresponding lattices.
The rescaled largest cluster size, $ (A_c(t,L)/L^{D_A}) \, \ell_d(t)^{-(2-D_A)}$,
reaches a plateau 
at a value that is very close to $0.6683$ (indicated by a black dashed line)
which is the value of the ratio $A_c/L^{D_A}$ for critical site percolation on the square lattice (with PBC). 
For the rescaled $l_c(t,L)$ instead, we do not observe the same type of agreement neither for wrapping nor for
non-wrapping hulls. On the honeycomb lattice the deviations are large. 
This may be explained by the fact that we are using $\ell_G$, the excess-energy growing length, as an estimate of $\ell_d$, while
there could be an ``hidden`` proportionality factor $\alpha$ between $\ell_G$ and the true $\ell_d$ (with value depending on the lattice geometry), 
as already stater in Sec.~\ref{subsec:LC-Kawasaki},
whose effect is more pronounced in the case of the scaling of $l_c$ because $\ell_d$ appears with a larger exponent, namely as $\ell_d(t)^{D_{\ell}-1}$.
On the triangular lattice instead, both the rescaled largest cluster size, $ (A_c/L^{D_A}) \, \ell_d(t)^{-(2-D_A)}$, 
and the rescaled length of wrapping hulls of the largest cluster, $ (l_c(t,L)/L^{D_\ell}) \, \ell_d(t)^{-(1-D_{\ell})}$, coincide with the corresponding
ratios at critical percolation at $t=0$ (not shown because of the logarithmic scale), 
and remain close to those values for a long period of time.

\vspace{0.5cm}

\vspace{0.5cm}

\subsection{Number density of domain areas}
\label{subsec:subsec:num_den_dom_NLK}

We now focus on the study of the number density of domain areas, ${\cal N}(A,t, L)$.
As done in Sec.~\ref{subsec:num_dens_dom_Ka} for the local Kawasaki dynamics, we need to single out the triangular lattice, with 
no pre-percolating regime, from the other lattices where we  observe the regime of approach
to critical percolation in the form of an additional length scale $\ell_p(t)$ that grows faster than $\ell_d(t)$ and describes 
the scaling of domains with much larger area than $\ell^2_d(t)$.

\subsubsection{Triangular lattice.}

\begin{figure}[h]
\begin{center}
 \includegraphics[scale=0.65]{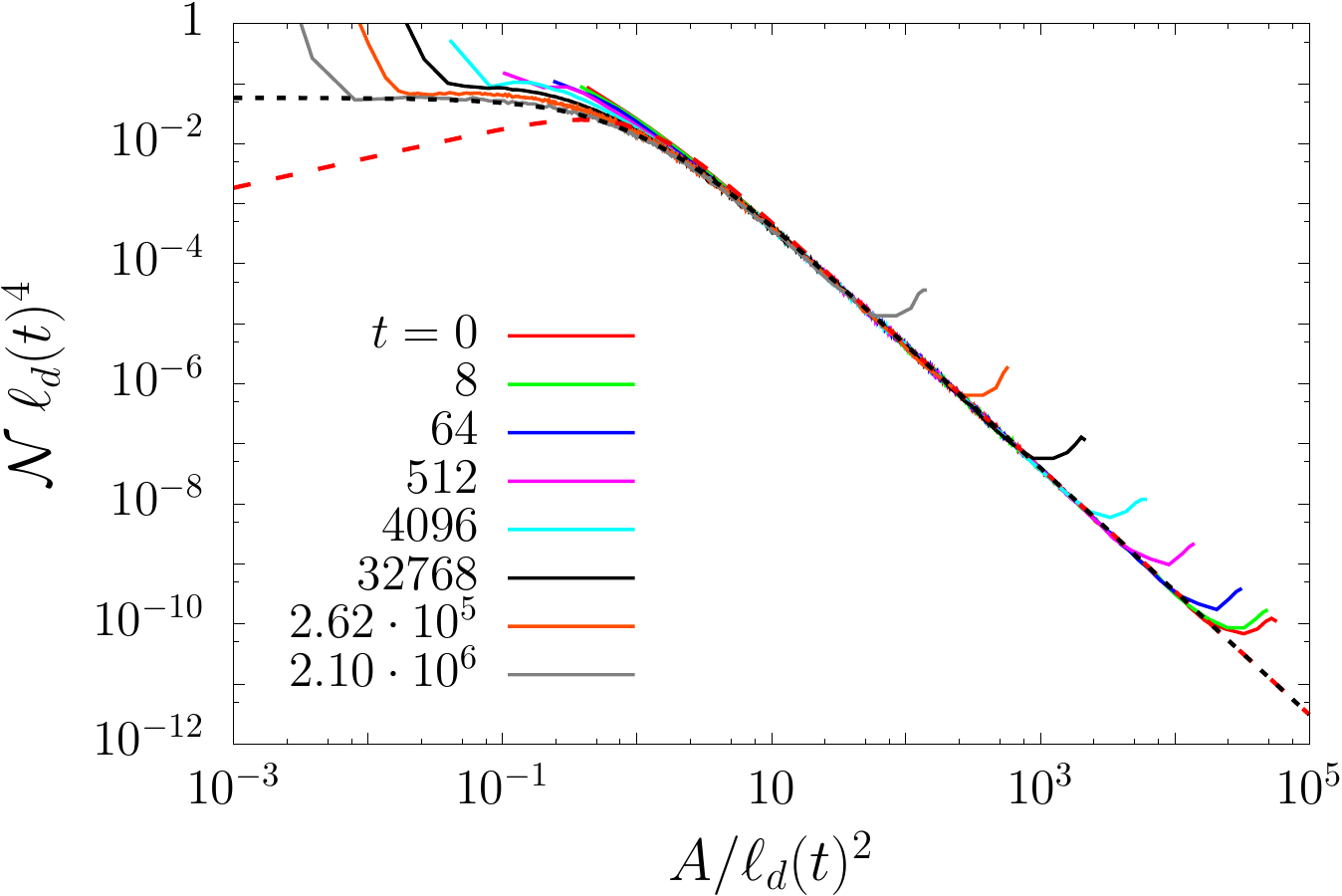}%
\end{center}
\caption{\small Scaling of the time-dependent number density of cluster areas  in a $2d$ Ising model on a 
triangular lattice evolved with the 
nonlocal Kawasaki dynamics at temperature $T_c/2$ ($L=640$).
We set  $\ell_d(t) \propto \ell_G(t)$ with $\ell_G(t)$ the inverse of the excess-energy. 
The proportionality constant is fixed to $1.66$.
The curves $f(x) = 2 \, c_d  \left( 1 +x \right)^{-\tau_A}$ and
$g(x) = 2 \, c_d \, x^{1/2} \, \left( 1 +x^{3/2} \right)^{-(2 \, \tau_A+1)/3}$
are represented by dashed lines.
The dataset for $t=2.10 \cdot 10^6$ matches the curve $f(x)$ (black dashed) in the region corresponding to $A/\ell_d(t)^2 < 1$,
for not too small domain areas (for which effects due to the discreteness of the system cause a deviation from the master curve).
}
\label{NLK_Tr_NA}
\end{figure}

We show in Fig.~\ref{NLK_Tr_NA}  the rescaled domain area number density, $\mathcal{N}(A,t) \, \ell^4_d(t)$
against $A/\ell^2_d(t)$, on the triangular lattice.
The characteristic length $\ell_d(t)$ is taken to be proportional to $\ell_G(t)$, $\ell_d(t) = \alpha \ \ell_G(t)$, 
with the constant $\alpha$ fixed by looking at the value that makes the rescaled data collapse onto a master curve,
$\alpha \simeq 1.66$. In the plot one can see two master curves.
The one of the NCOP class, 
$f(x) = 2 \, c_d \,  \left( 1 + x\right)^{-\tau_A}$,
and the one of the LCOP dynamics 
$g(x) = 2 \, c_d \, x^{1/2} \, \left( 1 + x^{3/2} \right)^{-(2\, \tau_A +1)/3}$ (see Eq.~(\ref{eq:NA_scaling_functions})).
These functions differ only at small values of the scaling variable $x$, yielding  $\sim 2 c_d$ and $\sim 2 c_d \ x^{1/2}$, respectively.
The data for the longest time shown, $t=2.10 \cdot 10^6$, matches very accurately the 
function $f(x) = 2 \, c_d \,  \left( 1 + x\right)^{-\tau_A}$, also shown in the plot in the 
full range of variation of the scaling variable. A similar result was found in the experiments in~\cite{Sicilia-etal08}.
This result confirms the fact that, for sufficiently high temperature (below $T_c$), the domain growth process caused
by nonlocal spin-exchange dynamics falls in the universality class of NCOP coarsening.

\vspace{0.5cm}

\subsubsection{Pre-percolating scaling}
\label{subsubsec:NLK_prepercolating}

\begin{figure}[h]
\begin{center}
 \includegraphics[scale=0.55]{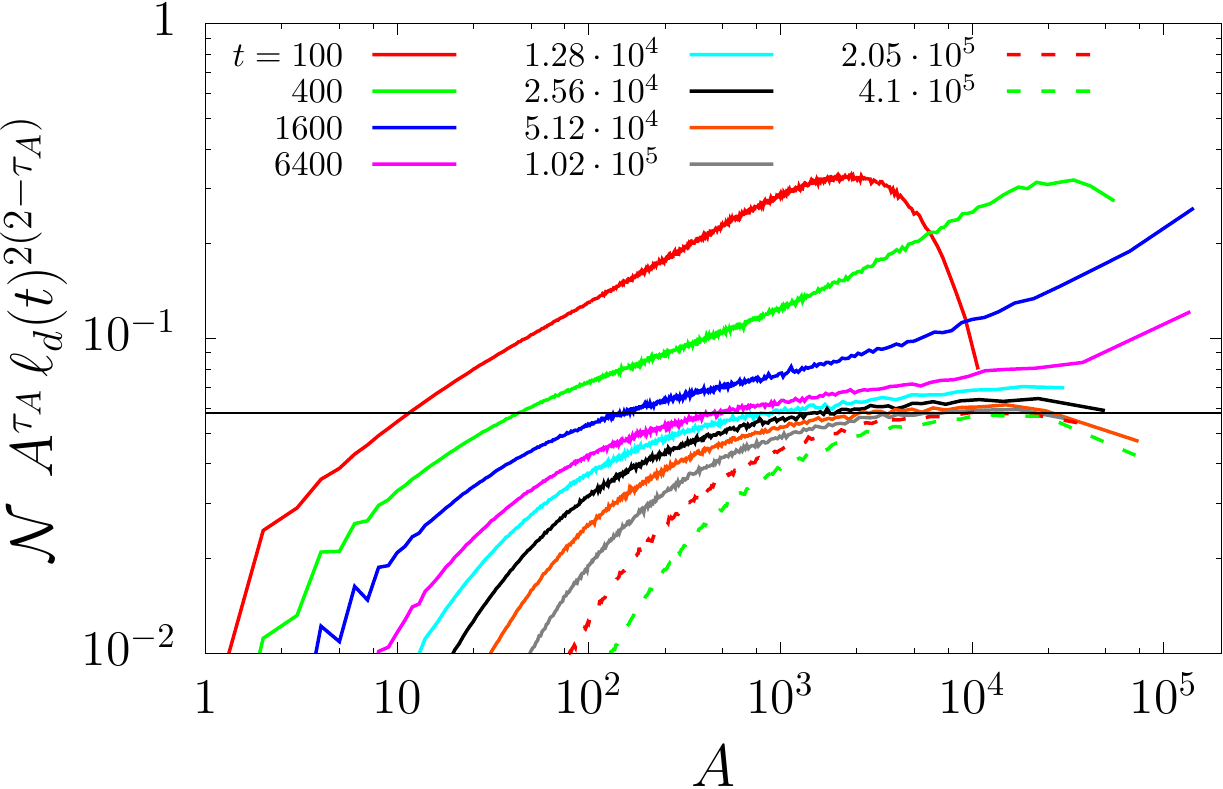}\quad%
 \includegraphics[scale=0.54]{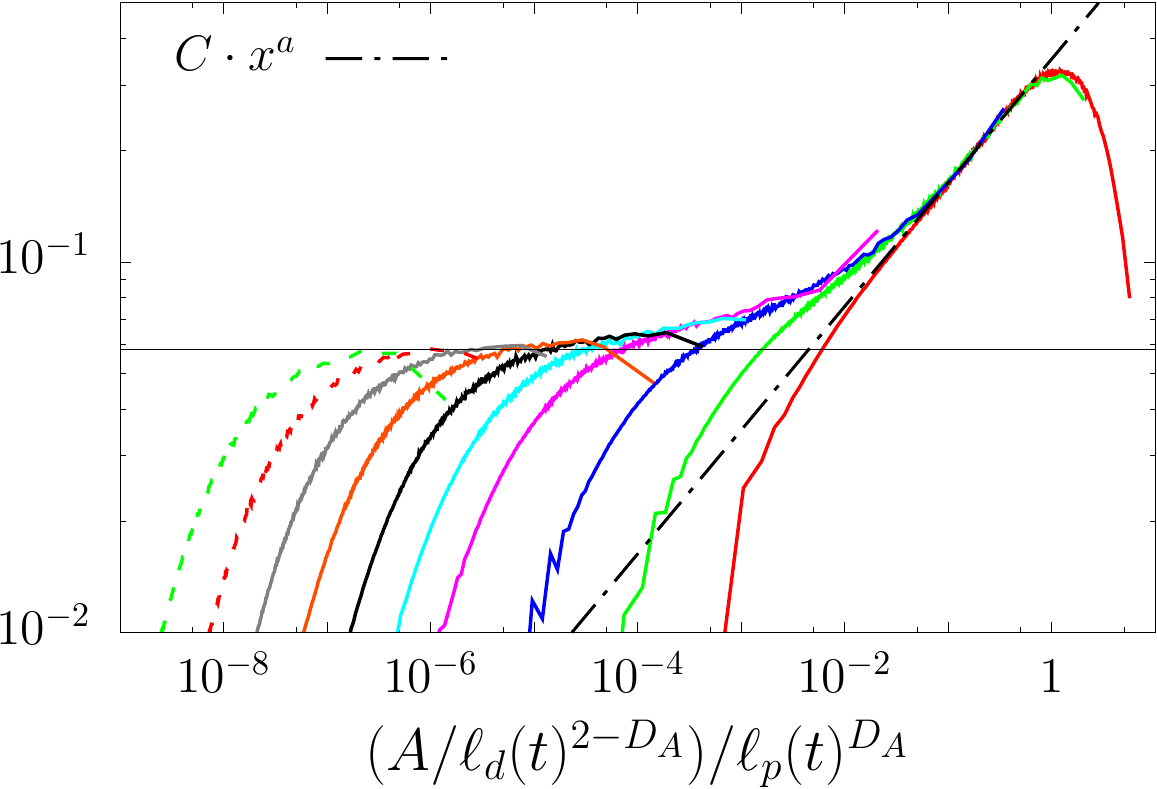}
\end{center}
\caption{\small 
Nonlocal Kawasaki dynamics on a square lattice with linear size $L=1280$, at $T_c/2$ and
50:50 up-down spin mixture. We show the number density of cluster areas, ${\mathcal N}(A,t, L)$, at different times given in the key in the left panel. 
In the left panel we plot
${\mathcal N}(A,t, L) \, A^{\tau_A} \; \ell_d(t)^{2(2- \tau_A)}$,
where $\ell_d(t) = \alpha \ell_G(t)$ and $\alpha \simeq 0.77$. The constant $2 c_d \approx 0.0580$ is indicated by the black horizontal line.
In the right panel, the same quantity is plotted against the rescaled area
$ \left( A / \ell_d(t)^{2-D_A} \right) / \ell_p(t)^{D_A}$, with $\ell_p(t) = \ell_d(t) \, t^{1/\zeta}$ with $\zeta \simeq 1.20$.
The function $\Phi(x) = C \, x^{a}$ has been fitted to the data at $t=100$ in the interval $[0.01,1]$ yielding $a \simeq 0.333$.
}
\label{FigNA2}
\end{figure}

In Fig.~\ref{FigNA2} we display the rescaled number density of domain areas,
$A^{\tau_A} \ \ell_d(t)^{2(2-\tau_A)} \; {\mathcal N}(A,t, L)$, against $A$ (left panel)
and  against the rescaled size $ ( A/ \ell_d(t)^{2-D_A}) \,/ \ell_p(t)^{D_A}$ (right panel), for the nonlocal Kawasaki dynamics
on a square lattice with $L=1280$, at  $T=T_c/2$.
As in Sec.~\ref{subsec:num_dens_dom_Ka}, for the case of the local version of the dynamics, the idea behind the scaling of the vertical axis
is to highlight the presence of the critical-percolation-like regime in the form of a plateau for sufficiently long $t$ and large $A$.
In the particular example shown in Fig.~\ref{FigNA2}, the onset of this regime occurs around $t=5 \cdot 10^4$.
We used $\ell_d(t) = \alpha \, \ell_G(t)$ with $\alpha$ a proportionality factor needed to match the height of the plateau at  
the latest time and the constant $2 c_d \approx 0.0580$ (indicated by the black horizontal line).

\vspace{0.5cm}

In the right panel of Fig.~\ref{FigNA2}, we use the relation $\ell_p(t) = \ell_d(t) \, t^{1/\zeta}$.
As we have already explained, 
scaling the cluster size $A$ as $ ( A/ \ell_d(t)^{2-D_A}) \,/ \ell_p(t)^{D_A} = A/\ell^2_d(t)\, \left( \ell_d(t)/\ell_p(t) \right)^{D_A}$
serves to highlight the presence of the pre-percolation regime, represented by the ``shoulder'' in the rescaled distribution. 
The exponent $\zeta$ was fixed by looking for the value that makes the datasets corresponding to different times collapse in this scaling region. The value
that gives the best result is $\zeta \simeq 1.20$ which, although not coincident with the one obtained through the scaling of the other observables, 
is not very far from it, taking into account that the  method used to estimate it is very rough.
Moreover, in the pre-percolating scaling regime the rescaled distribution can be fitted to 
$\Phi(x) = C\, x^a$ with $a \approx 0.333$ (indicated by a black dashed line in the right panel of Fig.~\ref{FigNA2}), 
a value that is very close to the one found for the other dynamic rules already studied.

\begin{figure}[h]
\begin{center}
 \includegraphics[scale=0.56]{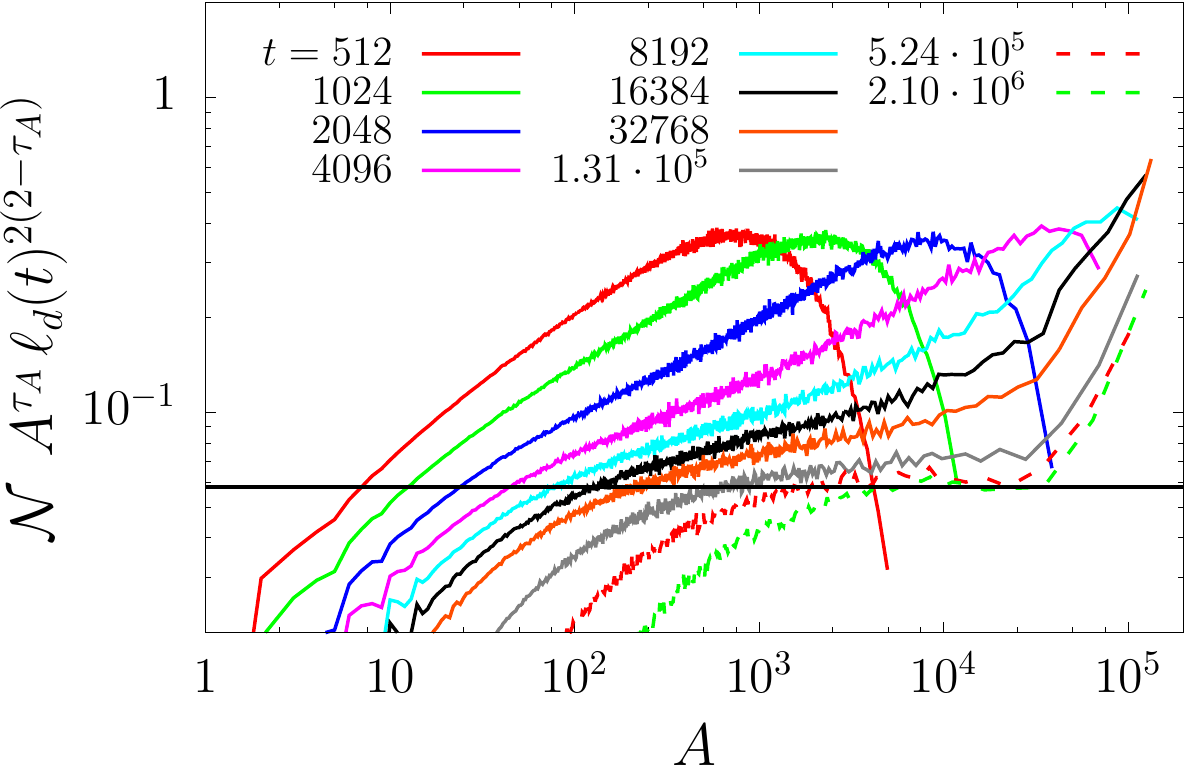}\quad%
 \includegraphics[scale=0.55]{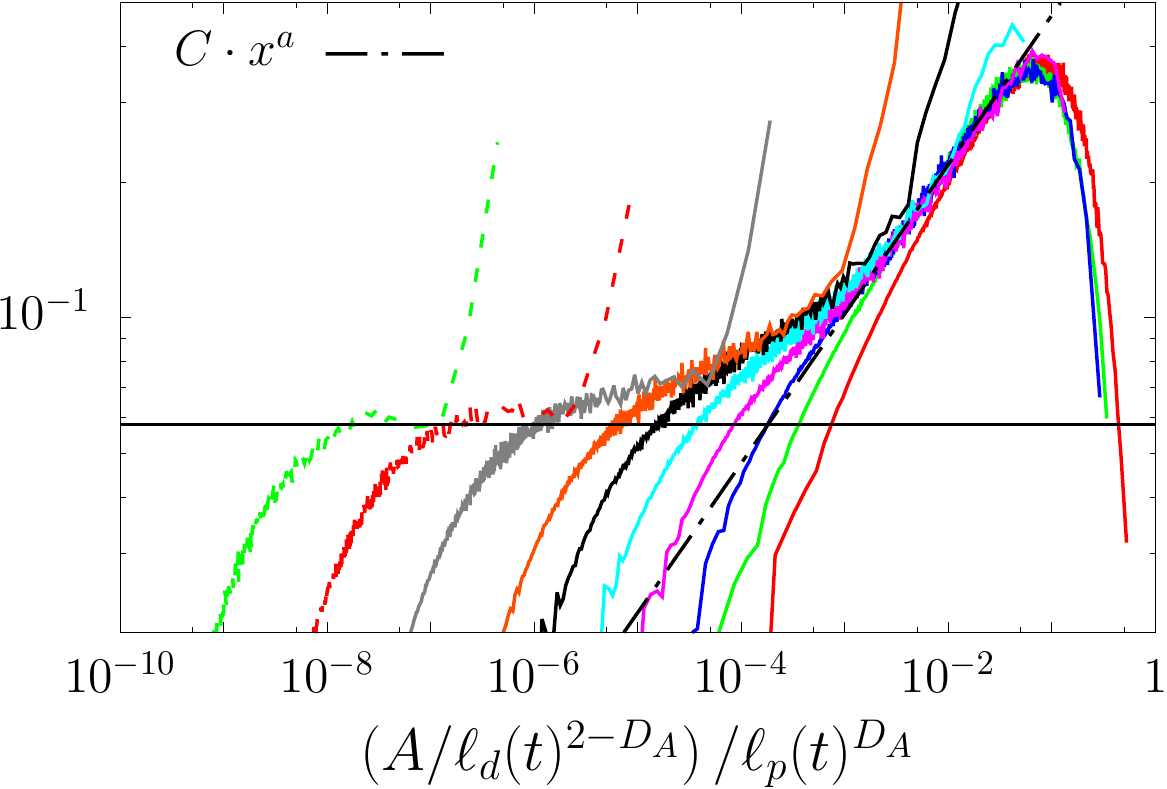}
\end{center}
\caption{\small 
Nonlocal Kawasaki dynamics on a honeycomb lattice with linear size $L=640$, at  $T_c/2$ and
50:50 up-down spin mixture. The number density of cluster areas, ${\mathcal N}(A,t, L)$, at different times given in the key in the left panel. 
In the left panel,
${\mathcal N}(A,t, L) \, A^{\tau_A} \; \ell_d(t)^{2(2- \tau_A)}$,
where $\ell_d(t) = \alpha \ell_G(t)$ and $\alpha \simeq 0.33$ against the area $A$. 
The constant $2 c_d \approx 0.058$ is indicated by the black horizontal line.
In the right panel, the same quantity against the rescaled area
$ \left( A / \ell_d(t)^{2-D_A} \right) / \ell_p(t)^{D_A}$, with $\ell_p(t) = \ell_d(t) \, t^{1/\zeta}$ with $\zeta \simeq 1.18$.
The function $\Phi(x) = C \, x^{a}$ has been fitted to the data at $t=1024$ in the interval $[10^{-3},10^{-2}]$ yielding $a \simeq 0.332$.
}
\label{NLK_Hon_NA}
\end{figure}

A similar behaviour is observed on the honeycomb lattice. 
In Fig.~\ref{NLK_Hon_NA} we show the rescaled number density,
$A^{\tau_A} \ \ell_d(t)^{2(2-\tau_A)} \; {\mathcal N}(A,t, L)$, for the nonlocal Kawasaki dynamics on a lattice with linear size $L=640$, at  $T=T_c/2$.
As in Fig.~\ref{FigNA2}, in the left panel the rescaled distribution is plotted against $A$, while in the right one, against the rescaled
size $ ( A/ \ell_d(t)^{2-D_A}) \,/ \ell_p(t)^{D_A}$. Here again $\ell_p(t) = \ell_d(t) \, t^{1/\zeta}$ and
$\ell_d(t) = \alpha \, \ell_G(t)$, with $\alpha \simeq 0.33$ ensuring that the plateau
appearing for large times ($t= 2.1 \times 10^6$ as a reference in the plot) matches the constant $2 \, c_d = 0.0580$.
The best collapse of the data in the ``shoulder'' region is achieved by using $\zeta \simeq 1.18$.
The function $\Phi(x) = C \, x^{a}$ was also fitted to the data at $t=1024$ in an interval inside the pre-percolating scaling region
yielding $a \simeq 0.332$.

\vspace{0.5cm}

\subsection{Summary}

We conclude by stating that the nonlocal spin-exchange moves do not change significantly the global picture that we are building. 

For the system sizes and longest time scales used
the effective exponent $z_{\rm eff}$ in the time-dependence of the excess energy growing length approaches $1/2$ at
$T_c/2$, it goes slightly above  ${1/3}$
at $T\geq T_c/4$, and it stays close to $1/3$ at $T_c/8$, 
within numerical accuracy. Only at $T_c/2$ it gets close to $t^{1/2}$, the law expected from renormalisation and 
approximate calculations of the corresponding field theory~\cite{Bray94b,Sire95,Rutenberg96}. 

These uncommon dynamics, in which the order parameter is conserved only 
globally and permits nonlocal energy transfer, also takes the spin configurations to the critical percolation ones. 
The scaling analysis of the wrapping probabilities, the size of the largest cluster, the length of the largest cluster hull
and the number density of cluster areas, indicates that
\begin{equation}
 \ell_p(t) \sim \ell_d(t) \, t^{1/\zeta} \; ,  \qquad \mbox{with} \quad \zeta \in [1.15,1.20] \;
\end{equation}
in the case of the dynamics on the square and honeycomb lattices.
The fact that the exponent takes approximately the same value for these two cases suggests that 
the nonlocality of the spin update rule is such that the characteristic length $\ell_p(t)$ is not affected much (apart from the 
factor $\ell_d(t)$ which depends on the geometry of the lattice, at least in the early time regime)
by the number of neighbours that each lattice site has, in contrast to what happens for the local version of the dynamics and
the NCOP dynamics, where different values of $\zeta$ are found.
On the triangular lattice no rescaling of this type is needed, since there is no pre-percolating regime, as already explained.
The winding angle variance scales as expected with $\ell_G(t)$ and,
for long curvilinear distances, it predicts a $\kappa$ parameter very close to the critical percolation one.
The number density of domain areas conforms to the scaling function of the NCOP dynamics, as in the
experimental realisation in~\cite{Sicilia-etal08}.

\section{Voter dynamics}
\label{sec:local-voter}

The voter model (VM)~\cite{CliffordSudbury73,HolleyLiggett75,Liggett99}
is a purely dynamical stochastic system, used to describe
the kinetics of catalytic reactions~\cite{Krapivsky92,Krapivsky92b,FrachebourgKrapivsky96} and as a prototype model of 
opinion and population dynamics~\cite{Vazquez03,FernandezGracia14,Korolev-et-al-2010}. 
In its simplest realisation, a bi-valued opinion variable, $s_i=\pm 1$, is 
assigned to each site on a lattice or graph with probability $1/2$ and an
equal number of one and the other state is enforced. 
At each subsequent microscopic time step, a variable chosen at random  
adopts the opinion of a randomly-chosen first neighbour. Therefore, the probability of the chosen spin to 
flip in a time step is simply given by the fraction of neighbours with opposite orientation. 
These moves mimic, in a very simple fashion, the influence of the neighbourhood on the individual opinion. 
The model is parameter free and invariant under global inversion of the spins. As a site 
surrounded by others sharing the same
opinion cannot fluctuate, there is no bulk noise and the dynamics are uniquely 
driven by interfacial fluctuations. The states of the system evolve through a coarsening 
process. In the long-time limit, $t_{\rm abs} \simeq L^2 \ln L$, the systems approach one of two 
absorbing states of full consensus.

Actually, the voter and ferromagnetic Ising models are two particular instances of a family of 
stochastic systems with bimodal variables, up/down symmetry, and isotropic and short-ranged 
interactions that determine the single spin flip updates.  This class of models is defined by two-parameter
dependent transition rates between configurations. (The two parameters basically control bulk and interfacial noise.)
The voter and Ising cases correspond to particular choices of these parameters~\cite{Oliveira,DrouffeGodreche99,GodrechePicco18}
and a remarkable difference is that the rates do not satisfy detailed balance in the voter model and they do in the Ising case.
It is then interesting to study these models side by side, as we do here, and to even go beyond this and analyse the behaviour of 
models in the full two-parameter family (this will be done in~\cite{GodrechePicco18}).

A thorough numerical investigation~\cite{TaCuPi15} of numerous macroscopic observables (interface density, 
space-time correlation, persistence, condensation time, {\it etc.}) of the square lattice VM 
corroborated the analytic predictions of the papers cited in the first paragraph.
The approach to critical percolation was also briefly discussed in~\cite{TaCuPi15} using a square lattice geometry. 
In this paper, we estimated (admittedly with a superficial analysis) that the appearance of percolation occurs at a time $t_p \simeq L^{z_p}$ with 
$ z_p \simeq 1.67$. This time was determined by considering the scaling of the total number of wrapping domains, the one of the largest area, 
$A_c$, and longest hull, $l_c$, only.

In this work, we reformulate our statements on the approach to percolation making them much more detailed and precise. This is 
achieved by the study of quantities which will allow us to characterise a critical intermediate state with considerable accuracy.  
In particular, the measure of the average squared winding angle will show that two criticalities characterise the interface lengths: 
a first one, that corresponds to percolation, appears after a very short  time span, 
and a second one, with different scaling properties, later conquers all the length scales growing from the short ones.
Importantly enough, the evaluation of the wrapping probabilities
will  prove that there is no long-lasting lapse over which critical percolation is established, contrary to what 
happens for the NCOP and COP dynamics of the $2d$IM.

\subsection{Snapshots}
\label{subsec:VM-snapshots}

\begin{figure}[h]
\begin{center}
  \subfloat[$t=0$]{\includegraphics[scale=0.33]{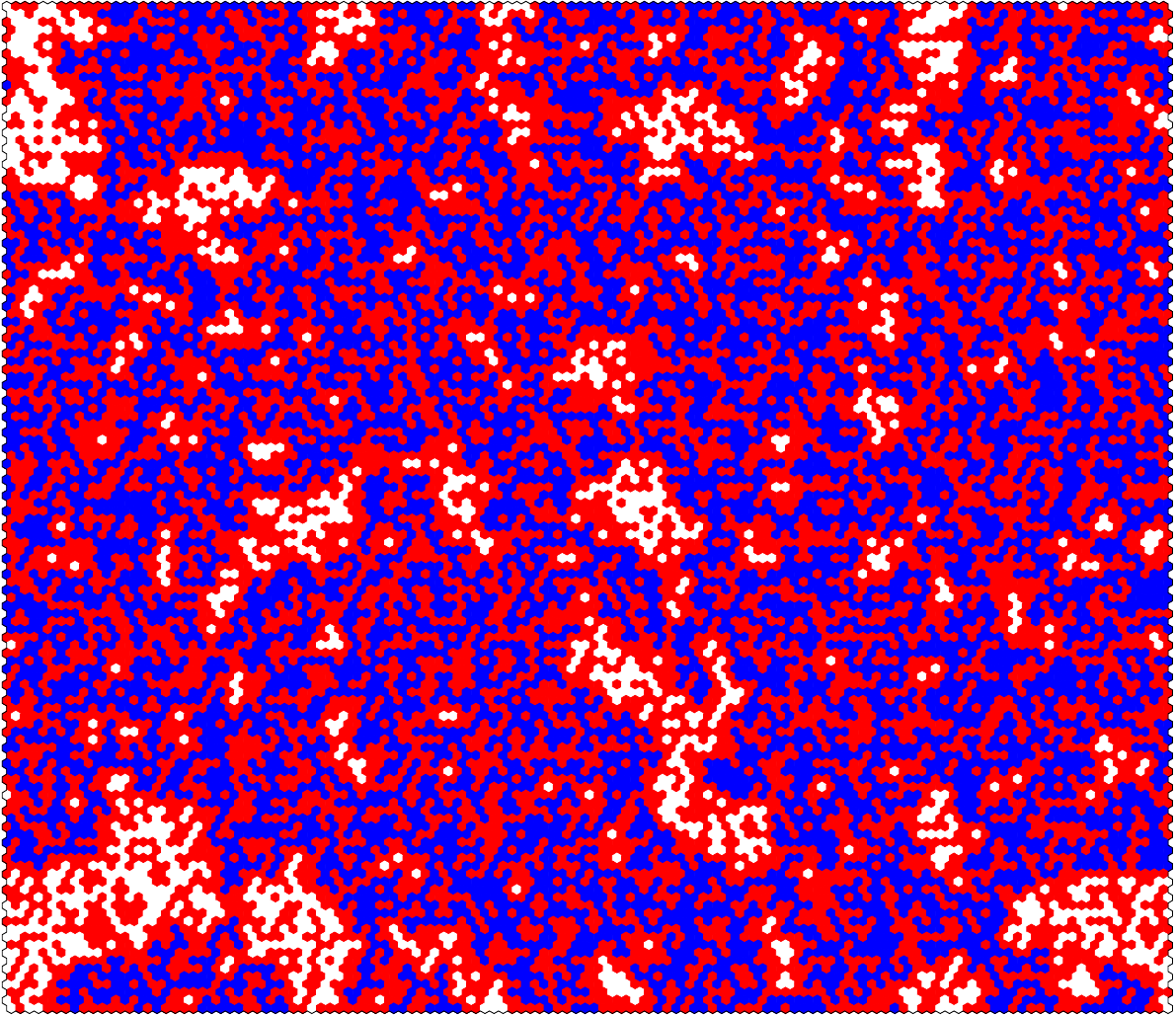}}\quad%
  \subfloat[$t=4$]{\includegraphics[scale=0.33]{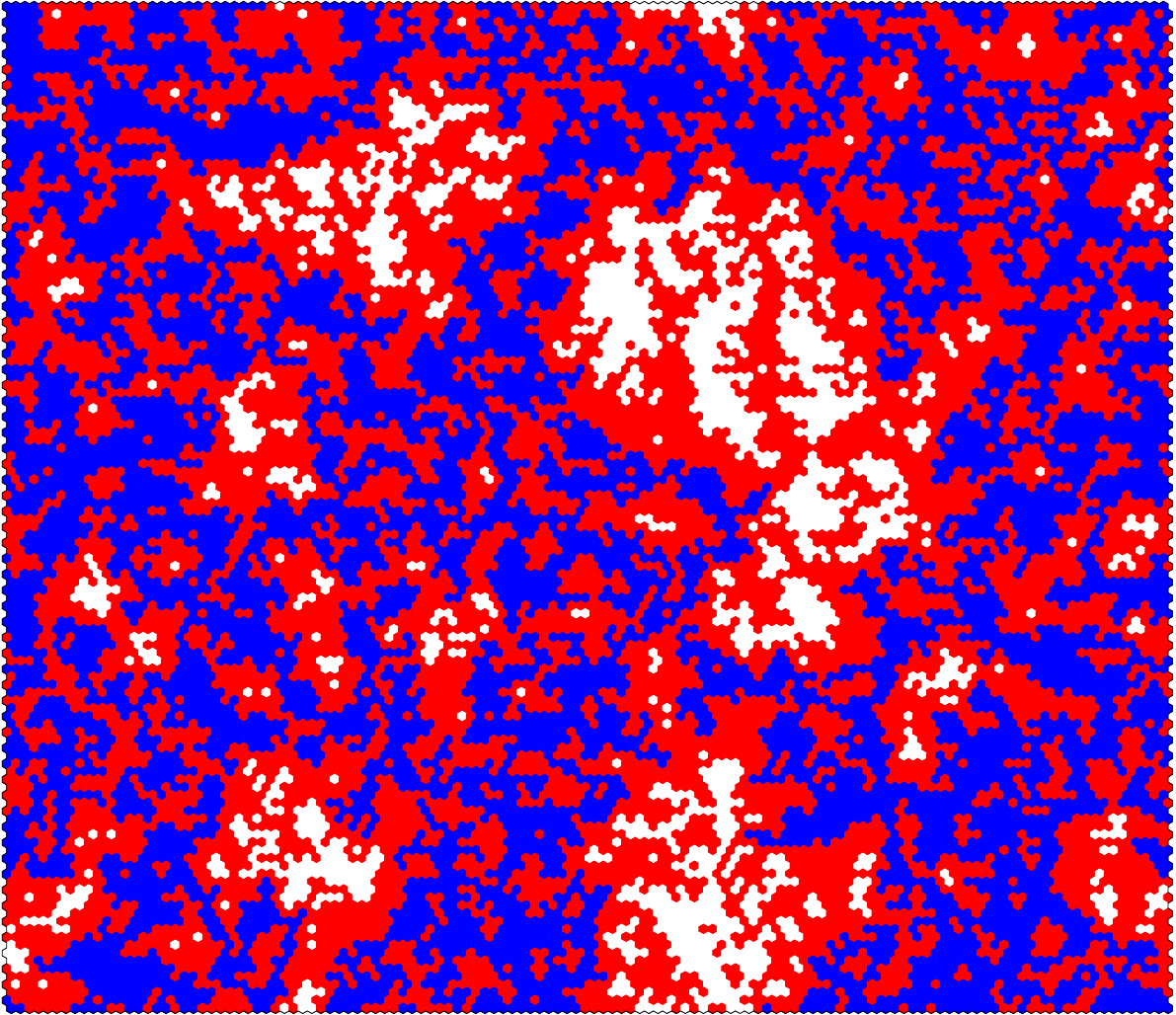}}\quad%
  \subfloat[$t=16$]{\includegraphics[scale=0.33]{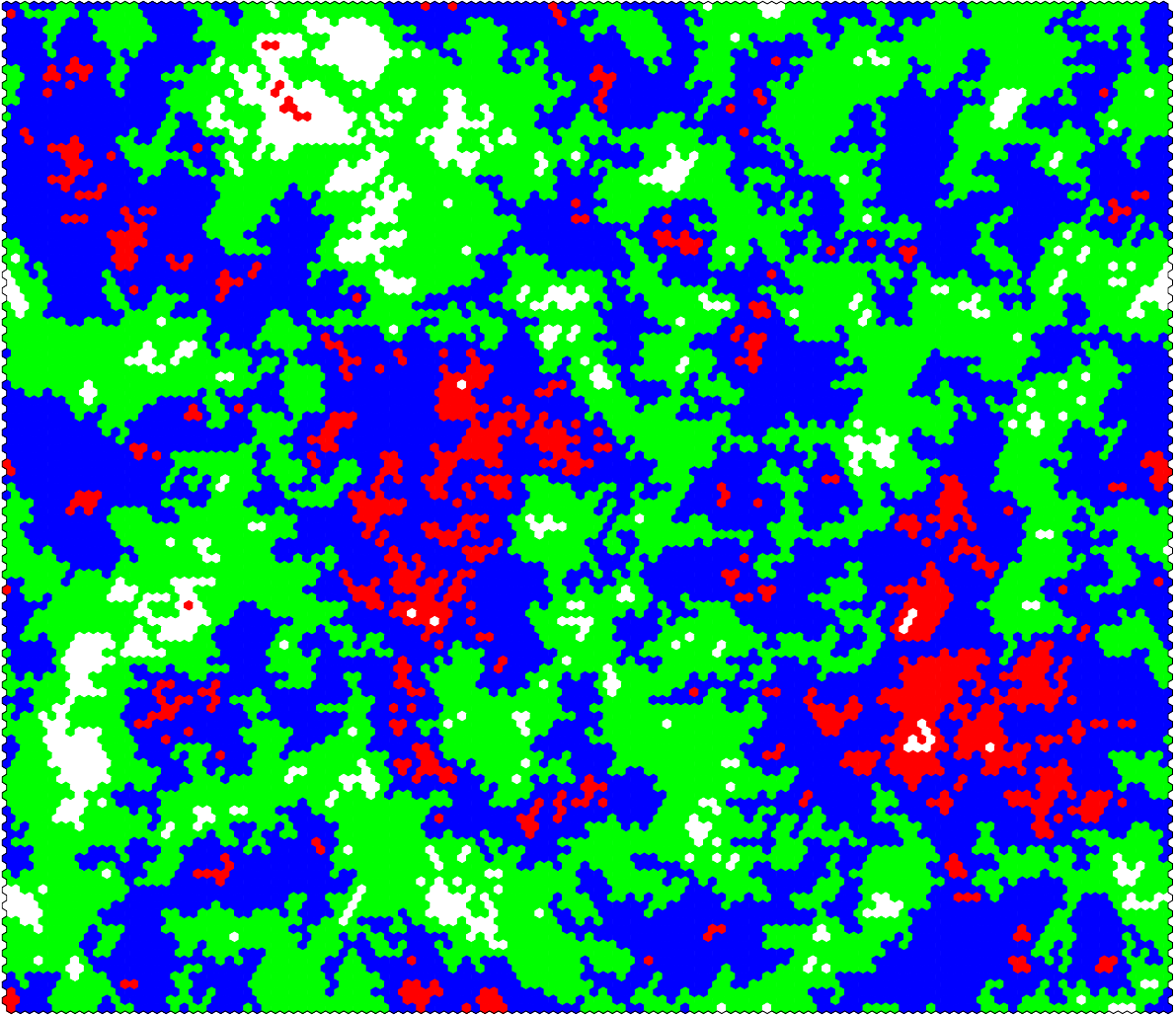}}

  \subfloat[$t=64$]{\includegraphics[scale=0.33]{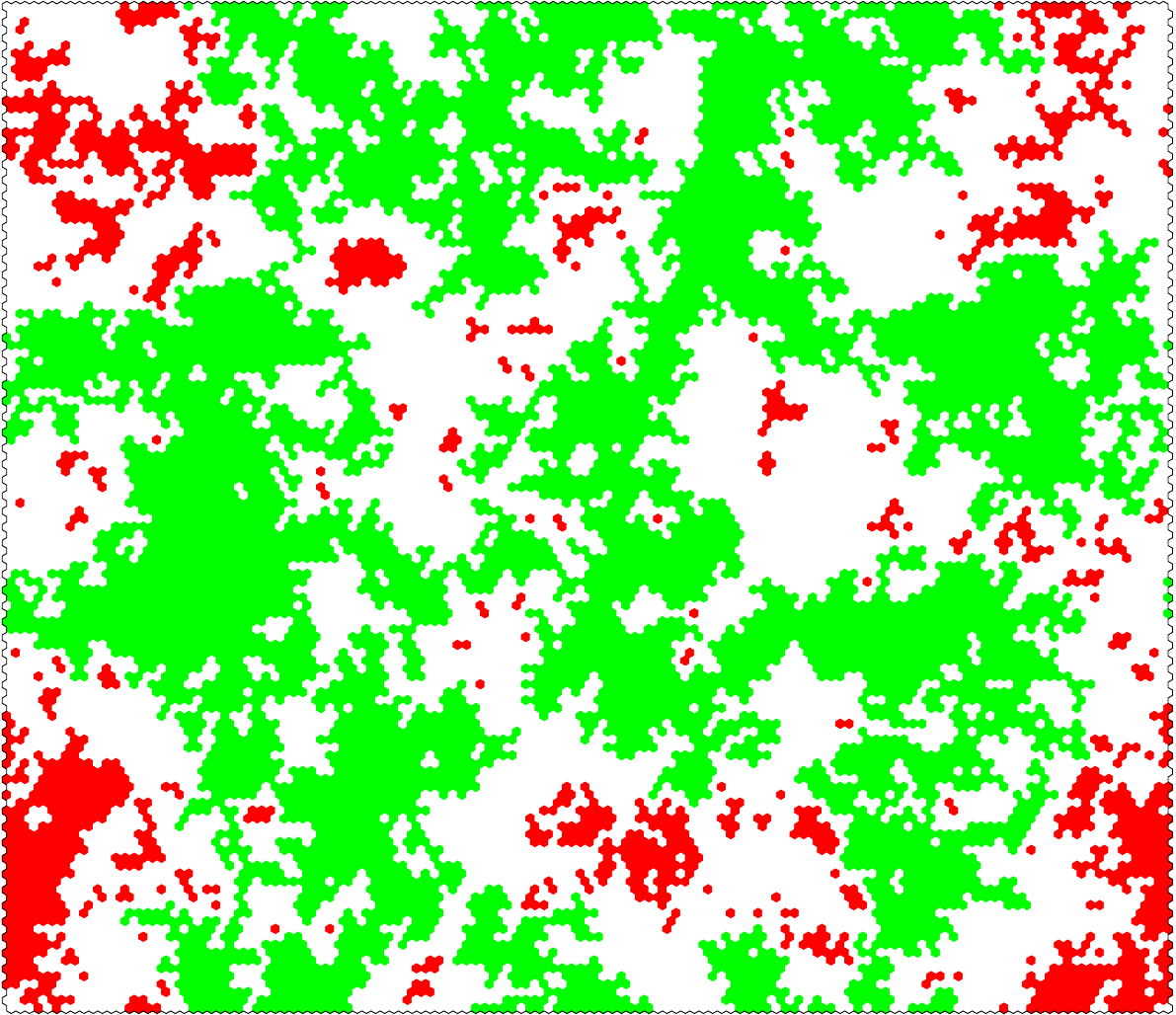}}\quad%
  \subfloat[$t=256$]{\includegraphics[scale=0.33]{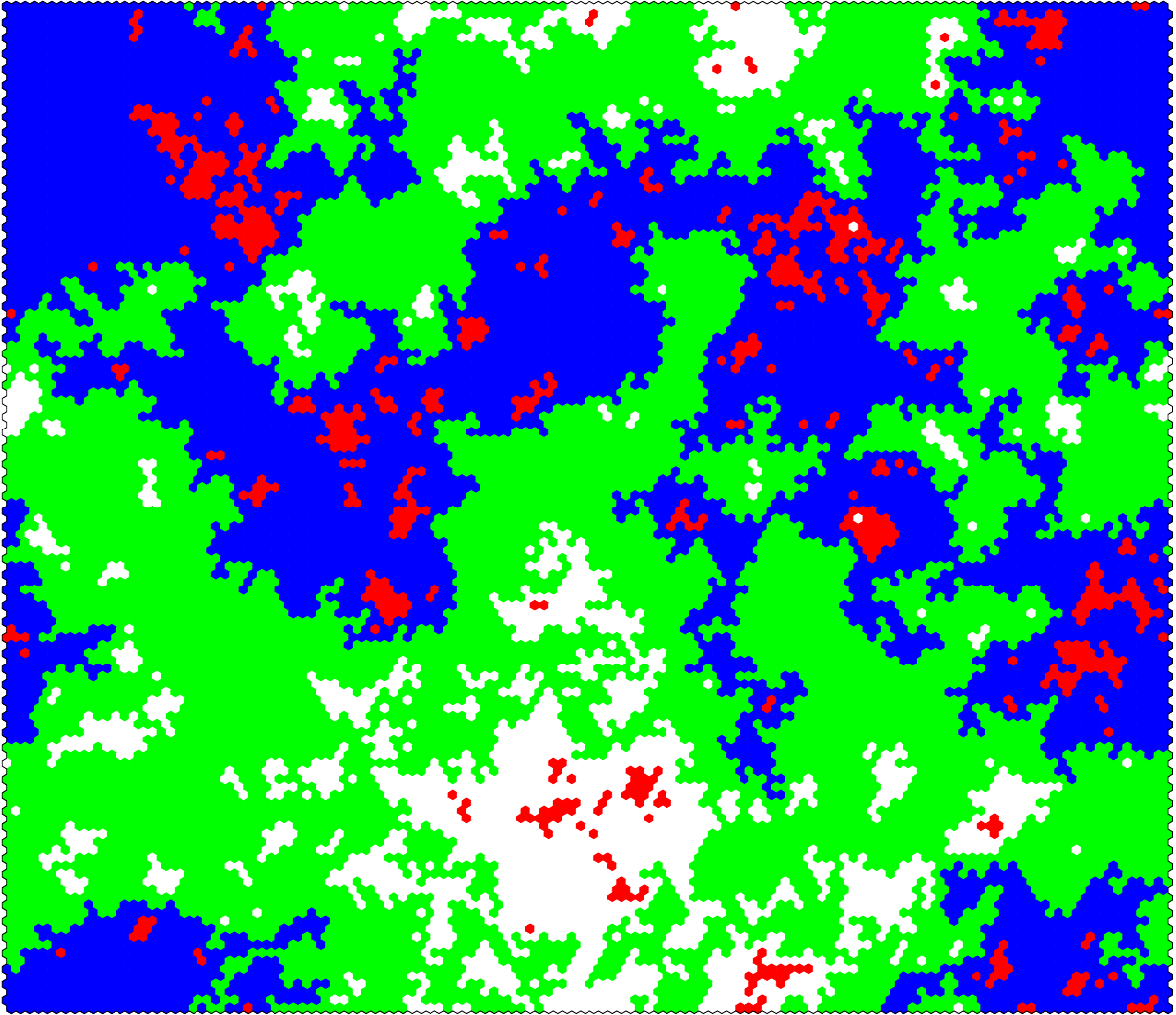}}\quad%
  \subfloat[$t=1024$]{\includegraphics[scale=0.33]{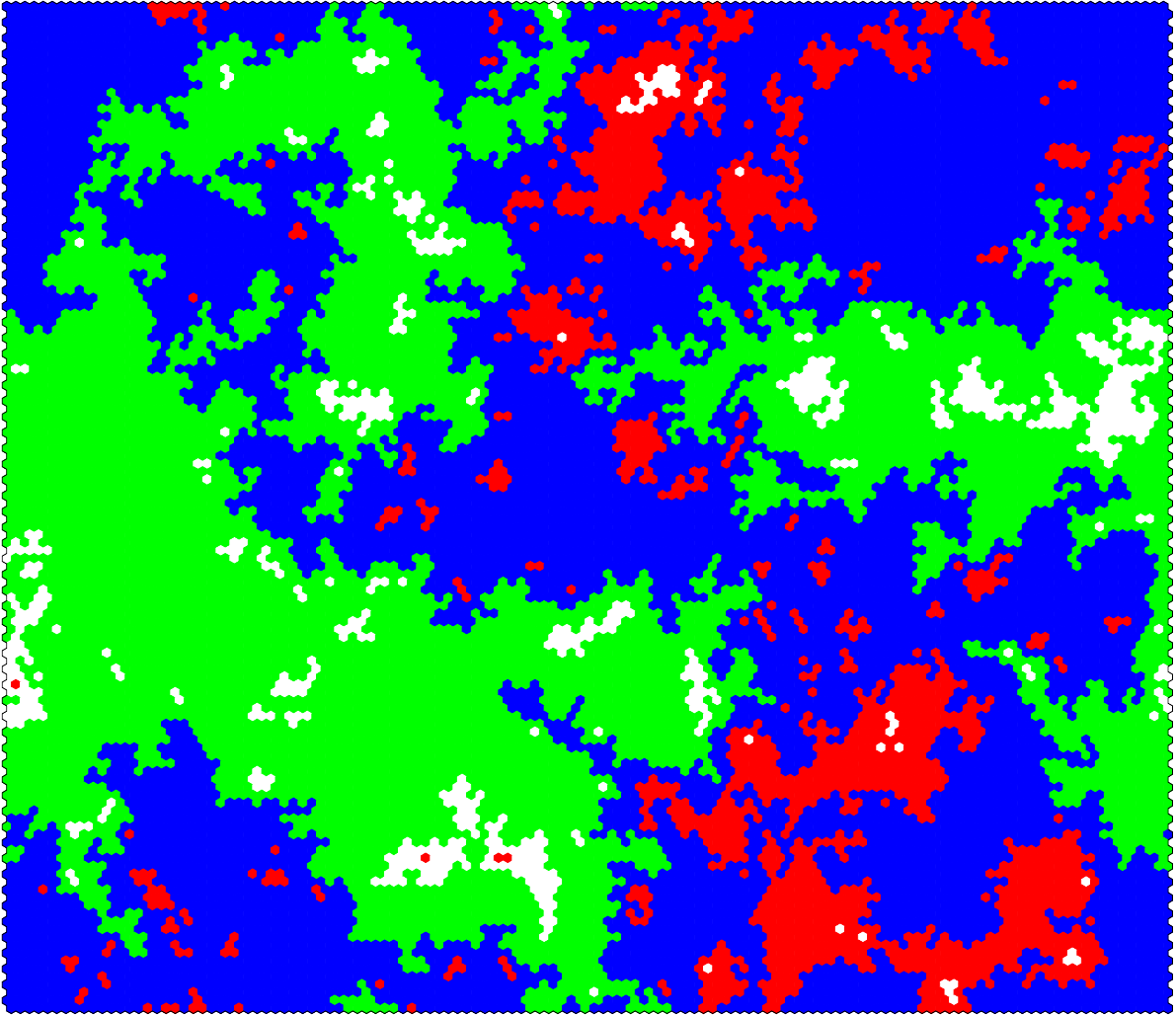}}
\end{center}
\caption{
\small Some snapshots of the evolution of a spin configuration under voter model dynamics on a triangular lattice with linear size $L=128$ and
PBC. Red cells and white cells represent $+1$ and $-1$ spins, respectively. Sites belonging to a cluster that wrap around the system
are highlighted in green for a $+1$ wrapping cluster and in blue for a $-1$ wrapping cluster.
}
\label{fig:voter-snapshots}
\end{figure}

To set the stage, 
in Fig.~\ref{fig:voter-snapshots} we show some snapshots of the evolution of a spin configuration under VM dynamics 
on a triangular lattice with linear size $L=128$ and PBC. The absence of bulk noise and the roughness
of the interfaces are clear in all images. The latter look very different from the ones that we exhibited for the $2d$IM. 
The initial state has a wrapping (blue, $-1$) cluster  which is later broken, 
see the snapshot at $t=64$. In the last image at $t=1024$ two spanning clusters having opposite phases are present. It is clear from the figures
that a coarsening process takes place although there is no energy function to minimise in this model. At much later times one of the two phases will 
predominate and conquer the sample.

\subsection{Average squared winding angle}
\label{subsec:winding-angle-vm}

As seen in Secs.~\ref{subsec:LK-winding-angle} and \ref{subsec:NLK-winding-angle}, 
the measurements of the average squared winding angle, $\langle \theta^2 \rangle$, can be used
to determine the type of criticality that the system is approaching during its evolution.
In the case of the VM, they reveal clearly that the system approaches, in the late stages of the dynamics,
a state in which the very large scale domain pattern has properties that are not the ones of $2d$ critical percolation, but of a distinct type.

\begin{figure}[h]
\begin{center}
        \includegraphics[scale=0.75]{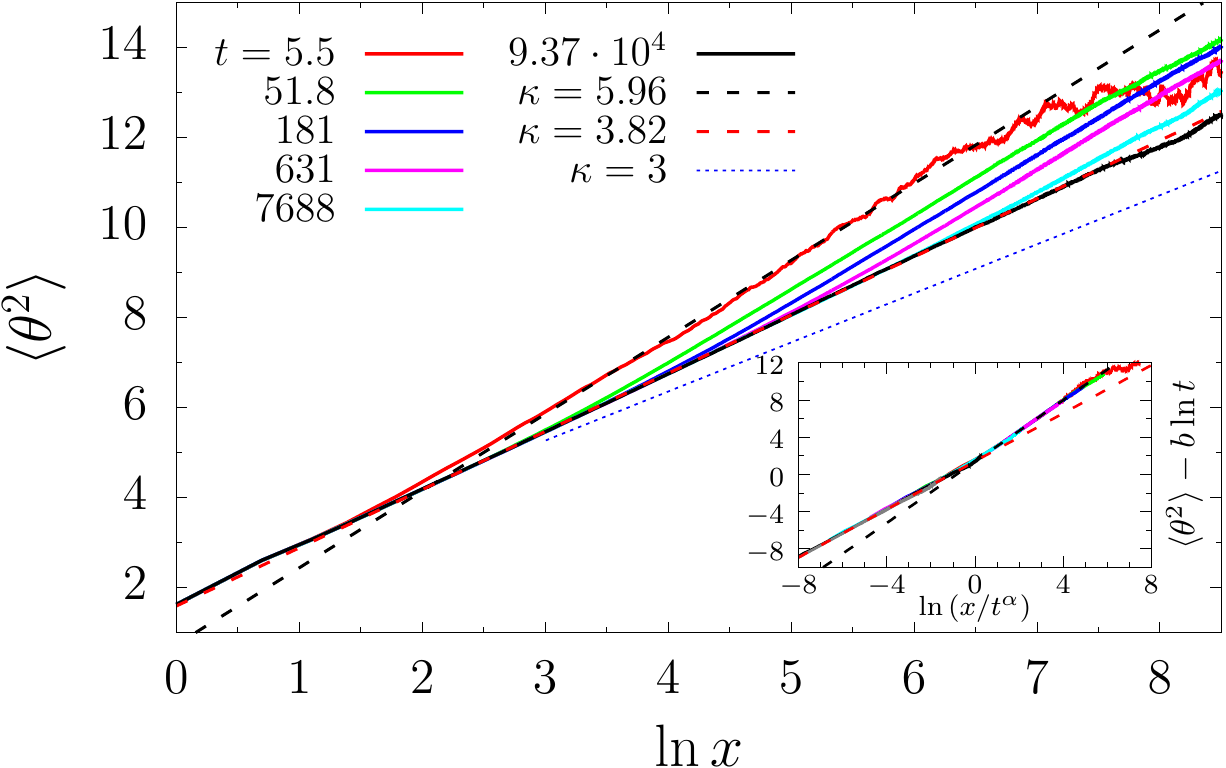}
\end{center}
\caption{\small 
Voter dynamics on the square lattice. Average squared winding angle $ \langle \theta^2 (x,t) \rangle $
for the wrapping hulls that form the largest cluster interface,
against the logarithm of the curvilinear coordinate $x$, at different times given in the key.
The lattice has linear size $L=640$.
A fit of $c + 4 \kappa/( 8 + \kappa ) \; \ln{x}$ \, to the data at $t=5.5$ yields $\kappa \simeq 5.96$
(black dashed line), while a fit to the data at late times, $t = 9.37 \cdot 10^4$, gives $\kappa = 3.82$ (red dashed line). 
We also show, for comparison, the slope associated to 
$\kappa=3$ with dashed-dotted blue line. In the inset, we collapse the dynamic data  by plotting $\langle \theta^2 \rangle - b \, \ln{t}$ 
against the rescaled curvilinear distance $x/t^{\alpha}$, with $\alpha = D_\ell/z_d = 0.74$ and $b=0.97$. See the text for 
an explanation of these parameters. 
}
\label{theta_SqVM}
\end{figure}

In Fig.~\ref{theta_SqVM} we report the average squared winding angle, $\langle \theta^2 \rangle$, for the hulls
of the largest cluster, against the logarithm of the curvilinear distance $x$, 
in the case of the VM on a square lattice with $L=640$.
The data shown refer to just the wrapping hulls. 
A spanning cluster appears at a very early time, $t \simeq 5$ for $L=640$.  
The ``critical'' function $c + 4 \kappa/( 8 + \kappa ) \; \ln{x}$ is fitted to the data corresponding to $t=5.5$, yielding
$\kappa \simeq 5.96$, a value which is very close to the $\kappa=6$ of critical percolation.
At later times, the long distance behaviour remains the same:
the slope of the curves is approximately the one at $t=5.5$ for distances $x>x_c(t)$, with $x_c$ a crossover
of the curvilinear distance. Meanwhile, for $x<x_c(t)$, the slope has changed but the behaviour is still
critical in the sense that $\langle \theta^2(x) \rangle \simeq \mathrm{const.}  + C \; \ln{x}$.
After a sufficiently late time, $\langle \theta^2 \rangle$ takes the form $\mathrm{const.}  + C \; \ln{x}$
over all the range of $x$ shown in the plot. 
By fitting the function $c + 4 \kappa/( 8 + \kappa ) \; \ln{x}$ to the data corresponding to the latest time 
we obtain $\kappa \simeq 3.82$. 
This result, associated to the interfaces, is very different from the one found for the $2d$IM with NCOP and COP 
dynamics, cases in which the interfaces are smooth and $\kappa=0$ at short  length scales.  

\vspace{0.5cm}

\begin{figure}[h]
\begin{center}
        \includegraphics[scale=0.57]{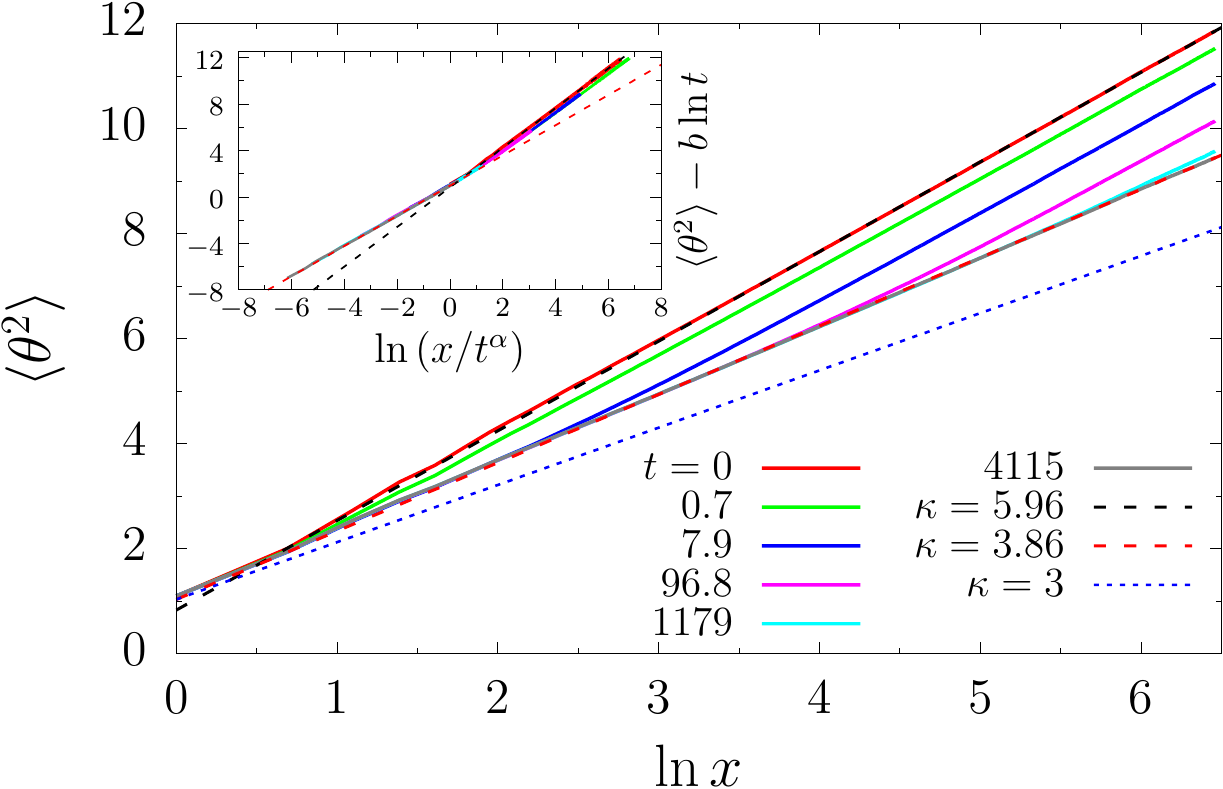}
        \includegraphics[scale=0.57]{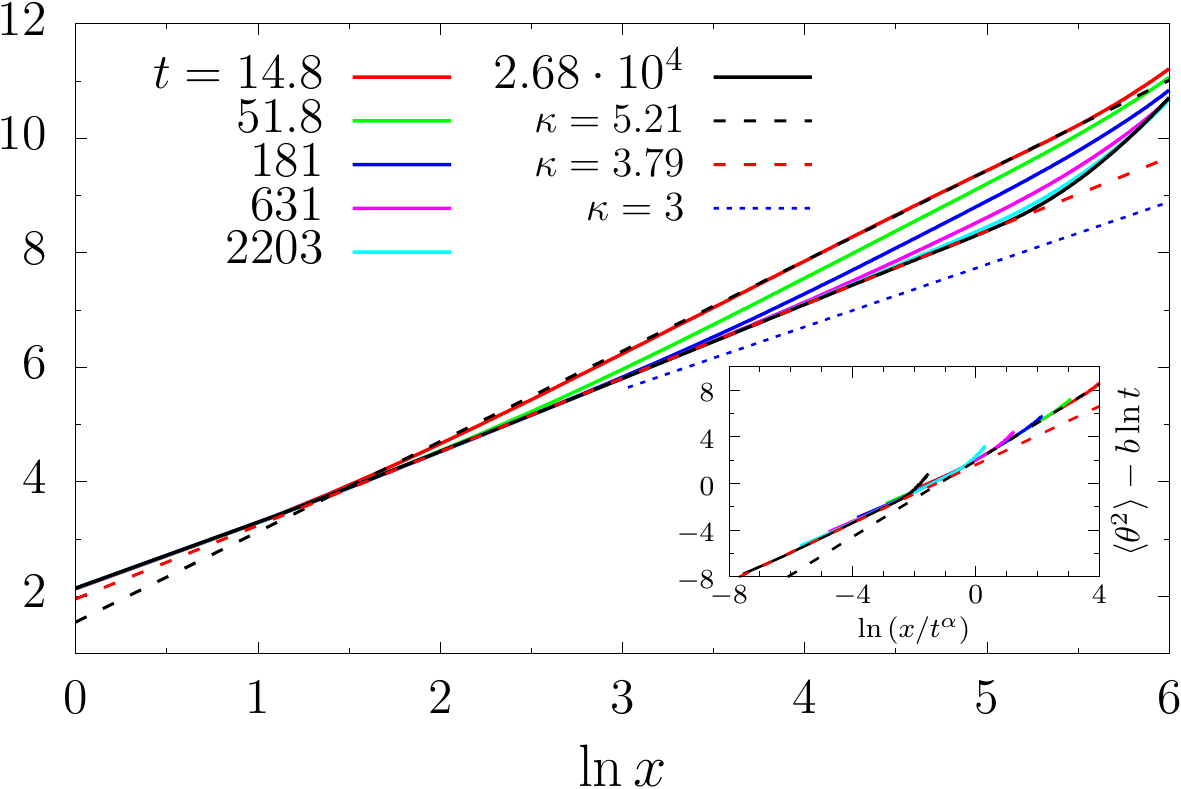}
\end{center}
\caption{\small 
Voter dynamics on the triangular (left) and honeycomb (right) lattice with $L=320$ and $L=160$, respectively.
Average squared winding angle $ \langle \theta^2 (x,t) \rangle $ for the
wrapping hulls of the largest clusters against
the logarithm of the curvilinear coordinate $x$, at different times given in the keys.
We show fits of the function $c + 4 \kappa/( 8 + \kappa ) \; \ln{x}$ with dashed lines.
In the left panel, a fit to the data at $t=0$ yields $\kappa \simeq 5.96$ (black dashed line),
while at $t=4115$, the value is reduced to $\kappa = 3.86$ (red dashed line).
In the right panel, $\kappa \simeq 5.21$ at $t=14.8$ and $\kappa = 3.79$ at $t=2.68 \cdot 10^4$. 
In the insets $\langle \theta^2 \rangle \, - b \ln{t} $ against $x/t^{\alpha}$,
with $\alpha = D_\ell/z_d  = 0.74$ and $b= 0.97$ on the triangular lattice, and $\alpha = 0.74 $ and $b= 0.95$ on the honeycomb lattice, 
yielding the best data collapse. See the text for an explanation of the meaning of the fitting parameters.
}
\label{theta_TrHonVM}
\end{figure}

In the inset of Fig.~\ref{theta_SqVM}, we show
$\langle \theta^2 \rangle \, - b \ln{t} $ \, against the logarithm of the rescaled curvilinear distance $x/t^{\alpha}$
with $\alpha = 0.74$ and $b=0.97$.  The justification for this scaling is the following. The unscaled data
show a crossover from the early critical percolation 
behaviour to a new kind of criticality at a time-dependent curvilinear distance $x_c(t) \propto t^{\alpha}$.
Visually, $x_c(t)$ corresponds to the point where  $\langle \theta^2 \rangle$ (when plotted against $\ln{x}$) 
changes from one slope to another.
This length scale is related to  $\ell_d(t)$ by 
$x_c(t) \sim \ell^{D_{\ell}}_d(t) \propto t^{D_\ell/z_d}$ with $D_\ell$ the fractal dimension
of the cluster hulls and $z_d$ the dynamical exponent.
Thus, $\alpha$ should be related to $\kappa$ and $z_d$ by $\alpha = D_\ell(\kappa)/z_d = \left( 1+\kappa/8 \right)/z_d$.
In the case indicated in Fig.~\ref{theta_SqVM}, as $t$ increases, $\langle \theta^2 \rangle$ approaches a functional form 
with $\kappa = 3.82$ (see the fitting function). Using now $z_d=2$ we deduce $D_\ell \simeq 1.48$ and
$\alpha \simeq 1.48/2 = 0.74$, which is the value used to scale $x$ in the inset.
The coefficient $b$ can be related to $\kappa$ and $z_d$ too. In fact, in order to obtain data collapse for different times 
we also need to subtract from $\langle \theta^2 \rangle$ the winding angle variance corresponding to a length $x_c(t) \propto t^{\alpha}$, that is
a quantity $4 \kappa / (8 + \kappa) \, \ln{t^{\alpha}}$. Then we have $b = 4 \, \alpha \, \kappa / (8 + \kappa) = \kappa/(2\, z_d) \simeq 0.97$.

For the other geometries, triangular and honeycomb lattices, similar fits yield
$3 \leq \kappa \leq 4$ at short length scales, see the caption of Fig.~\ref{theta_TrHonVM} for the precise values found in each case.
On the triangular lattice, as expected,  critical percolation properties are already present  at the moment of the 
quench, $t=0$. The insets in Fig.~\ref{theta_TrHonVM} display the time scaling  and 
confirm the arguments exposed in the previous paragraph.

The same scaling method was adopted by Blanchard \textit{et al.}~in~\cite{BlCuPi12}
in the study of quenches {\it between critical points}; more precisely, in the analysis of the $2d$IM on the triangular 
lattice evolving with NCOP dynamics at $T=T_c$ after a sudden quench from $T\to\infty$. 
These dynamics show a crossover from the initial critical percolation point to the Ising critical point 
controlled by the growing length $\ell_c(t)\simeq t^{1/z_c}$ with $z_c$ the critical dynamic exponent at $T_c$. In this case, the cluster hulls have 
the equilibrium geometrical features of the critical Ising point for $x<x_c(t)$, while  they obey the properties 
of critical percolation for $x>x_c(t)$. This double criticality is similar to what we observe for the interfaces of the 
VM on the triangular lattice. 

On the square lattice, the behaviour is also close to the one found after a quench to $T_c$~\cite{Ricateau}. In this case, 
a short time is needed to see the wrapping angles with $\kappa=6$ and only at a later time, 
the $\kappa=3$ of the  critical Ising point.
The crossover between the critical  percolation and the critical  Ising scaling is also controlled by 
 $\ell_c(t)\simeq t^{1/z_c}$ in this case. 

Obtaining a result that is similar to the one for a critical quench of the  Ising model should not come as a surprise. 
Indeed, as already announced in the introductory paragraphs to this Section, the VM and IM belong to the same
class of stochastic systems, distinguished by the updating rules. Moreover, 
the VM as well as the {\it critical} IM lie on a line of critical points, 
separating a paramagnetic phase from a ferromagnetic phase~\cite{Oliveira,DrouffeGodreche99}, in the 
bidimensional space of parameters that distinguish the microscopic updates. While it is believed that 
the low temperature phase has a universal behaviour (the one of the IM at zero temperature), the situation is 
less clear on this critical line. Our present results seem to indicate 
that a {\it new universality class}, at least for the interface behaviour, 
would exist for the VM (see~\cite{GodrechePicco18} for more details on this issue).

Summarising, with the analysis of the winding angle we have shown the following.
\begin{itemize}
\item On the triangular lattice, the VM evolves in time leaving the initial 
      random percolation criticality. The interfaces approach another kind of criticality, close to the 
      one of the equilibrium critical $2d$IM one, but not identical to it ($\kappa \simeq 3.86$ {\it vs.} $\kappa_{IM}=3$).
\item On the other lattices, the interfaces of the VM approach critical percolation properties
      in a relatively short span compared to the time required to reach the absorbing state. They then soon
      depart from this kind of geometry to enter 
      the coarsening regime with interfaces having critical properties  close but not identical to the ones of
      the critical Ising model (again, $\kappa \simeq 3.80$).
\end{itemize}
In both cases, the fractal dimension of the short length scales  
is approximately $D_\ell(\kappa=3.8) = 1.48$ and 
the temporal scaling is controlled by the usual coarsening length $\ell_d(t)\simeq t^{1/z_d}$ 
with $z_d=2$ and possible logarithmic corrections that we cannot control. A scaling of the winding
angle using $\ell_p(t)\simeq t^{1/z_p}$ with $z_p=1.67$ is not satisfactory (not shown).

Note also that the time needed to reach the second criticality {\it on all} length scales is very long, $t \simeq 10^5$ for a system 
with linear size $L=640$, while aspects of critical percolation geometry were already observed much earlier. Indeed, 
 at $t=5.5$, the interfaces (from an ordinary quench) were already characterised by $\kappa $ close to 6
 at long distances. 

\subsection{Initial configuration with a flat interface}

The results shown up to now were obtained using a 
paramagnetic initial state in equilibrium at infinite temperature. 
We are now going to consider a slab initial state with half of the spins taking a value $+1$ and the other half taking a value $-1$, 
ordered in such a way that there are two stripes of width $L/2$. 
Concretely, the spins at a position $x,y$ with 
$y \in [1,L/2]$ for any $x$ are $s(x,y)=1$ and the spins with  $y\in [L/2+1,L]$ for any $x$ are $s(x,y)=-1$.
Due to the PBC  the configuration contains, initially, two large clusters with volume $L^2/2$ 
separated by two straight interfaces with length $L$.  
This is a {\it fully stable} state for Glauber dynamics but it is not for the voter rule, as we will see below. 

\vspace{0.25cm}

\begin{figure}[h!]
\begin{center}
        \includegraphics[scale=0.5]{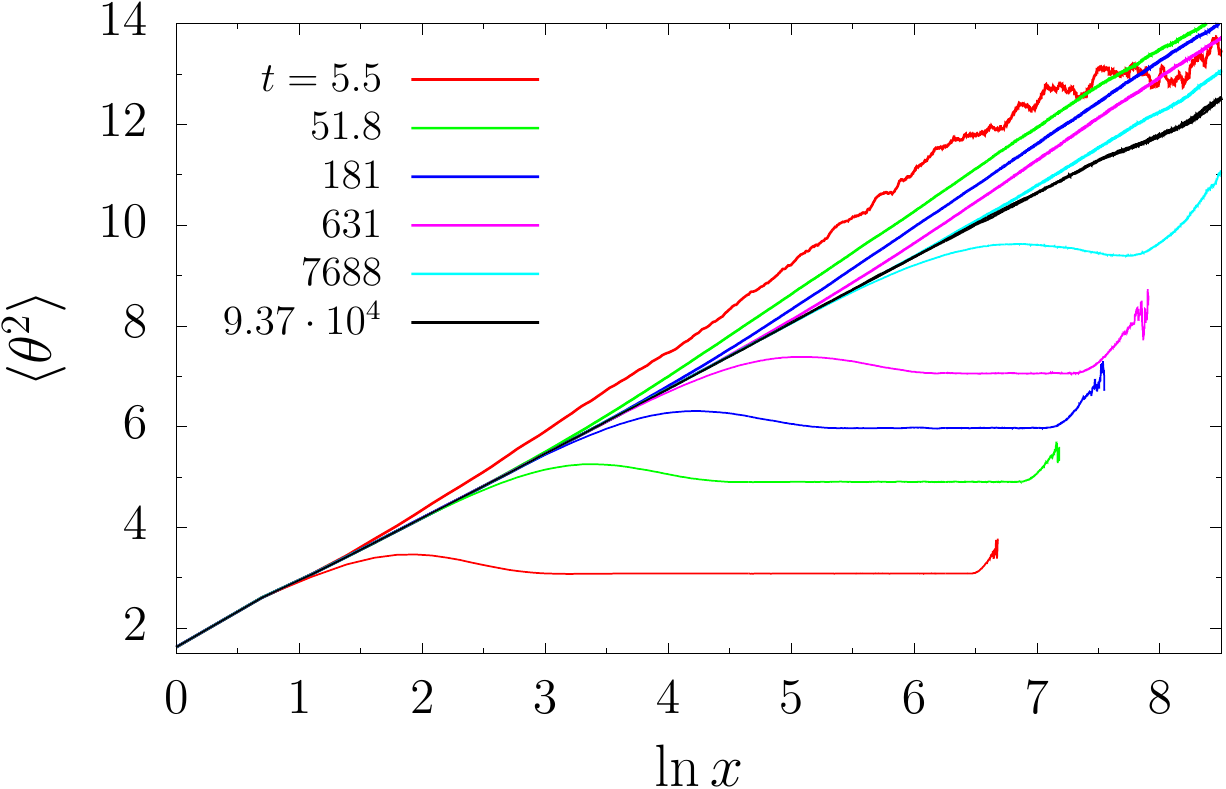}
        \hspace{0.25cm}
         \includegraphics[scale=0.5]{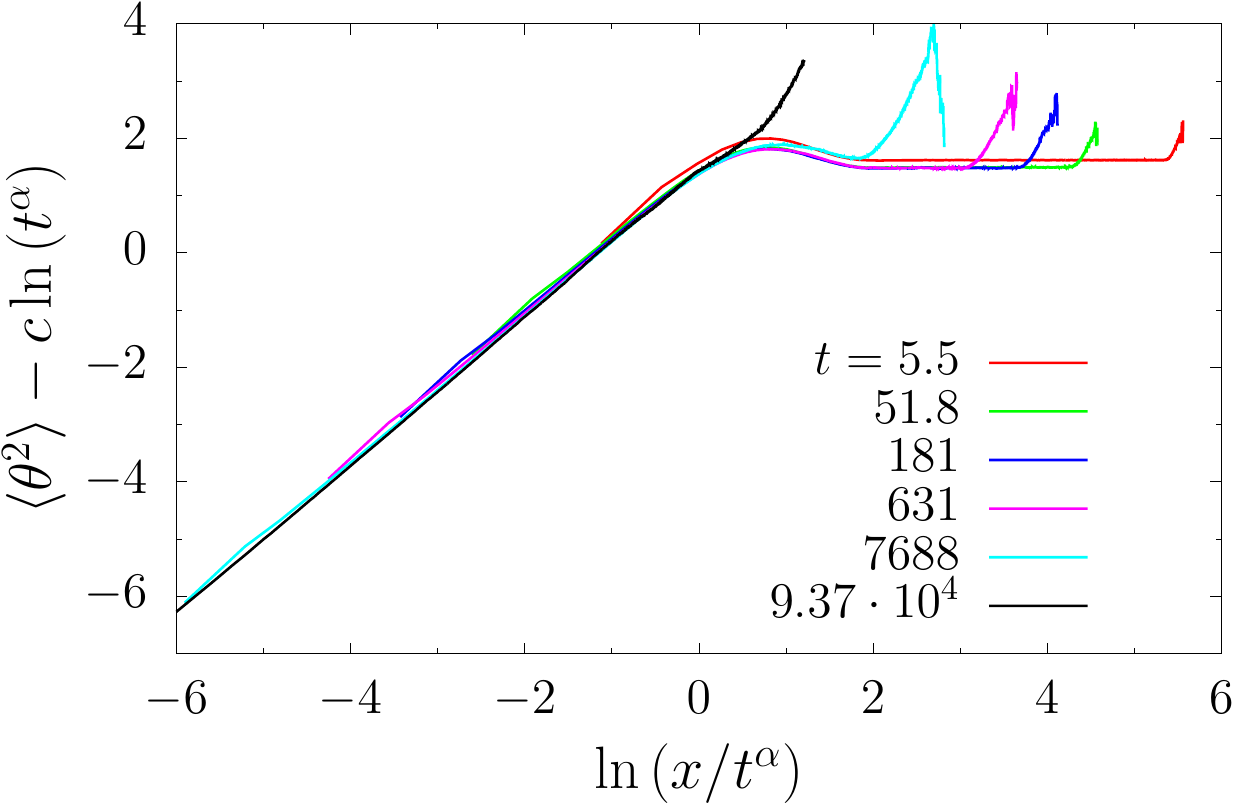}
\end{center}
\caption{\small 
Voter dynamics on the square lattice with $L=640$. 
Average squared winding angle $ \langle \theta^2 (x,t) \rangle $
for the wrapping hulls that form the largest cluster interface,
against the logarithm of the curvilinear coordinate $x$, at different times given in the key.
Left:  Raw data. The data above the black line were produced after a quench from
infinite temperature (same data as in Fig.~\ref{theta_SqVM}). The data below this line evolved from the slab initial state. 
Right: scaling of the data for the slab initial condition. We plot the quantity
$ \langle \theta^2 \rangle  - c \ln{(t^{\alpha})}$ against $\ln{ \left( x/t^\alpha \right)}$, where
$c = 4 \kappa/(8+\kappa)$ with $\kappa$ fixed to $3.85$ , and the exponent $\alpha = 0.67$ chosen to yield the best data collapse, as done in Fig.~\ref{theta_SqVM} and
~\ref{theta_TrHonVM}.
}
\label{theta_SqVM2}
\end{figure}

Under the voter dynamics, a flat interface  fluctuates. 
This is proven in Fig.~\ref{theta_SqVM2}, left panel, where we show the average square winding angle, $\langle \theta^2 \rangle$, 
for the (wrapping) hulls of the largest cluster on a square lattice with $L=640$, against the logarithm of the curvilinear distance $x$. 
The figure displays the same data as in Fig.~\ref{theta_SqVM}, for an instantaneous quench from infinite temperature 
(above the black curve), and data for the quench from the slab state  (below the black curve). 
The times at which the data are collected are the same in the two cases as indicated by the colour code and the key.
Concerning the latter set of data, at early  times, $t=5.5$, the interfaces are
curved only up to some short distance, while at longer distances, they remain flat as in the initial state ($\kappa=0$). 
As we let time pass,  the distance up to which the interfaces have a curvature increases
and, moreover, this curvature (or the corresponding $\kappa$) is equal, within numerical accuracy, to the one of the 
VM quenched from $T\to\infty$, {\it i.e.} $\kappa \simeq 3.85$. For the latest time shown,
$t=9.37  \cdot 10^4$, the data for the two kinds of initial states coincide.  
This shows that the ``second criticality'' is not related to critical percolation but it is inherent to the 
long term interfaces in the voter dynamics.

We identify a time-dependent crossover length, that we denote by $x_c(t)$. 
At curvilinear length scales larger than $x_c(t)$ the domain walls 
still retain the properties of the initial condition, that is, they are flat on average, while for smaller length scales the domain walls have fractal properties
associated to $\kappa\simeq 3.85$. As done in the previous cases, we suppose that $x_c(t) \sim t^{\alpha}$ with an exponent $\alpha$ to be determined
from the scaling. 
We show this scaling in the right panel of Fig.~\ref{theta_SqVM2},
where $ \langle \theta^2 \rangle  - c \ln{(t^{\alpha})}$ is plotted against $\ln{ \left( x/t^\alpha \right)}$, with
$c = 4 \kappa/(8+\kappa)$ and $\kappa$ fixed to $3.85$. The best collapse is achieved with $\alpha\simeq 0.67$.  Notice that this value
is close to the one obtained for the scaling of $\langle \theta^2 \rangle$ in the case of infinite temperature initial condition.

\subsection{Wrapping probabilities}

We can now ask whether the long-distance critical percolation properties of the interfaces truly represent a long-lasting critical percolation 
state. A way to address this question is to study the wrapping probabilities over the whole time
evolution going from $t=0$ to $t=t_{\rm abs}$.

\begin{figure}[h]
\begin{center}
  \includegraphics[scale=0.5]{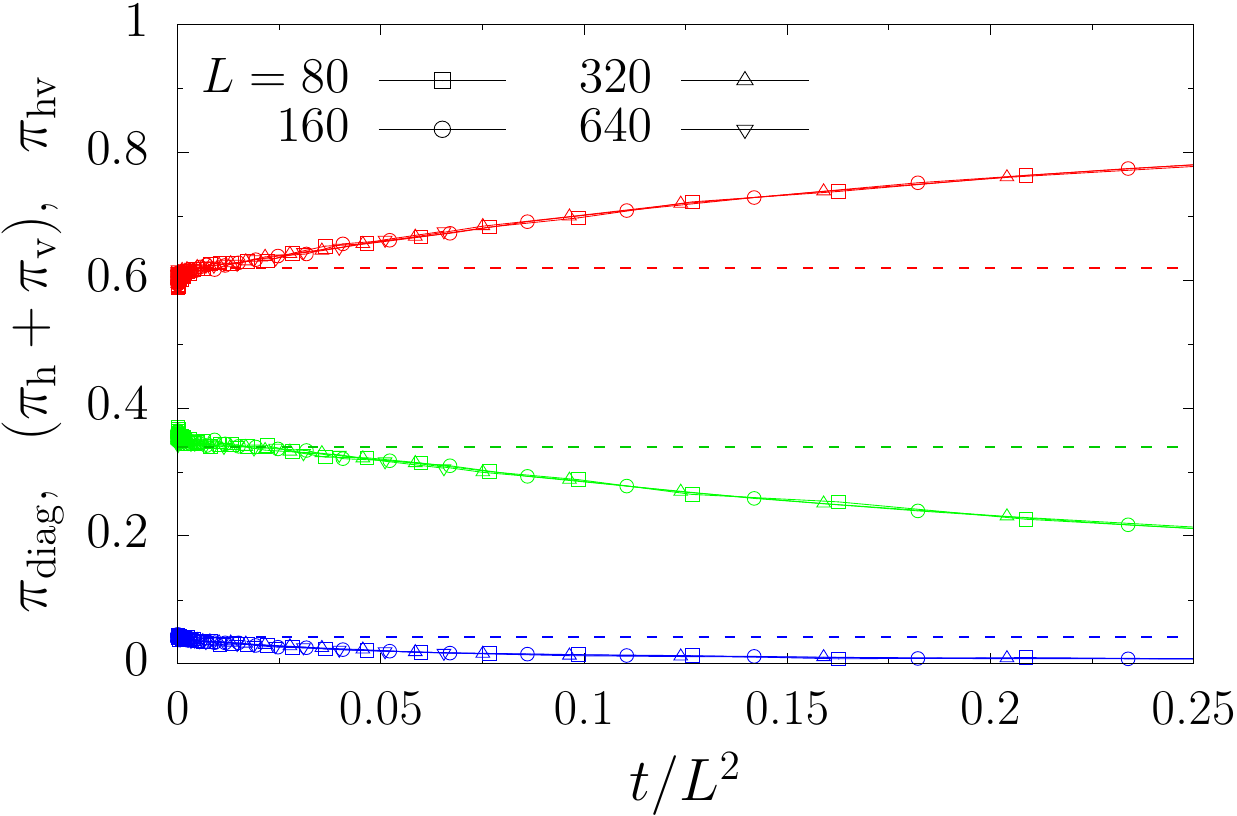}
    \includegraphics[scale=0.5]{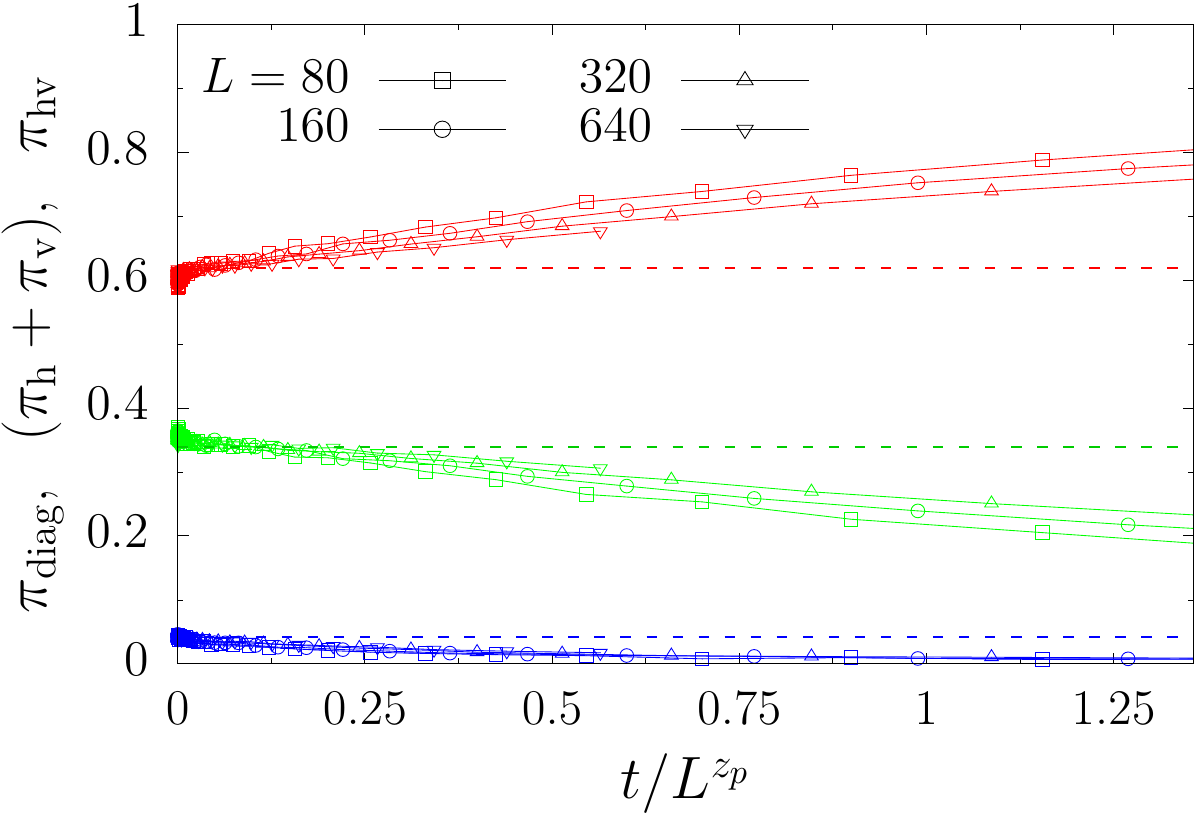}
\end{center}
\caption{\small
Voter model dynamics on the triangular lattice.
Probability that a cluster wraps  in both lattice directions with a cross topology ($\pi_{\rm hv}$, red points), 
along the horizontal or vertical direction ($\pi_{\rm h} + \pi_{\rm v}$, green points),
and along both directions but with a diagonal topology ($\pi_{\rm diag}$, blue points). Data for different values of $L$
are plotted against $t/L^{2}$ in the left panel (good scaling), and
against $t/L^{z_p}$ with $z_p \simeq 1.67$ in the right panel (failure of scaling).
The horizontal dashed lines indicate the wrapping probabilities $\pi_{\rm hv}$, 
$\pi_{\rm h} + \pi_{\rm v}$ and $\pi_{\rm diag}$ (from top to bottom) for $2d$ critical percolation.
}
\label{TrVM-WR}
\end{figure}

We assume that, for finite size systems, these probabilities obey
\begin{equation}
  \pi_{\alpha} (t,L) \sim \tilde{\pi}_{\alpha} \left( \frac{t}{L^z} \right)
 \label{eq:wp_scaling}
\end{equation}
with $\tilde{\pi}_{\alpha}$ a scaling function and the exponent $z$ to be determined.
Note that this scaling form is different from the one adopted for the spin-exchange dynamics, where 
we supposed that the $\pi$s depend on $t$ and $L$ through the rescaled time $t / (L/\ell_d(t))^{\zeta}$.
The form expressed in Eq.~(\ref{eq:wp_scaling}) is simpler.

In Fig.~\ref{TrVM-WR} - left panel, we report the cluster wrapping probabilities on a triangular lattice
against the rescaled time $t/L^2$. The value of $z$  that gives the best collapse 
coincides with $z_d=2$, the dynamical exponent.
The reason behind this result is that, on the triangular lattice, the initial spin configuration is already a realisation  of critical percolation.
The further evolution caused by the voter dynamics brings the system away from it.
Thus, the only relevant growing length is, in this case,  $\ell_d(t) \sim t^{1/z_d}$.
Had we used a logarithmic scale in the horizontal axis, we would have seen that for a  period of time lasting until $t/L^{2} \simeq 10^{-2}$ the probabilities remain almost constant
and very close to the exact values of $2d$ critical percolation, indicated by horizontal dashed lines.
At the characteristic time $t^{\star} \simeq 0.01 \cdot L^{2}$ they detach from these metastable values and in a bit less than two decades of the 
scaling variable they reach the values at the  absorbing state, namely $\pi_{\rm hv}=1$ and all 
other $\pi$'s equal zero. In the right panel of the same figure
we tried a scaling with $L^{z_p}$ and $z_p=1.67$ and we confirm that the data do not scale with this power.

\begin{figure}[h]
\begin{center}
  \includegraphics[scale=0.52]{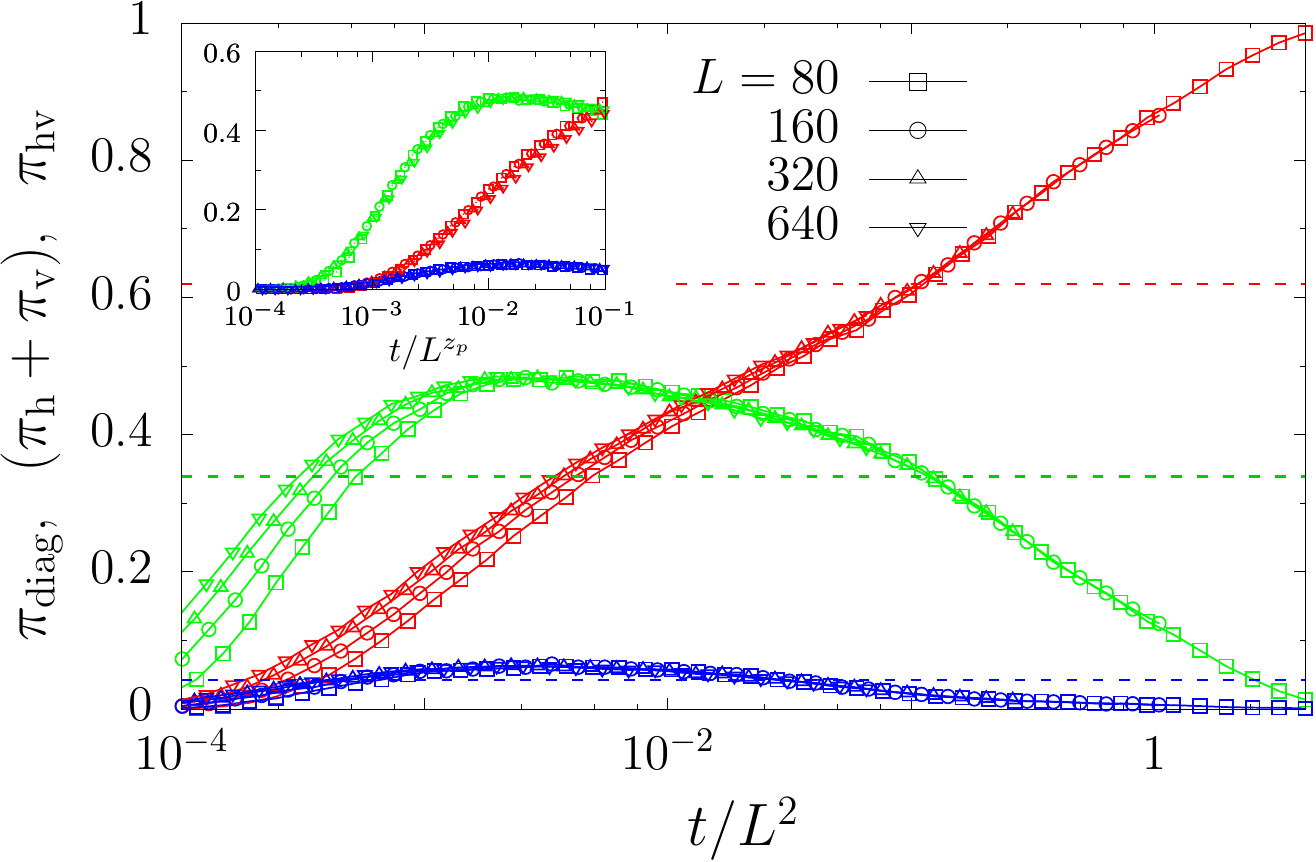}\quad%
  \includegraphics[scale=0.52]{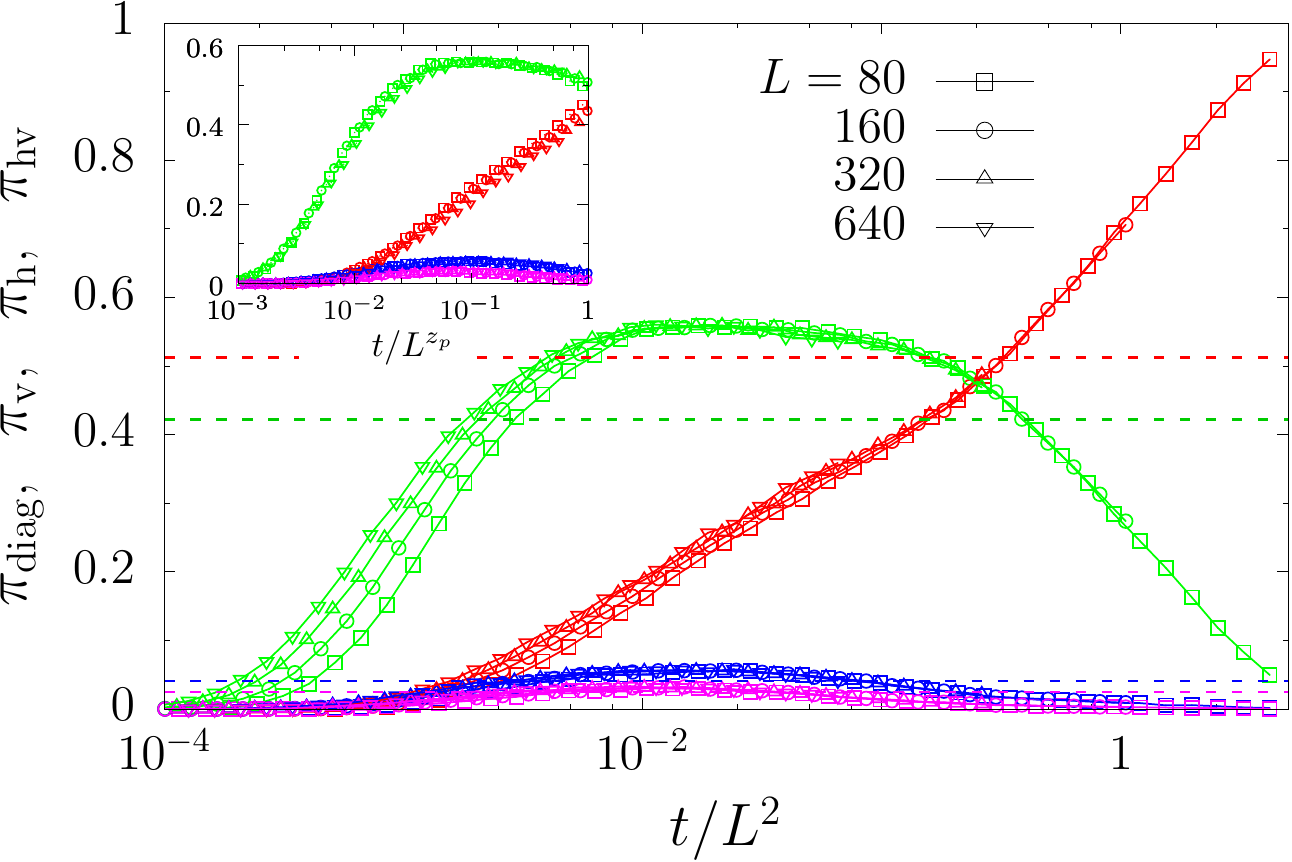}
\end{center}
\caption{ \small
The wrapping probabilities for the VM on a square lattice (left) and on the honeycomb lattice (right), 
for different values of $L$ indicated in the keys.
On the left we show $\pi_{\rm hv}$ (red points),   $\pi_{\rm h} + \pi_{\rm v}$ (green points),
and $\pi_{\rm diag}$ (blue points).
The horizontal dashed lines are $\pi_{\rm hv}$, 
$\pi_{\rm h} + \pi_{\rm v}$ and $\pi_{\rm diag}$ (from top to bottom) for $2d$ critical percolation on a lattice with unit aspect ratio.
On the right, $\pi_{\rm h}$ and $\pi_{\rm v}$ are separated and indicated by green and blue points respectively. 
The horizontal dashed lines are $\pi_{\rm hv}$, 
$\pi_{\rm h}$, $\pi_{\rm v}$ and $\pi_{\rm diag}$ (from top to bottom) for $2d$ critical percolation on a lattice with aspect ratio $\sqrt{3}$.
In the main plots, the wrapping probabilities are plotted against $t/L^2$.
In the insets, the same data are plotted against the rescaled time $t/L^{z_p}$, with $z_p \simeq 1.67$.
}
\label{VM-WR_Sq_Hon}
\end{figure}

In Fig.~\ref{VM-WR_Sq_Hon} we show the time evolution of the cluster wrapping probabilities on a
square lattice, in the left panel, and on a honeycomb lattice, in the right one. 
For the honeycomb case we need to separate $\pi_{\mathrm{h}}$ and $\pi_{\mathrm{v}}$
since the aspect ratio is not 1. The dashed horizontal lines
are the values at $2d$ critical percolation.
In the main panels, the data are plotted against the rescaled time $t/L^{z_d}$ with 
$z_d =  2$ that yields the best collapse for $t/L^{z_d} > 5 \cdot 10^{-2}$ in both cases, say.
In the insets a different scaling variables is used, $t/L^z$ with 
$z \simeq1.67$ that yields the best collapse for $t/L^{z} < 0.5$, approximately.
The curves {\it do not} show a long-lasting metastable state with the properties of critical percolation
(contrary to what happens for NCOP and COP dynamics).
The $\pi$s approach the critical percolation values ($\pi_{\rm hv}$ from below, while
$\pi_{\rm h}$, $\pi_{\rm v}$ and $\pi_{\rm diag}$ from above after the short transient in which they reached their maximum value)
around a time $t^{\star} \simeq 0.1 \, L^{z_d}$ (on the square lattice) and $t^{\star} \simeq 0.5 \, L^{z_d}$ (on the honeycomb lattice), 
but later they continue to evolve towards the asymptotic values 
dictated by the absorbing states ($\pi_{\rm hv} = 1$ and  $\pi_{\rm h} = \pi_{\rm v} = \pi_{\rm diag} = 0$).
This is very different from what is shown, {\it e.g.}, in Fig.~\ref{LKaWrapping_b2} for Kawasaki dynamics.

These measurements suggest that, in the cases in which the process does not start from a
critical percolation state, there is a crossover between two different scaling regimes. In the fist one, 
the relevant length scale is $\ell_p(t) \simeq t^{1/z_p}$ with $z_p=1.67$. 
The second and usual coarsening scaling regime, in which one observes the
characteristic length scale $\ell_d(t) \sim t^{1/2}$, sets in roughly after $\pi_h + \pi_v$ has reached its maximum
value.

\subsection{Largest cluster scaling}
\label{subsec:VM_LC_scaling}

The critical percolation criticality, $\kappa=6$, sets the fractal dimensions to be 
$D_A^{\rm cp} = 1.8958$ and $D_\ell^{\rm cp} = 1.75$.  
In Sec.~\ref{subsec:winding-angle-vm} we established that at short length scales  the interfaces are critical 
with $\kappa\simeq 3.85$. If we could use this $\kappa$ to characterise the 
area and length fractal properties we would get 
\begin{equation}
D_A(\kappa=3.85)  \simeq 1.8804 \; , \qquad\qquad D_\ell (\kappa=3.85) \simeq 1.4813
\; . 
\label{eq:fractal-dim-kappa3.85}
\end{equation}

In this Section we revisit the scaling properties of the size of the largest cluster, $A_c$, and the length
of the hulls forming its interface, $l_c$.
We assume  that $A_c$ and $l_c$ have the scaling  behaviour
\begin{equation}
  A_c(t,L) \, \sim \, L^{D^{\star}_A} \; \tilde{A}_c\left( \frac{t}{L^{z}} \right) \, , \qquad
  l_c(t,L) \, \sim \, L^{D^{\star}_{\ell}} \; \tilde{l}_c\left( \frac{t}{L^{z}} \right) \, , 
 \label{eq:scaling_Ac_lc_VM}
\end{equation}
with $\tilde{A}_c$ and $\tilde{l}_c$ some scaling functions, 
$ D^{\star}_A$ and $D^{\star}_{\ell}$ the fractal dimensions of the largest cluster size and 
hull length, respectively, and $z$ a dynamic exponent, and we use
$ D^{\star}_A$, $D^{\star}_{\ell}$ and $z$ as free parameters to be 
adjusted together.

\vspace{0.5cm}

\begin{figure}[h]
\begin{center}
        \includegraphics[scale=0.55]{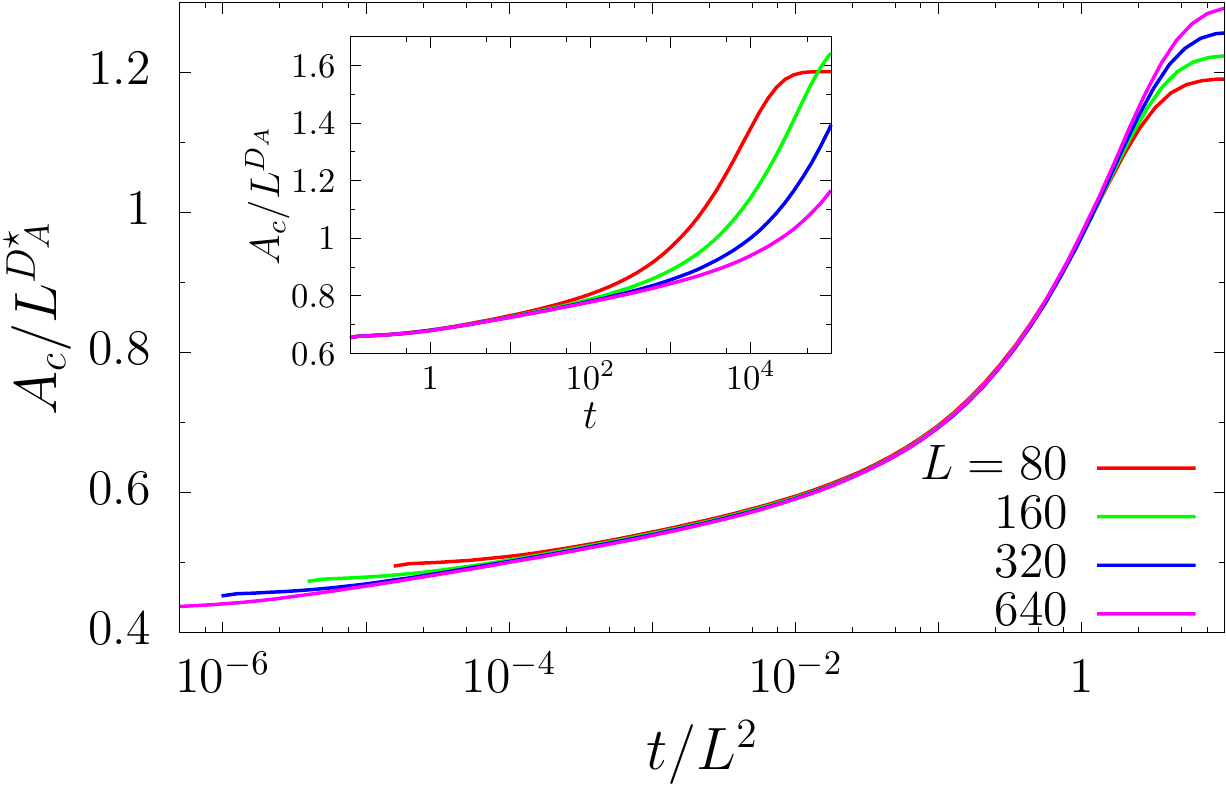}\quad%
        \includegraphics[scale=0.55]{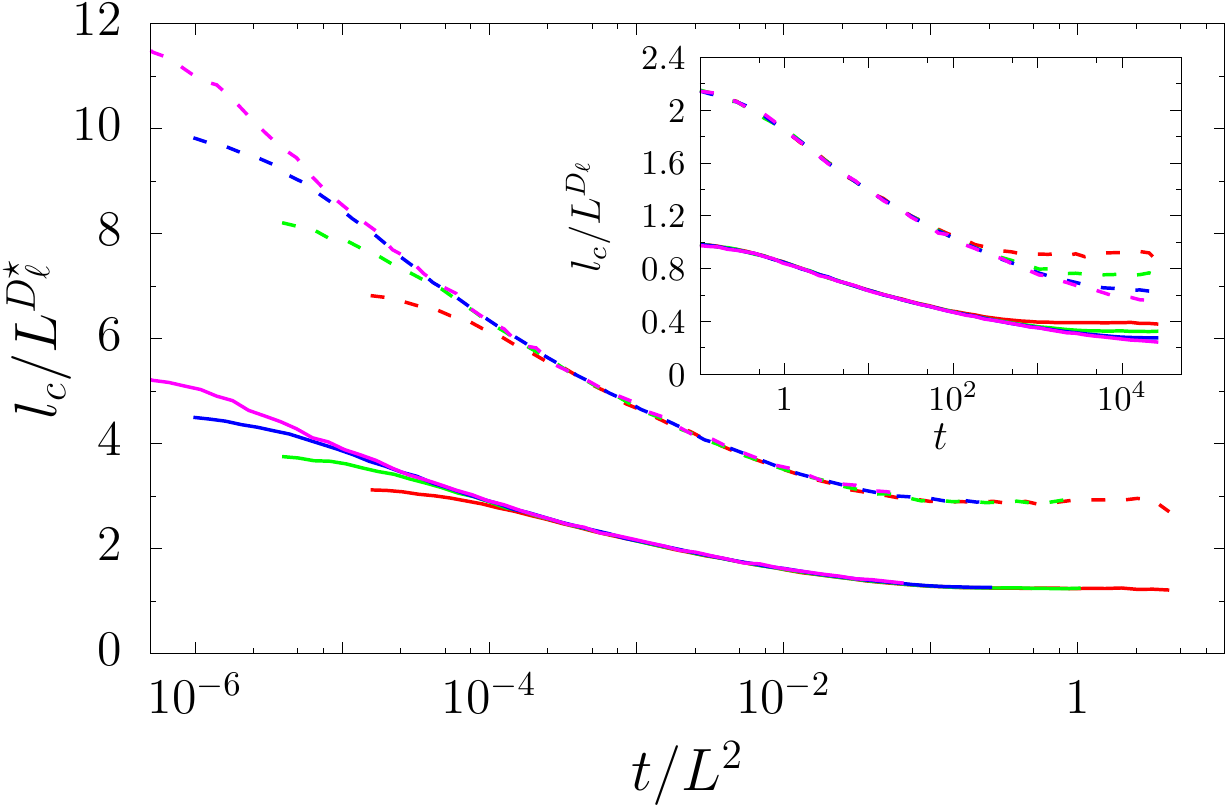}
\end{center}
\caption{\small 
The scaling of the geometric properties of the largest cluster for the VM with sizes in the key on a triangular lattice. 
In the left panel, the size of the largest cluster $A_c$ divided by 
$L^{D^{\star}_A}$, with $D^{\star}_A = 1.96 $, while in the right panel, the length of its external hull $l_c$ divided by 
$L^{D^{\star}_{\ell}}$, with $D^{\star}_{\ell}=1.49$.
Regarding $l_c$, we have separated the contribution of the  wrapping hulls (continuous lines) from the one 
of the non-wrapping ones (dashed lines). All the quantities are plotted against the rescaled time $t/L^{2}$.
In the insets we scale the data using the fractal dimensions of $2d$ critical percolation,
$D_A = 91/48$ and $D_{\ell} = 7/4$, and plot them against time $t$.
}
\label{lc_TrVM}
\end{figure}

In Fig.~\ref{lc_TrVM} we show the rescaled largest cluster size, $A_c(t,L)/L^{D^{\star}_A}$ (left), and the rescaled largest cluster
hull length separating the wrapping from the non-wrapping contributions, $l_c(t,L)/L^{D^{\star}_{\ell}}$ (right),  
on a triangular lattice. The above scaling relations are satisfied if 
we plot both quantities against $t/L^2$, and we use the fractal
dimensions $D^{\star}_A \simeq 1.96$ and $D^{\star}_{\ell} \simeq 1.49$.
The time scaling confirms that on the triangular lattice the   
coarsening of already percolating large clusters is controlled by the growing length
$\ell_d(t) \sim t^{1/2}$. Concerning the fractal dimensions found with the fit,
the first one is quite far from the one that would correspond to criticality at 
 $\kappa=3.85$ while the latter is very close to the one for $\kappa=3.85$, see Eq.~(\ref{eq:fractal-dim-kappa3.85}).

In the insets of Fig.~\ref{lc_TrVM} we present the same data as a function of time $t$,
but scaled with the fractal dimensions of critical percolation, $D_A = 91/48$ and $D_{\ell} = 7/4$. At short times, 
the data collapse is perfect since the critical percolation initial states survive 
in the early stages of the dynamics on length scales longer than $\ell_d(t)$. At longer times, the scaling breaks down, 
since the system evolves towards a regime with different characteristics, with the dimensions used 
in the main panels.

\vspace{0.5cm}

\begin{figure}[h]
\begin{center}
        \includegraphics[scale=0.55]{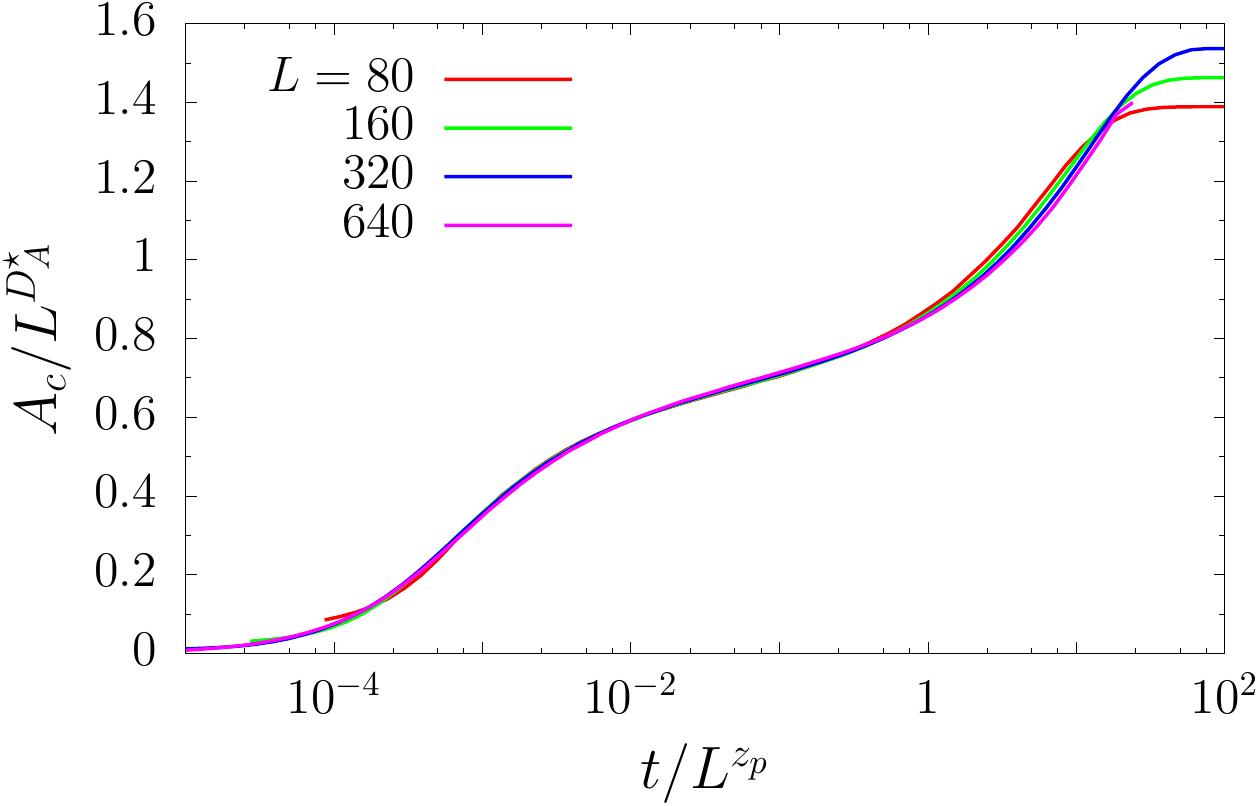}\quad%
        \includegraphics[scale=0.55]{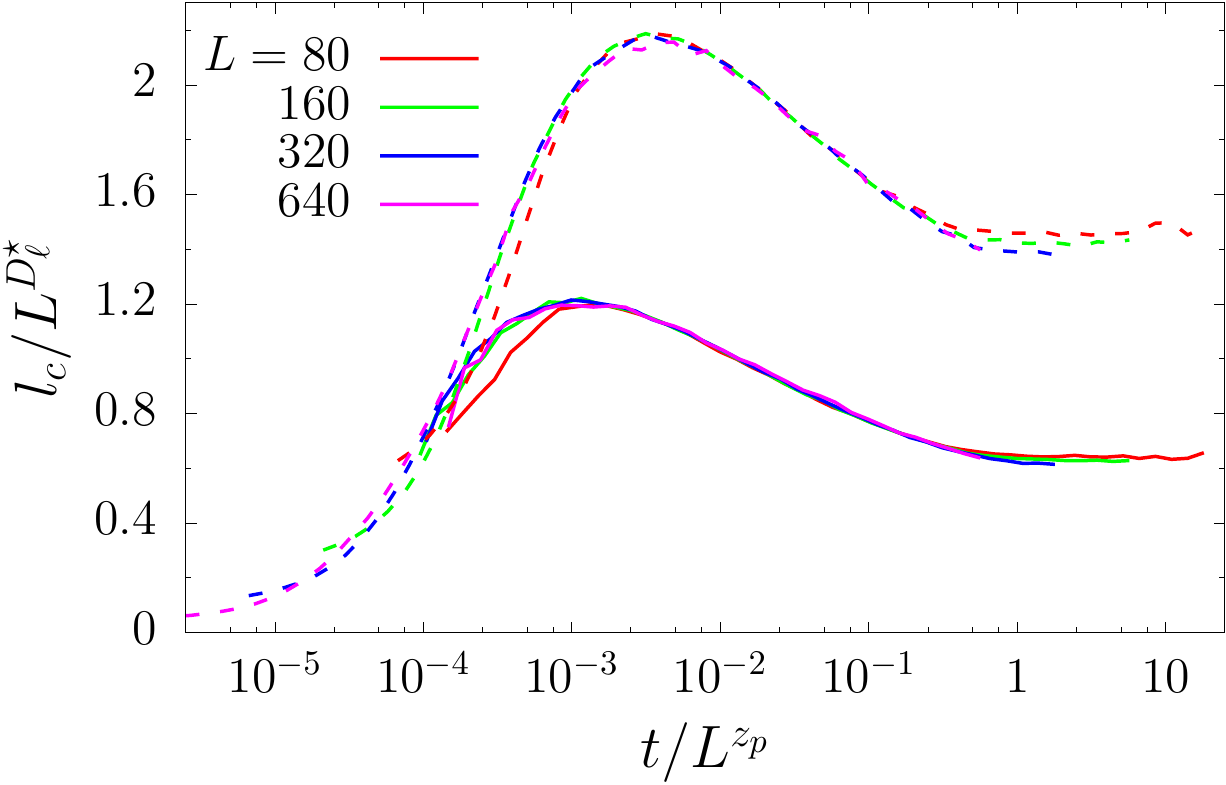}
\end{center}
\caption{\small 
The scaling of the geometric properties of the largest cluster for the VM dynamics on a square lattice. 
The size of the largest cluster $A_c$ divided by 
$L^{D^{\star}_A}$, with $D^{\star}_A = 1.93$ (left), and the length of its external hull $l_c$ divided by 
$L^{D^{\star}_{\ell}}$, with $D^{\star}_{\ell}=1.53$ (right).
As in Fig.~\ref{lc_TrVM}, the contribution of the wrapping hulls to $l_c$ is shown with continuous lines 
while the one of non-wrapping hulls is shown with dashed lines. All the quantities are plotted against the rescaled time $t/L^{z_p}$, with $z_p=1.67$.
}
\label{lc_SqVM}
\end{figure}

\begin{figure}[h]
\begin{center}
        \includegraphics[scale=0.55]{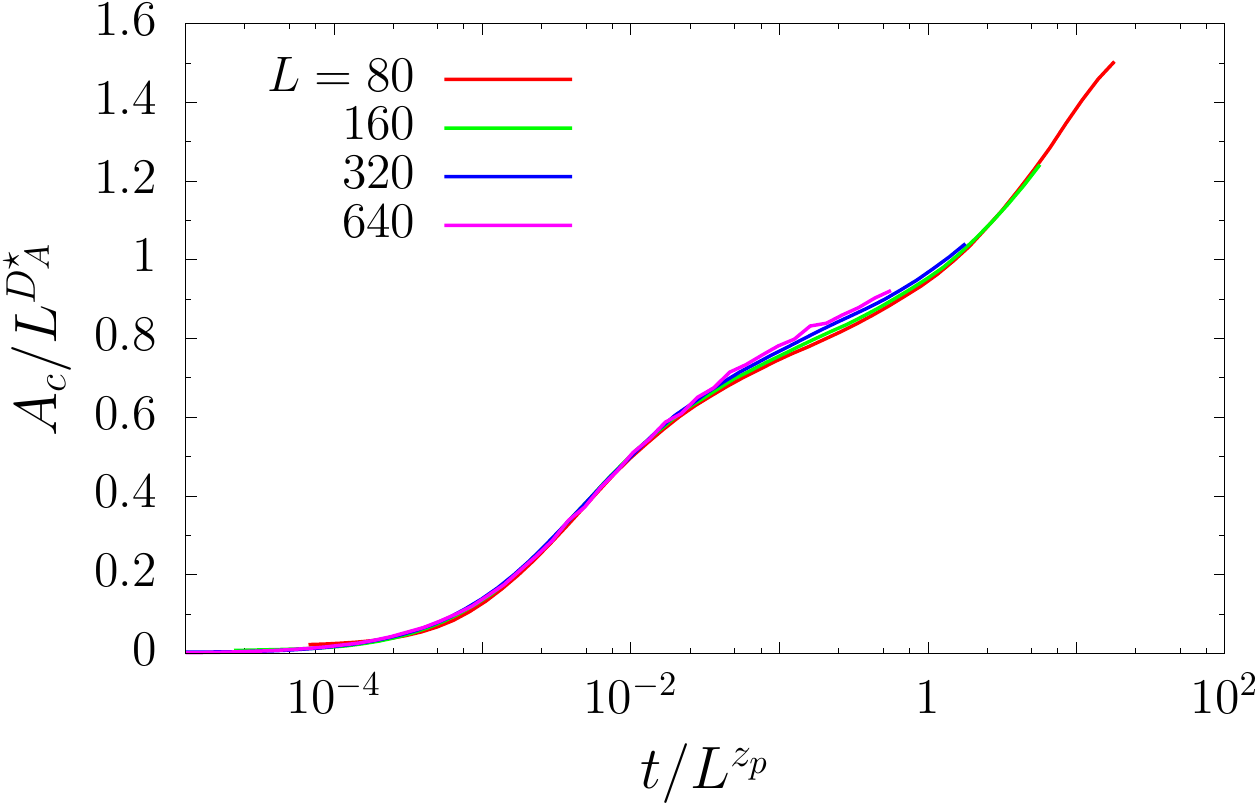}\quad%
        \includegraphics[scale=0.55]{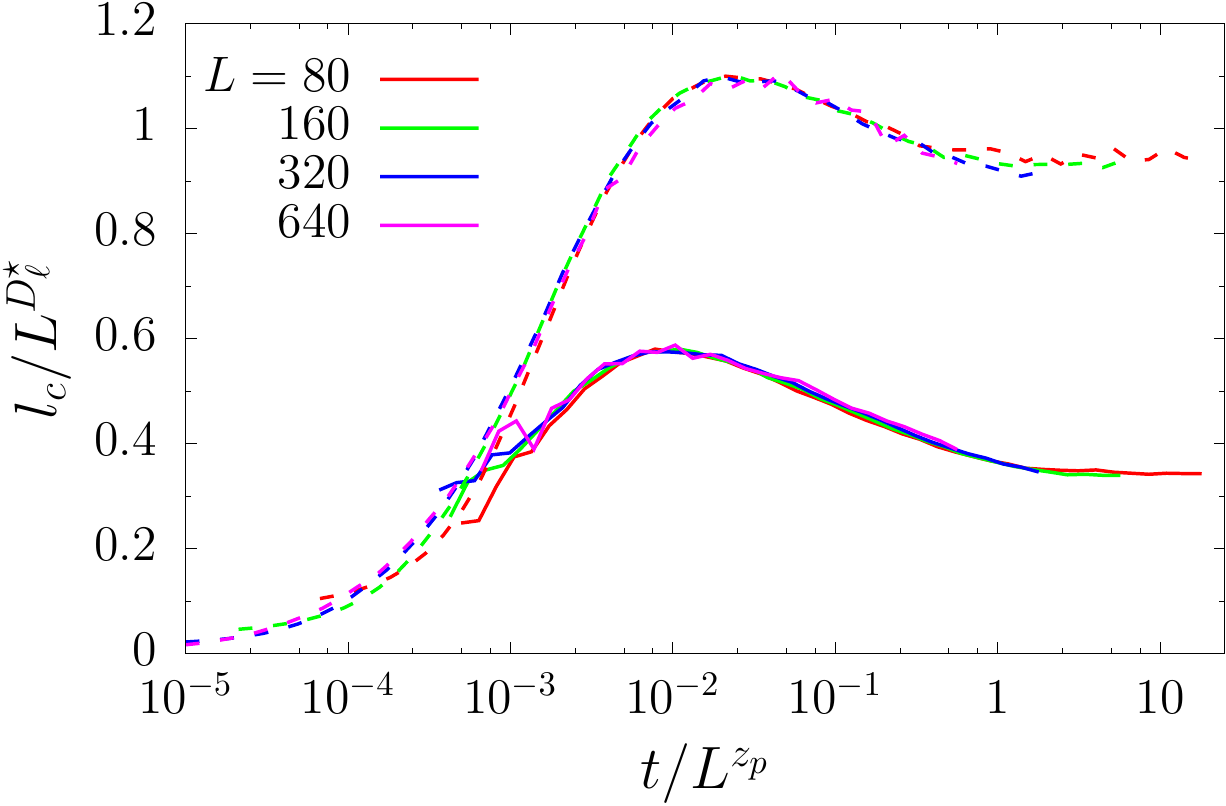}
\end{center}
\caption{\small 
The scaling of the geometric properties of the largest cluster for the VM on a honeycomb lattice. 
The size of the largest cluster $A_c$ divided by 
$L^{D^{\star}_A}$, with $D^{\star}_A = 1.90$ (left), and the length of its external hull $l_c$ divided by 
$L^{D^{\star}_{\ell}}$, with $D^{\star}_{\ell}=1.52$ (right).
As in Fig.~\ref{lc_TrVM}, the contribution of the wrapping hulls to $l_c$ is shown with continuous lines 
while the one of non-wrapping hulls is shown with dashed lines. All the quantities are plotted against the rescaled time $t/L^{z_p}$, with $z_p=1.67$.
}
\label{lc_HonVM}
\end{figure}

In Fig.~\ref{lc_SqVM} we display the same type of scaling for the VM dynamics on a square lattice.
The best scaling is achieved by using $D^{\star}_A \simeq 1.93$ and $D^{\star}_{\ell} \simeq 1.53$
for the largest cluster size and the length of its hull, respectively, and $z_p\simeq 1.67$. We confirm what is found for the
scaling of the wrapping probabilities, that is the existence of a time scale $t_p(L) \sim L^{z_p}$ that marks the 
end of a first scaling regime and the entrance into the last coarsening one. We find similar results on a honeycomb lattice, 
see Fig.~\ref{lc_HonVM}, with $D^{\star}_A \simeq 1.90$, $D^{\star}_{\ell} \simeq 1.52$ and $z_p\simeq 1.67$. The value of the
parameter $\kappa$ derived from $D^{\star}_A$ is significantly different from the one
derived from $D^{\star}_{\ell}$, $\kappa \simeq 3.20$ and $\kappa\simeq 4.23$, respectively, so that we 
can not associate both scaling behaviours to 
a single universality class. 

To conclude, on the triangular lattice the initial condition is one of critical percolation and the 
dynamics soon complies with the scaling properties of the coarsening regime, namely $z_d=2$. 
On the square and honeycomb lattices, instead, we found that the observables scale with a first typical time scale 
$t_p \sim L^{z_p}$ with $z_p\simeq 1.67$ before entering the long-term coarsening process. During the transient regime the dynamics seem to approach 
a state with fractal scaling properties, different from the ones of critical percolation, but not associated 
to any well-known criticality. For the numerical accuracy we have,  the $\kappa$ associated to interfaces and bulk 
properties are different. This fact holds also for the triangular lattice model. Naively, we could argue that the 
VM has the peculiarity of having only interfacial noise, a fact that makes the interfaces behave very differently from the 
bulk properties.

\subsection{Number density of domain areas}

In~\cite{TaCuPi15} we presented an analysis of the scaling properties of 
the number density of domain areas for the voter model dynamics on a square lattice and we showed that,
for sufficiently late times $t$ and large domain size $A$, it develops an algebraic decay, 
$\mathcal{N}(A,t) \simeq C(t) \, A^{-\tau}$, with exponent $\tau \approx 1.98$ for lattice sizes up to $L=640$.
This exponent has to become larger than $2$ in the infinite 
size limit and asymptotically in time to ensure normalisation of the number density.
In the pre-asymptotic regime this quantity behaved in a very similar manner 
to the one shown for the NCOP dynamics~\cite{BlCuPiTa-17}.


\begin{figure}[h]
\begin{center}
\includegraphics[scale=0.6]{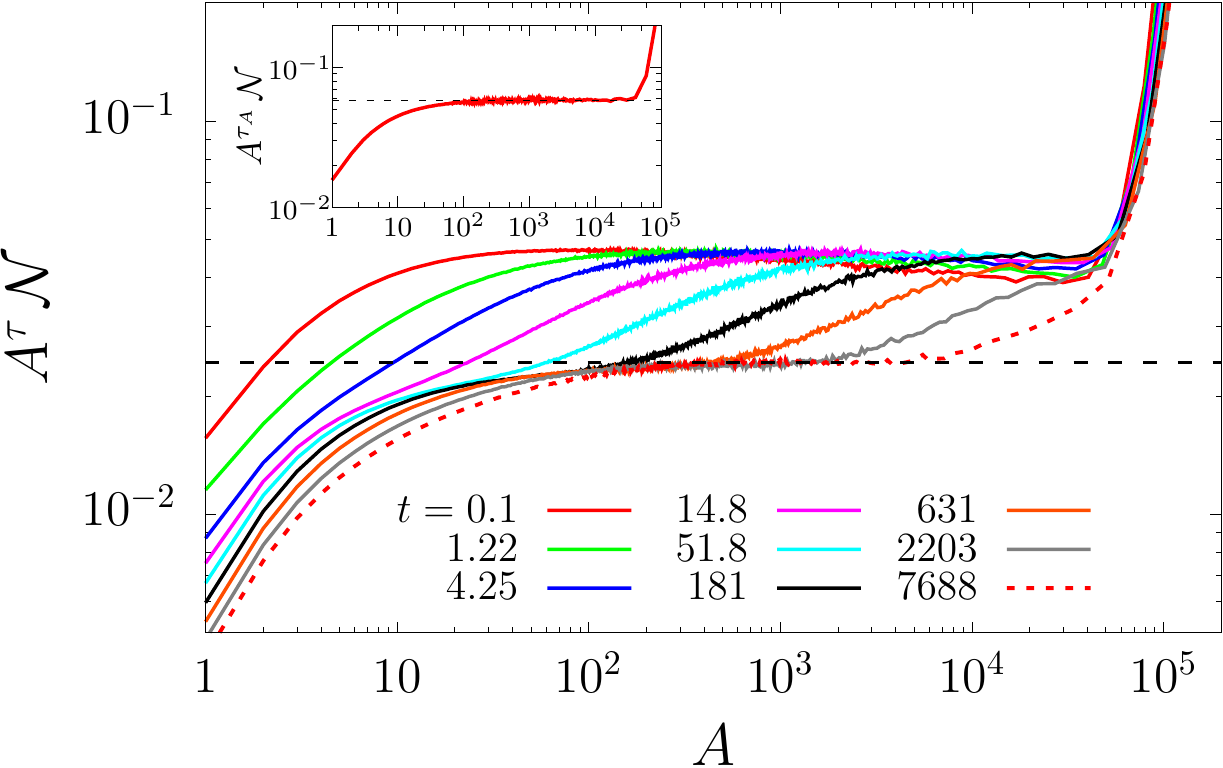}\quad%
\includegraphics[scale=0.58]{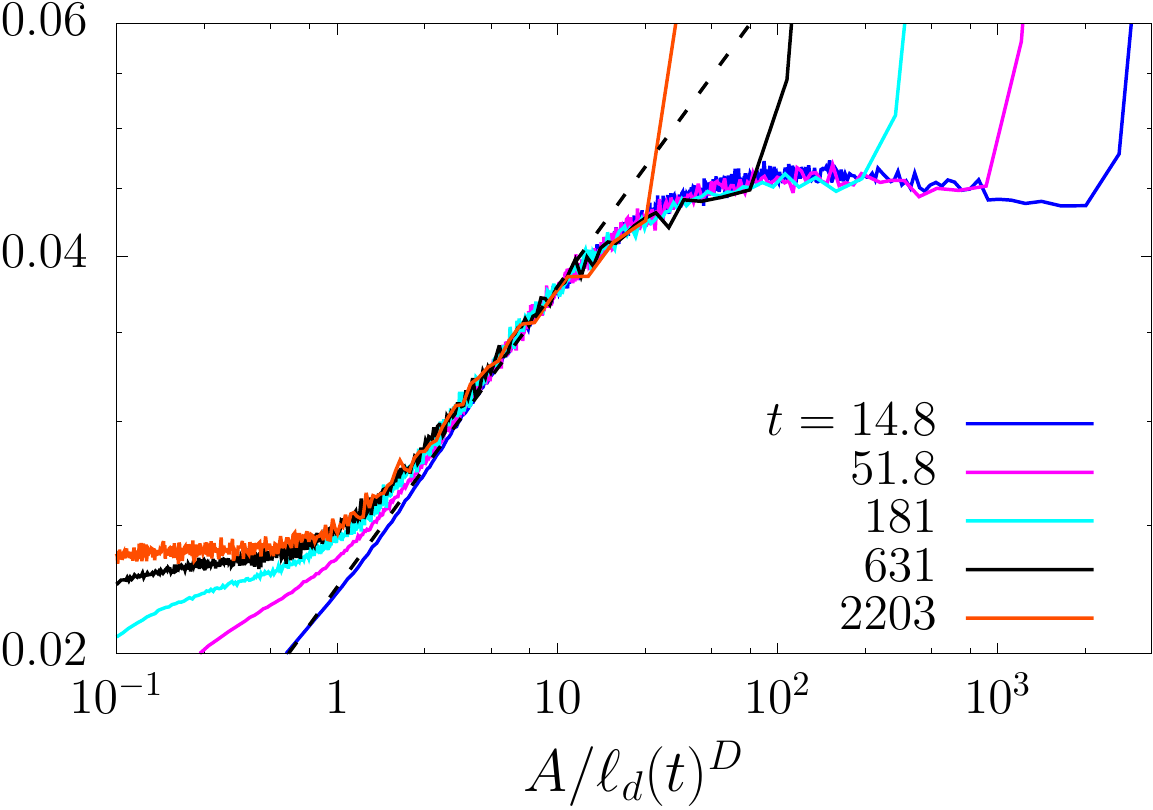}
\end{center}
\caption{\small The VM on a triangular lattice with linear size $L=640$. 
In the left panel we show  $A^{\tau} \, {\cal N}(A, t, L)$ against $A$, with $\tau \simeq 2.016$,
at different times indicated in the key.
The value of  $\tau$ was found by fitting the function $C \, A^{-\tau}$ to the
data relative to $t=7688$, in the interval $[10^2,\,10^4]$.
The constant $C \simeq 0.024$  is shown with a black dashed line.
In the inset,  $A^{\tau_A} \, {\cal N}(A, 0, L)$,
$\tau_A = 187/91$ the Fisher exponent for $2d$ critical percolation, against $A$.
In the right panel, $A^{\tau} \, {\cal N}(A, t, L)$ is plotted against the rescaled size $A/\ell_d(t)^{D}$, 
with $\tau \simeq 2.016$ as in the left panel, $\ell_d(t) = (1 + t)^{\frac{1}{2}}$ the dynamical characteristic length
and $D=2/(\tau - 1) \simeq 1.97$.
The function $C' \, x^{a}$, fitted at
$t=14.8$ and in the interval $[2.5,10]$ of the rescaled area
$x=A/t^{\xi}$, yields $a \simeq 0.228$ and is shown by an inclined dashed line.
}
\label{NATrVM-2}
\end{figure}

We repeat  the same type of analysis of the 
number density of areas already performed for  Kawasaki dynamics and its nonlocal version,
now for the VM on the triangular and honeycomb lattices, cases that 
were not included in~\cite{TaCuPi15}.
In particular, we examine the  triangular lattice case with the peculiarity that, as already stated multiple times
before, the initial spin configuration has a percolating cluster.
A simple scaling argument implies that the domain area number density should satisfy 
$\mathcal{N}(A,t) \sim \, \ell^{-4}_d(t) \, f\left( A / \ell^2_d(t) \right)$, with $\ell_d(t)$ the characteristic length
associated to coarsening and $f(x) \simeq 2c_d x^{-\tau_A}$ at large $x$ with $\tau_A = 187/91 \approx 2.0550$ and $2 \, c_d \simeq 0.0580$.

We observe that
the cluster size distribution goes from the critical percolation algebraic decay to a new power-law,
$C \, A^{-\tau}$, characterised by a different exponent $\tau$ and normalisation constant $C$. 
From a fit to the data collected at the latest time available, $t=7688$, for areas $A$ in the interval $[10^2, 10^4]$, we find 
$C \simeq 0.024$ and $\tau \simeq 2.016$.
To highlight the different critical behaviour that the system is approaching at late times, in Fig.~\ref{NATrVM-2}  we
plotted $A^{\tau} \, {\cal N}$ against $A$, with $\tau = 2.016$, the value obtained by the fit
(the plateau in the interval $[10^2, 10^4]$ is highlighted by a discontinuous black line which corresponds to the value of the constant $C$ obtained through the fit).

Note that since $D_A$ in this regime, as estimated in the previous Subsection, 
is larger than the one for critical percolation, $\tau = 1 + 2/D_A \simeq 2.016$ 
consistently takes a smaller value than $2.0550$, the value of $\tau_A$ expected for critical percolation.
Moreover, the data at $t=0$ complies with the critical percolation scenario, with an algebraic
decay, $\mathcal{N}(A,0) \simeq 2 \, c_d \, A^{-\tau_A}$, 
with $\tau_A=187/91$ and prefactor $2c_d\simeq 0.0580$ (see the inset).

There is a `smooth' crossover between the initial cluster size distribution satisfying the statistics of random critical percolation
and the new distribution in the late time scaling regime.
The crossover \textit{ramps} corresponding to different values of $t$ can be collapsed one onto the other 
by properly rescaling the area $A$ by $t^{D/z_d}$, with $D\simeq 1.97$ the domain fractal dimension associated with a Fisher
exponent $\tau \simeq 2.016$ and $z_d=2$ (Recall that on the triangular lattice we only found this dynamic exponent, 
so there is no ambiguity of choice between $z_d$ and $z_p$ on this lattice.)

We zoom over the crossover region, after having rescaled the area by $\ell_d(t)^{D}$,  in the right panel in Fig.~\ref{NATrVM-2}.
We fitted the function $C' \, x^{a}$ to the distribution at $t=14.8$, in the interval $[2.5,10]$ of the rescaled area
$x=A/t^{\xi}$, obtaining  $a \simeq 0.228$. The fitting curve is shown by an inclined dashed line in the plot.
(Recall that for the triangular lattice this is not a pre-percolating regime.)

\begin{figure}[h]
\begin{center}
\includegraphics[scale=0.55]{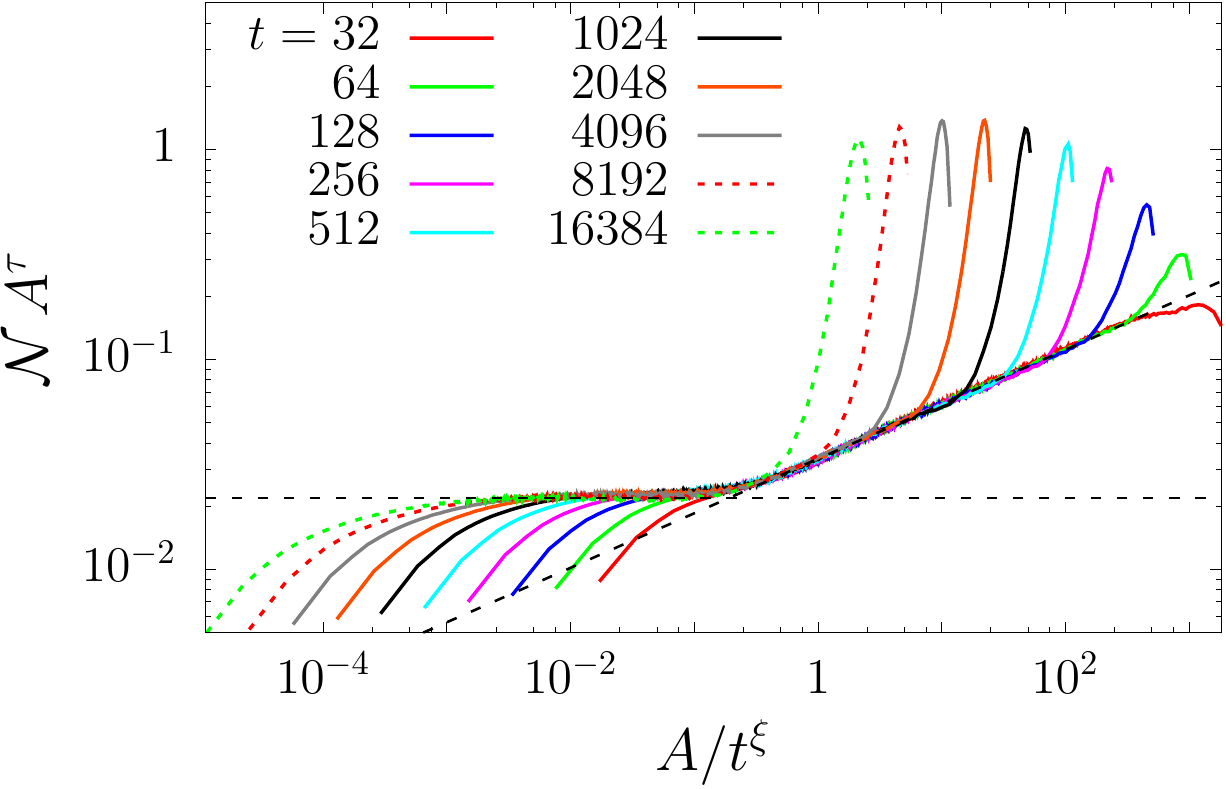}\quad%
\includegraphics[scale=0.55]{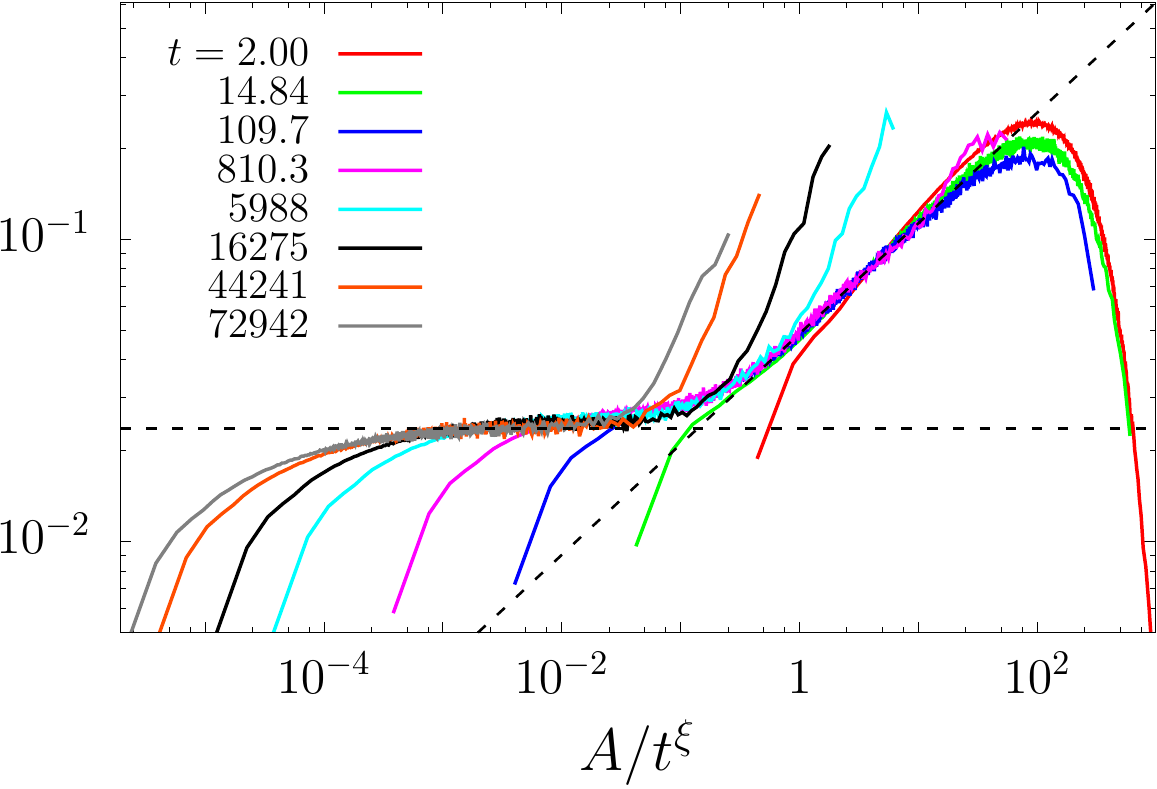}
\end{center}
\caption{\small
Scaling of the cluster size number density, ${\cal N}(A, t, L)$, for the VM
on a square lattice (left) and on a honeycomb lattice (right), both with $L=640$. 
The quantity $A^{\tau} \, {\cal N}(A, t, L)$ is plotted against 
the rescaled size $A/t^{\xi}$, at different times indicated in the keys.
As explained in the main text, the value of the exponent $\tau$ is not the one of critical percolation, but it has
been estimated numerically by fitting the function $C \, A^{-\tau}$ to the unscaled distribution ${\cal N}(A, t, L)$ 
at the latest time $t$ reached by our simulations. For the case of the dynamics on the square lattice, we found
$\tau \simeq 1.983$, while for the honeycomb lattice case, $\tau \simeq 1.984$.
The horizontal dashed lines correspond to $C \simeq 0.022$ for the square lattice case (left panel) and $C \simeq 0.024$ for the honeycomb lattice case (right panel).
In both cases, the value of the exponent $\xi$ was chosen to obtain collapse
in the ``ramp'' region, finding $\xi \simeq 1.18$ for the square lattice and 
$\xi \simeq 1.17$ for the honeycomb one.
The function $\Phi(x) = C' \, x^{a}$ has been fitted to the rescaled distribution at $t=64$ and in the interval $[1,10^{2}]$ 
of the rescaled area
$x=A/t^{\xi}$ in the square lattice case, while it has been fitted to the rescaled distribution at $t=109.7$ and in the 
interval $[1,10]$ of the rescaled area in the honeycomb lattice case, yielding
$a \simeq 0.260$ and $a \simeq 0.366 $, respectively. The fitting curve is shown by an inclined dashed line.
}
\label{NA-VM_Sq_Hon}
\end{figure}

A similar analysis was performed on the square and honeycomb lattices, see Fig.~\ref{NA-VM_Sq_Hon}.
In these cases we observe strong finite-size effects on the value of the exponent $\tau$ dictating the algebraic decay 
of ${\cal N}(A,t,L)$. In fact, on the square lattice, a fit of the data  for $L=640$ at $t=16384$ to the function
 $C \, A^{-\tau}$, taking  $A \in [10^2,\,10^4]$, yields
$\tau \simeq 1.983$ and $C \simeq 0.022$. This confirms the inconvenience that  
we already stated in~\cite{TaCuPi15}, that is $\tau < 2$ for the lattice sizes $L$ used 
and the time scales reached by our simulations.
On a honeycomb lattice with $L=640$, a similar fit of the
data at  $t=72942$ taking  $A \in [10^2,\,2 \times 10^3]$  yields $\tau \simeq 1.984$ and $C \simeq 0.024$.

The scaling behaviour of $\mathcal{N}(A,t)$ is similar
to what is observed in the case of the NCOP dynamics~\cite{BlCuPiTa-17} and the two versions of spin-exchange dynamics presented earlier in this paper.
In fact, as one can see from Fig.~\ref{NA-VM_Sq_Hon}, there is a characteristic size $A_c \sim t^{\xi}$ that separates
the plateau from the ramp-like region, which corresponds to an algebraic decay $ A^{-\tau^{\prime}}$
with $\tau^{\prime}> \tau$ (the so-called pre-asymptotic region).
By plotting $\mathcal{N}(A,t) \, A^{\tau}$ against the rescaled area $A/t^{\xi}$ it is thus possible to collapse the data corresponding to different times
onto the same master curve, save for the finite-size effects occuring for too small $A$ (due to the discreteness of the lattice) and
the late time  deviations, in correspondence with areas  of the order of the system size  (the large percolating clusters
whose statistics do not obey the law $\mathcal{N}(A) \sim A^{-\tau}$). 
The best collapse is achieved by using $\xi = 1.18$ on the square lattice, and
$\xi = 1.17$ on the honeycomb lattice.

If we assume that the characteristic size $t^{\xi}$ is given by $\ell^D_p(t)$, 
with $\ell_p(t) \sim t^{1/z_p}$ the first growing length and $D$ 
the fractal dimension of clusters in this regime, we then get 
$D \simeq 1.18 \cdot 1.67 \simeq 1.97 $ on the square lattice, and 
$D \simeq 1.17 \cdot 1.67 \simeq 1.96$ on the honeycomb lattice.
These estimates are compatible with the values of $D_A$ found by scaling the largest cluster size in Sec.~\ref{subsec:VM_LC_scaling}.

We also fitted the power law
$\Phi(x) = C' \, x^{a}$ to the rescaled distribution, $\mathcal{N}(A,t) \, A^{\tau}$, in the ramp-like region
as already done for the other types of dynamics shown in this paper.
On the square lattice, the fit was performed at
$t=64$, in the interval $[1,10^{2}]$ of the rescaled area, yielding $a \simeq 0.260$, while on the honeycomb lattice
the fit was done to the data corresponding to $t=109.7$ and in the interval $[1,10]$ of the rescaled area, yielding
$a \simeq 0.366 $. The fitting curves are shown by inclined dashed lines in Fig.~\ref{NA-VM_Sq_Hon}.
Note that the exponent $a$ estimated in this way is in the same range of values observed for the other types of dynamics,
see Sec.~\ref{subsubsec:Ka_prepercolating} and Sec.~\ref{subsubsec:NLK_prepercolating}.

\vspace{0.5cm}

\subsection{Summary}

We now summarise the picture that emerged form this study, a picture that completes (and is slightly different from) the
one that  we published in~\cite{TaCuPi15}.

We have pretty much confirmed the most significant result of our previous paper~\cite{TaCuPi15},
namely, that the coarsening process induced by the voter rule is characterised
by a crossover between an early regime in which the system's configurations are close to the ones of the 
initial disordered state and share certain aspects of critical percolation and a later regime with spin clusters with 
geometrical properties of an unknown universality class, before finally approaching the stable absorbing state. 

More precisely, the VM either starts from critical percolation (on the triangular lattice) or gets very close to 
such criticality in a very short time span (for the other lattices). Differently from what happens with the NCOP
and COP dynamics of the $2d$IM, the percolation state is not stable over a long period of time.

The further evolution from the percolating state takes the system to a regime in which 
interfaces are still critical with fractal properties that are close to the ones of the critical Ising
model, though still different from them. We see, for example, the $\kappa$ parameter
that we extract from the analysis of the domain boundaries, 
changing from a value that is very close to 6 to another value that is very close to $3.8$ in our simulations 
(see~\cite{GodrechePicco18} for a more accurate measurement of this parameter).

The time-scaling of all observables measured on the triangular lattice are controlled by the 
growing length $\ell_d(t) \simeq t^{1/z_d}$ with $z_d=2$. Instead, when the model is defined on 
other lattices that are not exactly at the critical percolation point initially, we find an early time regime 
in which the finite size scaling is given by $L^{z_p}$ and $z_p < 2$.
 
As the voter dynamics do not really care about the local connectivity of the 
lattice one may argue that $z_p$ should not depend on the  lattice geometry.
Admitting that our numerical accuracy is limited to make a strong statement,
our data for the square and honeycomb lattices are compatible with the same 
$z_p \simeq 1.67$ controlling the first time regime. 

In~\cite{TaCuPi15} we ascribed this time scale to the time needed to reach a critical percolation  state; however,  we 
proved in this paper that certain aspects of critical percolation, like $\kappa \simeq 6$ characterising the 
variance of the interface winding angle at long curvilinear distances, establish earlier and, moreover, 
there is no stable critical percolation geometry lasting over a long period of time. After the early regime 
the dynamics rapidly crosses over to the usual coarsening regime with $z_d=2$ for all lattice geometries.
Having said this, the two time scales 
$t_p(L) \sim L^{1.67}$ and $t_{\mathrm{abs}} \simeq L^2 \ln L$ are not very well separated
and this makes the distinction between the processes linked to the first and second regime difficult to disentangle.
 
\section{Conclusions}

We studied the early stages of the dynamics of three  bidimensional kinetic spin models.
On the one hand, we analysed the evolution of the $2d$ ferromagnetic Ising model (IM) quenched from infinite to 
a sub-critical temperature and evolving with local (Kawasaki) and nonlocal  
spin exchanges. On the other hand, we studied the purely dynamical voter model (VM). In both cases the 
initial state was chosen to have equal number of up and down spins, imposed exactly on the former and 
on average on the latter. These configurations are typical of equilibrium at infinite temperature.

For the $2d$IM on the square and honeycomb lattices we confirmed,  or showed for the first time, that during the early stages of its evolution the
system goes towards a critical percolation state, to only much later enter the usual dynamic scaling regime that eventually takes it to  
thermal equilibrium. A signature of this behaviour is given by the properties of the variance of the winding angle
relative to the cluster domain walls, the observable that we denoted by $\langle \theta^2 \rangle$.
As explained in Sec.~\ref{subsec:observables}, this observable allows us to easily determine the type of criticality,  or $\mathrm{SLE}$ family, 
that the system is reaching through the parameter $\kappa$ that characterises the slope in its linear dependence
on the logarithm of the curvilinear length along the interface. 
We found that, after a transient, the long length scale dependence of $\langle \theta^2\rangle$ yields  a
$\kappa$  value that is very close to $6$. This 
 let us conclude that both the local Kawasaki dynamics and its nonlocal version approach a critical-percolation-like regime.
 Moreover, the large scale properties  
 remain stable during the ensuing domain growth process that takes the system to equilibrium. Coarsening changes the geometric and statistical properties of 
 the objects, be them areas or interfaces, that are smaller or shorter than the typical ones $\ell_d^2(t)$ and $\ell_d(t)$, respectively,
but not those beyond these scales that keep the critical percolation geometry.

The transient between the initial completely disordered state and this critical-percolation-like scaling regime is 
governed by a characteristic length scale $\ell_p(t) \sim \ell_d(t) \, t^{1/\zeta}$, which grows faster than the one associated
to usual coarsening, $\ell_d(t) \sim t^{1/z_d}$, as it was already observed for NCOP dynamics~\cite{BlCoCuPi14,BlCuPiTa-17}.
The exponent $\zeta$ depends on the particular microscopic update rule and the lattice geometry.
We found:
\begin{itemize}
 \item $\zeta = 2.00$ on the square lattice and $\zeta = 1.15$ on the honeycomb lattice with local Kawasaki dynamics;
 \item $\zeta \in [1.15, 1.20]$ approximately on both the square and honeycomb lattices with nonlocal spin-exchange  dynamics.
\end{itemize}

In the NCOP dynamics of the $2d$IM $\ell_d(t)\simeq t^{1/z_d}$ and $\ell_p(t)$ is just another power law, $\ell_p(t)\simeq t^{1/z_p}$. In the COP
problem, $\ell_d(t)$ does not have a pure algebraic dependence on $t$ in the (short) times during which the approach to 
critical percolation takes place. Therefore, we proposed to use the more general functional form $\ell_p(t) \sim \ell_d(t) \, t^{1/\zeta}$ 
for $\ell_p(t)$.

The dynamics on a triangular lattice is a bit tricky since the spin configuration is already critical at $t=0$ and, with the analysis in 
this paper,  we did not observe a characteristic length scale $\ell_p(t)$ as in the other cases. A clarifying statement with 
respect to what we wrote in~\cite{BlCoCuPi14} is now in order. In this reference we 
identified a time scale $t_p \sim L^{z_p}$, with $z_p \simeq 1/3$, from the scaling of the crossing number correlation 
and the behaviour of the overlap between two copies of the system
in the $T=0$ NCOP dynamics. These two observables yield a time, that we called $t_p$, that represents 
the time at which the critical percolation state (already present at $t=0$) becomes ``stable'', in the sense
that its topology (the number of percolating clusters and their orientation) remains unchanged 
for the rest of the evolution.

In the case of the VM  on the square and honeycomb lattices we also found 
a transient between the initial disordered state and a  state in which spin clusters display long-distance fractal properties
as the ones at critical percolation. However, this state
is not as stable and long-lived as the one observed for the ``Hamiltonian'' microscopic dynamics. This 
strong difference can be confirmed by confronting, for example,  the data for the wrapping probabilities in Fig.~\ref{FigWP-NLK}
(Ising model) with those in Fig.~\ref{VM-WR_Sq_Hon} (Voter model) that look completely different with, essentially, no constant behaviour.
Having said this, from the scaling of these  
curves, and other observables, a typical time scale $t_p \simeq L^{z_p}$ with $z_p = 1.67$
can still be extracted.

The separation between the state reached at $t_p \simeq L^{z_p}$ and the ensuing coarsening regime is not 
sharp in the VM. Moreover, the domains being formed do not have smooth interfaces. Their boundaries, at 
short length scales, also look critical with $\kappa \simeq 3.85$. More details on this fact will be 
given in~\cite{GodrechePicco18}.

\vspace{0.5cm}

\begin{figure}[h!]
\begin{center}
        \includegraphics[scale=0.7]{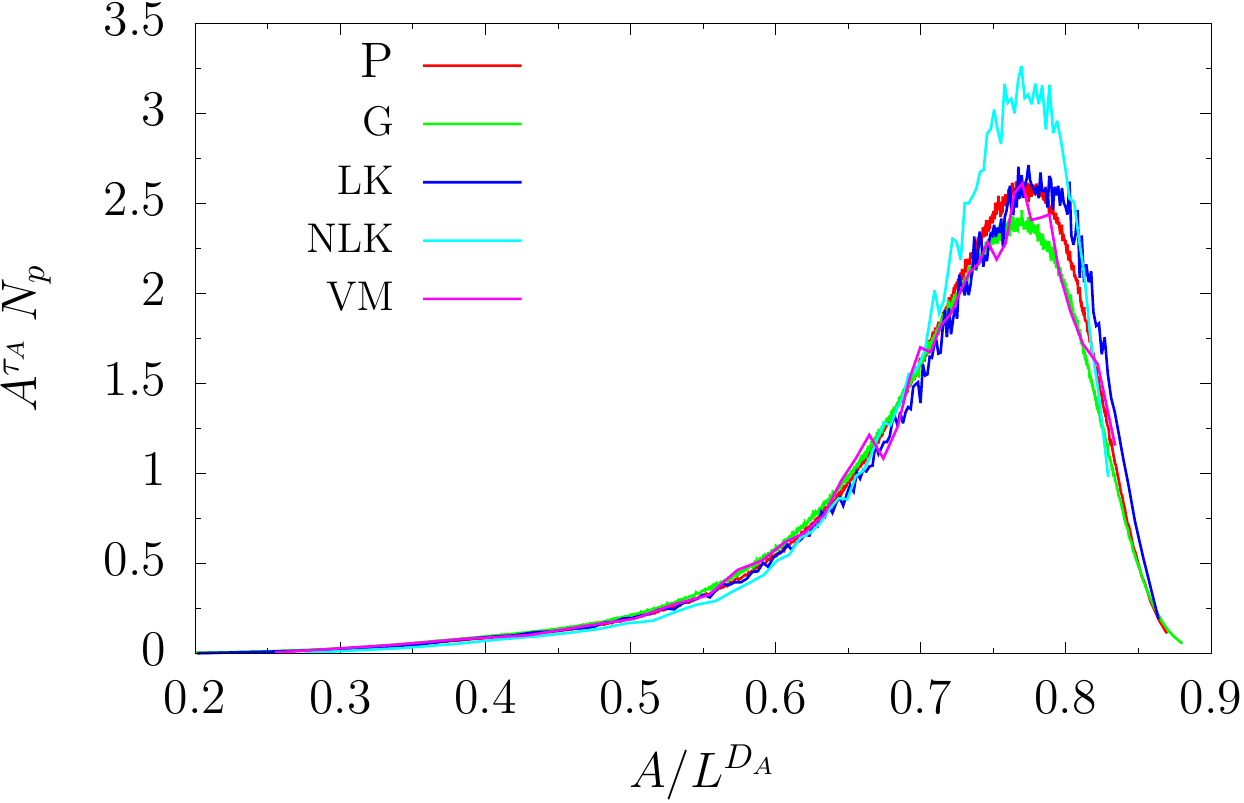}\quad%
\end{center}
\caption{\small
The figure shows $A^{\tau_A} \, N_p$ against $A/L^{D_A}$ for random site percolation on a square lattice with
$L=320$ at the critical threshold (P) and for the four different types of dynamics explored:
$T=0$ NCOP dynamics (G), local Kawasaki dynamics (LK) at $T=T_c/2$, nonlocal Kawasaki dynamics (NLK) at $T=T_c/2$
and voter model (VM), all of them on a square lattice with $L=320$ and PBC.
For the dynamical models the data correspond to times $t=8$ for G, 
$t = 8192$ for LK, $t = 23170$ for NLK and $t = 52 $ for VM, in units of MC steps (the same for all types of dynamics).
These times are close to the onset of the critical-percolation-like regime in each case.
}
\label{LCdist_cmp}
\end{figure}

Another nice way to exhibit the fact that all the microscopic spin update rules that we discussed in this paper,
at a certain point of the dynamics, take the system to a state in which there are large domains that
have the statistical and geometrical properties of the ones at the critical percolation point is to study the
distribution of percolating spin clusters, $N_p$, at times close to $t_p$, and compare it (after proper rescaling)
to the one relative to $2d$ critical percolation.
This is shown in Fig.~\ref{LCdist_cmp}, where $A^{\tau_A} \, N_p$ is plotted against $A/L^{D_A}$
for NCOP, local Kawasaki, nonlocal Kawasaki and VM dynamics, 
together with the data relative to site percolation at the threshold occupation probability, $p_c$.
Apart from deviations in the vicinity of the peak of the distribution (which are larger in the case of the nonlocal Kawasaki dynamics),
the five curves fall on top of each other quite convincingly.

The typical time $t_p$ at which the critical percolation structure establishes varies appreciably with the microscopic 
dynamics. Although we cannot get a sharp prediction for $t_p$, the analysis of the time-dependent 
wrapping probabilities using scaling arguments gives us orders of magnitude.
Concretely, for a system with $L = 320$, square lattice geometry and microscopic updates 
satisfying detailed balance, we found the 
values in Table~\ref{ref:table-tp} where we just kept the order of magnitude. (We recall that for the 
VM there is no stable time lapse over which critical percolation is seen, so we omitted this case.)
Notice that, in the case of the nonlocal Kawasaki dynamics, there is a discrepancy between the order of magnitude of $t_p$
at which the distribution of percolating spin clusters, $N_p$, reaches the form of critical percolation
and the estimate of $t_p$ (for a lattice of the same size) given in Sec.~\ref{subsec:WP-NLK}, which is based on a scaling argument
on the wrapping probabilities. In particular, we reckon that the value of $t_p$ given here is shorter than the one found by analysing
the wrapping probabilities. 
However, we must stress the fact that neither of them represent a precise measurement of $t_p$, but just a rough estimate.

\vspace{0.5cm}

\begin{table}[h!]
\begin{center}
\begin{tabular}{|c|c|c|c|}
\hline
& \mbox{NCOP} & \mbox{Kawasaki} & \mbox{Nonlocal Spin Exchange} 
\\
\hline
$t_p$ & $10$ &  $10^4$ & $10^5$ 
\\
\hline
\end{tabular}
\end{center}
\caption{\small Order of magnitude of $t_p$ for different spin updates satisfying detailed balance 
on a square lattice with linear size $L=320$.}
\label{ref:table-tp}
\end{table}

This paper is the partner of Ref.~\cite{BlCuPiTa-17} where a similarly detailed analysis of the early time 
dynamics of the $2d$IM with NCOP dynamics was performed. Some effects of weak quenched randomness
were addressed in~\cite{InCoCuPi16} where the $2d$ random bond and random field IMs were considered. 
Many issues remain to be studied in more detail and we 
mention some of them before closing the paper.

The interplay between critical percolation as we see it appearing under the time evolution
and dilution effects in the lattice itself (of either site or bond type) are the focus of some of our present
efforts in this context.

Other microscopic dynamic rules could also be used to address similar issues. 
An example of single spin-flip relaxation dynamics that breaks the full ``nearest-neighbour'' symmetry 
was recently studied~\cite{GodrechePleimling-18}.
Possibly, the most interesting case would be to investigate 
percolation properties in fluid dynamics~\cite{GonnellaYeomans09,Tanaka12,Das-etal15} for its many applications.

Temperature effects, both on the initial condition and the working set up were analysed in~\cite{Ricateau} 
and~\cite{BlCuPiTa-17}, respectively. Particularly interesting could be to understand in more detail the 
effect of temperature on the interface conformation and dynamics~\cite{PhysRevE.78.011109}
at the early stages of evolution for all kinds of microscopic updates.

In the present setting we only focused on the instantaneous quenches from infinite temperature to 
the working sub-critical one. In the last decade, great interest was set on the influence of slow cooling 
on coarsening phenomena, and more precisely on the number of topological effects 
encountered at a given time after crossing a critical point in field theories for cosmology and 
quantum models for cold atom, but also in statistical 
physics models~\cite{Biroli10,Krapivsky10,Jelic11,Liu-etal14,Priyanka,XuWuRuKaSa17,RuXuSa17}.
The influence of a slow cooling process on the critical percolation properties of the $2d$IM with NCOP 
appeared in~\cite{Ricateau}. This analysis could be extended to the system with other microscopic dynamics.

Finally, putting the VM in contact with the critical Ising model is an interesting problem that is currently
being studied~\cite{GodrechePicco18}.

\vspace{0.75cm}

\noindent
{\bf Acknowledgements.}
L. F. C. is a member of Institut Universitaire de France and thanks the KITP University of California at 
Santa Barbara for hospitality during part of the preparation of this work. 
We thank very useful discussions with F. Corberi, M. Esposito, C. Godr\`eche, F. Insalata and H. Ricateau.

\vspace{1cm}

\bibliographystyle{phaip}
\bibliography{coarsening}

\end{document}